\documentclass[apj,twocolumn]{openjournal}

\usepackage{xcolor}
\usepackage[utf8]{inputenc}
\usepackage[english]{babel}

\usepackage{orcidlink}
\usepackage{makecell}
\usepackage{amsmath}
\usepackage{booktabs}
\usepackage{multirow}
\usepackage{placeins}

\newcommand{\lt}{\left}
\newcommand{\rt}{\right}

\newcommand{\CXO}{\textit{Chandra}}
\newcommand{\XMM}{XMM-\textit{Newton}}

\usepackage{siunitx}
\DeclareSIUnit{\kms}{\kilo\meter\per\second}
\DeclareSIUnit{\erg}{erg}
\DeclareSIUnit{\ergs}{\erg\per\second}
\DeclareSIUnit{\keV}{\kilo\electronvolt}
\DeclareSIUnit \parsec {pc}
\DeclareSIUnit{\kpc}{\kilo\parsec}
\DeclareSIUnit{\mpc}{\mega\parsec}
\definecolor{linkcolor}{rgb}{0.0,0.3,0.5}
\usepackage{hyperref}
\hypersetup{
    unicode, 
    colorlinks=true,
    linkcolor=linkcolor,
    citecolor=linkcolor,
    filecolor=linkcolor,
    urlcolor=linkcolor,
}
\usepackage{cleveref}

\crefname{figure}{Figure}{Figures}
\crefname{equation}{Equation}{Equations}
\crefname{table}{Table}{Tables}
\crefname{section}{Section}{Sections}
\crefname{subsection}{Subsection}{Subsections}

\crefformat{figure}{Figure #2#1#3}
\crefformat{table}{Table #2#1#3}
\crefformat{equation}{Equation #2#1#3}
\crefformat{section}{Section #2#1#3}
\crefformat{subsection}{Subsection #2#1#3}

\Crefformat{figure}{Figure #2#1#3}
\Crefformat{table}{Table #2#1#3}
\Crefformat{equation}{Equation #2#1#3}
\Crefformat{section}{Section #2#1#3}
\Crefformat{subsection}{Subsection #2#1#3}

\newcommand{\uat}[2]{\href{https://astrothesaurus.org/uat/#2}{#1\ (#2)}}

\newcommand{\usetup}[3]{ #1 & #2 & #3}

\makeatletter

\renewcommand{\@tablenotes}{}

\newcommand{\tablenotesreset}{\gdef\@tablenotes{}}

\renewcommand{\tablenotemark}[1]{\textsuperscript{#1}}

\makeatother

\newcommand{\tablenoterow}[2]{% #1 = number of columns, #2 = note text
  \addlinespace[2pt]%
  \multicolumn{#1}{@{}l@{}}{%
    \footnotesize\raggedright\setlength{\parindent}{0pt}%
    Note. -- #2%
  }\\%
}



\newcommand{\tablenotelastrow}[2]{% #1 = number of columns, #2 = note text
  \addlinespace[1pt]%
  \multicolumn{#1}{@{}l@{}}{%
    \footnotesize\raggedright\setlength{\parindent}{0pt}%
    #2%
  }\\%
}

\makeatletter

\@ifundefined{LaTeXML}{%
  \renewcommand{\tablenotetext}[2]{%
  \g@addto@macro\@tablenotes{\makebox[0.5em][l]{\textsuperscript{\vphantom{b}#1}}#2\newline}%
}

\newcommand{\printtablenotes}[1]{%
  \noalign{\vskip 2pt}%
  \multicolumn{#1}{@{}l@{}}{%
    \parbox[t]{\linewidth}{%
      \footnotesize\raggedright\setlength{\parindent}{0pt}%
      \@tablenotes
    }%
  }\\%
}
}{%
  \renewcommand{\arcmin}{\ensuremath{^\prime}}
  \renewcommand{\arcsec}{\ensuremath{^{\prime\prime}}}

  \renewcommand{\tablenotetext}[2]{%
  \g@addto@macro\@tablenotes{\textsuperscript{#1}#2\newline}%
}

\newcommand{\printtablenotes}[1]{%
  \noalign{\vskip 2pt}%
  \multicolumn{#1}{@{}l@{}}{%
    \begin{minipage}{\linewidth}%
      \footnotesize\raggedright\setlength{\parindent}{0pt}%
      \@tablenotes
    \end{minipage}%
  }\\%
}

}
\makeatother

\makeatletter

\makeatletter
\newcommand{\pdffig}[1]{\@ifundefined{LaTeXML}{#1}{}}
\newcommand{\htmlfig}[1]{\@ifundefined{LaTeXML}{}{#1}}
\makeatother

\begin{document}

\title{X-SORTER (X-ray Survey Of meRging clusTErs in Redmapper): X-ray and Spectroscopic Characterization of 12 Optically Selected Galaxy Cluster Merger Candidates}

\makeatletter
\@ifundefined{LaTeXML}{%
  % =========================
  % Normal PDF build (pdfLaTeX)
  % =========================
  \author{
    {Christopher Hopp$^1$\orcidlink{0000-0001-8559-4538}}, 
    {David Wittman$^1$\orcidlink{0000-0002-0813-5888}},
    {Rodrigo Stancioli$^1$\orcidlink{0000-0002-6217-4861}},
    {Zhuoran Gao$^1$},
    {Faik Bouhrik$^{1,2}$\orcidlink{0009-0007-5074-5595}}, and
    {Scott Adler$^3$\orcidlink{0009-0004-6902-4649}}
  }
  \email{chopp@ucdavis.edu}
  \affiliation{$^1$Department of Physics \& Astronomy, University of California, Davis, CA 95616}
  \affiliation{$^2$Department of Physics \& Astronomy, California State University, Sacramento, CA 95819}
  \affiliation{$^3$School of Physics \& Astronomy, Rochester Institute of Technology, Rochester, NY 14623}
}{%
  % =========================
  % LaTeXML HTML build
  % =========================

  % 2) Feed LaTeXML separate author entries (it handles this MUCH better).
  \author{Christopher Hopp\orcidlink{0000-0001-8559-4538}}
  \affiliation{Department of Physics \& Astronomy, University of California, Davis, CA 95616}
  \email{chopp@ucdavis.edu}

  \author{David Wittman\orcidlink{0000-0002-0813-5888}}
  \affiliation{Department of Physics \& Astronomy, University of California, Davis, CA 95616}

  \author{Rodrigo Stancioli\orcidlink{0000-0002-6217-4861}}
  \affiliation{Department of Physics \& Astronomy, University of California, Davis, CA 95616}

  \author{Zhuoran Gao}
  \affiliation{Department of Physics \& Astronomy, University of California, Davis, CA 95616}

      \author{Faik Bouhrik\orcidlink{0009-0007-5074-5595}}
  \affiliation{Department of Physics \& Astronomy, University of California, Davis, CA 95616}
  \affiliation{Department of Physics \& Astronomy, California State University, Sacramento, CA 95819}

  \author{Scott Adler\orcidlink{0009-0004-6902-4649}}
  \affiliation{School of Physics \& Astronomy, Rochester Institute of Technology, Rochester, NY 14623}

}
\makeatother

\begin{abstract}
Merging galaxy clusters offer a unique probe of dark matter (DM) interactions through the spatial offsets between galaxies, the intracluster medium, and the DM halo. Systems that are binary, near the plane of the sky, and observed shortly after first pericenter provide the cleanest constraints on the DM self-interaction cross-section. The \textbf{X-SORTER} (X-ray Survey Of meRging clusTErs in redMaPPer) program aims to systematically identify such mergers using optical indicators of binarity in the redMaPPer cluster catalog and to follow up promising candidates with X-ray and spectroscopic observations. We select massive clusters where the top redMaPPer brightest cluster galaxy (BCG) probability is below 0.98, the top two BCGs are separated by at least 0.95\arcmin, and the optical richness exceeds $\lambda = 120$. We present \XMM\ and Keck/DEIMOS observations of twelve clusters with no previous \XMM\ or \CXO\ archival data meeting these criteria. The X-ray data reveal that most targets are morphologically disturbed, with several clear post-pericenter, dissociative systems exhibiting X-ray peaks between the BCGs. Spectroscopy confirms cluster membership and rules out foreground or background contamination. Together, these results demonstrate that optical BCG properties provide an efficient means of identifying dynamically active clusters, including clean dissociative mergers suitable for detailed, multi-wavelength studies of dark matter and cluster evolution.
\end{abstract}


\maketitle

% %%%%%%%%%%%%%%%%%%%%%%%%%%%%%%%%%%%%%%%%%%%%%%%%%%%%%%%%%%%%%%%%%%%%%%%%%%%%%%%%%%%%%%%%%%%%%%
% %%%%%%%%%%%%%%%%%%%%%%%%%%%%%%%%%                            %%%%%%%%%%%%%%%%%%%%%%%%%%%%%%%%%
% %%%%%%%%%%%%%%%%%%%%%%%%%%%%%%%%%         Introduction       %%%%%%%%%%%%%%%%%%%%%%%%%%%%%%%%%
% %%%%%%%%%%%%%%%%%%%%%%%%%%%%%%%%%                            %%%%%%%%%%%%%%%%%%%%%%%%%%%%%%%%%
% %%%%%%%%%%%%%%%%%%%%%%%%%%%%%%%%%%%%%%%%%%%%%%%%%%%%%%%%%%%%%%%%%%%%%%%%%%%%%%%%%%%%%%%%%%%%%%

\section{Introduction}\label{sec:Introduction}

Merging galaxy clusters are powerful laboratories for constraining the properties of dark matter (DM). As two (or more) clusters collide, their constituent components---galaxies, intracluster medium (ICM), and dark matter---are provided ample opportunity to interact. During these events, the galaxies act as effectively collisionless particles, passing through the merger with minimal interaction. The hot gas of the ICM, by contrast, is highly collisional and interacts through ram pressure, causing it to lag behind the collisionless components. Dark matter, inferred from gravitational lensing, behaves more similarly to the collisionless galaxies than the collisional gas, suggesting it is also nearly collisionless. Quantifying the spatial offsets between components in these dissociative mergers offers a way to constrain the interaction properties of dark matter, with a lag between the DM halo and galaxies favoring a DM model with a non-zero self-interaction cross-section \citep{markevitch2004direct,clowe2006direct, randall2008constraints,kim2017wake,golovich2019merging1}. Because self-interactions cannot be probed in the laboratory, astrophysical systems like these remain the only way to test such models. In particular, merging clusters probe DM interactions at relative velocities exceeding \SI{1000}{\kms} (compared to $\sim \SI{10}{\kms}$ at dwarf galaxy scales and $\sim \SI{200}{\kms}$ at galaxy scales \citep{tulin2018dark}), providing constraints in a velocity regime inaccessible to lower-mass systems.

While cluster mergers provide a powerful method to study dark matter, not all mergers are equally useful. First, the offset between DM and galaxies is predicted to be small, $\sim$10--\SI{40}{\kpc} for self-interaction cross-sections of \SI{1}{\centi\meter\squared\per\gram} \citep{kim2017wake,robertson2019observable}, which may be further diminished by projection effects. Additionally, systems with complex substructure are prohibitively complicated to model with sufficient accuracy to constrain the nature of dark matter. Finally, the displacement between galaxies, DM, and the ICM depends on the impact parameter and dynamics of the merger. Accurate models must consider the merger stage, infall velocity, and time since pericenter passage \citep{kim2017wake, golovich2019merging1}. Thus, the ideal scenario is a head-on binary merger in the plane of the sky observed shortly after the first pericenter passage. These simple, well-modeled systems have utility beyond the study of dark matter. Studies of cluster dynamics, the ICM, and galaxy evolution all benefit from clean merger models.

The quintessential example of a high-quality merger scenario is the Bullet Cluster \citep{clowe2006direct,robertson2016does,cha2025high}. Other well-studied systems include the Musket Ball Cluster \citep{dawson2012discovery}, El Gordo \citep{jee2014weighing}, and Baby Bullet \citep{bradavc2008revealing}, all of which have produced constraints on the self-interacting dark matter (SIDM) cross-section. From these individual clusters, the tightest upper limit on the SIDM cross-section is $\sigma / m < \SI{0.7}{\centi\meter\squared\per\gram}$ \citep{randall2008constraints}. While these constraints are useful, there is still a wide swath of parameter space available, meriting further study (for reference, these current limits are approximately that of the largest cross-sections in the Standard Model). To further tighten these constraints with mergers, one approach is to consider a larger, statistical ensemble of merging systems \citep{harvey2015nongravitational}. However, care must be taken to select an ensemble from clusters in the same merger phase, as the offset between DM and galaxies changes sign once the subhalo is returning from apocenter \citep{kim2017wake, wittman2018mismeasure}. 

The current sample of such ideal mergers is small, historically discovered serendipitously by their disturbed X-ray morphologies. Extensive work has focused on identifying merging clusters systematically through X-ray \citep{mann2012x, arendt2024identifying} or radio \citep{golovich2019merging2} data. These methods typically identify disturbed systems but do not preferentially select for bimodal mergers with separated subclusters. Other attempts have been made to develop algorithms to identify clusters in optical wavelengths to leverage the high-quality imaging now available across much of the sky from surveys such as SDSS \citep{york2000sloan}, DES \citep{dark2016dark}, Pan-STARRS \citep{kaiser2002pan}, DESI \citep{dey2019overview}, and eventually LSST \citep{ivezic2019lsst}. One approach, using a Friends-of-Friends algorithm, has shown promise in identifying merging structures \citep{tempel2017merging}. More recently, \citet{wen2024catalogue} had considerable success in identifying merging systems by locating galaxy overdensities between identified brightest cluster galaxies (BCGs). They identified 3446 systems with post-collision merging features, including two bullet-like clusters.

In this work, we pursue an optical selection strategy using the redMaPPer catalog \citep{rykoff2014redmapper} to identify clean binary mergers from BCG positions alone. RedMaPPer identifies galaxy clusters from SDSS \citep{york2000sloan} and, for each cluster, ranks five candidate BCGs by their probability of being the true BCG. In relaxed clusters, the top-ranked BCG is overwhelmingly likely to be the true BCG, whereas in a merger, we expect a secondary BCG with a non-negligible probability.

This strategy forms the basis of the \textbf{X-SORTER} (X-ray Survey Of meRging clusTErs in Redmapper) program, which uses optical BCG information to systematically identify merging clusters for targeted X-ray and spectroscopic observations. In this paper, we apply the X-SORTER selection method by targeting clusters where the top BCG probability is less than $98\%$ and the two most likely BCGs have a projected separation of ${>}0.95\arcmin$, criteria designed to identify potential binary mergers. While the full program spans both archival and new observations, we focus here on 12 galaxy clusters lacking previous \XMM\ or \CXO\ coverage,\footnote{RM J121917.6+505432.8 did have very shallow \CXO\ data, insufficient for detailed analysis.} selected for follow-up with \XMM. We complement the X-ray data with Keck/DEIMOS spectroscopy to identify cluster members and rule out foreground and background contamination. These observations are intended as a first-look assessment and do not uniquely constrain the merger scenario of each candidate. However, they provide sufficient information to evaluate the viability of each system for multi-wavelength follow-up.

The paper is structured as follows: in \S\ref{sec:Methods}, we discuss our selection criteria and data reduction processes. In \S\ref{sec:Results}, we cover results and discussion of our selection algorithm as well as specific analysis of each of the 12 clusters targeted. Finally, we summarize and conclude in \S\ref{sec:Conclusion}. Throughout this paper, we assume a flat $\Lambda$CDM cosmology with $H_0 = \SI{70.0}{\km\per\second\per\mega\parsec}$ and $\Omega_\mathrm{m} = 0.3$.

% %%%%%%%%%%%%%%%%%%%%%%%%%%%%%%%%%%%%%%%%%%%%%%%%%%%%%%%%%%%%%%%%%%%%%%%%%%%%%%%%%%%%%%%%%%%%%%
% %%%%%%%%%%%%%%%%%%%%%%%%%%%%%%%%%                            %%%%%%%%%%%%%%%%%%%%%%%%%%%%%%%%%
% %%%%%%%%%%%%%%%%%%%%%%%%%%%%%%%%%         Methods            %%%%%%%%%%%%%%%%%%%%%%%%%%%%%%%%%
% %%%%%%%%%%%%%%%%%%%%%%%%%%%%%%%%%                            %%%%%%%%%%%%%%%%%%%%%%%%%%%%%%%%%
% %%%%%%%%%%%%%%%%%%%%%%%%%%%%%%%%%%%%%%%%%%%%%%%%%%%%%%%%%%%%%%%%%%%%%%%%%%%%%%%%%%%%%%%%%%%%%%

\section{Methods}\label{sec:Methods}

% %%%%%%%%%%%%%%%%%%%%%%%%%%%%%%%%%%%%%%%%%%%%%%%%%%%%%%%%%%%%%%%%%%%%%%%%%%%%%%%%%%%%%%%%%%%%%%
% %%%%%%%%%%%%%%%%%%%%%%%%%%%%%%%%%      Target Selection      %%%%%%%%%%%%%%%%%%%%%%%%%%%%%%%%%
% %%%%%%%%%%%%%%%%%%%%%%%%%%%%%%%%%%%%%%%%%%%%%%%%%%%%%%%%%%%%%%%%%%%%%%%%%%%%%%%%%%%%%%%%%%%%%%

\subsection{Target Selection}\label{subsec:Target Selection}

We aim to identify binary mergers with a subcluster separation favorable for dark matter follow-up studies (\textit{i.e.}, post-pericenter mergers with small impact parameter and a separation vector mostly in the plane of the sky) using optical photometric properties for the initial selection. Optical properties are well suited to expose bimodality in the projected spatial distribution of the galaxies, but less so for inferring whether the two subclusters are heading toward pericenter (which we will refer to as \textit{premerger}), outbound after first pericenter (\textit{outbound}), or returning after first apocenter (\textit{returning}). See \citet{wen2024catalogue}, however, for promising recent work using optical properties to test the merger stage. Our approach relies on X-ray follow-up to identify candidates where the X-ray surface brightness (XSB) peak is between the BCGs, indicating that the ICM has been displaced. These \textit{dissociative} mergers are most likely to have had a small impact parameter and be seen after first pericenter \citep{dawson2012discovery, McDonald2022}.

Because our approach requires X-ray follow-up, we focus on richer clusters, which are more likely to yield strong X-ray signals. In turn, because the number of clusters steeply declines with richness, the sample size is much smaller than that of \citet{wen2024catalogue}. The sample is defined by cuts on the following three attributes listed in, or derivable from, the redMaPPer catalog:
\begin{itemize}
\item \textbf{Optical richness} ($\lambda$), redMaPPer's estimate of the number of galaxies in the cluster within a set magnitude range of the BCG magnitude.
\item \textbf{Projected separation} between the top two BCG candidates.
\item \textbf{Dominance of the top BCG candidate}, quantified by the redMaPPer attribute $P0$, the probability of the top BCG candidate being the BCG.  
\end{itemize}

%%%%%%%%%%%%%%%%%%%%%%%%%% Figure 1: RM Pair Plot %%%%%%%%%%%%%%%%%%%%%%%%%%
\begin{figure*}
  \centering
  \includegraphics[clip, trim=0.0cm 1.0cm 1.0cm 1.0cm, width=0.888\textwidth]{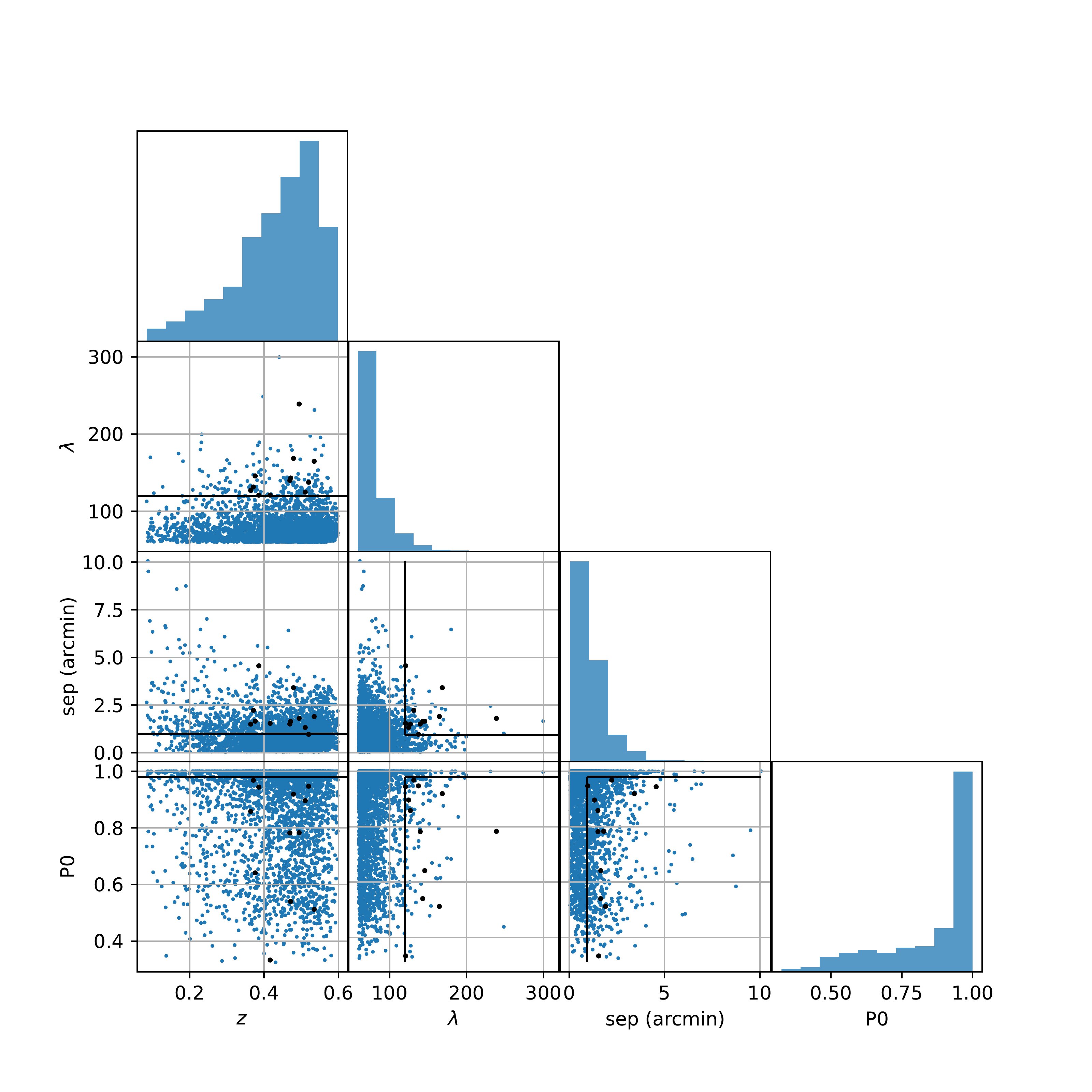}
  \caption{Pair plot for the redMaPPer attributes of redshift, richness $\lambda$, separation between the top two BCG candidates, and $P0$ (probability of the top BCG candidate being the actual BCG). This plot includes only clusters with $\lambda>60$ to reduce crowding. Black lines illustrate our cuts, and black points are clusters studied in this paper.}
  \label{fig-RMpairplot}
\end{figure*}

\cref{fig-RMpairplot} shows the joint distribution of these three attributes, along with the redshift $z$ which redMaPPer estimates from photometric properties, with a typical uncertainty of ${\approx}0.01$. We cut on these properties as follows:
\begin{itemize}
\item The bottom right panel of \cref{fig-RMpairplot} shows that the modal value of P0 is near unity, so we impose a cut of $P0 < 0.98$ to remove a large number of presumably relaxed clusters. This cut by itself removes only about half the clusters, but we found (based on the procedure described below) that good merger candidates can be found even with $P0 \approx 0.9$, so we chose to make other cuts more restrictive.
\item We wanted subclusters to be separated by several times the \XMM\ angular resolution of 10\arcsec\ so that the location of the X-ray peak relative to the galaxy subclusters could be determined with relatively short X-ray exposures. Therefore, we seek separations of $\gtrsim1\arcmin$ between the top two BCG candidates; we settled on a cut of $\ge0.95\arcmin$ to obtain a sample size of ${\approx}50$. A fixed angular cut has the possible disadvantage of yielding a redshift-dependent cut on the physical separation, thus favoring low-redshift candidates, but in practice, we had no excess of low-redshift candidates (likely due to the richness cut explained below). Over the redshift range 0.2--0.55, the 0.95\arcmin\ cut ranges from 190--370 kpc in physical separation.
\item Rich candidates are more practical for follow-up, so we imposed a richness cut of $\lambda \ge 120$, to yield a small number of rich candidates. \cref{fig-RMpairplot} shows that richness is largely uncorrelated with the other attributes, except for the unavoidable lack of rare rich clusters at low $z$ due to the small volume probed there. Therefore, the richness cut reduces the sample size with the side effect of disfavoring low-redshift candidates and preferentially selecting massive clusters. Given the opposing redshift effects of the richness and separation cuts, the redshifts of the resulting candidates range from 0.1--0.55, similar to the parent catalog. 
\end{itemize}

%%%%%%%%%%%%%%%%%%%%%%%%%% Figure 2: Decision Tree %%%%%%%%%%%%%%%%%%%%%%%%%%
\pdffig{
\begin{figure*}
  \centering
  \includegraphics[width=0.98\textwidth]{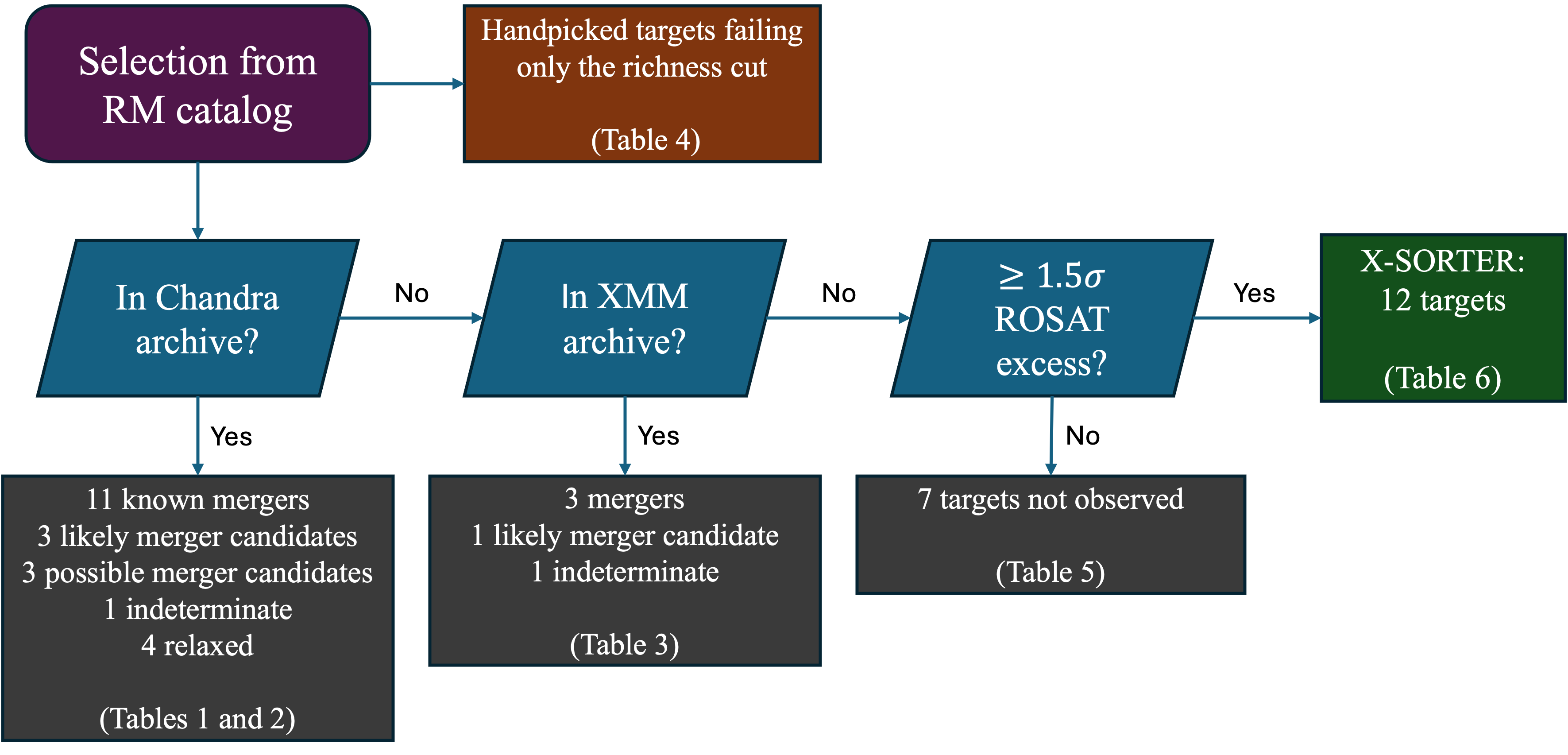}
  \caption{Decision tree for X-SORTER targets. The green box on the right is the focus of this paper.}
  \label{fig-flowchart}
\end{figure*}}

We found two cases in which redMaPPer failed to identify an overwhelmingly dominant BCG despite the presence of a single clear BCG: RM J073220.3+314120.7 (Abell 586) and RM J140100.5+025149.7. We suspect the underlying photometry was at fault, for example, if the BCG was saturated or misclassified due to AGN activity. Therefore, our final cut is a visual inspection to rule out this failure mode. To limit the role of subjective human judgment, we considered no other aspects of redMaPPer performance in this inspection.

The resulting target list was evaluated using the decision tree illustrated in \cref{fig-flowchart}. First, we performed a \CXO\ archival search (in mid-2020) to determine how well the selection criteria were yielding systems with an XSB peak between the BCG candidates.  This allowed for some experimentation with the cuts. With the final set of cuts, there were 46 targets, with 22 in the \CXO\ archive, 17 of which were also in the \XMM\ archive.

\htmlfig{
\begin{figure*}
  \centering
  \includegraphics[width=0.98\textwidth]{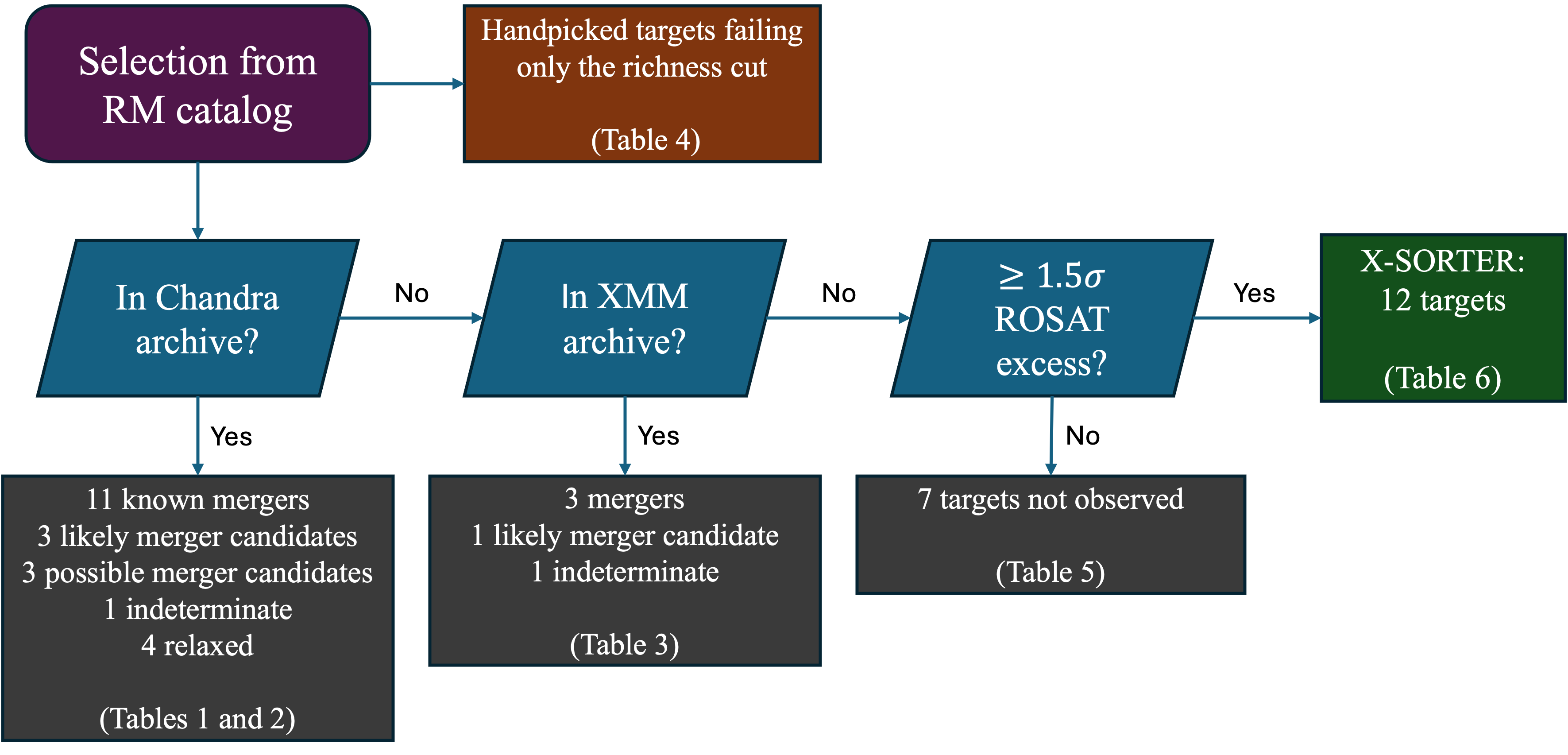}
  \caption{Decision tree for X-SORTER targets. The green box on the right is the focus of this paper.}
  \label{fig-flowchart}
\end{figure*}}

A summary of these 22 systems is given in \cref{tab:CXO_and_XMM,tab:CXO}, with dynamical classifications assigned according to \S\ref{subsec:Classification}. The classification given to each system is based on both the observed X-ray morphology and published results in the literature. Of the 22 systems in the \CXO\ archive, we find 10 previously-known mergers and 7 merger candidates. For the remaining 24 targets, we searched the \XMM\ archive and found five that had been previously observed, three of which are clear mergers (\cref{tab:XMM_archive}).

%%%%%%%%%%%%%%%%%%%%%%%%%% Table 1: Candidates in Both Archives %%%%%%%%%%%%%%%%%%%%%%%%%%
\begin{table*}
    \centering
    \caption{Candidates in both the {\CXO} and {\XMM} Archives}
    \label{tab:CXO_and_XMM}
    \tablenotetext{a}{As estimated by \citet{rykoff2016redmapper}.}
    \tablenotetext{b}{Cluster morphology parameter reported from \citet{yuan2022dynamical}.}
\begin{tabular*}{\textwidth}{@{\extracolsep{\fill}} cccccc} \toprule\toprule
        \hspace{1em} Name\hspace{1em}            & \hspace{1em}Redshift\tablenotemark{a}\hspace{1em}  & \hspace{1em}Richness\tablenotemark{a}\hspace{1em}  & \hspace{1em} $\delta$\tablenotemark{b}\hspace{1em}  & \hspace{2em}Alt. name(s)\hspace{2em} & \hspace{2em}Comment\hspace{2em} \\\toprule
RM J004629.3+202804.8   & 0.104                     & 123                       & $0.85$                    & \makecell{Abell 98\\PSZ1 G121.35-42.47}                           & \makecell{Merger\\ \citep{Paterno-Mahler2014}\\ \citep{sarkar2022shock}\\\citep{Sarkar2023}}\\ \midrule
RM J015949.3$-$084958.9 & 0.408                     & 168                       & $-0.13$                   & \makecell{MACS J0159.8-0849\\RXC J0159.8-0850\\
                                                                                                                        PSZ2 G167.66-65.59\\ACT-CL J0159.8-0849}                & \makecell{Relaxed\\ \citep{cibirka2018relics}}\\ \midrule
RM J082529.1+470800.9   & 0.128                     & 132                       & $0.50$                    & \makecell{Abell 655\\RXC J0825.5+4707\\PSZ2 G172.63+35.15}        & \makecell{Likely merger\\ \citep{zhang2023planck}\\ \citep{groeneveld2025serendipitous}}\\ \midrule
RM J092052.5+302740.3   & 0.294                     & 129                       & $0.92$                    & Abell 781                                                         & \makecell{Merger\\ \citep{Sehgal2008Abell781}\\ \citep{BotteonA781}}\\ \midrule
RM J100214.1+203216.6   & 0.324                     & 151                       & $0.98$                    & \makecell{Abell 913\\PSZ2 G213.30+50.99\\WHL J100226.8+203101}    & \makecell{Merger\\ \citep{geXray2019}}\\ \midrule
RM J101734.3+593339.8   & 0.289                     & 124                       & $0.60$                    & \makecell{Abell 959\\RXC J1017.5+5934\\PSZ2 G151.19+48.27}        & \makecell{Merger\\ \citep{Boschin2009Abell959}\\ \citep{Birzan2019Abell959}}\\ \midrule
RM J104729.0+151402.1   & 0.209                     & 128                       & $1.06$                    & \makecell{Abell 1095\\RXC J1047.5+1513\\PSZ1 G228.98+58.89}       & \makecell{Merger\\ \citep{ge2016baryon}\\ \citep{Burchett2018Abell1095}}\\ \midrule
RM J113613.0+400235.8   & 0.295                     & 155                       & $0.13$                    & \makecell{Abell 1319\\PSZ2 G168.33+69.73}                         & \makecell{Merger\\ \citep{Wilber2019A}}\\ \midrule
RM J114935.7+222354.6   & 0.536                     & 231                       & $0.86$                    & \makecell{MACS J1149.6+2223\\RXC J1149.5+2223\\PSZ2 G228.16+75.20}& \makecell{Merger\\ \citep{ogrean2016frontier}\\ \citep{golovichDynamical2016}\\\citep{Finner2025}}\\\midrule
RM J115914.9+494748.4   & 0.363                     & 134                       & $0.01$                    & \makecell{Abell 1430\\RXC J1159.2+4947\\PSZ2 G143.26+65.24}       & \makecell{Merger\\ \citep{Hoeft2021}}\\\midrule
RM J121218.5+273255.1   & 0.354                     & 158                       & $0.39$                    & \makecell{Abell 1489\\RXC J1212.3+2733\\PSZ2 G207.88+81.31}       & \makecell{Possible merger\\\citep{zitrin2020strong}\\\citep{rines2022spectroscopic}}\\\midrule
RM J155820.0+271400.3   & 0.095                     & 170                       & $0.01$                    & \makecell{Abell 2142\\RXC J1558.3+2713\\PSZ2 G044.20+48.66}       & \makecell{Merger\\ \citep{Liu2018Abell2142}}\\\midrule
RM J172227.2+320757.2   & 0.229                     & 180                       & $-0.45$                   & \makecell{Abell 2261\\RXC J1722.4+3208\\PSZ2 G055.59+31.85}       & \makecell{Likely relaxed\\\citep{kim2022a2261}\\\citep{dutta2025weak}} \\\midrule
RM J221145.9$-$034944.5 & 0.426                     & 159                       & $-0.09$                   & \makecell{MACS J2211.7$-$0349\\RXC J2211.7$-$0349\\
                                                                                                                        PSZ2 G057.25$-$45.34\\ACT-CL J2211.7$-$0349}            & \makecell{Likely relaxed\\\citep{cerny2018relics}\\\citep{george2021imaging}}\\\midrule
RM J222826.8+203852.4   & 0.389                     & 121                       & $0.25$                    & \makecell{MACS J2228.5+2036\\RXC J2228.6+2036\\PSZ2 G083.29-31.03}& \makecell{Merger\\ \citep{Soucail2015}\\\citep{stuardi2025radio}}\\\midrule
RM J224319.8$-$093530.9 & 0.441                     & 299                       & $0.73$                    & \makecell{MACS J2243.3$-$0935\\RXC J2243.3$-$0935\\
                                                                                                                        PSZ2 G056.93$-$55.08\\ACT-CL J2243.3$-$0935}            & \makecell{Merger\\ \citep{Cantwell2016}\\\citep{schirmer2011macsj2243}}\\ \midrule
RM J231147.6+034107.6   & 0.301                     & 166                       & $-0.15$                   & \makecell{Abell 2552\\MACS J2311.5+0338\\RXC J2311.5+0338\\
                                                                                                                        PSZ2 G081.00-50.93\\ACT-CL J2311.5+0338}                & \makecell{Likely relaxed \\\citep{kale2015extended}}\\\bottomrule
\printtablenotes{6}
\end{tabular*}
  \tablenotesreset
\end{table*}

%%%%%%%%%%%%%%%%%%%%%%%%%% Table 2: Candidates in Chandra Archive %%%%%%%%%%%%%%%%%%%%%%%%%%
\begin{table*}
    \centering
    \caption{Candidates Exclusively in the {\CXO} Archive}
    \label{tab:CXO}
    \tablenotetext{a}{As estimated by \citet{rykoff2016redmapper}.}
    \tablenotetext{b}{Cluster morphology parameter reported from \citet{yuan2022dynamical}.}
\begin{tabular*}{\textwidth}{@{\extracolsep{\fill}} cccccc} \toprule\toprule
        \hspace{1em} Name\hspace{1em}            & \hspace{1em}Redshift\tablenotemark{a}\hspace{1em}  & \hspace{1em}Richness\tablenotemark{a}\hspace{1em}  & \hspace{1em} $\delta$\tablenotemark{b}\hspace{1em}  & \hspace{4em}Alt. name(s)\hspace{4em} & Comment \\\toprule
RM J105417.5+143904.2   & 0.297                     & 125                       & $1.48$                     & \makecell{Abell 1127\\PSZ2 G231.56+60.03\\ACT-CL J1054.2+1438}    & \makecell{Likely merger\\\citep{duchesne2020murchison}}\\ \midrule
RM J124922.1+494742.1   & 0.287                     & 129                       & $1.10$                     & \makecell{Abell 1622\\PSZ2 G123.66+67.25}                         & \makecell{Likely merger\\\citep{van2021lofar}\\\citep{paul2021ugmrt}}\\ \midrule
RM J160319.0+031644.6   & 0.237                     & 132                       & $1.05$                     & \makecell{PSZ2 G014.09+38.38\\ACT-CL J1603.3+0317}                & Possible merger\\ \midrule
RM J211849.1+003337.2   & 0.265                     & 120                       & $1.50$                     & PSZ2 G052.35-31.98                                                & Indeterminate; low S/N \\ \midrule
RM J230707.5+163246.1   & 0.248                     & 130                       & $1.62$                     & \makecell{RXC J2307.0+1631\\PSZ2 G089.81-39.56\\ACT-CL J2307.0+1631}& Possible merger \\ \bottomrule
\printtablenotes{6}
\end{tabular*}
  \tablenotesreset
\end{table*}

%%%%%%%%%%%%%%%%%%%%%%%%%% Table 3: Candidates in XMM Archive %%%%%%%%%%%%%%%%%%%%%%%%%%
\begin{table*}
    \centering
    \caption{Candidates Exclusively in the {\XMM} Archive}
    \tablenotetext{a}{As estimated by \citet{rykoff2016redmapper}.}
    \label{tab:XMM_archive}
\begin{tabular*}{\textwidth}{@{\extracolsep{\fill}} ccccc} \toprule\toprule
        \hspace{1.5em} Name\hspace{1.5em}            & \hspace{1.5em}Redshift\tablenotemark{a}\hspace{1.5em}  & \hspace{1.5em}Richness\tablenotemark{a}\hspace{1.5em}  & \hspace{4em}Alt. name(s)\hspace{4em} & Comment \\\toprule
RM J003208.2+180625.3   & 0.398                     & 248                       & \makecell{MACS J0032.1+1808\\RXCJ 0032.1+1808\\
                                                                                            PSZ2 G116.50-44.47\\ACT-CL J0032.2+1808}        & \makecell{Merger\\\citep{acebron2020relics}\\\citep{gargantua2026inprep}} \\ \midrule
RM J003353.1$-$075210.4 & 0.304                     & 128                       & \makecell{Abell 56\\MACS J0033.8-0751\\RXC J0033.8-0750\\
                                                                                            PSZ2 G109.99-70.28\\ACT-CL J0033.8-0751}        & \makecell{Merger \\\citep{wittman2023new}\\ \citep{albuquerque2024unravelling}}\\ \midrule
RM J123416.1+151508.4   & 0.280                     & 137                       & ACT-CL J1234.2+1515                                       & Indeterminate; on edge of XMM field\\ \midrule
RM J150822.0+575515.2   & 0.544                     & 153                       & PSZ2 G094.56+51.03                                        & \makecell{Merger\\ \citep{stancioli2024new}}\\ \midrule
RM J224158.1+173303.7   & 0.323                     & 126                       & \makecell{Abell 2472\\MACS J2241.8+1732\\RXC J2241.8+1732
                                                                                            \\PSZ2 G084.13-35.41\\ACT-CL J2241.9+1732}      & \makecell{Likely merger\\\citep{botteon2022planck}}\\ \bottomrule
\printtablenotes{5}
\end{tabular*}
  \tablenotesreset
\end{table*}

 For the 19 targets lacking archival X-ray data, we estimated the \XMM\ exposure time required to record 3500 photons and thus locate the XSB peak to $\lesssim 10 \arcsec$, even for the highly disturbed morphologies typical of known merging clusters. This estimate required some prior X-ray detection to anchor the calculation, so we used the clusters that were in the \XMM\ archive to create a mapping between ROSAT All Sky Survey (RASS) flux \citep{voges1999rosat} and \XMM\ counts.  None of the 19 unobserved clusters were cataloged as sources in RASS, but generally had 1.5--3$\sigma$ flux excesses. To ensure the proposed exposures would be grounded in plausible X-ray signal rather than noise, we opted to drop the seven targets with ${<}1.5\sigma$ RASS excesses\footnote{For reference, all archival systems clear the same $1.5\sigma$ threshold with the exception of RM J100214.1+203216.6 (Abell 913; \citealt{geXray2019}), which sits at ${\sim}1\sigma$. Because this criterion was only used in prioritizing the pointed XMM observations, it was not applied to the archival systems as a selection cut.}. We recognize that with such low signal-to-noise (S/N) in RASS, there is randomness in whether a target appeared above or below $1.5\sigma$. Therefore, the seven dropped targets may be fruitful in the future. We also identified a few additional promising clusters that are closely related to this project. These are clusters that came to our attention while developing the cuts; had a wealth of archival data worthy of study; and ultimately failed only the richness cut. We list these clusters in \cref{tab:handpicked} only to clarify the relationship of this paper to other papers, and do not consider them further in this work. The seven dropped targets are listed separately in \cref{tab:notobserved}.

%%%%%%%%%%%%%%%%%%%%%%%%%% Table 4: Related Clusters %%%%%%%%%%%%%%%%%%%%%%%%%%
\begin{table*}
    \centering
    \caption{Merging Clusters in Related Publications}
    \tablenotetext{a}{As estimated by \citet{rykoff2016redmapper}.}
    \label{tab:handpicked}
\begin{tabular*}{\textwidth}{@{\extracolsep{\fill}} ccccc} \toprule\toprule
        \hspace{1.5em} Name\hspace{1.5em}            & \hspace{1.5em}Redshift\tablenotemark{a}\hspace{1.5em}  & \hspace{1.5em}Richness\tablenotemark{a}\hspace{1.5em}  & \hspace{4em}Alt. name(s)\hspace{4em} & Comment \\\toprule
RM J001938.7+033557.3   & 0.277                     & 109                       & \makecell{ZwCl 0017.0+0320\\PSZ2 G107.66-58.32\\ACT-CL J0019.6+0336}  & \makecell{\citep{pillay2021multiwavelength}\\ (Stancioli  \textit{et al.} in prep.)} \\\midrule
RM J024808.3$-$021637.2 & 0.235                     & 97                        & \makecell{Abell 384\\RXC J0248.2$-$0216\\PSZ2 G176.25$-$52.57\\
                                                                                            ACT-CL J0248.1$-$0216}                                      & \makecell{\citep{chatterjee2025exploring}\\ (Stancioli  \textit{et al.} in prep.)} \\ \midrule
RM J130558.9+263048.4   & 0.316                     & 70                        & \makecell{ZWCL 1303.6+2647\\RXC J1306.0+2630\\PSZ2 G023.17+86.71}     & \citet{Bouhrik2025Champagne} \\ \midrule
RM J213518.8+012527.0   & 0.234                     & 109                       & \makecell{Abell 2355\\RXC J2135.2+0125\\PSZ2 G055.95-34.89\\ACT-CL J2135.2+0125}& Bouhrik \textit{et  al.} in prep. \\ \bottomrule
\tablenoterow{5}{Failing the richness cut, these clusters were followed up outside the scope of this paper.}
\printtablenotes{5}
\end{tabular*}
  \tablenotesreset
\end{table*}

%%%%%%%%%%%%%%%%%%%%%%%%%% Table 5: Low ROSAT Flux %%%%%%%%%%%%%%%%%%%%%%%%%%
\htmlfig{
\begin{table}
    \centering
    \caption{Unobserved Candidates with Low ROSAT Flux}
    \tablenotetext{a}{As estimated by \citet{rykoff2016redmapper}.}
    \label{tab:notobserved}
\begin{tabular*}{\columnwidth}{@{\extracolsep{\fill}} ccc} \toprule\toprule
        Name            & \hspace{1.5em} Redshift\tablenotemark{a}\hspace{1.5em}  & Richness\tablenotemark{a} \\ \toprule
RM J005119.7$-$104940.8 & 0.502&128\\
RM J005936.8+131020.8 & 0.492 &148\\
RM J020824.7+245522.0 & 0.485 & 131\\
RM J112529.9+101339.1 & 0.502 &129\\
RM J222921.6+105809.0 & 0.517 &129\\
RM J231211.3+151320.7 & 0.449 &152\\
RM J233344.5+244324.4  & 0.499 &131 \\ \bottomrule
\printtablenotes{3}
\end{tabular*}
  \tablenotesreset
\end{table}}

The remaining 12 targets with ${\geq}1.5\sigma$ RASS excesses, listed in \cref{tab:clusters}, form the basis of the \XMM\ survey described in this paper. Henceforth, we refer to clusters by truncating their redMaPPer designation to the first four digits. All 12 systems are minimally characterized in the literature and none has been the subject of a dedicated multiwavelength analysis. Three clusters (RMJ0829, RMJ1219, and RMJ2321) were included in a LOFAR analysis of Planck SZ-selected clusters by \citet{botteon2022planck}, while several others appear in large-area optical or SZ cluster catalogs, but without detailed study. These 12 clusters therefore comprise a newly defined sample of optically identified merger candidates whose X-ray properties have not been systematically examined.

%%%%%%%%%%%%%%%%%%%%%%%%%% Table 5: Low ROSAT Flux %%%%%%%%%%%%%%%%%%%%%%%%%%
\pdffig{
\begin{table}
    \centering
    \caption{Unobserved Candidates with Low ROSAT Flux}
    \tablenotetext{a}{As estimated by \citet{rykoff2016redmapper}.}
    \label{tab:notobserved}
\begin{tabular*}{\columnwidth}{@{\extracolsep{\fill}} ccc} \toprule\toprule
        Name            & \hspace{1.5em} Redshift\tablenotemark{a}\hspace{1.5em}  & Richness\tablenotemark{a} \\ \toprule
RM J005119.7$-$104940.8 & 0.502&128\\
RM J005936.8+131020.8 & 0.492 &148\\
RM J020824.7+245522.0 & 0.485 & 131\\
RM J112529.9+101339.1 & 0.502 &129\\
RM J222921.6+105809.0 & 0.517 &129\\
RM J231211.3+151320.7 & 0.449 &152\\
RM J233344.5+244324.4  & 0.499 &131 \\ \bottomrule
\printtablenotes{3}
\end{tabular*}
  \tablenotesreset
\end{table}}

%%%%%%%%%%%%%%%%%%%%%%%%%% Table 6: Cluster Info %%%%%%%%%%%%%%%%%%%%%%%%%%
\begin{table*}
    \centering
    \caption{RedMaPPer Properties of the 12 X-SORTER Clusters Studied in This Work}
    \label{tab:clusters}
\begin{tabular*}{\textwidth}{@{\extracolsep{\fill}}cccccccc} \toprule\toprule
                        Name       & Redshift & $\sigma_z$  & Richness  & $\sigma_\lambda$  &  RA  {[}deg{]}     & Dec {[}deg{]}        & Alt. name(s) \\ \toprule
    \textbf{RM J000343.8+100123.8} & 0.372    & 0.014       & 131       & 6                 & \phantom{00}0.9325 & 10.0233              & ACT-CL J0003.7+1001 \\ \midrule
    \textbf{RM J010934.2+330301.0} & 0.479    & 0.012       & 168       & 17                & \phantom{0}17.3923 & 33.0503              & ...\\ \midrule
    \textbf{RM J021952.2+012952.2} & 0.364    & 0.014       & 127       & 6                 & \phantom{0}34.9673 & \phantom{0}1.4978    & ACT-CL J0219.9+0130 \\ \midrule
    \textbf{RM J080135.3+362807.5} & 0.511    & 0.014       & 125       & 17                & 120.3971           & 36.4688              & ... \\ \midrule
    \textbf{RM J082944.9+382754.4} & 0.376    & 0.013       & 146       & 6                 & 127.4372           & 38.4651              & PSZ2 G183.30+34.98 \\ \midrule
    \textbf{RM J092647.3+050004.0} & 0.469    & 0.011       & 140       & 13                & 141.6970           & \phantom{0}5.0011    & \makecell{PSZ2 G227.89+36.58\\ACT-CL J0926.7+0500}\\ \midrule
    \textbf{RM J104311.1+150151.9} & 0.417    & 0.012       & 121       & 8                 & 160.7962           & 15.0311              & ...\\ \midrule
    \textbf{RM J121917.6+505432.8} & 0.535    & 0.015       & 165       & 23                & 184.8232           & 50.9091              & PSZ2 G135.17+65.43 \\ \midrule
    \textbf{RM J125725.9+365429.4} & 0.520    & 0.013       & 138       & 17                & 194.3580           & 36.9082              & ...\\ \midrule
    \textbf{RM J132724.2+534656.5} & 0.386    & 0.013       & 121       & 6                 & 201.8506           & 53.7823              & ... \\ \midrule
    \textbf{RM J163509.2+152951.5} & 0.472    & 0.011       & 143       & 13                & 248.7885           & 15.4976              & ACT-CL J1635.1+1529 \\ \midrule
    \textbf{RM J232104.1+291134.5} & 0.494    & 0.011       & 239       & 22                & 350.2673           & 29.1929              & PSZ2 G100.22-29.64 \\ \bottomrule
    \tablenoterow{8}{All properties from \citet{rykoff2016redmapper}.}
\end{tabular*}
\end{table*}

% %%%%%%%%%%%%%%%%%%%%%%%%%%%%%%%%%%%%%%%%%%%%%%%%%%%%%%%%%%%%%%%%%%%%%%%%%%%%%%%%%%%%%%%%%%%%%%
% %%%%%%%%%%%%%%%%%%%%%%%%%%%%%%%%  Classification Framework   %%%%%%%%%%%%%%%%%%%%%%%%%%%%%%%%%
% %%%%%%%%%%%%%%%%%%%%%%%%%%%%%%%%%%%%%%%%%%%%%%%%%%%%%%%%%%%%%%%%%%%%%%%%%%%%%%%%%%%%%%%%%%%%%%
\subsection{Classification Framework}\label{subsec:Classification}

To compare merger candidates across the program in a uniform way, we adopt a single classification scheme that applies whether a system has been characterized in the published literature or with the new data presented in this work.

We assign each candidate to one of six tiers:
\begin{itemize}
    \item \textbf{Merger:} Disturbed X-ray morphology with substructure independently confirmed by multiple corroborating diagnostics (e.g., bimodal galaxy luminosity distribution, weak lensing mass structure, spectroscopic velocity substructure, or radio relic emission).
    \item \textbf{Likely Merger:} Disturbed X-ray morphology supported by at least one corroborating diagnostic, but with residual uncertainty in geometry, mass ratio, or component identification.
    \item \textbf{Possible Merger:} Disturbed or asymmetric X-ray morphology without corroborating evidence; a single dynamically complex halo cannot be ruled out.
    \item \textbf{Indeterminate:} Data quality insufficient for classification (e.g., low signal-to-noise or off-axis placement on the detector).
    \item \textbf{Likely Relaxed:} Regular X-ray morphology with at least one corroborating diagnostic suggesting equilibrium.
    \item \textbf{Relaxed:} Regular X-ray morphology with multiple corroborating diagnostics.
\end{itemize}

The morphological parameter $\delta$, defined by \citet{yuan2020dynamical}, is listed for the systems in \cref{tab:CXO_and_XMM,tab:CXO} (the remaining systems lack a decisive match). This composite index combines two adaptive measurements taken within the best-fit elliptical region of a cluster's X-ray emission: a profile parameter $\kappa$, which characterizes the surface brightness profile, and an asymmetry factor $\alpha$, which quantifies departures from elliptical symmetry. The index is constructed to correlate with the traditional dynamical-state indicators of concentration, centroid shift, and power ratio ($P_3/P_0$), with $\delta > 0$ corresponding to dynamically active systems and $\delta < 0$ to relaxed systems. We adopt values from \citet{yuan2022dynamical}, which extends the original \CXO\ catalog with \XMM\ archival imaging. We treat $\delta$ as one possible corroborating diagnostic but do not consider it sufficient on its own to elevate a candidate above Possible Merger.

For the 12 X-SORTER targets presented in this work, uniform optical luminosity density, X-ray, and spectroscopic data permit an additional, orthogonal assessment of binarity. \textit{Binary} systems show two clear, comparably significant components with limited additional substructure. \textit{Likely Binary} systems have two dominant components but with some uncertainty about minor substructure or geometry. \textit{Possibly Binary} systems show two prominent components in some diagnostics but contain confounding additional structure. \textit{Not Binary} systems display multi-component or filamentary configurations that are not well described by a two-body interaction. We do not assess binarity for archival systems, since the uniform optical and spectroscopic analysis required is not available across that sample.

Merger and binarity classifications for each target are reported in their respective per-cluster summaries in \S\ref{subsec:Individual Clusters} and aggregated in \cref{fig:results}.

% %%%%%%%%%%%%%%%%%%%%%%%%%%%%%%%%%%%%%%%%%%%%%%%%%%%%%%%%%%%%%%%%%%%%%%%%%%%%%%%%%%%%%%%%%%%%%%
% %%%%%%%%%%%%%%%%%%%%%%%%%%%%%%%%%       XMM Data Reduction   %%%%%%%%%%%%%%%%%%%%%%%%%%%%%%%%%
% %%%%%%%%%%%%%%%%%%%%%%%%%%%%%%%%%%%%%%%%%%%%%%%%%%%%%%%%%%%%%%%%%%%%%%%%%%%%%%%%%%%%%%%%%%%%%%
\subsection{XMM Data Reduction}\label{subsec:XMM}
The targets were observed across the AO20-22 programs using the \XMM\ European Photon Imaging Camera (EPIC) and reduced using the Science Analysis System (SAS) Version 22.1.0 \citep{gabriel2004astronomical} in conjunction with the Extended Source Analysis Software (ESAS), largely following the ESAS Cookbook \citep{snowden2014cookbook}. The ESAS tasks were executed via a custom bash pipeline that streamlines the standard reduction workflow and is made publicly available.\footnote{\url{https://github.com/chrishopp12/xmm-esas-pipeline}}

For each detector (MOS1, MOS2, and PN), the data were first filtered of proton flaring events by identifying good time intervals in the 10--12 keV band, removing all events during the contaminated intervals. The exact filtering level depends on the observation, but was generally around 0.3 counts s$^{-1}$ for the MOS detectors and 0.5 counts s$^{-1}$ for the PN. Next, the exposures were checked for CCDs in anomalous (broken) states, which were removed from further processing. Point sources were subtracted with the \texttt{cheese} routine, with a manual inspection for any missing or wrongfully removed point sources. 

Background contamination was removed following the double-subtraction method of \citet{arnaud2002xmm}, with blank-sky files \citep{carter2007xmm} to account for spatial variability in the backgrounds across each detector. First, source and background regions with a typical radius of ${\sim}1\arcmin$--$1.5\arcmin$ were defined, with the background region selected to avoid bright sources, dead pixels, chip gaps, and anomalous CCDs. After correcting for vignetting effects in both the observation and blank-sky files with the routine \texttt{evigweight}, the effective area (A) of each region was calculated, as was the total count rate in the 10--12 keV band (R) for each detector.

A scaled blank-sky subtraction was then performed on the spectra (S) of each region,
\begin{align}
    S_\text{Source} &= A_\text{Obs, Src} \lt( \frac{S_\text{Obs, Src}}{A_\text{Obs, Src}} -Q \frac{S_\text{Blank, Src}}{A_\text{Blank, Src}}\rt), \\
    S_\text{Bkg} &= A_\text{Obs, Src} \lt(\frac{S_\text{Obs, Bkg}}{A_\text{Obs, Bkg}} - Q\frac{S_\text{Blank, Bkg}}{A_\text{Blank, Bkg}}\rt),
\end{align}
where Q is the ratio in count rates, 
\begin{equation}
    Q = \frac{R_\text{Obs}}{R_\text{Blank}}.
\end{equation}
The subtracted and scaled spectra were then subtracted a second time to produce the final spectra,
\begin{equation}
    S = S_\text{Source} - S_\text{Bkg}.
\end{equation}

The resulting spectrum for each detector was then grouped to a minimum of 30 counts per bin and fit simultaneously in XSPEC \citep{arnaud1996xspec}. An \texttt{apec} model for continuum plasma X-ray emission was combined with a \texttt{phabs} model for galactic absorption. To account for cross-calibration differences between detectors, a \texttt{const} component was added, with the value allowed to float on both MOS detectors. All detectors were fit simultaneously to produce a best-fit temperature for each cluster.

The best-fit temperature was then used to recalculate a source region with $r_{500}$ defined by the relationship from \citet{arnaud2005structural} to match the procedure from \citet{upsdell2023xmm}:
\begin{equation}\label{eq:ez}
    E(z)r_{500} = B_{500}\lt(\frac{T_X}{\SI{5}{\keV}}\rt)^\beta,
\end{equation}
where $E(z)=\sqrt{\Omega_M(1+z)^3 + \Omega_\Lambda}$, and constants $B_{500}=\SI{1104}{\kpc}$ and $\beta=0.57$. The double-subtraction and XSPEC fitting process was then repeated using the new source region defined by $r_{500}$. After fitting in the 0.3--\SI{10.0}{\keV} band, the cross-calibration and temperature components were frozen and the \texttt{clumin} convolution model was used to calculate the luminosities in 0.1--2.4 and  0.5--\SI{2.0}{\keV} bands.

Image processing was done with ESAS by first implementing \texttt{binadapt} with a bin factor of four and adaptively smoothing with \texttt{combimage} using a kernel sized to enclose 50 counts to produce a point-source-subtracted, exposure-corrected, smoothed image in the $0.4-\SI{1.1}{\keV}$ band. Residual holes in the image from point-source masking were iteratively filled using the \texttt{inpaint\_biharmonic} routine in \texttt{skimage.restoration} \citep{van2014scikit}. Surface brightness contours varied in spacing depending on the observation, but were generally set to include 12 linearly spaced levels starting at ${\sim}0.5 \sigma$, where $\sigma$ is the standard deviation of the XSB, and extending up to the maximum pixel value in the field.

Additional image processing was performed to qualitatively inspect sharp features in the XSB. The ESAS-processed image was reprojected onto an $800 \times 800$ pixel grid using \texttt{reproject.exact}, which conserves flux \citep{robitaille2020reproject}. After infilling holes in the image due to point source masking, the image was smoothed with a 10\arcsec\ Gaussian kernel and one-dimensional profiles were extracted across any suspected features. A map of the Gaussian gradient magnitude was also generated using the same 10\arcsec\ kernel to highlight sharp intensity gradients. Finally, an unsharp mask was produced by subtracting an image smoothed with a 50\arcsec\ kernel from the image smoothed at 10\arcsec.

% %%%%%%%%%%%%%%%%%%%%%%%%%%%%%%%%%%%%%%%%%%%%%%%%%%%%%%%%%%%%%%%%%%%%%%%%%%%%%%%%%%%%%%%%%%%%%%
% %%%%%%%%%%%%%%%%%%%%%%%%%%%%%  Member Galaxy Redshift Survey %%%%%%%%%%%%%%%%%%%%%%%%%%%%%%%%%
% %%%%%%%%%%%%%%%%%%%%%%%%%%%%%%%%%%%%%%%%%%%%%%%%%%%%%%%%%%%%%%%%%%%%%%%%%%%%%%%%%%%%%%%%%%%%%%
\subsection{Member Galaxy Redshift Survey} \label{sec:deimos}
Spectroscopic observations were carried out using the DEIMOS multi-object spectrograph \citep{faber2003deimos} on the Keck II telescope between 2023 and 2025 as part of a multi-semester program. For each cluster, we designed two slitmasks to maximize spectroscopic coverage of likely cluster members and potential foreground and background structures. Targets for each mask were selected using Pan-STARRS photometric data, with priorities based on both brightness and likelihood of cluster membership. Specifically, we assigned each galaxy a weight proportional to
\begin{equation}
    \frac{24 - r}{\sigma_{\mathrm{PS}}} \exp\left( \frac{-(z_{\mathrm{PS}} - z_{\mathrm{cl}})^2}{2 \sigma_{\mathrm{PS}}^2} \right),
\end{equation}
where $r$ is the $r$-band apparent magnitude and $z_{\mathrm{PS}}$ and $\sigma_{\mathrm{PS}}$ are the photometric redshift and its uncertainty from the Pan-STARRS catalog, respectively. Because the Pan-STARRS photometric redshifts have substantial uncertainty (${\sim}0.15$), this weighting allows for some focus on the cluster redshift while retaining sensitivity to foreground and background structures. The purpose of the magnitude weighting is to prioritize brighter galaxies (those more likely to successfully yield a redshift) within the target cluster. A side effect of the magnitude weighting is that foreground galaxies are more likely to be targeted than background galaxies. These weights were input into the \texttt{dsimulator} software to prioritize slit placement. We excluded stars based on Pan-STARRS morphology and avoided targeting galaxies with existing spectroscopic redshifts, except on the outskirts of the field of view, where redundancy was used for consistency checks.

%%%%%%%%%%%%%%%%%%%%%%%%%% Table 7: Deimos Config %%%%%%%%%%%%%%%%%%%%%%%%%%
\begin{table}
    \centering
    \caption{DEIMOS Configurations}
    \label{tab:deimos}
    \begin{tabular*}{\columnwidth}{@{\extracolsep{\fill}}lclcc} \toprule\toprule
                         Cluster  & Mask & Date (UT)        & Exp [s]   & Wavelength [\AA] \\ \toprule
\multirow{2}{*}{\textbf{RMJ0003}} & A    & 2023 Sept 8      & 2520      &  4680-7320 \\ 
                                  & B    & 2023 Sept 14     & 2700      &  4680-7320 \\ \midrule
\multirow{2}{*}{\textbf{RMJ0109}} & A    & 2023 Sept 14     & 2160      &  5180-7820 \\ 
                                  & B    & 2023 Sept 14     & 2160      &  5180-7820 \\ \midrule
\multirow{2}{*}{\textbf{RMJ0219}} & A    & 2023 Sept 14     & 2700      &  4680-7320 \\
                                  & B    & 2023 Sept 14     & 2700      &  4680-7320 \\ \midrule
\multirow{2}{*}{\textbf{RMJ0801}} & A    & 2023 April 4     & 1800      &  5370-8010 \\
                                  & B    & 2023 April 4     & 1800      &  5370-8010 \\ \midrule
\multirow{2}{*}{\textbf{RMJ0829}} & A    & 2025 Feb 23, 24  & 6300      &  4780-7420 \\
                                  & B    & 2025 Feb 23      & 2700      &  4780-7420 \\  \midrule
 \multirow{2}{*}{\textbf{RMJ0926}}& A    & 2025 Feb 23      & 2700      &  5080-7720 \\
                                  & B    & 2025 Feb 23      & 2700      &  5080-7720 \\  \midrule   
\multirow{2}{*}{\textbf{RMJ1043}} & A    & 2025 Feb 23      & 2700      &  4930-7570 \\
                                  & B    & 2025 Feb 23      & 2580      &  4930-7570 \\  \midrule
\multirow{2}{*}{\textbf{RMJ1219}} & A    & 2023 April 23    & 1800      &  5370-8010 \\
                                  & B    & 2023 April 23    & 1800      &  5370-8010 \\ \midrule
\textbf{RMJ1257}                  & A    & 2022 July 1      & 2340      &  5370-8010 \\ \midrule
\multirow{2}{*}{\textbf{RMJ1327}} & A    & 2025 Feb 24      & 2280      &  4780-7420 \\
                                  & B    & 2025 Feb 23      & 2700      &  4780-7420 \\ \midrule
\multirow{3}{*}{\textbf{RMJ1635}} & A    & 2023 April 12    & 3240      &  5180-7820 \\
                                  & B    & 2023 April 23    & 3240      &  5370-8010 \\
                                  & C    & 2023 Sept 8      & 2160      &  5370-8010 \\ \midrule
\multirow{2}{*}{\textbf{RMJ2321}} & A    & 2023 Sept 8      & 2160      &  5180-7820 \\
                                  &B     & 2023 Sept 14     & 2700      &  5180-7820 \\ \bottomrule
    \end{tabular*}

\end{table}

\cref{tab:deimos} lists the masks and corresponding DEIMOS configurations. While the specifics of each observation are dependent on cluster redshift and observing conditions, the same general approach was applied to all clusters. All masks utilized the 1200G grating, giving a dispersion of ${\sim}0.33$ \AA\ pixel$^{-1}$ and a spectral resolution of ${\sim}1$\AA, corresponding to a velocity resolution of ${\sim}40$ km s$^{-1}$. The wavelength coverage varied with cluster redshift so that it consistently included rest-frame features from the 3727\AA\ [O~\textsc{ii}] doublet to the 5177\AA\ magnesium line. (Within a mask, the exact wavelength coverage depends on the slit position, but always covers these lines for galaxies in the target cluster.)

The baseline plan was for three 900-second exposures on each mask, but exposure times were shortened (and in a few cases lengthened) as needed to complete the masks within the allotted night. The ``B" mask for RMJ1257 is omitted from \cref{tab:deimos} because it received almost no exposure time and yielded no redshifts. Conversely, a third mask was designed for RMJ1635 to fill a gap in target scheduling, and RMJ0829 mask A was observed with approximately double the standard exposure time over two nights for a similar reason. The mask name listed in the Keck Science Archive can be constructed from \cref{tab:deimos} by combining the four digits in the cluster name with the mask letter, for example \textit{0926B} or \textit{1635C}.

We reduced the data using \texttt{PypeIt} \citep{prochaska2020pypeit}, which performs wavelength calibration, sky subtraction, and extraction of 1D spectra. Redshifts were measured with a custom cross-correlation algorithm that compares each spectrum to a set of redshifted galaxy templates, following the procedure detailed in \citet{wittman2023new}, with typical uncertainties of $ \lesssim 10^{-4}$.

Each mask was designed with ${\sim}100$ slits and yielded ${\sim}70$ secure redshifts. These new redshifts were combined with archival results in a 10\arcmin\ radius circle around the cluster from SDSS \citep{kollmeier2019sdss}, BOSS \citep{dawson2012baryon}, and DESI \citep{abdul2025data}. For each cluster, we used the combined redshift catalog to identify cluster members and assess substructure following a consistent procedure. We first inspected the full redshift distribution to define the cluster redshift range. Within that range, we fit Gaussian mixture models (GMMs) with one, two, and three components and used the Bayesian Information Criterion (BIC) to select the preferred number of components. The cluster-range distribution, and individual GMM components where multiple were preferred, was tested for Gaussianity using the Kolmogorov--Smirnov (KS) and Anderson--Darling (AD) tests, with AD $p$-values from the Monte Carlo option in \texttt{scipy.stats.anderson}. GMM components outside the cluster redshift range identified foreground and background structures, whose spatial distributions were inspected for any overlap with the X-ray emission or galaxy density.

To assess substructure within each cluster, the field was divided into regions defined by the bisectors between candidate BCGs and a fixed radius from each BCG (typically $2.5\arcmin$), and velocity dispersion and mean redshift were computed for the spectroscopic members in each region.

% %%%%%%%%%%%%%%%%%%%%%%%%%%%%%%%%%%%%%%%%%%%%%%%%%%%%%%%%%%%%%%%%%%%%%%%%%%%%%%%%%%%%%%%%%%%%%%
% %%%%%%%%%%%%%%%%%%% Red Sequence Selection and Luminosity Density %%%%%%%%%%%%%%%%%%%%%%%%%%%%
% %%%%%%%%%%%%%%%%%%%%%%%%%%%%%%%%%%%%%%%%%%%%%%%%%%%%%%%%%%%%%%%%%%%%%%%%%%%%%%%%%%%%%%%%%%%%%%
\subsection{Red Sequence Selection and Luminosity Density}
\label{subsec:redseq_contours}

We identified red sequence galaxies using DESI Legacy Survey \citep{dey2019overview} photometry in the \textit{g} and \textit{r} bands. For each cluster, we extracted sources within a 10\arcmin\ radius around the cluster center and constructed a color-magnitude diagram (CMD) using $ g-r$ color versus $r$-band magnitude. Sources with missing or invalid photometry were removed.

We cross-matched the photometric catalog with available spectroscopic redshifts to isolate confirmed cluster members. A linear red sequence was then fit in $g-r$ versus $r$ using these members with iterative $\sigma$-clipping to remove outliers and obtain a robust slope and intercept. We defined red sequence candidates as photometric sources lying within $\pm0.2$ mag of the best-fit relation and added them to our catalog of cluster members identified by spectroscopy.

Each cluster member was then assigned a luminosity weight using its $r$-band magnitude:
\begin{equation}
    w_L = 10^{-0.4r},
\end{equation}
which serves as a proxy for stellar luminosity. These weights were then used to construct a smoothed luminosity-weighted spatial density map, generated by applying a Gaussian kernel density estimator with \texttt{scipy.stats.gaussian\_kde} \citep{virtanen2020scipy}. Density contours varied in spacing depending on the observation, but were typically constructed with 12 linearly spaced levels from the minimum to maximum weighted density values, with the lowest contour omitted for clarity.

% %%%%%%%%%%%%%%%%%%%%%%%%%%%%%%%%%%%%%%%%%%%%%%%%%%%%%%%%%%%%%%%%%%%%%%%%%%%%%%%%%%%%%%%%%%%%%%
% %%%%%%%%%%%%%%%%%%%%%%%%%%%%%%%%%                            %%%%%%%%%%%%%%%%%%%%%%%%%%%%%%%%%
% %%%%%%%%%%%%%%%%%%%%%%%%%%%%%%%%%   Results and Discussion   %%%%%%%%%%%%%%%%%%%%%%%%%%%%%%%%%
% %%%%%%%%%%%%%%%%%%%%%%%%%%%%%%%%%                            %%%%%%%%%%%%%%%%%%%%%%%%%%%%%%%%%
% %%%%%%%%%%%%%%%%%%%%%%%%%%%%%%%%%%%%%%%%%%%%%%%%%%%%%%%%%%%%%%%%%%%%%%%%%%%%%%%%%%%%%%%%%%%%%%
\section{Results and Discussion}\label{sec:Results}

%%%%%%%%%%%%%%%%%%%%%%%%%% Table 8: XMM Results %%%%%%%%%%%%%%%%%%%%%%%%%%
\pdffig{
\begin{table*}
    \centering
    \caption{\XMM\ Observation Results}
    \label{tab:XMM}
\begin{tabular*}{\textwidth}{@{\extracolsep{\fill}} c c cccc c c cc} \toprule\toprule
            Name        &  Obs ID   & \multicolumn{4}{c}{Exposure [ks]}                                 & $r_{500}$ [kpc] & $T_X$ [keV]           & \multicolumn{2}{c}{$L_X$ [$10^{44}$ erg s$^{-1}$]} \\ \cmidrule(lr){3-6}\cmidrule(lr){9-10}
                        &           & \scriptsize Nominal &  \multicolumn{3}{c}{\scriptsize Usable}   &  &                     & \scriptsize 0.5--2.0 keV & \scriptsize 0.1--2.4 keV \\ \cmidrule(lr){4-6}
                        &           &                     & \scriptsize MOS1 & \scriptsize MOS2 & \scriptsize PN                            &  &                     & & \\\toprule

\textbf{RMJ0003} & 0881900801 &  38.9 & \usetup{21.6}{21.9}{13.8} & 1160 & \phantom{0}$7.65 \pm 1.70$ & $4.09$ {\raisebox{0.5ex}{\tiny$\substack{+0.063 \\ -0.062}$}} & \phantom{0}$6.60$ {\raisebox{0.5ex}{\tiny$\substack{+0.093 \\ -0.092}$}} \\ \midrule
\textbf{RMJ0109} & 0922150401 &  28.9 & \usetup{17.2}{17.3}{14.8} & \phantom{0}980 & \phantom{0}$6.38 \pm 0.48$ & $4.19$ {\raisebox{0.5ex}{\tiny$\substack{+0.063 \\ -0.062}$}} & \phantom{0}$6.78$ {\raisebox{0.5ex}{\tiny$\substack{+0.101 \\ -0.100}$}} \\ \midrule
\textbf{RMJ0219} & 0922150601 &  36.6 & \usetup{24.6}{24.6}{21.7} & \phantom{0}860 & \phantom{0}$4.50 \pm 0.26$ & $1.93$ {\raisebox{0.5ex}{\tiny$\substack{+0.026 \\ -0.025}$}} & \phantom{0}$3.13$ {\raisebox{0.5ex}{\tiny$\substack{+0.041 \\ -0.041}$}}  \\ \midrule
\textbf{RMJ0801} & 0881901001 &  24.0 & \usetup{22.1}{22.3}{20.4} & 1150 & \phantom{0}$8.65 \pm 1.67$ & $4.51$ {\raisebox{0.5ex}{\tiny$\substack{+0.062 \\ -0.061}$}}& \phantom{0}$7.26$ {\raisebox{0.5ex}{\tiny$\substack{+0.100 \\ -0.098}$}}  \\ \midrule
\textbf{RMJ0829} & 0901870901 &  28.0 & \usetup{26.7}{25.2}{24.5} & 1090 & $\phantom{0}6.95 \pm 0.26$ & $5.50$ {\raisebox{0.5ex}{\tiny$\substack{+0.035 \\ -0.035}$}}& \phantom{0}$8.89$ {\raisebox{0.5ex}{\tiny$\substack{+0.057 \\ -0.057}$}} \\ \midrule
\textbf{RMJ0926} & 0901870201 &  19.0 & \usetup{\phantom{0}7.9}{\phantom{0}8.0}{\phantom{0}4.9} & 1320 & $10.55 \pm 1.62$ & $6.84$ {\raisebox{0.5ex}{\tiny$\substack{+0.120 \\ -0.118}$}}& $10.99$ {\raisebox{0.5ex}{\tiny$\substack{+0.192 \\ -0.189}$}}  \\ \midrule
\textbf{RMJ1043} & 0922150301 &  21.0 & \usetup{19.7}{19.7}{17.5} & \phantom{0}630 & \phantom{0}$2.77 \pm 0.44$ & $0.56$ {\raisebox{0.5ex}{\tiny$\substack{+0.024 \\ -0.023}$}}& \phantom{0}$0.91$ {\raisebox{0.5ex}{\tiny$\substack{+0.039 \\ -0.038}$}} \\ \midrule
\textbf{RMJ1219} & 0881900301 &  30.0 & \usetup{17.5}{18.6}{10.1} & 1020 & \phantom{0}$7.27 \pm 0.83$ & $5.32$ {\raisebox{0.5ex}{\tiny$\substack{+0.103 \\ -0.101}$}}& \phantom{0}$8.58$ {\raisebox{0.5ex}{\tiny$\substack{+0.166 \\ -0.163}$}}  \\ \midrule
\textbf{RMJ1257} & 0881900701 &  14.0 & \usetup{12.6}{12.4}{\phantom{0}1.1} & \phantom{0}740 & \phantom{0}$4.04 \pm 0.68$ & $3.04$ {\raisebox{0.5ex}{\tiny$\substack{+0.123 \\ -0.118}$}}& \phantom{0}$4.95$ {\raisebox{0.5ex}{\tiny$\substack{+0.200 \\ -0.193}$}} \\ \midrule
\textbf{RMJ1327} & 0881901201 &  28.7 & \usetup{10.9}{10.3}{12.4} & \phantom{0}830 & \phantom{0}$4.31 \pm 0.85$ & $1.21$ {\raisebox{0.5ex}{\tiny$\substack{+0.048 \\ -0.046}$}}& \phantom{0}$1.97$ {\raisebox{0.5ex}{\tiny$\substack{+0.078 \\ -0.075}$}} \\ \midrule
\textbf{RMJ1635} & 0881900501 &  18.8 & \usetup{15.6}{16.3}{\phantom{0}4.0} & \phantom{0}930 & \phantom{0}$5.74 \pm 1.14$ & $4.09$ {\raisebox{0.5ex}{\tiny$\substack{+0.151 \\ -0.145}$}}& \phantom{0}$6.62$ {\raisebox{0.5ex}{\tiny$\substack{+0.244 \\ -0.235}$}} \\ \midrule
\textbf{RMJ2321} & 0922150101 &  17.9 & \usetup{13.9}{13.8}{11.8} & 1150 & \phantom{0}$8.62 \pm 1.01$ & $7.20$ {\raisebox{0.5ex}{\tiny$\substack{+0.108 \\ -0.106}$}}& $11.59$ {\raisebox{0.5ex}{\tiny$\substack{+0.173 \\ -0.171}$}} \\ \bottomrule
    \tablenoterow{10}{RMJ1257 was fit using only MOS detectors due to proton flaring.}
\end{tabular*}
\end{table*}}

% %%%%%%%%%%%%%%%%%%%%%%%%%%%%%%%%%%%%%%%%%%%%%%%%%%%%%%%%%%%%%%%%%%%%%%%%%%%%%%%%%%%%%%%%%%%%%%
% %%%%%%%%%%%%%%%%%%%%%%%%%%%%%         X-Ray Analysis         %%%%%%%%%%%%%%%%%%%%%%%%%%%%%%%%%
% %%%%%%%%%%%%%%%%%%%%%%%%%%%%%%%%%%%%%%%%%%%%%%%%%%%%%%%%%%%%%%%%%%%%%%%%%%%%%%%%%%%%%%%%%%%%%%
\subsection{X-Ray Analysis}\label{subsec:Xray}

X-ray temperatures and luminosities were determined following the procedure described in \S\ref{subsec:XMM}, with results shown in \cref{tab:XMM}. Luminosities are measured within $r_{500}$, computed from the scaling relation in \cref{eq:ez} using the fitted $T_X$; the uncertainty in $r_{500}$ is not propagated into the reported $L_X$ uncertainties. We use the scaling relations provided in \citet{upsdell2023xmm} to compare fitted temperatures and luminosities to expected values at the given richness, assuming self-similar evolution. These relations do not explicitly account for selection effects. However, we use relations derived from an optically selected sample, which includes X-ray faint sources (those not detected in the XMM Cluster Survey, possibly due to low exposure time or off-axis positioning on the detector) at an estimated upper-limit luminosity to reduce bias towards X-ray bright sources. Significant positive deviations from these relations can indicate ongoing merger activity, since cluster mergers shock-heat and compress the intracluster medium during pericenter passage, transiently boosting both $T_X$ and $L_X$ above the equilibrium values expected for a relaxed system of the same mass \citep{ricker2001off, poole2007impact}.

%%%%%%%%%%%%%%%%%%%%%%%%%% Table 8: XMM Results %%%%%%%%%%%%%%%%%%%%%%%%%%
\htmlfig{
\begin{table*}
    \centering
    \caption{\XMM\ Observation Results}
    \label{tab:XMM}
\begin{tabular*}{\textwidth}{@{\extracolsep{\fill}} c c cccc c c cc} \toprule\toprule
            Name        &  Obs ID   & \multicolumn{4}{c}{Exposure [ks]}                                 & $r_{500}$ [kpc] & $T_X$ [keV]           & \multicolumn{2}{c}{$L_X$ [$10^{44}$ erg s$^{-1}$]} \\ \cmidrule(lr){3-6}\cmidrule(lr){9-10}
                        &           & \scriptsize Nominal &  \multicolumn{3}{c}{\scriptsize Usable}   &  &                     & \scriptsize 0.5--2.0 keV & \scriptsize 0.1--2.4 keV \\ \cmidrule(lr){4-6}
                        &           &                     & \scriptsize MOS1 & \scriptsize MOS2 & \scriptsize PN                            &  &                     & & \\\toprule

\textbf{RMJ0003} & 0881900801 &  38.9 & \usetup{21.6}{21.9}{13.8} & 1160 & \phantom{0}$7.65 \pm 1.70$ & $4.09$ {\raisebox{0.5ex}{\tiny$\substack{+0.063 \\ -0.062}$}} & \phantom{0}$6.60$ {\raisebox{0.5ex}{\tiny$\substack{+0.093 \\ -0.092}$}} \\ \midrule
\textbf{RMJ0109} & 0922150401 &  28.9 & \usetup{17.2}{17.3}{14.8} & \phantom{0}980 & \phantom{0}$6.38 \pm 0.48$ & $4.19$ {\raisebox{0.5ex}{\tiny$\substack{+0.063 \\ -0.062}$}} & \phantom{0}$6.78$ {\raisebox{0.5ex}{\tiny$\substack{+0.101 \\ -0.100}$}} \\ \midrule
\textbf{RMJ0219} & 0922150601 &  36.6 & \usetup{24.6}{24.6}{21.7} & \phantom{0}860 & \phantom{0}$4.50 \pm 0.26$ & $1.93$ {\raisebox{0.5ex}{\tiny$\substack{+0.026 \\ -0.025}$}} & \phantom{0}$3.13$ {\raisebox{0.5ex}{\tiny$\substack{+0.041 \\ -0.041}$}}  \\ \midrule
\textbf{RMJ0801} & 0881901001 &  24.0 & \usetup{22.1}{22.3}{20.4} & 1150 & \phantom{0}$8.65 \pm 1.67$ & $4.51$ {\raisebox{0.5ex}{\tiny$\substack{+0.062 \\ -0.061}$}}& \phantom{0}$7.26$ {\raisebox{0.5ex}{\tiny$\substack{+0.100 \\ -0.098}$}}  \\ \midrule
\textbf{RMJ0829} & 0901870901 &  28.0 & \usetup{26.7}{25.2}{24.5} & 1090 & $\phantom{0}6.95 \pm 0.26$ & $5.50$ {\raisebox{0.5ex}{\tiny$\substack{+0.035 \\ -0.035}$}}& \phantom{0}$8.89$ {\raisebox{0.5ex}{\tiny$\substack{+0.057 \\ -0.057}$}} \\ \midrule
\textbf{RMJ0926} & 0901870201 &  19.0 & \usetup{\phantom{0}7.9}{\phantom{0}8.0}{\phantom{0}4.9} & 1320 & $10.55 \pm 1.62$ & $6.84$ {\raisebox{0.5ex}{\tiny$\substack{+0.120 \\ -0.118}$}}& $10.99$ {\raisebox{0.5ex}{\tiny$\substack{+0.192 \\ -0.189}$}}  \\ \midrule
\textbf{RMJ1043} & 0922150301 &  21.0 & \usetup{19.7}{19.7}{17.5} & \phantom{0}630 & \phantom{0}$2.77 \pm 0.44$ & $0.56$ {\raisebox{0.5ex}{\tiny$\substack{+0.024 \\ -0.023}$}}& \phantom{0}$0.91$ {\raisebox{0.5ex}{\tiny$\substack{+0.039 \\ -0.038}$}} \\ \midrule
\textbf{RMJ1219} & 0881900301 &  30.0 & \usetup{17.5}{18.6}{10.1} & 1020 & \phantom{0}$7.27 \pm 0.83$ & $5.32$ {\raisebox{0.5ex}{\tiny$\substack{+0.103 \\ -0.101}$}}& \phantom{0}$8.58$ {\raisebox{0.5ex}{\tiny$\substack{+0.166 \\ -0.163}$}}  \\ \midrule
\textbf{RMJ1257} & 0881900701 &  14.0 & \usetup{12.6}{12.4}{\phantom{0}1.1} & \phantom{0}740 & \phantom{0}$4.04 \pm 0.68$ & $3.04$ {\raisebox{0.5ex}{\tiny$\substack{+0.123 \\ -0.118}$}}& \phantom{0}$4.95$ {\raisebox{0.5ex}{\tiny$\substack{+0.200 \\ -0.193}$}} \\ \midrule
\textbf{RMJ1327} & 0881901201 &  28.7 & \usetup{10.9}{10.3}{12.4} & \phantom{0}830 & \phantom{0}$4.31 \pm 0.85$ & $1.21$ {\raisebox{0.5ex}{\tiny$\substack{+0.048 \\ -0.046}$}}& \phantom{0}$1.97$ {\raisebox{0.5ex}{\tiny$\substack{+0.078 \\ -0.075}$}} \\ \midrule
\textbf{RMJ1635} & 0881900501 &  18.8 & \usetup{15.6}{16.3}{\phantom{0}4.0} & \phantom{0}930 & \phantom{0}$5.74 \pm 1.14$ & $4.09$ {\raisebox{0.5ex}{\tiny$\substack{+0.151 \\ -0.145}$}}& \phantom{0}$6.62$ {\raisebox{0.5ex}{\tiny$\substack{+0.244 \\ -0.235}$}} \\ \midrule
\textbf{RMJ2321} & 0922150101 &  17.9 & \usetup{13.9}{13.8}{11.8} & 1150 & \phantom{0}$8.62 \pm 1.01$ & $7.20$ {\raisebox{0.5ex}{\tiny$\substack{+0.108 \\ -0.106}$}}& $11.59$ {\raisebox{0.5ex}{\tiny$\substack{+0.173 \\ -0.171}$}} \\ \bottomrule
    \tablenoterow{10}{RMJ1257 was fit using only MOS detectors due to proton flaring.}
\end{tabular*}
\end{table*}}

%%%%%%%%%%%%%%%%%%%%%%%%%% Figure 3: Temp vs Richness %%%%%%%%%%%%%%%%%%%%%%%%%%
\pdffig{
\begin{figure}[b]
    \centering
    \includegraphics[width=\columnwidth]{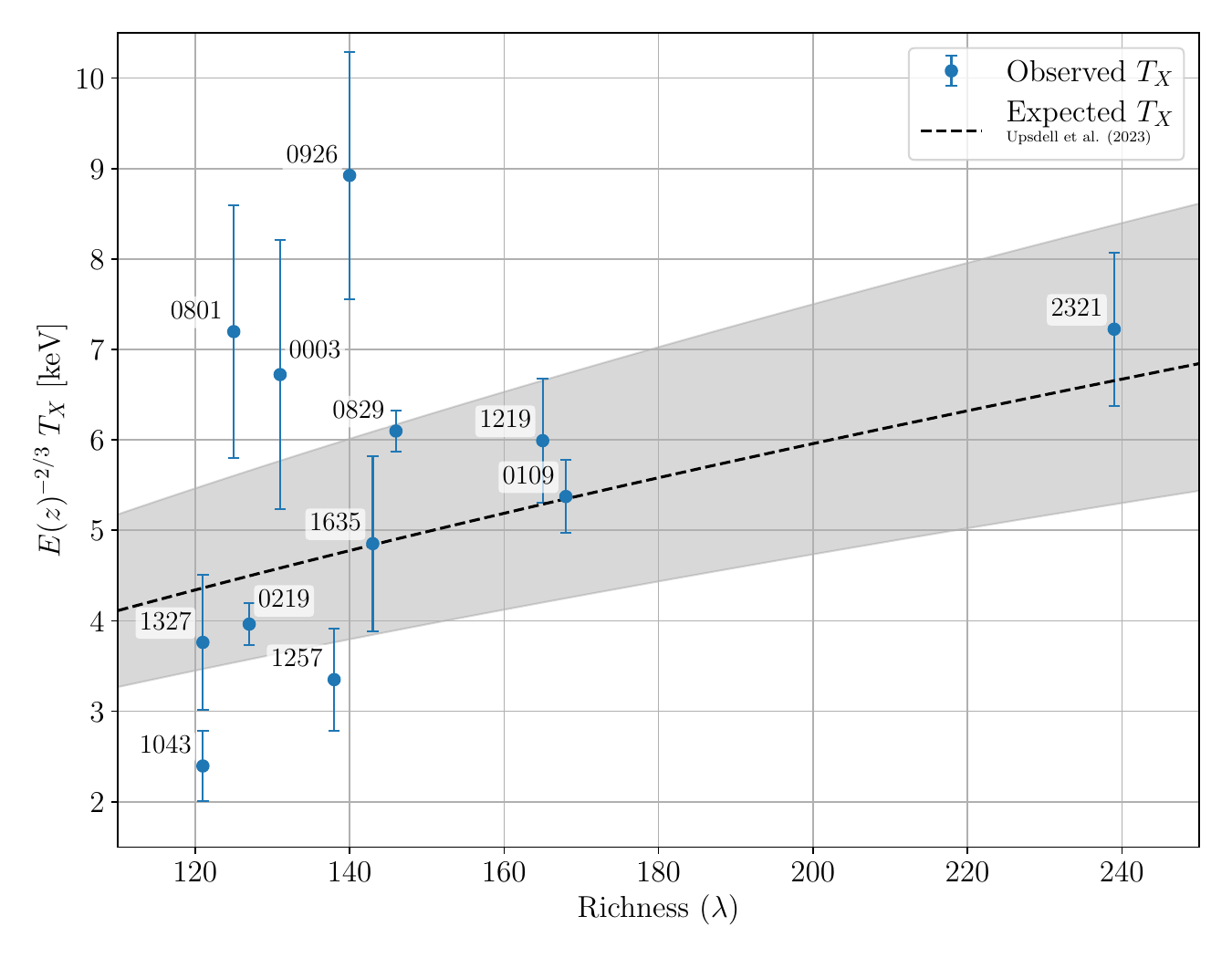}
    \caption{X-ray temperature, with self-similar correction $E(z)^{-2/3}$, versus richness with scaling relation given in \cref{eq:T}. The shaded region gives $1\sigma$ uncertainty in the scaling relation by \citet{upsdell2023xmm}.}
    \label{fig:rich_temp}
\end{figure}}

Looking at the richness--temperature scaling relation from \citet{upsdell2023xmm},
\begin{equation}\label{eq:T}
    \log\lt(\frac{T_X}{E(z)^{2/3}T_0}\rt) = \log(A_T) + B_T \log\lt(\frac{\lambda}{\lambda_0}\rt) \pm \sigma_T,
\end{equation}
with $E(z)$ given by \cref{eq:ez} and fitting parameters $A_T=1.13$, $B_T=0.62$, $T_0=\SI{2.5}{\keV}$, $\lambda_0=60.0$, and $\sigma_T=0.23$ is given in natural log space, we would expect temperatures of ${\sim}5$--$\SI{6}{\keV}$ for clusters of this richness and redshift (except for RMJ2321, which has an expected temperature of $\SI{7.9}{\keV}$), shown in \cref{fig:rich_temp}.

%%%%%%%%%%%%%%%%%%%%%%%%%% Figure 3: Temp vs Richness %%%%%%%%%%%%%%%%%%%%%%%%%%
\htmlfig{
\begin{figure}[b]
    \centering
    \includegraphics[width=\columnwidth]{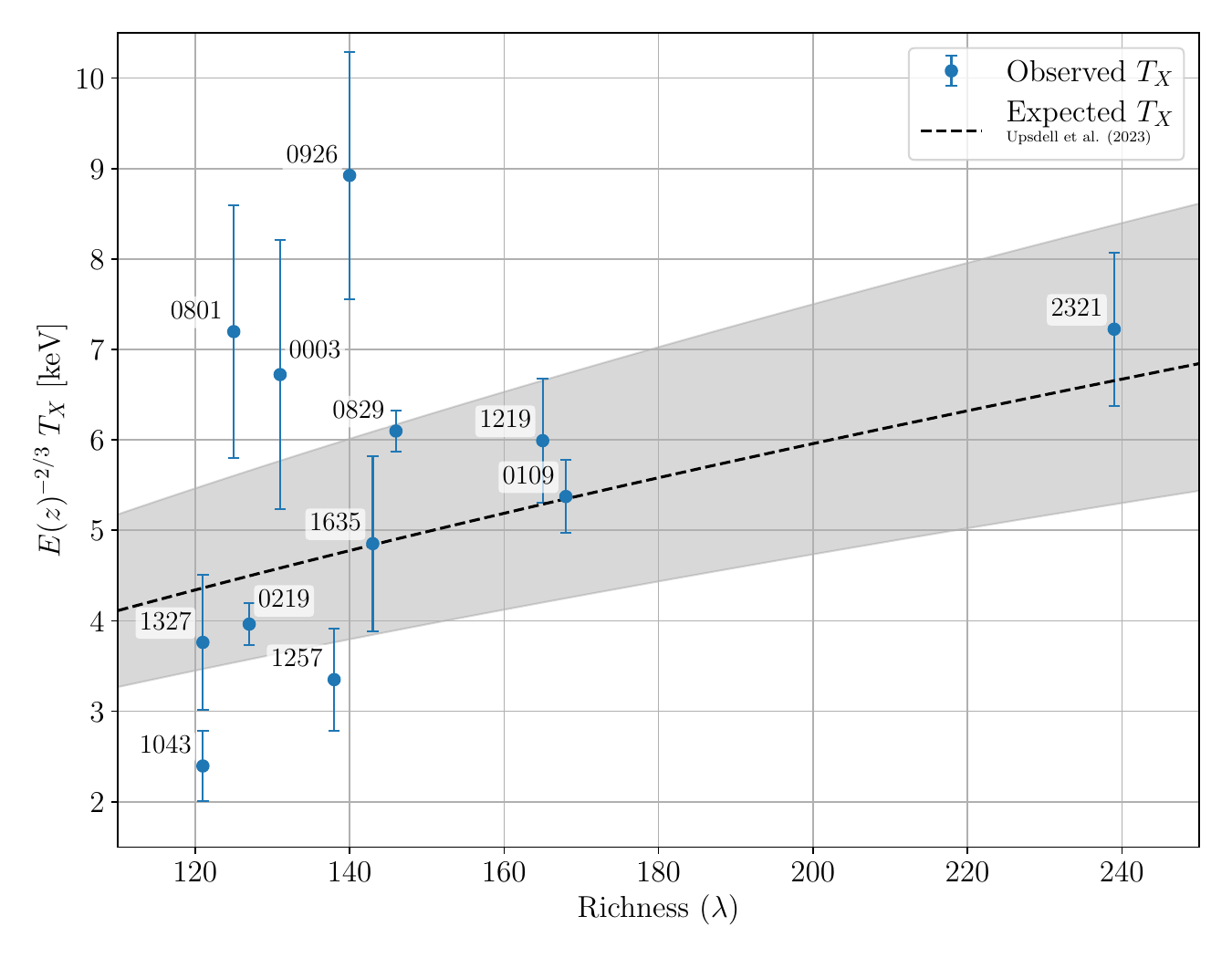}
    \caption{X-ray temperature, with self-similar correction $E(z)^{-2/3}$, versus richness with scaling relation given in \cref{eq:T}. The shaded region gives $1\sigma$ uncertainty in the scaling relation by \citet{upsdell2023xmm}.}
    \label{fig:rich_temp}
\end{figure}}

To assess the significance of deviations from the relation, we compare the logarithmic offset between the observed and expected temperatures to the quadrature sum of the intrinsic scatter and the observational uncertainty,
\begin{equation}
    \sigma = \sqrt{\sigma_{\mathrm{ln, obs}}^2 + \sigma_T^2},
\end{equation} where $\sigma_{\mathrm{ln, obs}}=\sigma_{\mathrm{obs}}/T_X$ is the observational uncertainty in natural log space (\textit{i.e.}, the fractional uncertainty). Three clusters (RMJ0003, RMJ0801, and RMJ0926) have temperatures more than $1\sigma$ greater than predicted by the scaling relation, while two clusters (RMJ1043 and RMJ0219) lie more than $1\sigma$ below the predicted value.

%%%%%%%%%%%%%%%%%%%%%%%%%% Figure 4: Luminosity vs Richness %%%%%%%%%%%%%%%%%%%%%%%%%%
\begin{figure}[b]
    \centering
    \includegraphics[width=\columnwidth]{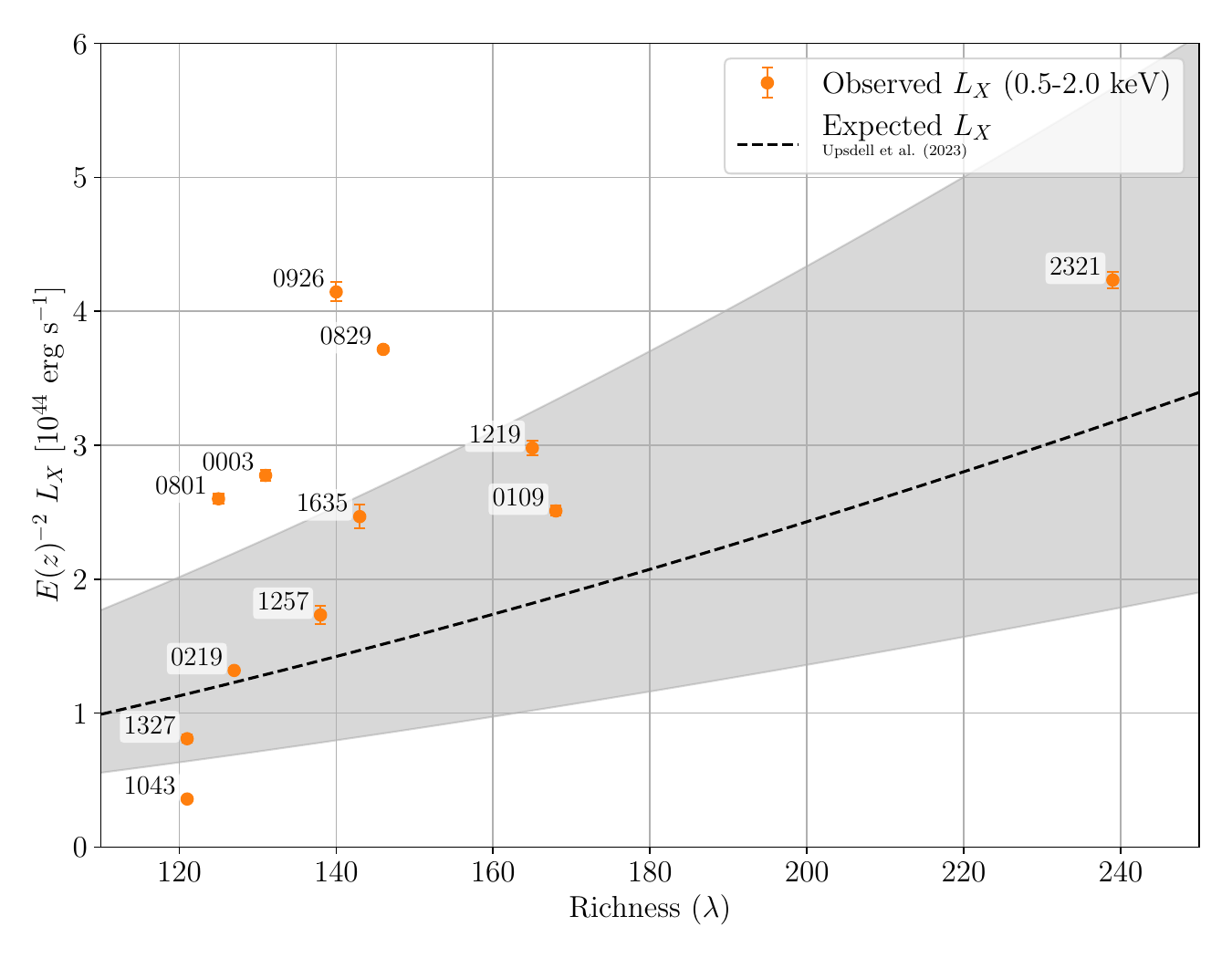}
    \caption{X-ray luminosity in the 0.5--\SI{2.0}{\keV} band, with self-similar correction $E(z)^{-2}$, versus richness with scaling relation given in \cref{eq:L}. The shaded region gives $1\sigma$ uncertainty in the scaling relation by \citet{upsdell2023xmm}.}
    \label{fig:rich_lum}
\end{figure}

Fitted luminosities in both the 0.1--2.4 and 0.5--\SI{2.0}{\keV} bands are listed in \cref{tab:XMM}. The richness--luminosity scaling relation used in \citet{upsdell2023xmm}, 
\begin{equation}\label{eq:L}
    \log\lt(\frac{L_X}{E(z)^{2}L_0}\rt) = \log(A_L) + B_L \log\lt(\frac{\lambda}{\lambda_0}\rt) \pm \sigma_L,
\end{equation}
with $A_L=0.57$, $B_L=1.5$, $L_0=\SI{0.7e44}{\ergs}$, $\lambda_0=60.0$, and $\sigma_L=0.58$,
utilizes the 0.5--\SI{2.0}{\keV} band and is shown in \cref{fig:rich_lum}. Only one of our clusters, RMJ1043, has a luminosity significantly below the prediction, while four (RMJ0003, RMJ0801, RMJ0829, and RMJ0926) lie more than $1\sigma$ above, where the uncertainty is again given in natural log space, $\sigma = \sqrt{\lt(\sigma_{\mathrm{obs}}/L_X\rt)^2 + \sigma_L^2}$. RMJ0926 exhibits the largest positive deviation from the expected relations in both luminosity and temperature, with offsets of $1.8\sigma$ and $2.7\sigma$ respectively, and appears to be a prime candidate for a binary dissociative merger. All of these systems are discussed at length in \S\ref{subsec:Individual Clusters}.

% %%%%%%%%%%%%%%%%%%%%%%%%%%%%%%%%%%%%%%%%%%%%%%%%%%%%%%%%%%%%%%%%%%%%%%%%%%%%%%%%%%%%%%%%%%%%%%
% %%%%%%%%%%%%%%%%%%%%%%%%  Redshift and Cluster Member Analysis  %%%%%%%%%%%%%%%%%%%%%%%%%%%%%%
% %%%%%%%%%%%%%%%%%%%%%%%%%%%%%%%%%%%%%%%%%%%%%%%%%%%%%%%%%%%%%%%%%%%%%%%%%%%%%%%%%%%%%%%%%%%%%%
\subsection{Redshift and Cluster Member Analysis}\label{subsec:Redshifts and Members}

%%%%%%%%%%%%%%%%%%%%%%%%%% Table 9: Redshift Summary %%%%%%%%%%%%%%%%%%%%%%%%%%
\begin{table}
    \centering
    \caption{Redshift and Membership Summary}
    \label{tab:Redshifts}
\begin{tabular*}{\columnwidth}{@{\extracolsep{\fill}}cccccc} \toprule\toprule
        Name & Redshift Range & \multicolumn{2}{c}{Redshifts} & \multicolumn{2}{c}{Members} \\
        \cmidrule(lr){3-4}\cmidrule(lr){5-6} & & Arch. & New & Spec. & CMD \\ \toprule
    \textbf{RMJ0003} & $0.360 \leq z \leq 0.385$ & 113 & 101 & 91 & 468 \\ \midrule
    \textbf{RMJ0109} & $0.449 \leq z \leq 0.470$ & 22 & 123 & 71 & 371 \\ \midrule
    \textbf{RMJ0219} & $0.355 \leq z \leq 0.375$ & 178 & 119 & 108 & 1110 \\ \midrule
    \textbf{RMJ0801} & $0.485 \leq z \leq 0.520$ & 153 & 96 & 62 & 479 \\ \midrule
    \textbf{RMJ0829} & $0.380 \leq z \leq 0.410$ & 110 & 147 & 119 & 588 \\ \midrule
    \textbf{RMJ0926} & $0.445 \leq z \leq 0.475$ & 208 & 143 & 89 & 585 \\ \midrule
    \textbf{RMJ1043} & $0.420 \leq z \leq 0.440$ & 65 & 154 & 61 & 631 \\ \midrule
    \textbf{RMJ1219} & $0.535 \leq z \leq 0.566$ & 126 & 118 & 75 & 432 \\ \midrule
    \textbf{RMJ1257} & $0.520 \leq z \leq 0.540$ & 139 & 39 & 31 & 320 \\ \midrule
    \textbf{RMJ1327} & $0.380 \leq z \leq 0.420$ & 151 & 171 & 124 & 616 \\ \midrule
    \textbf{RMJ1635} & $0.465 \leq z \leq 0.485$ & 83 & 143 & 76 & 792 \\ \midrule
    \textbf{RMJ2321} & $0.480 \leq z \leq 0.510$ & 95 & 115 & 113 & 775 \\ \bottomrule
    \end{tabular*}
\end{table}

A summary of redshift results is shown in \cref{tab:Redshifts}, including the number of archival members within 10\arcmin\ of the cluster and the number of new spectroscopic results from DEIMOS. The cluster redshift range was determined from the redshift distribution following the procedure described in \S\ref{sec:deimos}, using both archival and new results. That redshift range was then used to define a red sequence and identify additional cluster members from Legacy photometry as described in \S\ref{subsec:redseq_contours}. The full redshift catalog is available in machine-readable form; a portion is shown in Appendix \ref{sec:Catalog} (\cref{tab:redshift_table}).

%%%%%%%%%%%%%%%%%%%%%%%%%%%%%%%%%%%%%%%%%%%%%%%%%%%%%%%%%%%%%%%%%%%%%%%%%%%%%%%%%%%%%%%%%%%%%%
%%%%%%%%%%%%%%%%%%%%%%%%%%%%%%%%%                            %%%%%%%%%%%%%%%%%%%%%%%%%%%%%%%%%
%%%%%%%%%%%%%%%%%%%%%%%%%%%%%%%%% Individual Cluster Results %%%%%%%%%%%%%%%%%%%%%%%%%%%%%%%%%
%%%%%%%%%%%%%%%%%%%%%%%%%%%%%%%%%                            %%%%%%%%%%%%%%%%%%%%%%%%%%%%%%%%%
%%%%%%%%%%%%%%%%%%%%%%%%%%%%%%%%%%%%%%%%%%%%%%%%%%%%%%%%%%%%%%%%%%%%%%%%%%%%%%%%%%%%%%%%%%%%%%

%%%%%%%%%%%%%%%%%%%%%%%%%%%%%%%%%%%%%%%%%%%%%%%%%%%%%%%%%%%%%%%%%%%%%%%%%%%%%%%%%%%%%%%%%%%%%%
%%%%%%%%%%%%%%%%%%%%%%%%%%%%%%%%%          RMJ 0003          %%%%%%%%%%%%%%%%%%%%%%%%%%%%%%%%%
%%%%%%%%%%%%%%%%%%%%%%%%%%%%%%%%%%%%%%%%%%%%%%%%%%%%%%%%%%%%%%%%%%%%%%%%%%%%%%%%%%%%%%%%%%%%%%
\subsection{Individual Cluster Analysis}\label{subsec:Individual Clusters}
\subsubsection{RM J000343.8+100123.8}\label{subsubsec:RMJ0003}

%%%%%%%%%%%%%%%%%%%%%%%%%% RMJ0003: Table - BCGs %%%%%%%%%%%%%%%%%%%%%%%%%%
\begin{table}
    \centering
    \caption{RMJ0003 BCG Information}
    \label{tab:RMJ0003_BCG}
    \tablenotetext{a}{SDSS DR18 \citep{almeida2023eighteenth}}
    \tablenotetext{b}{DESI DR1 \citep{abdul2025data}}
    \tablenotetext{c}{DEIMOS (This work)}
    \tablenotetext{d}{DESI Legacy Survey DR10 \citep{dey2019overview}}
\begin{tabular*}{\columnwidth}{@{\extracolsep{\fill}}cccccc} \toprule\toprule
    BCG & Probability & Redshift & $r$-mag\tablenotemark{\footnotesize{d}} & RA {[}deg{]} & Dec {[}deg{]} \\ \toprule
    1  & 0.9680 & 0.374\tablenotemark{\footnotesize{a}} & 18.12 & 0.93249 & 10.02327 \\ \midrule
    2  & 0.0293 & 0.377\tablenotemark{\footnotesize{b}} & 18.49 & 0.95326 & 10.05429 \\ \midrule
    3  & 0.0020 & 0.364\tablenotemark{\footnotesize{b}} & 18.60 & 0.92447 & 10.01528 \\ \midrule
    4  & 0.0007 & 0.372\tablenotemark{\footnotesize{b}} & 19.15 & 0.95263 & 10.03755 \\ \midrule
    5  & 0.0000 & 0.369\tablenotemark{\footnotesize{c}} & 19.60 & 0.92024 & 10.02599 \\ \bottomrule
\printtablenotes{6}
\end{tabular*}
  \tablenotesreset
\end{table}

%%%%%%%%%%%%%%%%%%%%%%%%%% RMJ0003: Figure - Optical %%%%%%%%%%%%%%%%%%%%%%%%%%
\begin{figure}[t]
    \centering
    \includegraphics[clip, trim=0.0cm 1.25cm 0.0cm 0.0cm, width=\columnwidth]{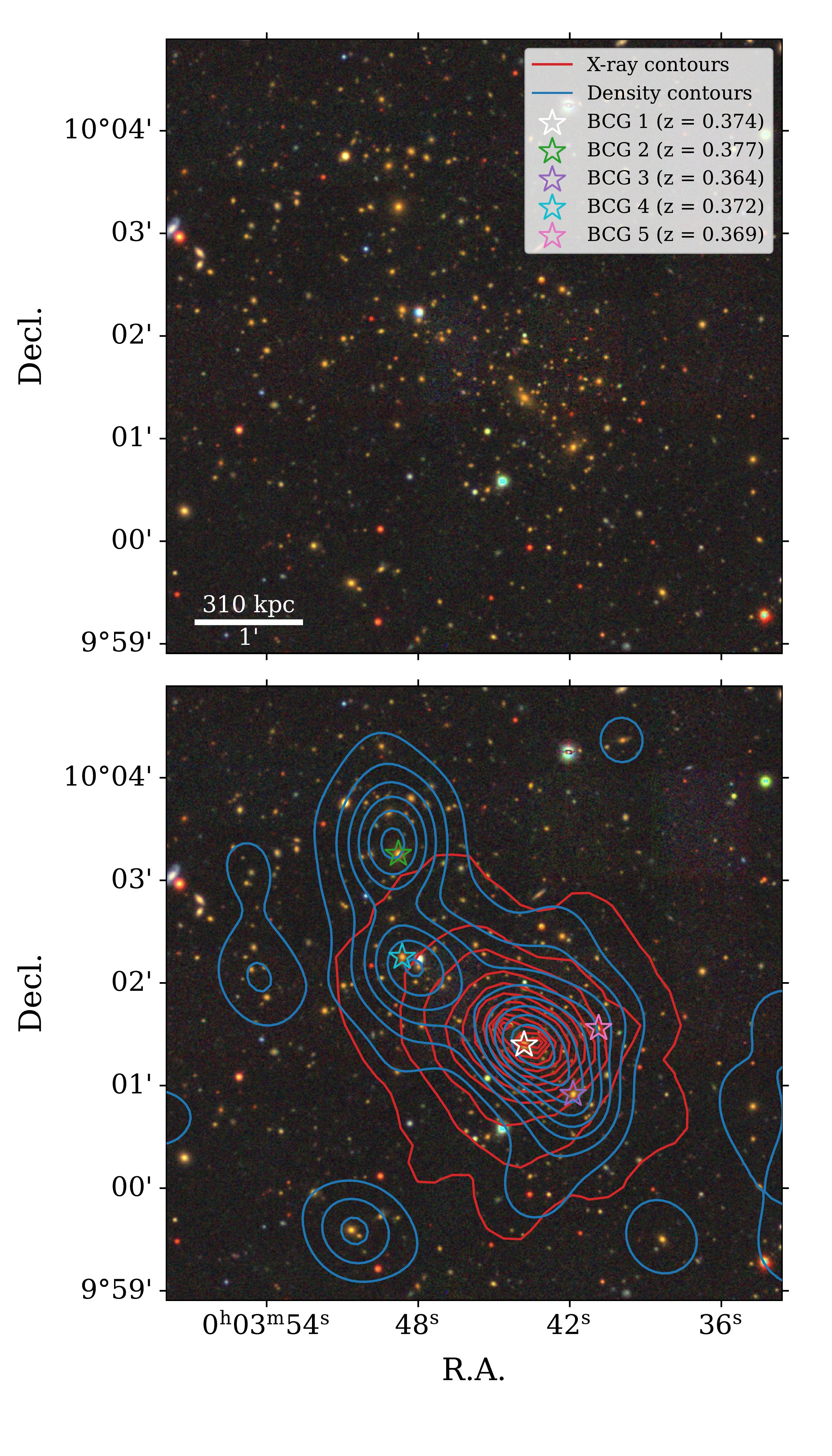}
    \caption{Optical image of RMJ0003. Top panel: 6\arcmin\ by 6\arcmin\ (1.86 by 1.86 Mpc) Legacy Survey image. Bottom panel: red sequence density contours (blue), X-ray surface brightness contours (red), and BCG candidates identified by redMaPPer.}
    \label{fig:RMJ0003_optical}
\end{figure}

%%%%%%%%%%%%%%%%%%%%%%%%%%%%%%%%%        RMJ 0003 Text       %%%%%%%%%%%%%%%%%%%%%%%%%%%%%%%%%

Our first cluster, RMJ0003, is the second-lowest redshift cluster in the sample at $z=0.372$ (\cref{tab:clusters}). The top priority BCG from redMaPPer barely meets our 0.98 probability threshold at $0.9680$, shown in \cref{tab:RMJ0003_BCG}. In \cref{fig:RMJ0003_optical}, we can see that the X-ray surface brightness is largely centered on BCG 1 and elongated towards BCG 4. The X-ray emission also shows a second peak between BCGs 1 and 4, offset ${\approx}\SI{100}{kpc}$ from BCG 1. Luminosity-weighted galaxy density contours show three individual structures, each collocated with one of the identified BCGs.

%%%%%%%%%%%%%%%%%%%%%%%%%% RMJ0003: Figure - Redshift Heatmap %%%%%%%%%%%%%%%%%%%%%%%%%%
\begin{figure}[t]
    \centering
    \includegraphics[clip, trim=0.0cm 0.1cm 0.0cm 0.0cm, width=\columnwidth]{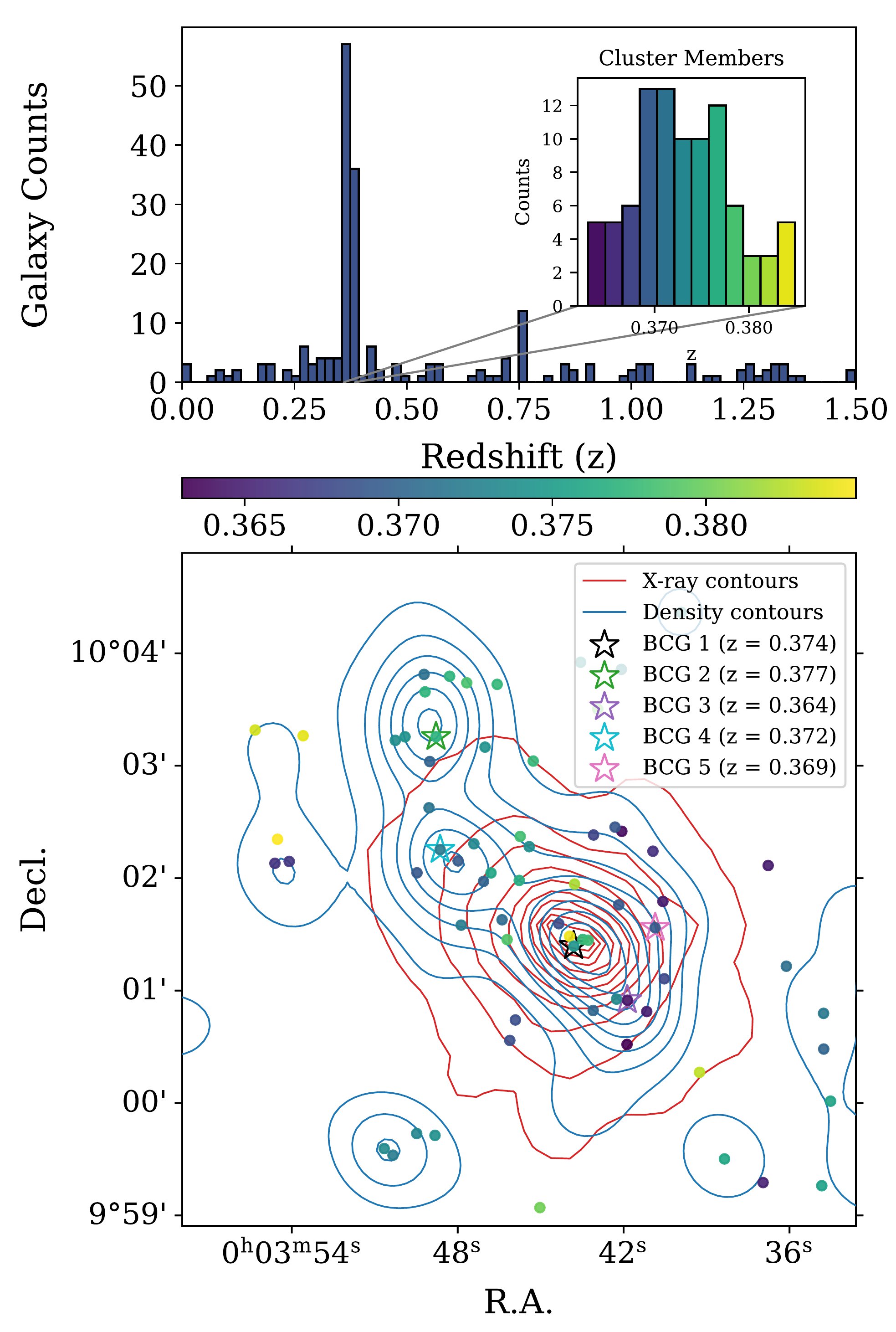}
    \caption{Redshift distribution of RMJ0003. Top panel: histogram of spectroscopic redshifts, combining archival data within 10\arcmin\ of the cluster with new observations from DEIMOS. The inset highlights galaxies in the redshift range $0.360 \leq z \leq 0.385$, which are classified as cluster members. Bottom panel: spatial distribution of those same members.}
    \label{fig:RMJ0003_redshifts}
\end{figure}

Looking at the spatial redshift distribution in the bottom panel of \cref{fig:RMJ0003_redshifts}, there is some clustering of lower redshift members near BCG 1 and higher redshift members near BCG 2, indicating possible separate subclusters. The redshift distribution of all galaxies in a 10\arcmin\ radius is shown in the top panel, with the inset showing all members between $0.360 \leq z \leq 0.385$ in the same field. There is a small background structure at $z\approx0.75$ with 12 spectroscopic members, which are all located southwest of our cluster and do not contaminate the observed X-ray emission. To determine if there are distinct subcluster distributions in redshift-space, we fit a Gaussian mixture model (GMM) with one, two, and three components to the cluster redshift distribution. A single-component model is preferred, with a Bayesian Information Criterion $\mathrm{BIC} = -690$ compared to $-679$ and $-668$ for two- and three-component models, respectively. We find the same one-component fit consistent with a Gaussian distribution according to both the Kolmogorov--Smirnov (KS) ($D=0.070$, $p=0.73$) and Anderson--Darling\footnote{For the KS test, the statistic $D$ is the maximum absolute difference between the empirical cumulative distribution function and that of a reference Gaussian distribution, and is primarily sensitive to global deviations. The Anderson--Darling statistic, $A^2$, places greater weight on deviations in the distribution tails. $A^2$ is typically reported in reference to tabulated critical values where $A^2 \lesssim 0.55$ are consistent with a Gaussian distribution at better than a $15\%$ significance level, while values $\gtrsim1$ strongly disfavor a normal distribution \citep{stephens1974edf}. Reported AD p-values are computed using the Monte Carlo option in \texttt{scipy.stats.anderson}.} (AD) ($A^2=0.46$, $p=0.26$) tests.

%%%%%%%%%%%%%%%%%%%%%%%%%% RMJ0003: Figure - Subclusters %%%%%%%%%%%%%%%%%%%%%%%%%%
\pdffig{
\begin{figure*}
    \centering
    \includegraphics[clip, trim=0.0cm 0.75cm 0.5cm 0.5cm, width=\textwidth]{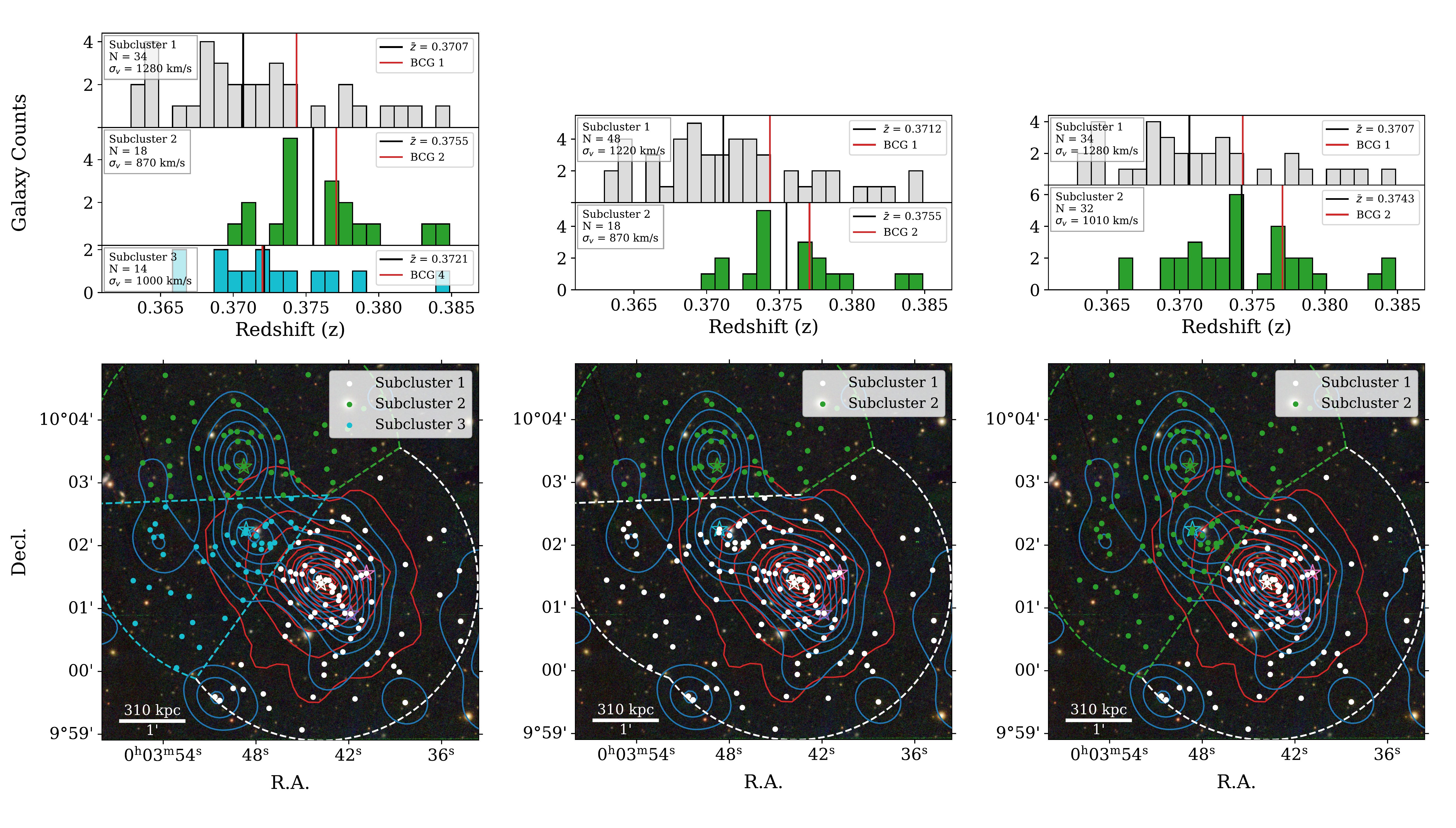}
    \caption{Subclustering analysis of RMJ0003. Bottom panels: spatial distribution of subcluster members, both spectroscopic and photometric, identified within regions defined by the bisector between BCG pairs and a radial distance of 2.5\arcmin\ from the BCG in each region. Top panels: redshift distribution of the spectroscopic members. The two right panels consider the subcluster associated with BCG 4 to be part of Subcluster 1 or 2.}
    \label{fig:RMJ0003_sub_1}
\end{figure*}}

To further explore the possible substructure in the cluster, we divided the cluster into three regions, each containing one of the BCGs associated with a density peak. The boundaries of these regions were defined by the bisectors between BCGs 1, 2, and 4 and a 2.5\arcmin\ radius from the associated BCG, shown in the bottom left panel of \cref{fig:RMJ0003_sub_1}. Individual members, both spectroscopic and those identified through red-sequence, were determined from spatial location without additional redshift cuts. The top left panel shows the redshift distribution of the spectroscopic members. Subclusters 1 and 2 (corresponding to BCGs 1 and 2) show a distinct separation in redshift ($\Delta v_{\mathrm{LOS}}=\SI{1430}{\kms}$ between subcluster means and \SI{590}{\kms} between BCGs), while Subcluster 3 (BCG 4) appears more uniformly distributed across the cluster redshift range. Each subcluster is individually consistent with a normal distribution by both KS and AD tests. 

%%%%%%%%%%%%%%%%%%%%%%%%%% RMJ0003: Figure - Subclusters %%%%%%%%%%%%%%%%%%%%%%%%%%
\htmlfig{
\begin{figure*}
    \centering
    \includegraphics[clip, trim=0.0cm 0.75cm 0.5cm 0.5cm, width=\textwidth]{Figures/RMJ_0003_all.jpg}
    \caption{Subclustering analysis of RMJ0003. Bottom panels: spatial distribution of subcluster members, both spectroscopic and photometric, identified within regions defined by the bisector between BCG pairs and a radial distance of 2.5\arcmin\ from the BCG in each region. Top panels: redshift distribution of the spectroscopic members. The two right panels consider the subcluster associated with BCG 4 to be part of Subcluster 1 or 2.}
    \label{fig:RMJ0003_sub_1}
\end{figure*}}

Given the difference in mean redshift between Subclusters 1 and 2 (\cref{tab:RMJ0003_subclusters}), we consider them to be distinct groups. To explore Subcluster 3 further, we considered scenarios that separately combine Subclusters 1 and 3 and Subclusters 2 and 3, shown in the center and right panels of \cref{fig:RMJ0003_sub_1}, respectively. Grouping Subclusters 1 and 3 together reduces the velocity dispersion of Subcluster 1 to \SI{1220}{\kms} (from \SI{1280}{\kms}) while the grouping of 1 and 2 increases the dispersion of Subcluster 2 to \SI{1010}{\kms}. The similarity in redshift distribution between Subclusters 1 and 3, along with an elevated temperature and luminosity (0.5--\SI{2.0}{\keV}) of \SI{7.65}{\keV} and \SI{4.09e44}{\ergs} (\cref{tab:XMM}) versus expected values of \SI{5.22}{\keV} and \SI{1.90e44}{\ergs}, indicates those subclusters are likely involved in an active merger near the plane of the sky.

%%%%%%%%%%%%%%%%%%%%%%%%%% RMJ0003: Table - Subclusters %%%%%%%%%%%%%%%%%%%%%%%%%%
\begin{table}
    \centering
    \caption{RMJ0003 Subcluster Properties}
    \label{tab:RMJ0003_subclusters}
\begin{tabular*}{\columnwidth}{@{\extracolsep{\fill}}cccccc} \toprule\toprule
    Subcluster & $N$ & BCG & BCG $z$ & Mean $z$ & $\sigma_v$ [km s$^{-1}$] \\ \toprule
All & 91 & ... & ... & 0.3731 & 1160 \\
1 & 34 & 1 & 0.374 & 0.3707 & 1280  \\
2 & 18 & 2 & 0.377 & 0.3755 & \phantom{0}870  \\
3 & 14 & 4 & 0.372 & 0.3721 & 1000  \\
\bottomrule
\end{tabular*}
\end{table}

The role of Subcluster 2 is more ambiguous. The lack of X-ray emission suggests gas-stripping and a low impact parameter post-pericenter scenario, or a low-mass subcluster. In a low impact parameter scenario, the subclusters would be on a radial trajectory where the line-of-sight (LOS) velocity component between subclusters and an on-sky separation of \SI{700}{\kpc} between BCGs 1 and 2 would produce a 3D separation of ${\approx}1$ Mpc. A significant LOS velocity at that separation would imply a moderate or low mass ratio merger while the complete stripping of the gas would favor a larger mass ratio between subclusters.

We believe the more likely scenario is a low-mass Subcluster 2. The luminosity density contours in the region are heavily dominated by BCG 2. Using unweighted density, the peak nearly vanishes while peaks at Subclusters 1 and 3 remain. Further, there is a large foreground galaxy that is likely obscuring members of Subcluster 3, falsely diminishing the local density. If the involvement with Subcluster 2 is minimal, the interaction between Subclusters 1 and 3 could still be well-modeled as a binary merger. This ambiguity, along with the elongated morphology and second X-ray peak between BCGs 1 and 4, leads us to not discount this cluster for possible future follow-up. We classify this system as a likely merger and possibly binary. Weak lensing would provide the clearest picture, but additional spectroscopy targeted on Subcluster 2 would help resolve any LOS velocity component and provide a mass estimate for that subcluster. 
%%%%%%%%%%%%%%%%%%%%%%%%%%%%%%%%%%%%%%%%%%%%%%%%%%%%%%%%%%%%%%%%%%%%%%%%%%%%%%%%%%%%%%%%%%%%%%
%%%%%%%%%%%%%%%%%%%%%%%%%%%%%%%%%          RMJ 0109          %%%%%%%%%%%%%%%%%%%%%%%%%%%%%%%%%
%%%%%%%%%%%%%%%%%%%%%%%%%%%%%%%%%%%%%%%%%%%%%%%%%%%%%%%%%%%%%%%%%%%%%%%%%%%%%%%%%%%%%%%%%%%%%%

\subsubsection{RM J010934.2+330301.0}\label{subsubsec:RMJ0109}

\htmlfig{
\begin{table}
    \centering
    \caption{RMJ0109 BCG Information}
    \label{tab:RMJ0109_BCG}
    \tablenotetext{a}{DEIMOS (This work)}
    \tablenotetext{b}{SDSS DR18 \citep{almeida2023eighteenth}}
    \tablenotetext{c}{DESI Legacy Survey DR10 \citep{dey2019overview}}
\begin{tabular*}{\columnwidth}{@{\extracolsep{\fill}}cccccc} \toprule\toprule
    BCG & Probability & Redshift & $r$-mag\tablenotemark{\footnotesize{c}} & RA {[}deg{]} & Dec {[}deg{]} \\ \toprule
    1  & 0.9195 & 0.459\tablenotemark{\footnotesize{a}} & 18.62 & 17.39234 & 33.05028 \\ \midrule
    2  & 0.0750 & 0.328\tablenotemark{\footnotesize{a}} & 18.88 & 17.33332 & 33.02213 \\ \midrule
    3  & 0.0031 & 0.465\tablenotemark{\footnotesize{b}} & 19.64 & 17.36767 & 33.02857 \\ \midrule
    4  & 0.0020 & 0.456\tablenotemark{\footnotesize{a}} & 19.99 & 17.35695 & 33.02684 \\ \midrule
    5  & 0.0004 & 0.465\tablenotemark{\footnotesize{a}} & 19.74 & 17.37253 & 33.03254 \\ \midrule
    A  & ...    & 0.449\tablenotemark{\footnotesize{b}} & 18.56 & 17.38853 & 33.03734 \\ \midrule
    B  & ...    & 0.460\tablenotemark{\footnotesize{b}} & 19.04 & 17.36798 & 33.03092 \\ \bottomrule
\printtablenotes{6}
\end{tabular*}
  \tablenotesreset
\end{table}}

%%%%%%%%%%%%%%%%%%%%%%%%%% RMJ0109: Figure - Optical %%%%%%%%%%%%%%%%%%%%%%%%%%
\begin{figure}[t]
    \centering
    \includegraphics[clip, trim=0.0cm 1.25cm 0.0cm 0.0cm, width=\columnwidth]{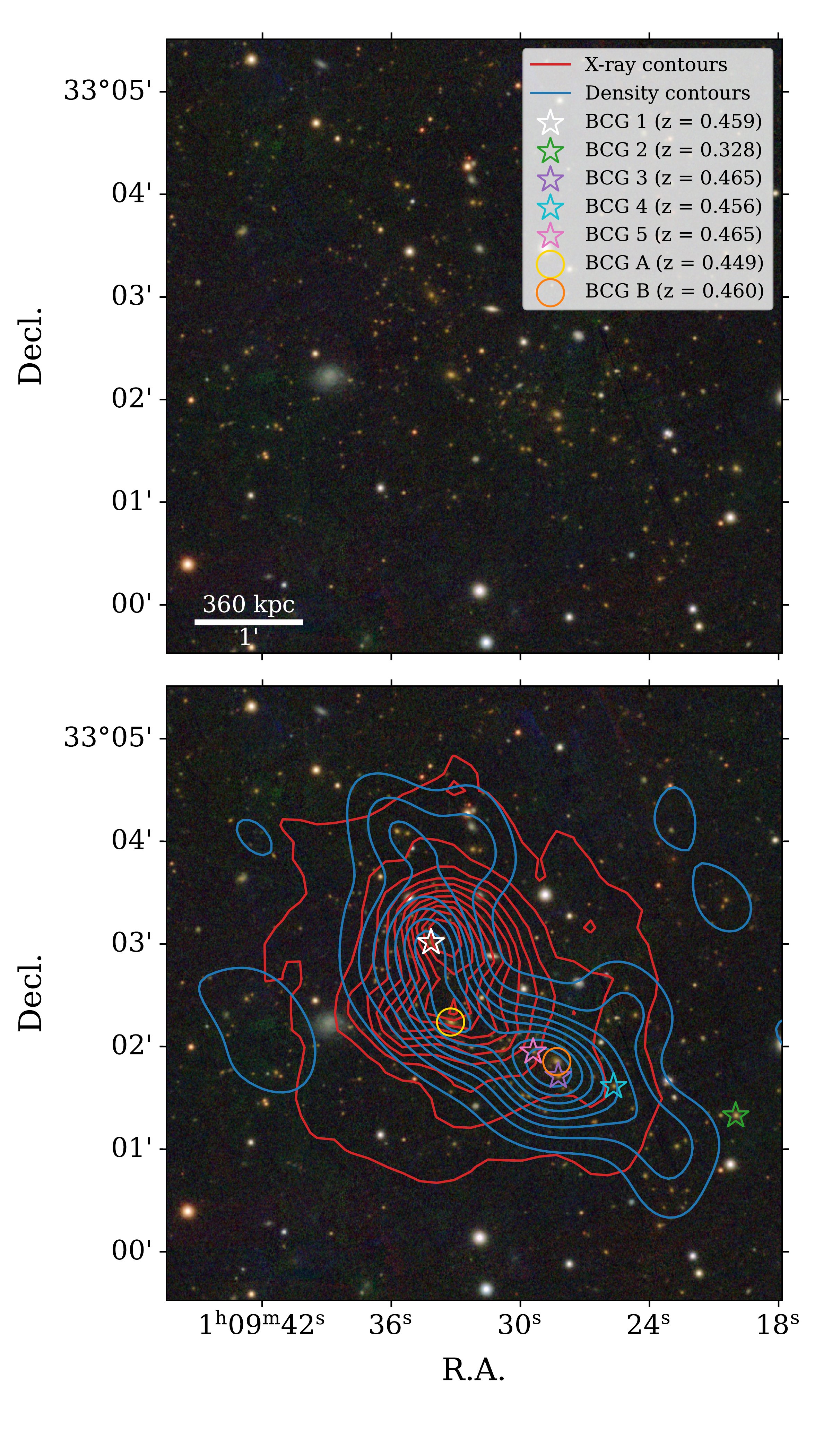}
    \caption{Optical image of RMJ0109. Top panel: 6\arcmin\ by 6\arcmin\ (2.16 by 2.16 Mpc) Pan-STARRS image. Bottom panel: red sequence density contours (blue), X-ray surface brightness contours (red), and BCG candidates identified by redMaPPer and this work.}
    \label{fig:RMJ0109_optical}
\end{figure}

%%%%%%%%%%%%%%%%%%%%%%%%%% RMJ0109: Figure - Redshift Heatmap %%%%%%%%%%%%%%%%%%%%%%%%%%
\pdffig{
\begin{figure}[t]
    \centering
    \includegraphics[clip, trim=0.0cm 0.5cm 0.0cm 0.0cm, width=\columnwidth]{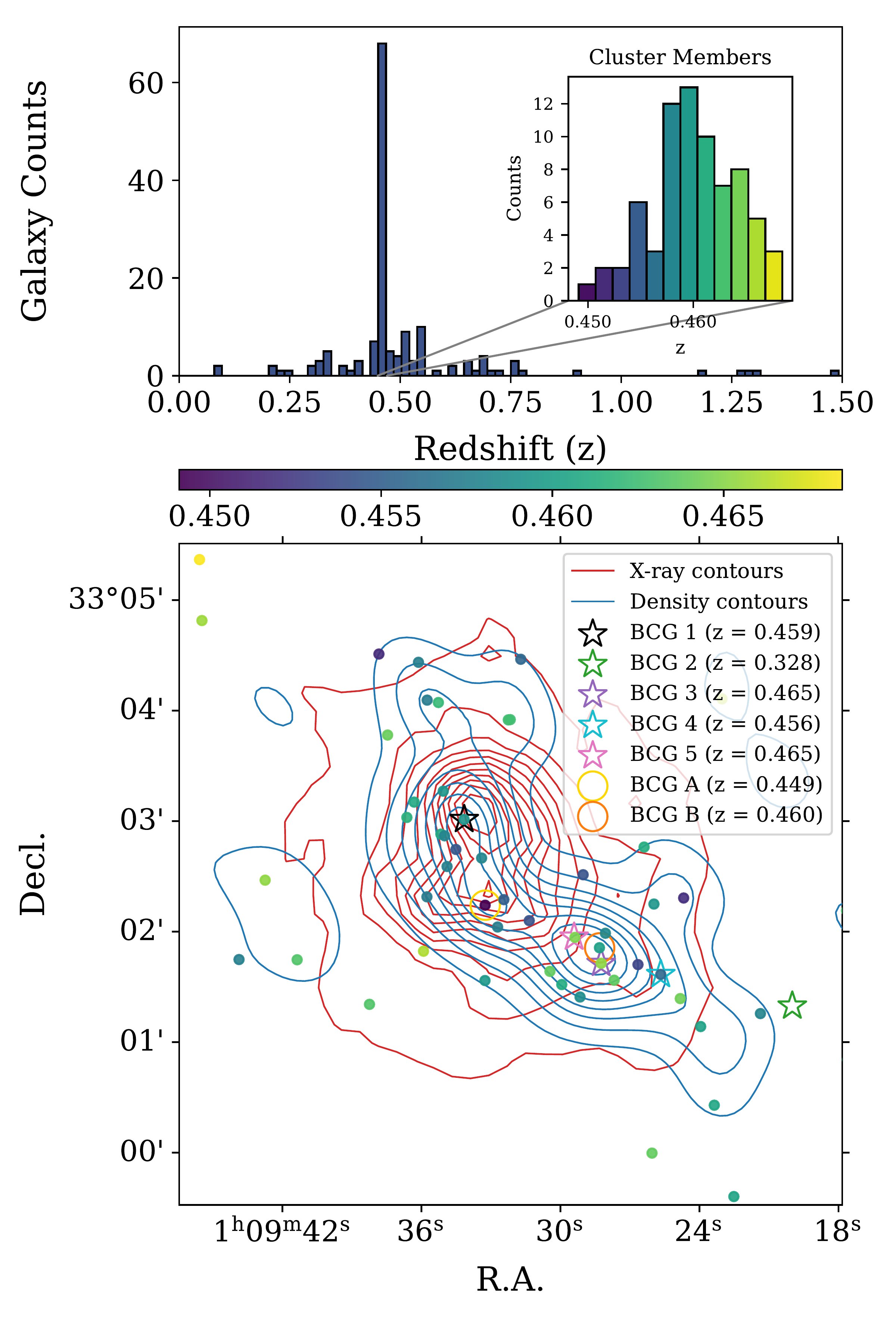}
    \caption{Redshift distribution of RMJ0109. Top panel: histogram of spectroscopic redshifts, combining archival data within 10\arcmin\ of the cluster with new observations from DEIMOS. The inset highlights galaxies in the redshift range $0.449 \leq z \leq 0.470$, which are classified as cluster members. Bottom panel: spatial distribution of those same members.}
    \label{fig:RMJ0109_redshifts}
\end{figure}

%%%%%%%%%%%%%%%%%%%%%%%%%% RMJ0109: Table - BCGs %%%%%%%%%%%%%%%%%%%%%%%%%%

\begin{table}
    \centering
    \caption{RMJ0109 BCG Information}
    \label{tab:RMJ0109_BCG}
    \tablenotetext{a}{DEIMOS (This work)}
    \tablenotetext{b}{SDSS DR18 \citep{almeida2023eighteenth}}
    \tablenotetext{c}{DESI Legacy Survey DR10 \citep{dey2019overview}}
\begin{tabular*}{\columnwidth}{@{\extracolsep{\fill}}cccccc} \toprule\toprule
    BCG & Probability & Redshift & $r$-mag\tablenotemark{\footnotesize{c}} & RA {[}deg{]} & Dec {[}deg{]} \\ \toprule
    1  & 0.9195 & 0.459\tablenotemark{\footnotesize{a}} & 18.62 & 17.39234 & 33.05028 \\ \midrule
    2  & 0.0750 & 0.328\tablenotemark{\footnotesize{a}} & 18.88 & 17.33332 & 33.02213 \\ \midrule
    3  & 0.0031 & 0.465\tablenotemark{\footnotesize{b}} & 19.64 & 17.36767 & 33.02857 \\ \midrule
    4  & 0.0020 & 0.456\tablenotemark{\footnotesize{a}} & 19.99 & 17.35695 & 33.02684 \\ \midrule
    5  & 0.0004 & 0.465\tablenotemark{\footnotesize{a}} & 19.74 & 17.37253 & 33.03254 \\ \midrule
    A  & ...    & 0.449\tablenotemark{\footnotesize{b}} & 18.56 & 17.38853 & 33.03734 \\ \midrule
    B  & ...    & 0.460\tablenotemark{\footnotesize{b}} & 19.04 & 17.36798 & 33.03092 \\ \bottomrule
\printtablenotes{6}
\end{tabular*}
  \tablenotesreset
\end{table}

%%%%%%%%%%%%%%%%%%%%%%%%%% RMJ0109: Figure - Subclusters %%%%%%%%%%%%%%%%%%%%%%%%%%
\begin{figure*}
    \centering
    \includegraphics[clip, trim=0.0cm 0.75cm 0.5cm 0.5cm, width=\textwidth]{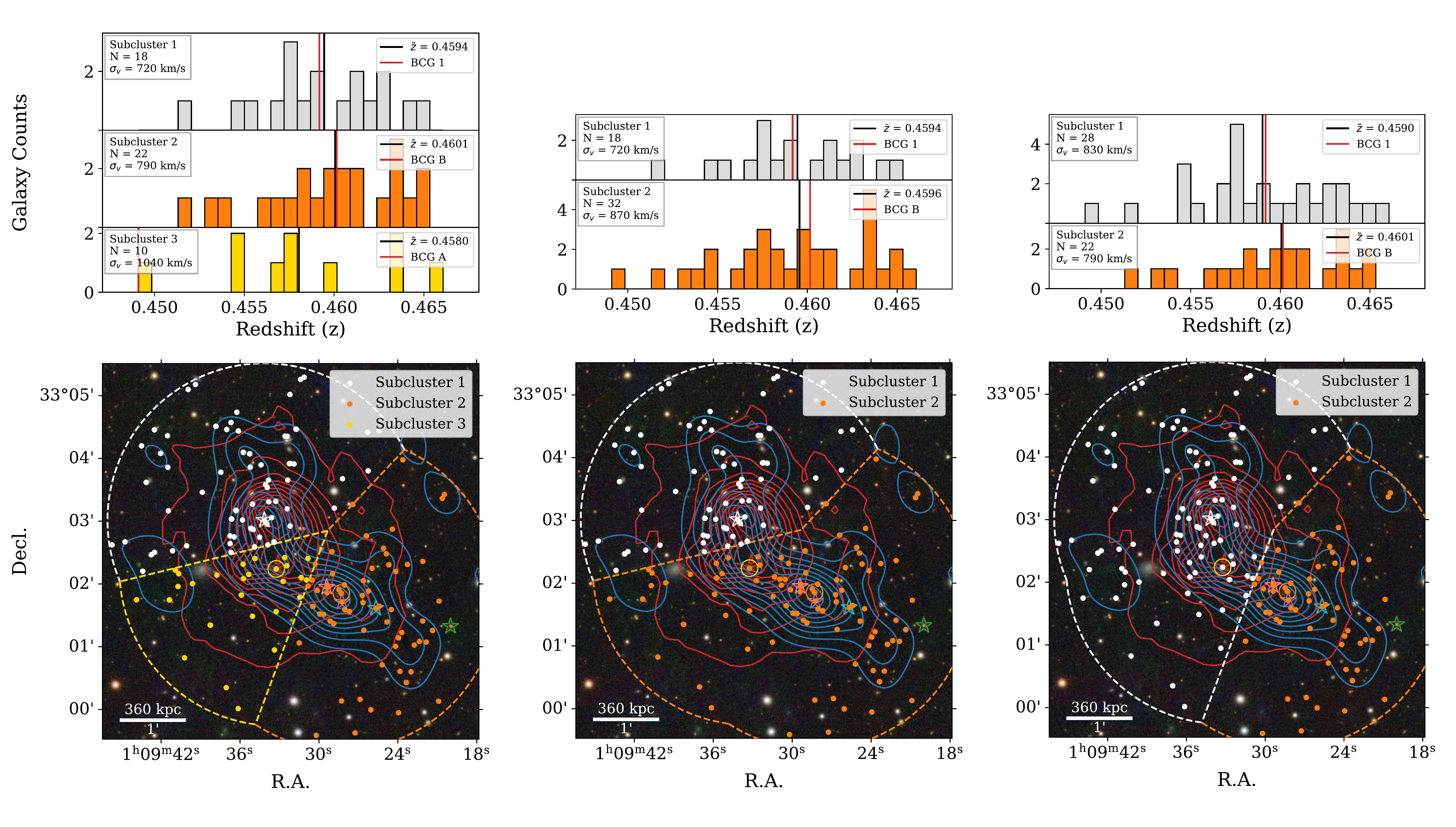}
    \caption{Subclustering analysis of RMJ0109. Bottom panels: spatial distribution of subcluster members, both spectroscopic and photometric, identified within regions defined by the bisector between BCG pairs and a radial distance of 2.5\arcmin\ from the BCG in each region. Top panels: redshift distribution of the spectroscopic members. The two right panels consider the subcluster associated with BCG A to be part of Subcluster 1 or 2.}
    \label{fig:RMJ0109_sub_1}
\end{figure*}
}

%%%%%%%%%%%%%%%%%%%%%%%%%%%%%%%%%        RMJ 0109 Text       %%%%%%%%%%%%%%%%%%%%%%%%%%%%%%%%%
Our second cluster, RMJ0109, illustrates some issues with redMaPPer BCG selection that are relevant to a few of our other targets as well. RedMaPPer identifies two primary BCG candidates with probabilities of $0.9195$ and $0.0750$ (\cref{tab:RMJ0109_BCG}). However, a visual inspection suggests that the brightest galaxy consistent with the cluster red sequence is the one near the center of \cref{fig:RMJ0109_optical}, just south of the small X-ray contour. This galaxy has an $r$-band magnitude of 18.56 as compared to the 18.62 for BCG 1 and SDSS spectroscopy identifies the galaxy at $z=0.44910$, making it the lowest redshift galaxy we define as in the cluster. We have marked this galaxy as an additional possible BCG, labeled BCG A, and adopted a convention of using letters to signify BCGs not classified by redMaPPer. It is not immediately clear to us why redMaPPer does not include BCG A in its list of top five BCG candidates. The redMaPPer algorithm incorporates local galaxy density when identifying potential BCG candidates. Although BCG A lies near the cluster center and appears on an extension of the primary peak in the luminosity-weighted galaxy density map, the unweighted galaxy number density in its immediate vicinity is relatively low, suggesting a possible explanation. 

Another bright galaxy ($r=19.04$), labeled BCG B in \cref{fig:RMJ0109_optical}, is also not listed among the BCG candidates. However, in this case, the reason may be that it is blended with a bluer foreground galaxy. The SDSS spectrum\footnote{Available from \url{https://skyserver.sdss.org/dr19/VisualTools/explore/summary?id=1237680316595175986}} shows a blend of early and late type spectra at different redshifts; the SDSS pipeline correctly identified only the strong emission lines at $z=0.17895$. We model the spectrum as the sum of this SDSS model (with freely varying amplitude) plus an early-type model with freely varying amplitude, redshift, and dust reddening. We find the redshift of the early type galaxy to be $0.46015 \pm 0.00008$. The composite model fits the data substantially better, with $\Delta\chi^2=166$ for three additional parameters. We therefore consider this galaxy to be in the cluster. Note that this new redshift does not appear in \cref{tab:redshift_table} because that table pertains only to Keck/DEIMOS redshifts.

Finally, the galaxy identified by redMaPPer as BCG 2 is actually a foreground galaxy, spectroscopically confirmed by SDSS at $z=0.32846$. Because redMaPPer does not use spectroscopic information, the color of the galaxy is similar enough to be misidentified. Of note, this galaxy was not classified as a member of our red sequence fit, being bluer than the fit by 0.216 mag (accepted members lie within $\pm0.2$ mag).

%%%%%%%%%%%%%%%%%%%%%%%%%% RMJ0109: Table - Subclusters %%%%%%%%%%%%%%%%%%%%%%%%%%
\pdffig{
\begin{table}
    \centering
    \caption{RMJ0109 Subcluster Properties}
    \label{tab:RMJ0109_subclusters}
\begin{tabular*}{\columnwidth}{@{\extracolsep{\fill}}cccccc} \toprule\toprule
    Subcluster & $N$ & BCG & BCG $z$ & Mean $z$ & $\sigma_v$ [km s$^{-1}$] \\ \toprule
All & 71 & ... & ... & 0.4601 & \phantom{0}870 \\
1 & 18 & 1 & 0.459 & 0.4594 & \phantom{0}720  \\
2 & 22 & B & 0.460 & 0.4601 & \phantom{0}790  \\
3 & 10 & A & 0.449 & 0.4580 & 1040  \\
\bottomrule
\end{tabular*}
\end{table}}

To summarize the multiple BCG issues in reverse order: a foreground galaxy had similar enough color to be assigned as BCG 2; a galaxy with brighter apparent magnitude than BCG 3 was not identified as a BCG candidate, possibly due to blending with a foreground galaxy; and the brightest galaxy was overlooked with no definitive explanation. The blending issue is unique to this cluster, but the other two issues are relevant to some of our other targets. This suggests that a more robust selection approach could be developed based on additional inputs beyond redMaPPer's BCG ranking (for example, spectroscopic information or redMaPPer's complete list of possible member galaxies).

Turning our attention to BCGs 3, 4, and 5, we see in \cref{fig:RMJ0109_optical} that they likely reside in a subcluster in the southwest with BCG B. Galaxy luminosity density contours show a peak between BCGs 3 and 5, and the X-ray emission is extended to the southwest. All three of these BCGs have comparatively faint magnitudes (19.64--19.99) and redMaPPer probabilities less than 0.3\%. 

The distribution of galaxies in redshift space, shown in the top panel of \cref{fig:RMJ0109_redshifts}, is best-fit by a one-component Gaussian (BIC of -567 compared to -555 for a two-component model) in the cluster region. Across the full redshift range, a GMM identifies a background subcluster at $z\sim 0.52$. However, these background galaxies are located in two primary groups northeast and southwest of the cluster and do not spatially overlap with the merging system or contribute to the observed X-ray morphology. The spatial distribution of cluster members, shown in the bottom panel of \cref{fig:RMJ0109_redshifts}, does not show any clear gradient in redshift or separation between subclusters.

\htmlfig{
\begin{figure}[t]
    \centering
    \includegraphics[clip, trim=0.0cm 0.5cm 0.0cm 0.0cm, width=\columnwidth]{Figures/RMJ_0109_redshifts_hist_hmap.jpg}
    \caption{Redshift distribution of RMJ0109. Top panel: histogram of spectroscopic redshifts, combining archival data within 10\arcmin\ of the cluster with new observations from DEIMOS. The inset highlights galaxies in the redshift range $0.449 \leq z \leq 0.470$, which are classified as cluster members. Bottom panel: spatial distribution of those same members.}
    \label{fig:RMJ0109_redshifts}
\end{figure}}

Given the low redshift of BCG A ($\Delta v_{\mathrm{LOS}}=\SI{2200}{\kms}$ as compared with BCG 1), we examined the redshift distribution in three distinct regions corresponding with BCGs 1, A, and B (including BCGs 3--5 collectively) (\cref{fig:RMJ0109_sub_1}). Of these three potential subclusters, Subcluster 3 has the largest velocity dispersion and fewest members (\cref{tab:RMJ0109_subclusters}). Further, we find no other low redshift members in the vicinity of BCG A and thus explore the possibility of BCG A being part of either a southern subcluster with BCG B or the northern subcluster with BCG 1, shown in the center and right panels of \cref{fig:RMJ0109_sub_1} respectively. In either scenario, both subclusters have similar mean redshifts with $\Delta v_{\mathrm{LOS}}<\SI{230}{\kms}$ between subclusters and velocity dispersions of ${\sim}\SI{800}{\kms}$.

%%%%%%%%%%%%%%%%%%%%%%%%%% RMJ0109: Figure - Subclusters %%%%%%%%%%%%%%%%%%%%%%%%%%
\htmlfig{
\begin{figure*}
    \centering
    \includegraphics[clip, trim=0.0cm 0.75cm 0.5cm 0.5cm, width=\textwidth]{Figures/RMJ_0109_all.jpg}
    \caption{Subclustering analysis of RMJ0109. Bottom panels: spatial distribution of subcluster members, both spectroscopic and photometric, identified within regions defined by the bisector between BCG pairs and a radial distance of 2.5\arcmin\ from the BCG in each region. Top panels: redshift distribution of the spectroscopic members. The two right panels consider the subcluster associated with BCG A to be part of Subcluster 1 or 2.}
    \label{fig:RMJ0109_sub_1}
\end{figure*}}

%%%%%%%%%%%%%%%%%%%%%%%%%% RMJ0109: Table - Subclusters %%%%%%%%%%%%%%%%%%%%%%%%%%
\htmlfig{
\begin{table}
    \centering
    \caption{RMJ0109 Subcluster Properties}
    \label{tab:RMJ0109_subclusters}
\begin{tabular*}{\columnwidth}{@{\extracolsep{\fill}}cccccc} \toprule\toprule
    Subcluster & $N$ & BCG & BCG $z$ & Mean $z$ & $\sigma_v$ [km s$^{-1}$] \\ \toprule
All & 71 & ... & ... & 0.4601 & \phantom{0}870 \\
1 & 18 & 1 & 0.459 & 0.4594 & \phantom{0}720  \\
2 & 22 & B & 0.460 & 0.4601 & \phantom{0}790  \\
3 & 10 & A & 0.449 & 0.4580 & 1040  \\
\bottomrule
\end{tabular*}
\end{table}}

RMJ0109 is the second richest cluster in our sample, with a richness of 168, given in \cref{tab:clusters}. Given this richness, the temperature and luminosity of $6.38$ keV and $\SI{4.19e44}{\ergs}$, respectively, are not immediately indicative of a post-pericenter merger. However, the X-ray morphology is clearly disturbed and double-peaked and, despite the BCG confusion, this is a reasonable candidate for a bimodal, or more complicated, merger. The BCG candidates are strung out along a NE-SW axis, while the XSB is highly concentrated at the NE end of that string. Furthermore, the galaxy luminosity distribution shows two peaks. The merger action may be primarily between groups headed by BCG A and BCG 1, which span the region of highest XSB. In this scenario, BCGs 3--5 and B perhaps form an infalling group or highlight a filament but are not necessarily part of the current merger action. On the other hand, BCGs 1 and B are separated by ${>}0.95\arcmin$ and there could be an argument for a merger between groups headed by BCG 1 (perhaps including BCG A) and BCG B (including BCGs 3--5).

It should be noted that, had BCG A been identified by redMaPPer as one of the top two BCG candidates, this cluster would have been excluded from our sample based on a projected separation of ${<}0.95\arcmin$ between BCGs 1 and A. Furthermore, spectroscopy shows that BCG 2 is not a cluster member, rendering the separation between BCGs 1 and 2 physically irrelevant. While the inclusion of this system does not strictly follow our original intent of selecting clusters whose two most likely BCGs trace distinct subclusters, it highlights that ambiguity in the primary BCG identification alone can be indicative of merging activity.  

We classify this as a merger, possibly binary, and believe this cluster would be a good candidate for lensing follow-up to determine the morphology of the mass distribution. This would provide a firmer foundation for identifying the subclusters involved, a key issue not resolved by the BCG information. This would also be a good candidate for further spectroscopic study. Prior to this work, there were only 22 archival redshifts within 10\arcmin\ of the cluster, the least of any cluster in our sample. We increase that number to 145, with 71 spectroscopic members including 50 within 2.5\arcmin\ of BCGs 1, A, or B, but a more complete spectroscopic picture could help to disentangle the underlying structure.

%%%%%%%%%%%%%%%%%%%%%%%%%%%%%%%%%%%%%%%%%%%%%%%%%%%%%%%%%%%%%%%%%%%%%%%%%%%%%%%%%%%%%%%%%%%%%%
%%%%%%%%%%%%%%%%%%%%%%%%%%%%%%%%%          RMJ 0219         %%%%%%%%%%%%%%%%%%%%%%%%%%%%%%%%%
%%%%%%%%%%%%%%%%%%%%%%%%%%%%%%%%%%%%%%%%%%%%%%%%%%%%%%%%%%%%%%%%%%%%%%%%%%%%%%%%%%%%%%%%%%%%%%
\subsubsection{RM J021952.2+012952.2}\label{subsubsec:RMJ0219}

%%%%%%%%%%%%%%%%%%%%%%%%%% RMJ0219: Table - BCGs %%%%%%%%%%%%%%%%%%%%%%%%%%
\htmlfig{
\begin{table}[b]
    \centering
    \caption{RMJ0219 BCG Information}
    \label{tab:RMJ0219_BCG}
    \tablenotetext{a}{DESI DR1 \citep{abdul2025data}}
    \tablenotetext{b}{SDSS DR18 \citep{almeida2023eighteenth}}
    \tablenotetext{c}{DEIMOS (This work)}
    \tablenotetext{d}{DESI Legacy Survey DR10 \citep{dey2019overview}}
\begin{tabular*}{\columnwidth}{@{\extracolsep{\fill}}cccccc} \toprule\toprule
    BCG & Probability & Redshift & $r$-mag\tablenotemark{\footnotesize{d}} & RA {[}deg{]} & Dec {[}deg{]} \\ \toprule
    1  & 0.8577 & 0.365\tablenotemark{\footnotesize{a}} & 17.80 & 34.96731 & 1.49783 \\ \midrule
    2  & 0.0871 & 0.368\tablenotemark{\footnotesize{b}} & 18.01 & 34.98425 & 1.51626 \\ \midrule
    3  & 0.0548 & 0.363\tablenotemark{\footnotesize{c}} & 18.32 & 34.97850 & 1.50815 \\ \midrule
    4  & 0.0004 & 0.372\tablenotemark{\footnotesize{c}} & 18.94 & 34.98640 & 1.50205 \\ \midrule
    5  & 0.0001 & 0.370\tablenotemark{\footnotesize{b}} & 19.20 & 34.98922 & 1.50065 \\ \bottomrule
\printtablenotes{6}
\end{tabular*}
  \tablenotesreset
\end{table}}

%%%%%%%%%%%%%%%%%%%%%%%%%% RMJ0219: Figure - Optical %%%%%%%%%%%%%%%%%%%%%%%%%%
\begin{figure}[t]
    \centering
    \includegraphics[clip, trim=0.0cm 1.35cm 0.0cm 0.0cm, width=\columnwidth]{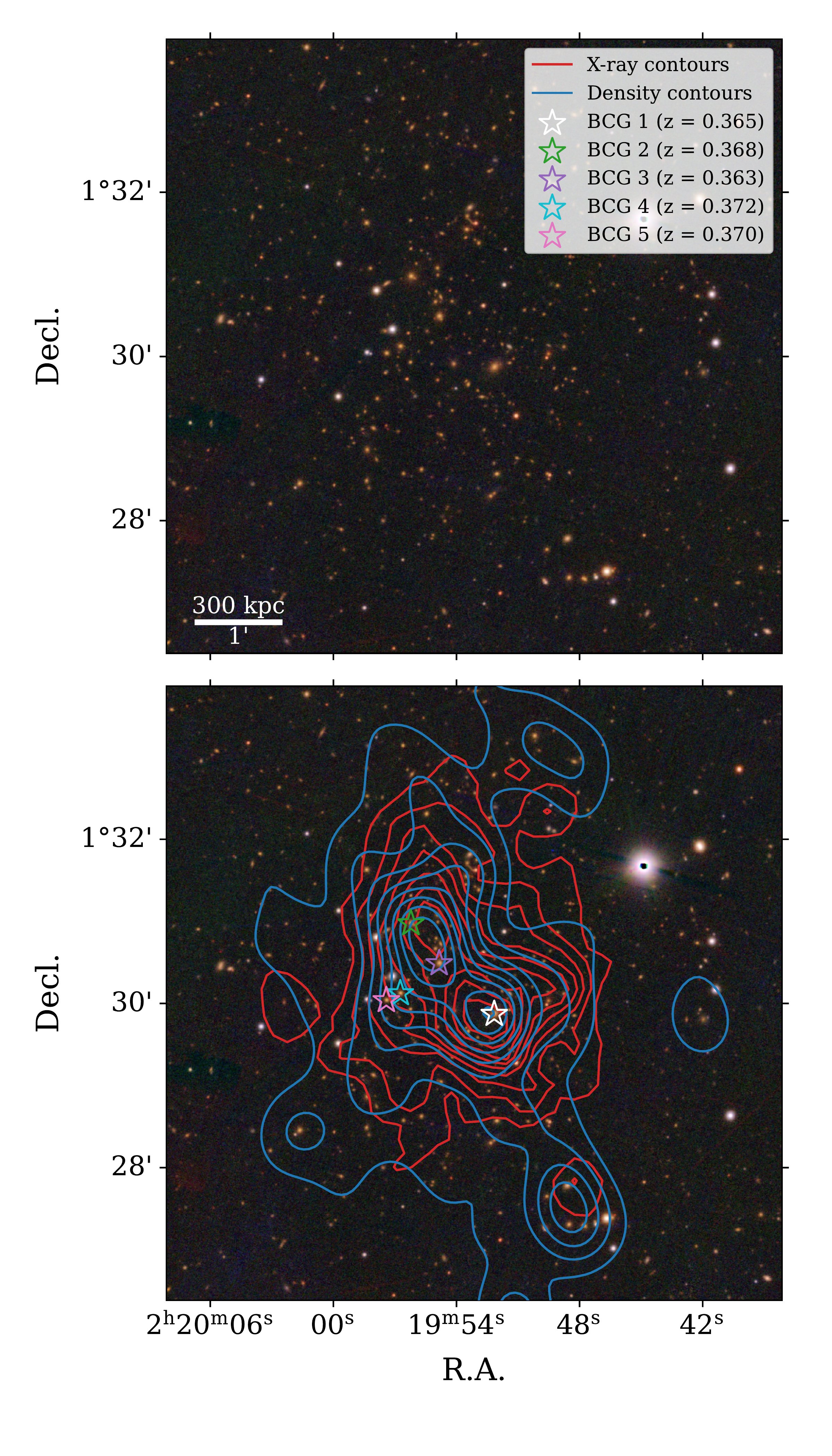}
    \caption{Optical image of RMJ0219. Top panel: 7.5\arcmin\ by 7.5\arcmin\ (2.25 by 2.25 Mpc) Pan-STARRS image. Bottom panel: red sequence density contours (blue), X-ray surface brightness contours (red), and BCG candidates identified by redMaPPer.}
    \label{fig:RMJ0219_optical}
\end{figure}

%%%%%%%%%%%%%%%%%%%%%%%%%% RMJ0219: Figure - Redshift Heatmap %%%%%%%%%%%%%%%%%%%%%%%%%%
\pdffig{
\begin{figure}[ht]
    \centering
    \includegraphics[clip, trim=0.0cm 0.88cm 0.0cm 0.0cm, width=0.985\columnwidth]{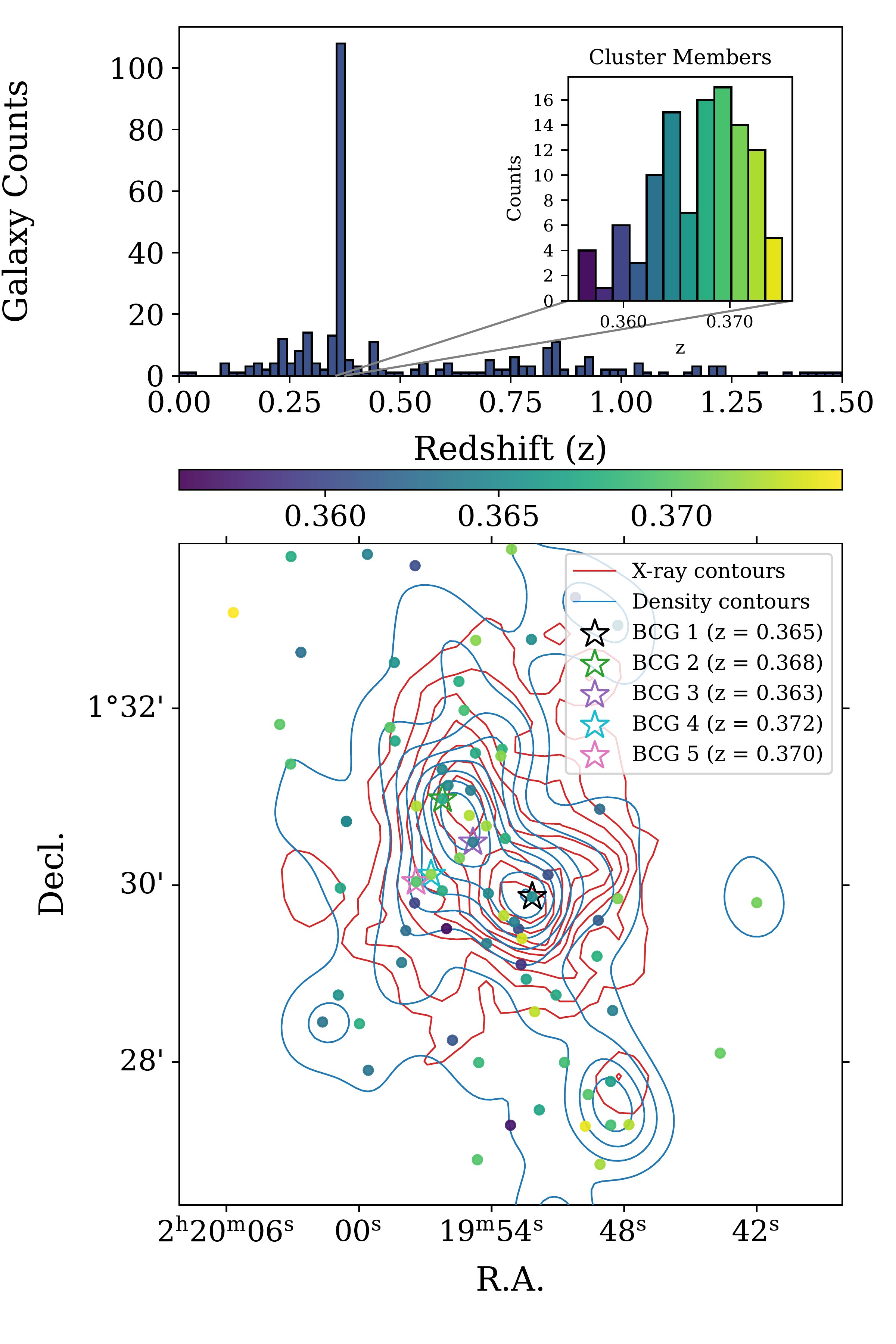}
    \caption{Redshift distribution of RMJ0219. Top panel: histogram of spectroscopic redshifts, combining archival data within 10\arcmin\ of the cluster with new observations from DEIMOS. The inset highlights galaxies in the redshift range $0.355 \leq z \leq 0.375$, which are classified as cluster members. Bottom panel: spatial distribution of those same members.}
    \label{fig:RMJ0219_redshifts}
\end{figure}}

%%%%%%%%%%%%%%%%%%%%%%%%%%%%%%%%%        RMJ 0219 Text       %%%%%%%%%%%%%%%%%%%%%%%%%%%%%%%%%
RMJ0219 displays a clear spatial separation between two subclusters with BCG 1 southwest of BCGs 2 and 3 (\cref{fig:RMJ0219_optical}). BCG 1 has a probability of $0.8577$, while BCGs 2 and 3 have lower, but non-negligible, probabilities of $0.0871$ and $0.0548$, respectively (\cref{tab:RMJ0219_BCG}). BCGs 4 and 5 are both on the eastern edge of the cluster. They are not associated with any overdensities of galaxies or X-ray emission, and are unlikely to be representative of any underlying substructure. Both the XSB and galaxy luminosity density are elongated along a NE--SW axis and double-peaked, with one peak at BCG 1 and the other between BCGs 2 and 3. There is a small group in the southwest with an overdensity in both galaxies and X-ray emission. This group may be underrepresented, with foreground galaxies obscuring some members. However, the overall contribution to the cluster dynamics does not appear significant.

\pdffig{
\begin{table}[b]
    \centering
    \caption{RMJ0219 BCG Information}
    \label{tab:RMJ0219_BCG}
    \tablenotetext{a}{DESI DR1 \citep{abdul2025data}}
    \tablenotetext{b}{SDSS DR18 \citep{almeida2023eighteenth}}
    \tablenotetext{c}{DEIMOS (This work)}
    \tablenotetext{d}{DESI Legacy Survey DR10 \citep{dey2019overview}}
\begin{tabular*}{\columnwidth}{@{\extracolsep{\fill}}cccccc} \toprule\toprule
    BCG & Probability & Redshift & $r$-mag\tablenotemark{\footnotesize{d}} & RA {[}deg{]} & Dec {[}deg{]} \\ \toprule
    1  & 0.8577 & 0.365\tablenotemark{\footnotesize{a}} & 17.80 & 34.96731 & 1.49783 \\ \midrule
    2  & 0.0871 & 0.368\tablenotemark{\footnotesize{b}} & 18.01 & 34.98425 & 1.51626 \\ \midrule
    3  & 0.0548 & 0.363\tablenotemark{\footnotesize{c}} & 18.32 & 34.97850 & 1.50815 \\ \midrule
    4  & 0.0004 & 0.372\tablenotemark{\footnotesize{c}} & 18.94 & 34.98640 & 1.50205 \\ \midrule
    5  & 0.0001 & 0.370\tablenotemark{\footnotesize{b}} & 19.20 & 34.98922 & 1.50065 \\ \bottomrule
\printtablenotes{6}
\end{tabular*}
  \tablenotesreset
\end{table}

%%%%%%%%%%%%%%%%%%%%%%%%%% RMJ0219: Figure - Subclusters %%%%%%%%%%%%%%%%%%%%%%%%%%
\begin{figure}[t]
    \centering
    \includegraphics[clip, trim=0.0cm 0.75cm 0.0cm 0.0cm, width=\columnwidth]{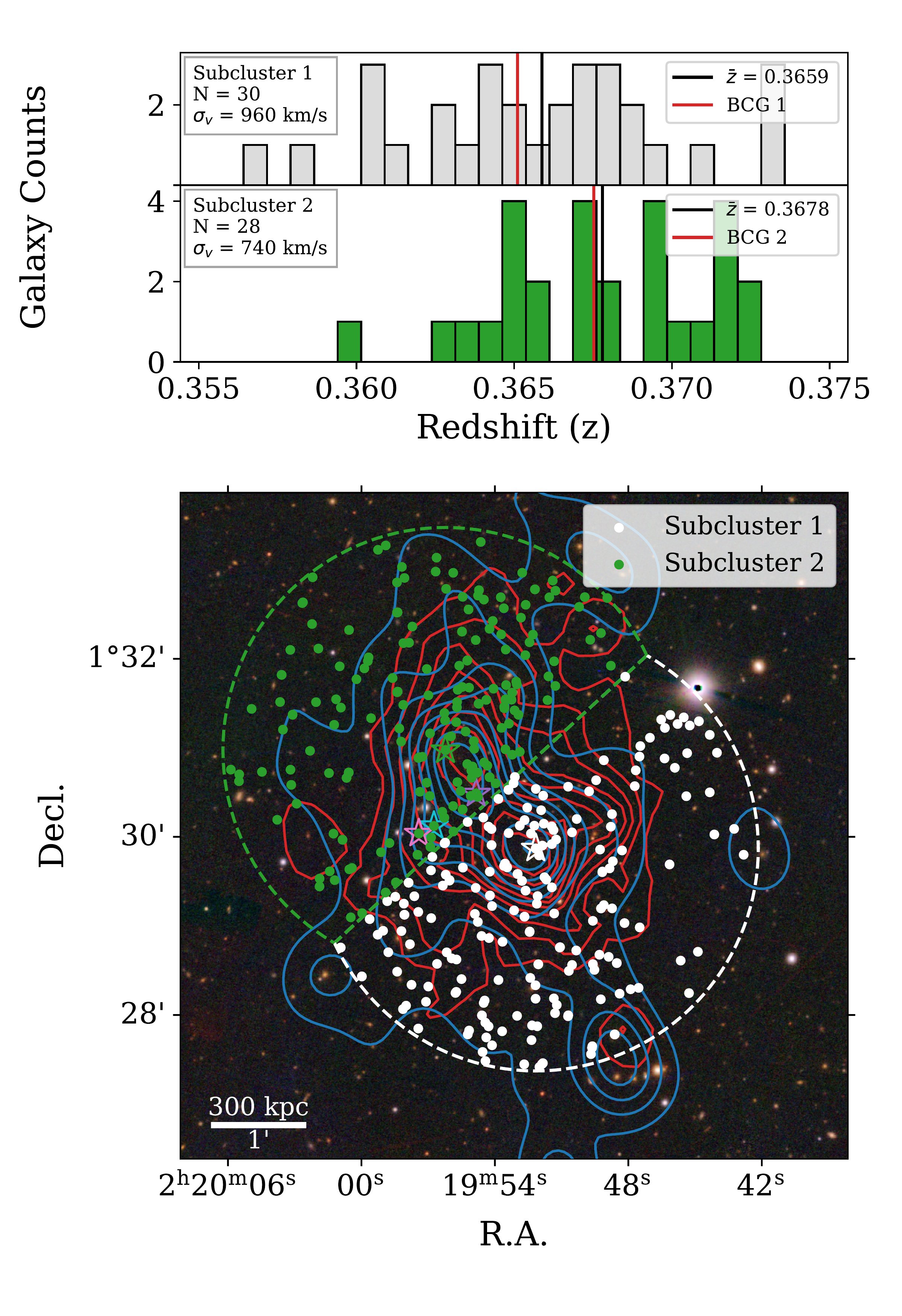}
    \caption{Subclustering analysis of RMJ0219. Bottom panel: spatial distribution of subcluster members, both spectroscopic and photometric, identified within regions defined by the bisector between BCG pairs and a radial distance of 2.5\arcmin\ from the BCG in each region. Top panel: redshift distribution of the spectroscopic members of those regions.}
    \label{fig:RMJ0219_sub_1}
\end{figure}

%%%%%%%%%%%%%%%%%%%%%%%%%% RMJ0219: Table - Subclusters %%%%%%%%%%%%%%%%%%%%%%%%%%
\begin{table}
    \centering
    \caption{RMJ0219 Subcluster Properties}
    \label{tab:RMJ0219_subclusters}
\begin{tabular*}{\columnwidth}{@{\extracolsep{\fill}}cccccc} \toprule\toprule
    Subcluster & $N$ & BCG & BCG $z$ & Mean $z$ & $\sigma_v$ [km s$^{-1}$] \\ \toprule
All & 110 & ... & ... & 0.3674 & 980 \\
1 & 30 & 1 & 0.365 & 0.3659 & 960  \\
2 & 28 & 2 & 0.368 & 0.3678 & 740  \\
\bottomrule
\end{tabular*}
\end{table}}

The redshift distribution, shown in \cref{fig:RMJ0219_redshifts}, shows no foreground or background structures. The inset histogram shows two apparent peaks within the cluster redshift range of  $0.355 \leq z \leq 0.375$, but this is likely a binning artifact and a single Gaussian fit is preferred. A one-component fit yields a lower BIC ($-873$ versus $-867$ for a two-component fit), and both KS ($D=0.068$, $p=0.63$) with AD ($A^2=0.63$, $p=0.10$) statistics are consistent with a Gaussian distribution. In the bottom panel of the figure, no gradient in redshifts is evident.

%%%%%%%%%%%%%%%%%%%%%%%%%% RMJ0219: Figure - Redshift Heatmap %%%%%%%%%%%%%%%%%%%%%%%%%%
\htmlfig{
\begin{figure}[ht]
    \centering
    \includegraphics[clip, trim=0.0cm 0.88cm 0.0cm 0.0cm, width=0.985\columnwidth]{Figures/RMJ_0219_redshifts_hist_hmap.jpg}
    \caption{Redshift distribution of RMJ0219. Top panel: histogram of spectroscopic redshifts, combining archival data within 10\arcmin\ of the cluster with new observations from DEIMOS. The inset highlights galaxies in the redshift range $0.355 \leq z \leq 0.375$, which are classified as cluster members. Bottom panel: spatial distribution of those same members.}
    \label{fig:RMJ0219_redshifts}
\end{figure}}

%%%%%%%%%%%%%%%%%%%%%%%%%% RMJ0219: Figure - Gauss %%%%%%%%%%%%%%%%%%%%%%%%%%
\pdffig{
\begin{figure}[t]
    \centering
    \includegraphics[clip, trim=0.0cm 0.25cm 0.0cm 0.0cm, width=\columnwidth]{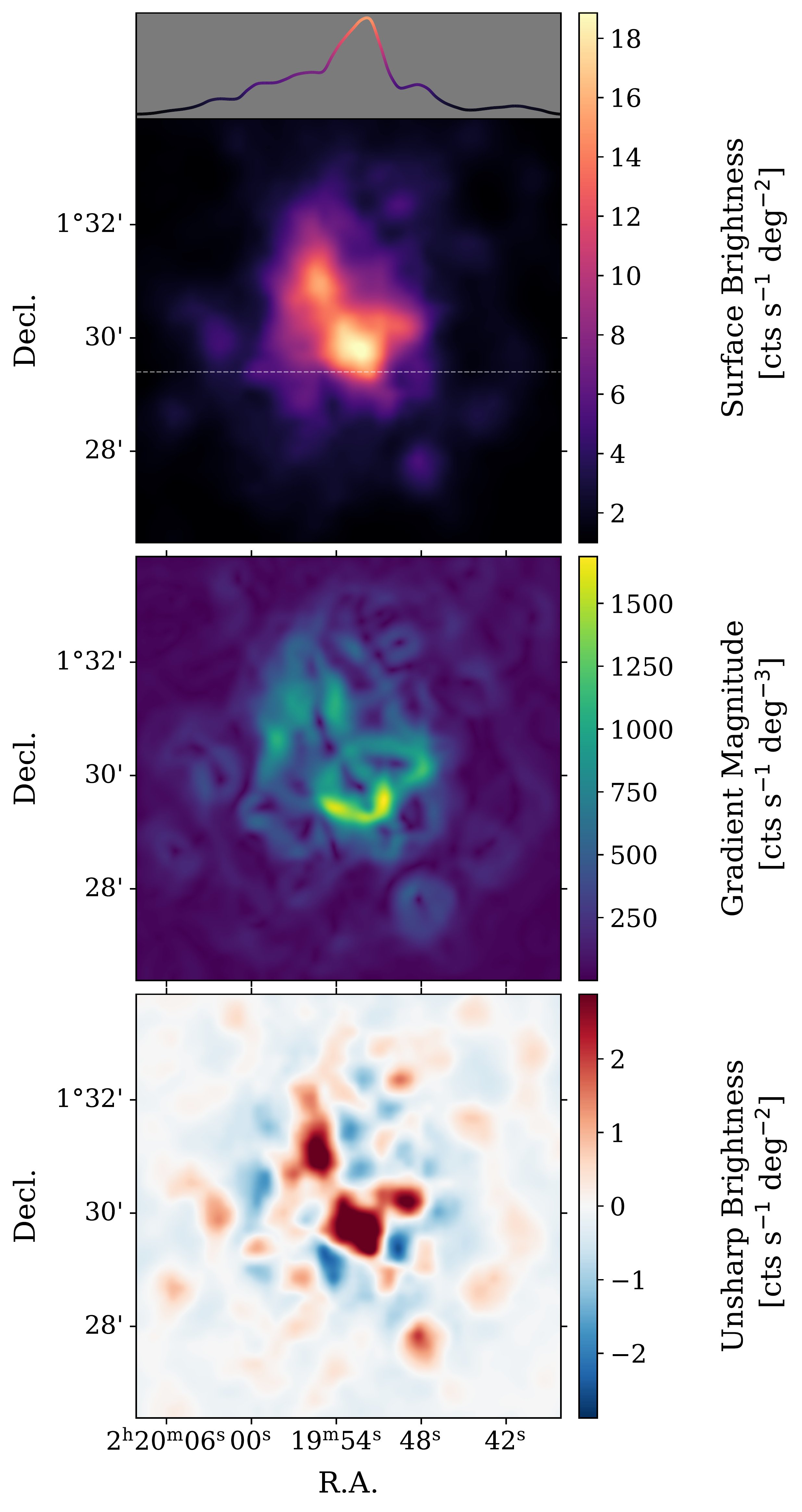}
    \caption{RMJ0219 X-ray features. Top panel: X-ray surface brightness smoothed with a 10\arcsec\ kernel and reprojected on an $800\times800$ pixel grid with the extracted 1D profile. The dashed line shows where the profile was extracted across the sharp feature. Center panel: gradient magnitude of the brightness. Bottom panel: residual features after unsharp masking, where an image smoothed with a kernel of 50\arcsec\ is subtracted from the image smoothed at 10\arcsec.}
    \label{fig:RMJ0219_gauss}
\end{figure}}

Dividing the cluster into regions defined by the bisector between BCGs 1 and 2 and a 2.5\arcmin\ radius from each, shown in \cref{fig:RMJ0219_sub_1}, also shows no evidence for bimodality in redshift space. The redshift distributions of both subclusters are consistent with Gaussianity, with a line-of-sight offset of $\Delta V_{\mathrm{LOS}} = \SI{420}{\kms}$ between subcluster means. Velocity dispersions for the two subclusters are similar at $960$ and $\SI{740}{\kms}$ (\cref{tab:RMJ0219_subclusters}). Overall, this system appears to be a clean, binary system near the plane of the sky.

\htmlfig{
\begin{figure}[t]
    \centering
    \includegraphics[clip, trim=0.0cm 0.75cm 0.0cm 0.0cm, width=\columnwidth]{Figures/RMJ_0219_subcluster_histograms_1_2.jpg}
    \caption{Subclustering analysis of RMJ0219. Bottom panel: spatial distribution of subcluster members, both spectroscopic and photometric, identified within regions defined by the bisector between BCG pairs and a radial distance of 2.5\arcmin\ from the BCG in each region. Top panel: redshift distribution of the spectroscopic members of those regions.}
    \label{fig:RMJ0219_sub_1}
\end{figure}

%%%%%%%%%%%%%%%%%%%%%%%%%% RMJ0219: Table - Subclusters %%%%%%%%%%%%%%%%%%%%%%%%%%
\begin{table}
    \centering
    \caption{RMJ0219 Subcluster Properties}
    \label{tab:RMJ0219_subclusters}
\begin{tabular*}{\columnwidth}{@{\extracolsep{\fill}}cccccc} \toprule\toprule
    Subcluster & $N$ & BCG & BCG $z$ & Mean $z$ & $\sigma_v$ [km s$^{-1}$] \\ \toprule
All & 110 & ... & ... & 0.3674 & 980 \\
1 & 30 & 1 & 0.365 & 0.3659 & 960  \\
2 & 28 & 2 & 0.368 & 0.3678 & 740  \\
\bottomrule
\end{tabular*}
\end{table}}

Assuming BCG 2 to be the true BCG of the northern subcluster (with magnitude 18.01 versus 18.32 for BCG 3), the associated X-ray peak is lagging slightly (${\sim}\SI{100}{\kpc}$) to the south. The X-ray surface brightness also exhibits a steep gradient and bullet-like morphology in the southwest, near BCG 1. To explore this structure further, X-ray images were processed following the procedure in \S\ref{subsec:XMM}. A one-dimensional profile extracted across the leading edge, shown in the top panel of \cref{fig:RMJ0219_gauss}, reveals a sharp change in surface brightness on both the eastern and western edges of the cluster. This morphology is further highlighted with a map of the Gaussian gradient magnitude, shown in the center panel of \cref{fig:RMJ0219_gauss}. An unsharp-masked image, shown in the bottom panel of \cref{fig:RMJ0219_gauss}, was created by subtracting an image smoothed with a 50\arcsec\ kernel from one smoothed at 10\arcsec, in which a bright edge tracing the leading boundary of the cluster is clearly visible.

\htmlfig{
\begin{figure}[t]
    \centering
    \includegraphics[clip, trim=0.0cm 0.25cm 0.0cm 0.0cm, width=\columnwidth]{Figures/RMJ0219_gaussian_gradient.jpg}
    \caption{RMJ0219 X-ray features. Top panel: X-ray surface brightness smoothed with a 10\arcsec\ kernel and reprojected on an $800\times800$ pixel grid with the extracted 1D profile. The dashed line shows where the profile was extracted across the sharp feature. Center panel: gradient magnitude of the brightness. Bottom panel: residual features after unsharp masking, where an image smoothed with a kernel of 50\arcsec\ is subtracted from the image smoothed at 10\arcsec.}
    \label{fig:RMJ0219_gauss}
\end{figure}}

From the X-ray morphology and galaxy distribution, we classify this as a binary merger. Additionally, this appears to be a post-pericenter merger occurring near the plane of the sky. However, the X-ray temperature ($\SI{4.50}{\keV}$) and luminosity ($\SI{1.93e44}{\ergs}$) are not elevated, casting doubt on a recent pericenter passage. The sharp surface-brightness edge in the south is indicative of merger activity and is morphologically consistent with a cold-front-like structure, which could be confirmed with spatially resolved temperature measurements. While evidence supports a post-pericenter merger, a returning scenario, with the galaxies already having fallen back through the ICM, is not ruled out. Even given the uncertainty in the merger stage, the binary nature of this system still makes it a valuable target for future study. Follow-up with lensing and radio observations could be useful in determining the subcluster positions and constraining the merger stage.

%%%%%%%%%%%%%%%%%%%%%%%%%%%%%%%%%%%%%%%%%%%%%%%%%%%%%%%%%%%%%%%%%%%%%%%%%%%%%%%%%%%%%%%%%%%%%%
%%%%%%%%%%%%%%%%%%%%%%%%%%%%%%%%%          RMJ 0801          %%%%%%%%%%%%%%%%%%%%%%%%%%%%%%%%%
%%%%%%%%%%%%%%%%%%%%%%%%%%%%%%%%%%%%%%%%%%%%%%%%%%%%%%%%%%%%%%%%%%%%%%%%%%%%%%%%%%%%%%%%%%%%%%
\subsubsection{RM J080135.3+362807.5}\label{subsubsec:RMJ0801}

%%%%%%%%%%%%%%%%%%%%%%%%%% RMJ0801: Table - BCGs %%%%%%%%%%%%%%%%%%%%%%%%%%

\begin{table}[b]
    \centering
    \caption{RMJ0801 BCG Information}
    \label{tab:RMJ0801_BCG}
    \tablenotetext{a}{DESI DR1 \citep{abdul2025data}}
    \tablenotetext{b}{SDSS DR18 \citep{almeida2023eighteenth}}
    \tablenotetext{c}{DEIMOS (This work)}
    \tablenotetext{d}{DESI Legacy Survey DR10 \citep{dey2019overview}}
\begin{tabular*}{\columnwidth}{@{\extracolsep{\fill}}cccccc} \toprule\toprule
    BCG & Probability & Redshift & $r$-mag\tablenotemark{\footnotesize{d}} & RA {[}deg{]} & Dec {[}deg{]} \\ \toprule
    1  & 0.8964 & 0.501\tablenotemark{\footnotesize{a}} & 19.11 & 120.39710 & 36.46876 \\ \midrule
    2  & 0.1015 & ...                                   & 19.23 & 120.37135 & 36.47645 \\ \midrule
    3  & 0.0019 & 0.510\tablenotemark{\footnotesize{b}} & 20.08 & 120.39008 & 36.47881 \\ \midrule
    4  & 0.0001 & 0.500\tablenotemark{\footnotesize{b}} & 20.12 & 120.36597 & 36.48641 \\ \midrule
    5  & 0.0001 & 0.498\tablenotemark{\footnotesize{c}} & 20.35 & 120.35158 & 36.48754 \\ \midrule
    A  & ... & 0.503\tablenotemark{\footnotesize{c}} & 19.70 & 120.42687 & 36.45626 \\ \bottomrule
\printtablenotes{6}
\end{tabular*}
  \tablenotesreset
\end{table}

%%%%%%%%%%%%%%%%%%%%%%%%%% RMJ0801: Figure - Optical %%%%%%%%%%%%%%%%%%%%%%%%%%
\begin{figure}[t]
    \centering
    \includegraphics[clip, trim=0.0cm 0.75cm 0.0cm 0.0cm, width=\columnwidth]{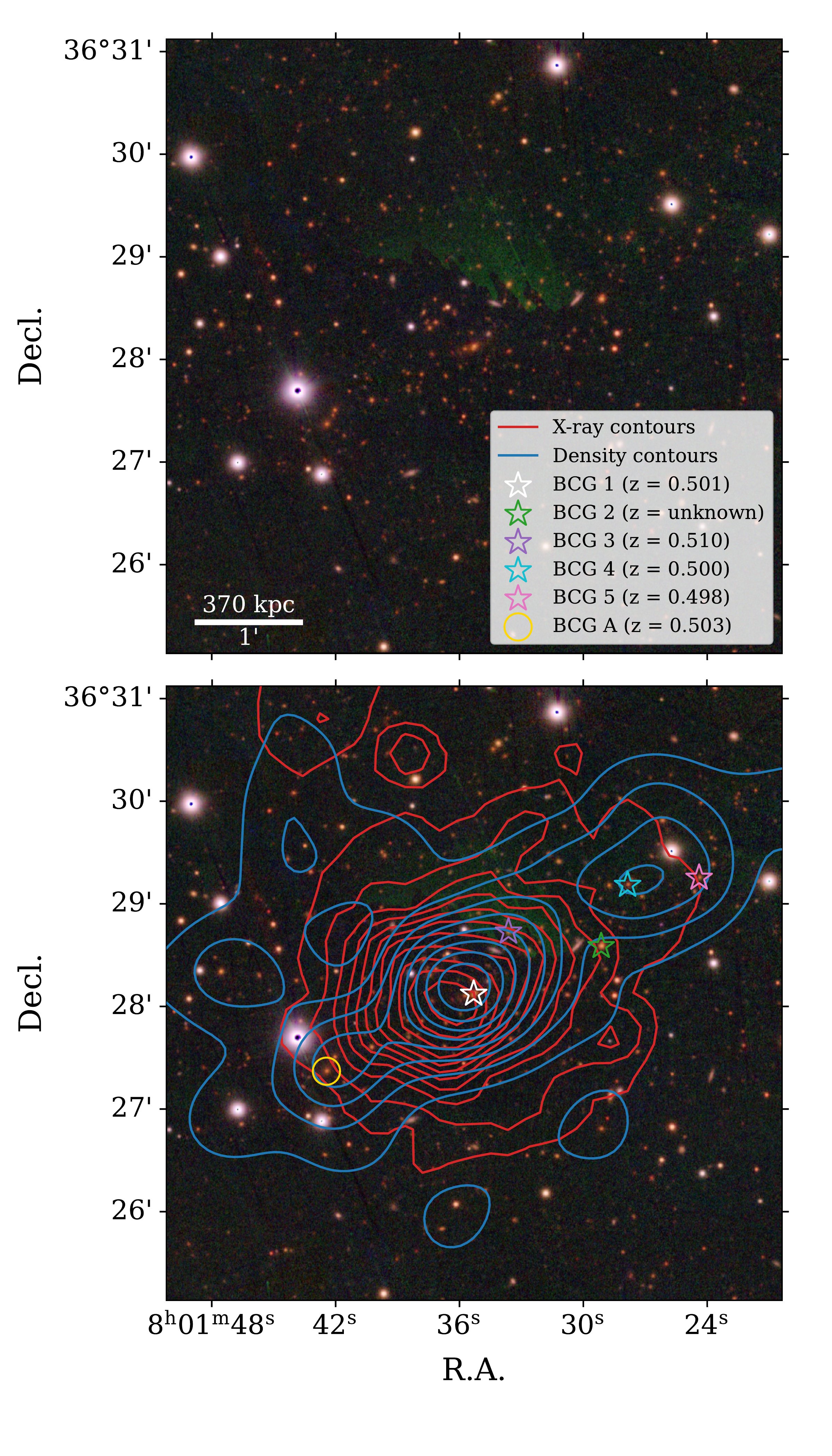}
    \caption{Optical image of RMJ0801. Top panel: 6\arcmin\ by 6\arcmin\ (2.22 by 2.22 Mpc) Pan-STARRS image. Bottom panel: red sequence density contours (blue), X-ray surface brightness contours (red), and BCG candidates identified by redMaPPer and this work. The diffuse green feature in the Pan-STARRS image is an imaging artifact and not associated with the cluster.}
    \label{fig:RMJ0801_optical}
\end{figure}
%%%%%%%%%%%%%%%%%%%%%%%%%% RMJ0801: Figure - Redshift Heatmap %%%%%%%%%%%%%%%%%%%%%%%%%%
\pdffig{
\begin{figure}[t]
    \centering
    \includegraphics[clip, trim=0.0cm 0.75cm 0.0cm 0.0cm, width=\columnwidth]{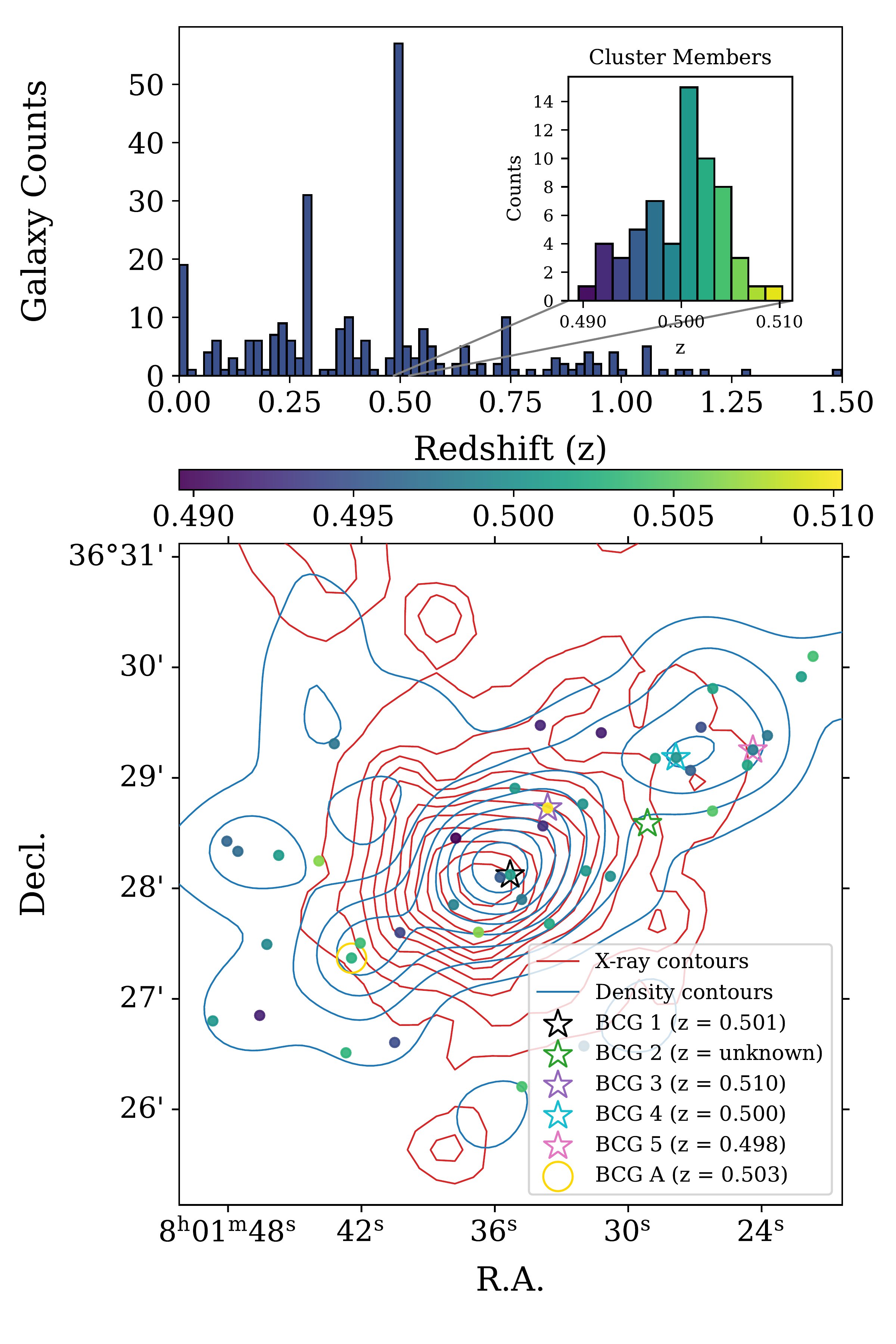}
    \caption{Redshift distribution of RMJ0801. Top panel: histogram of spectroscopic redshifts, combining archival data within 10\arcmin\ of the cluster with new observations from DEIMOS. The inset highlights galaxies in the redshift range $0.485 \leq z \leq 0.520$, which are classified as cluster members. Bottom panel: spatial distribution of those same members.}
    \label{fig:RMJ0801_redshifts}
\end{figure}}

%%%%%%%%%%%%%%%%%%%%%%%%%%%%%%%%%        RMJ 0801 Text       %%%%%%%%%%%%%%%%%%%%%%%%%%%%%%%%%
RedMaPPer lists the top two BCGs of RMJ0801 as having probabilities of $0.8964$ and $0.1015$, with all others having a probability less than 0.002. BCG 2 does not have a secure spectroscopic redshift, but it has been identified as a cluster member with a photometric redshift consistent with the cluster redshift \citep{szabo2011optical}. We also identified an additional possible BCG, which we label BCG A, with magnitude 19.70, placing it as the third brightest galaxy in the cluster (\cref{tab:RMJ0801_BCG}). We obtained a redshift of 0.503 with new DEIMOS spectroscopy, confirming it as a cluster member. BCG A also lies close to our fit of the cluster red sequence, with a difference of only 0.006 mag between the fit and actual $g-r$ color. Several bright foreground sources are visible in the optical image (\cref{fig:RMJ0801_optical}) possibly affecting the redMaPPer identification. However, unlike BCG B in RMJ0109, there is no foreground galaxy directly obscuring BCG A.

Galaxy density contours suggest three possible subclusters, with the main cluster centered on BCG 1, a second with BCG A, and a third to the northwest with BCGs 4 and 5. BCG 2 does not appear to be associated with any underlying substructure. The luminosity density near BCG A is likely understated due to the bright stars, while the northwestern subcluster contours are largely dominated by BCGs 4 and 5. 

The X-ray contours are elongated along a southeast-to-northwest axis, with a single peak southeast of BCG 1. While the X-ray peak is offset less than $\SI{100}{\kpc}$ from the BCG, the displacement along the axis through BCGs 1 and A is indicative of a merger. Both temperature and luminosity are elevated given a richness of $125$ (\cref{tab:clusters}), at $\SI{8.65}{\keV}$ and $\SI{4.51e44}{\ergs}$ (\cref{tab:XMM}), with the temperature $1.6\sigma$ above the predicted value of $\SI{5.35}{\keV}$, providing further support for a merger scenario.

%%%%%%%%%%%%%%%%%%%%%%%%%% RMJ0801: Figure - Subclusters %%%%%%%%%%%%%%%%%%%%%%%%%%
\pdffig{
\begin{figure*}
    \centering
    \includegraphics[clip, trim=0.0cm 0.75cm 0.5cm 0.5cm, width=\textwidth]{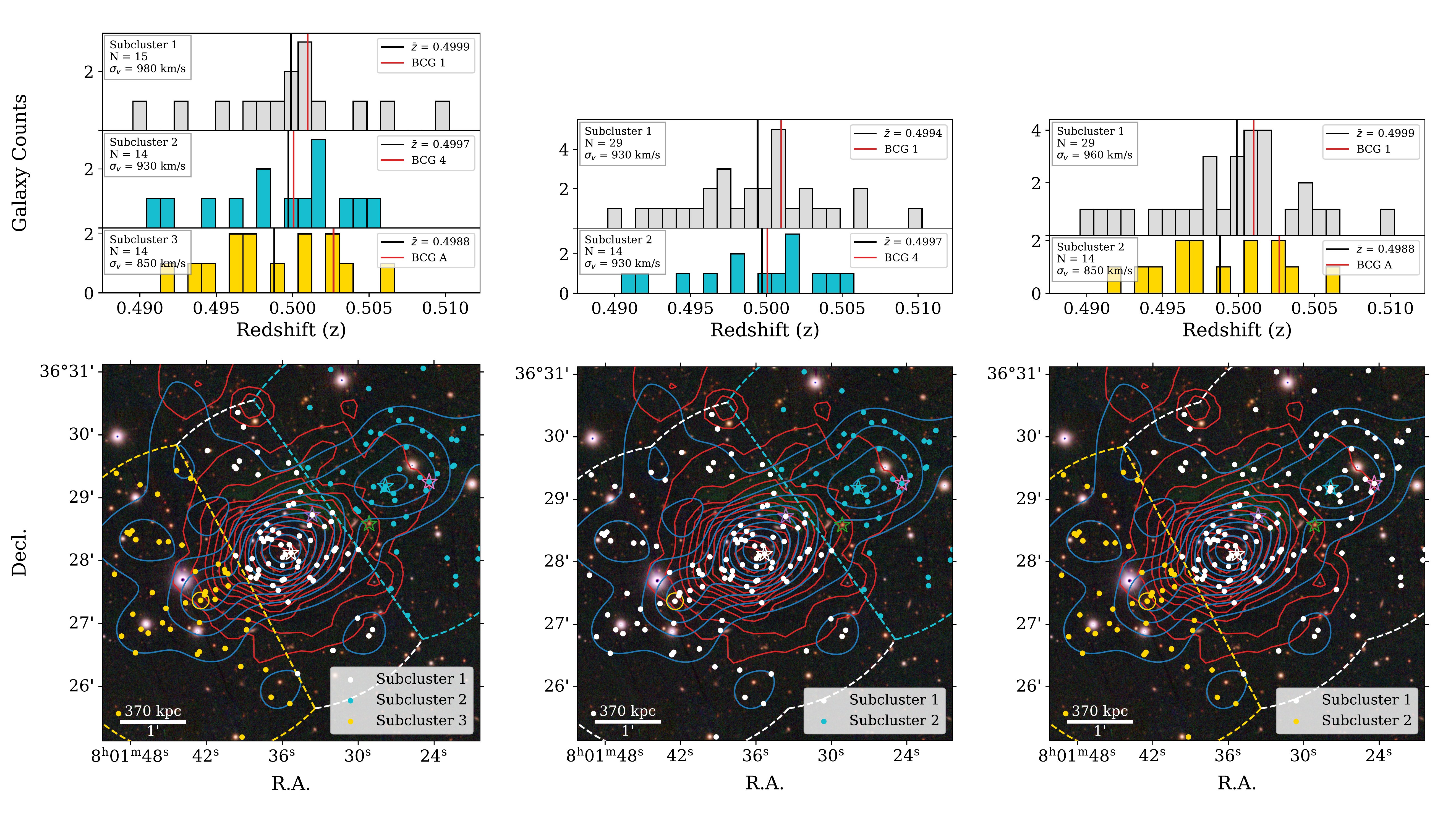}
    \caption{Subclustering analysis of RMJ0801. Bottom panels: spatial distribution of subcluster members, both spectroscopic and photometric, identified within regions defined by the bisector between BCG pairs and a radial distance of 2.5\arcmin\ from the BCG in each region. Top panels: redshift distribution of the spectroscopic members of those regions. The two right panels consider the subcluster associated with BCG A (center) and BCG 4 (right) to be part of Subcluster 1.}
    \label{fig:RMJ0801_sub_1}
\end{figure*}}

The redshift distribution, shown in \cref{fig:RMJ0801_redshifts}, shows two significant foreground structures, both identified as narrow components of the preferred GMM (five components total with three narrow). However, neither are in the immediate vicinity of the target cluster. The cluster at $z \sim 0.27$ is to the southwest out of the field of view shown. The ${\sim}20$ galaxies at $z\lesssim0.5$ are spread through the entire $10\arcmin \times 10\arcmin$ field. An additional smaller group at $z \sim 0.38$ is located due east of the cluster, also out of the field of view. The cluster redshift range was well fit by a single Gaussian and the spatial distribution shows no gradient or subclustering in redshift. 

\htmlfig{
\begin{figure}[t]
    \centering
    \includegraphics[clip, trim=0.0cm 0.75cm 0.0cm 0.0cm, width=\columnwidth]{Figures/RMJ_0801_redshifts_hist_hmap.jpg}
    \caption{Redshift distribution of RMJ0801. Top panel: histogram of spectroscopic redshifts, combining archival data within 10\arcmin\ of the cluster with new observations from DEIMOS. The inset highlights galaxies in the redshift range $0.485 \leq z \leq 0.520$, which are classified as cluster members. Bottom panel: spatial distribution of those same members.}
    \label{fig:RMJ0801_redshifts}
\end{figure}}

 We divided the cluster into three regions around BCGs 1, 3, and A, shown in the left panel of \cref{fig:RMJ0801_sub_1}, and find no evidence of a LOS velocity difference between subclusters. The largest difference in LOS velocity is between Subclusters 1 and 3 at \SI{220}{\kms} and all velocity dispersions lie in the range $850$--\SI{980}{\kms} (\cref{tab:RMJ0801_subclusters}). While the central subcluster is dominant, it is not clear whether subclusters in the northwest and southeast are truly distinct so we consider scenarios that combine BCGs 1 and A (center panel) and BCGs 1 and 3 (right panel). In the first scenario, the resulting subclusters have the same velocity dispersion at $\SI{930}{\kms}$ and $\Delta V_\mathrm{LOS} = \SI{60}{\kms}$ between subcluster means. The second scenario tells much the same story with Subcluster 1 velocity dispersion dropping to $\sigma_v=\SI{960}{\kms}$, and a $\SI{220}{\kms}$ LOS velocity difference between subclusters. 

\htmlfig{
\begin{figure*}
    \centering
    \includegraphics[clip, trim=0.0cm 0.75cm 0.5cm 0.5cm, width=\textwidth]{Figures/RMJ_0801_all.jpg}
    \caption{Subclustering analysis of RMJ0801. Bottom panels: spatial distribution of subcluster members, both spectroscopic and photometric, identified within regions defined by the bisector between BCG pairs and a radial distance of 2.5\arcmin\ from the BCG in each region. Top panels: redshift distribution of the spectroscopic members of those regions. The two right panels consider the subcluster associated with BCG A (center) and BCG 4 (right) to be part of Subcluster 1.}
    \label{fig:RMJ0801_sub_1}
\end{figure*}}

 %%%%%%%%%%%%%%%%%%%%%%%%%% RMJ0801: Table - Subclusters %%%%%%%%%%%%%%%%%%%%%%%%%%
\begin{table}[t]
    \centering
    \caption{RMJ0801 Subcluster Properties}
    \label{tab:RMJ0801_subclusters}
\begin{tabular*}{\columnwidth}{@{\extracolsep{\fill}}cccccc} \toprule\toprule
    Subcluster & $N$ & BCG & BCG $z$ & Mean $z$ & $\sigma_v$ [km s$^{-1}$] \\ \toprule
All & 62 & ... & ... & 0.5005 & 860 \\
1 & 15 & 1 & 0.501 & 0.4999 & 980  \\
2 & 14 & 4 & 0.500 & 0.4997 & 930  \\
3 & 14 & A & 0.503 & 0.4988 & 850  \\
\bottomrule
\end{tabular*}
\end{table}

Given the X-ray peak displaced toward BCG A along with the elevated temperature and luminosity, we classify this as possible merger and possibly binary between subclusters at BCGs 1 and A. The role of the northwestern group is more uncertain, but is likely a small infalling group. With a morphology that can possibly be well-modeled as a binary merger, we believe this is an excellent candidate for follow-up observations. Given the abundance of foreground sources contaminating the density contours, weak lensing would be particularly useful in determining the extent of the secondary subclusters.

%%%%%%%%%%%%%%%%%%%%%%%%%%%%%%%%%%%%%%%%%%%%%%%%%%%%%%%%%%%%%%%%%%%%%%%%%%%%%%%%%%%%%%%%%%%%%%
%%%%%%%%%%%%%%%%%%%%%%%%%%%%%%%%%          RMJ 0829          %%%%%%%%%%%%%%%%%%%%%%%%%%%%%%%%%
%%%%%%%%%%%%%%%%%%%%%%%%%%%%%%%%%%%%%%%%%%%%%%%%%%%%%%%%%%%%%%%%%%%%%%%%%%%%%%%%%%%%%%%%%%%%%%

\subsubsection{RM J082944.9+382754.4}\label{subsubsec:RMJ0829}

 %%%%%%%%%%%%%%%%%%%%%%%%%% RMJ0829: Table - BCGs %%%%%%%%%%%%%%%%%%%%%%%%%%

\begin{table}[b]
    \centering
    \caption{RMJ0829 BCG Information}
    \label{tab:RMJ0829_BCG}
    \tablenotetext{a}{SDSS DR18 \citep{almeida2023eighteenth}}
    \tablenotetext{b}{DEIMOS (This work)}
    \tablenotetext{c}{DESI DR1 \citep{abdul2025data}}
    \tablenotetext{d}{DESI Legacy Survey DR10 \citep{dey2019overview}}
\begin{tabular*}{\columnwidth}{@{\extracolsep{\fill}}cccccc} \toprule\toprule
    BCG & Probability & Redshift & $r$-mag\tablenotemark{\footnotesize{d}} & RA {[}deg{]} & Dec {[}deg{]} \\ \toprule
    1  & 0.6412 & 0.392\tablenotemark{\footnotesize{a}} & 18.42 & 127.43722 & 38.46511 \\ \midrule
    2  & 0.1930 & 0.342\tablenotemark{\footnotesize{b}} & 19.18 & 127.40229 & 38.46144 \\ \midrule
    3  & 0.1021 & 0.397\tablenotemark{\footnotesize{b}} & 19.54 & 127.46242 & 38.46859 \\ \midrule
    4  & 0.0359 & 0.395\tablenotemark{\footnotesize{b}} & 19.34 & 127.46410 & 38.47002 \\ \midrule
    5  & 0.0278 & 0.391\tablenotemark{\footnotesize{c}} & 18.29 & 127.40543 & 38.43779 \\ \bottomrule
\printtablenotes{6}
\end{tabular*}
  \tablenotesreset
\end{table}

%%%%%%%%%%%%%%%%%%%%%%%%%% RMJ0829: Figure - Optical %%%%%%%%%%%%%%%%%%%%%%%%%%
 \begin{figure}[t]
    \centering
    \includegraphics[clip, trim=0.0cm 1.25cm 0.0cm 0.0cm, width=\columnwidth]{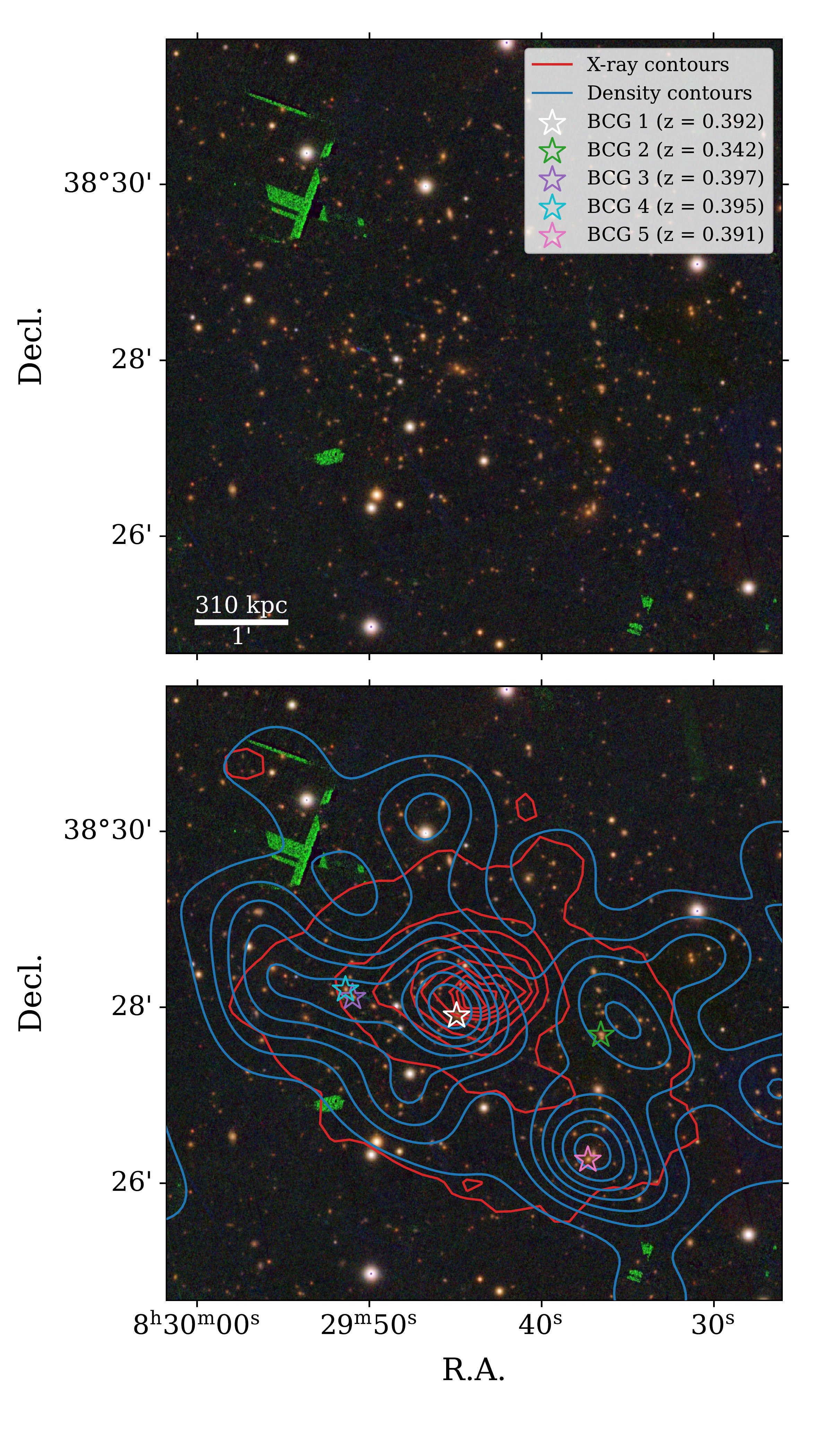}
    \caption{Optical image of RMJ0829. Top panel: 7\arcmin\ by 7\arcmin\ (2.17 by 2.17 Mpc) Pan-STARRS image. Bottom panel: red sequence density contours (blue), X-ray surface brightness contours (red), and BCG candidates identified by redMaPPer. Green features in the Pan-STARRS image are imaging artifacts and not associated with the cluster.}
    \label{fig:RMJ0829_optical}
\end{figure}

%%%%%%%%%%%%%%%%%%%%%%%%%% RMJ0829: Figure - Redshift Heatmap %%%%%%%%%%%%%%%%%%%%%%%%%%
\pdffig{
\begin{figure}[t]
    \centering
    \includegraphics[clip, trim=0.0cm 0.75cm 0.0cm 0.0cm, width=\columnwidth]{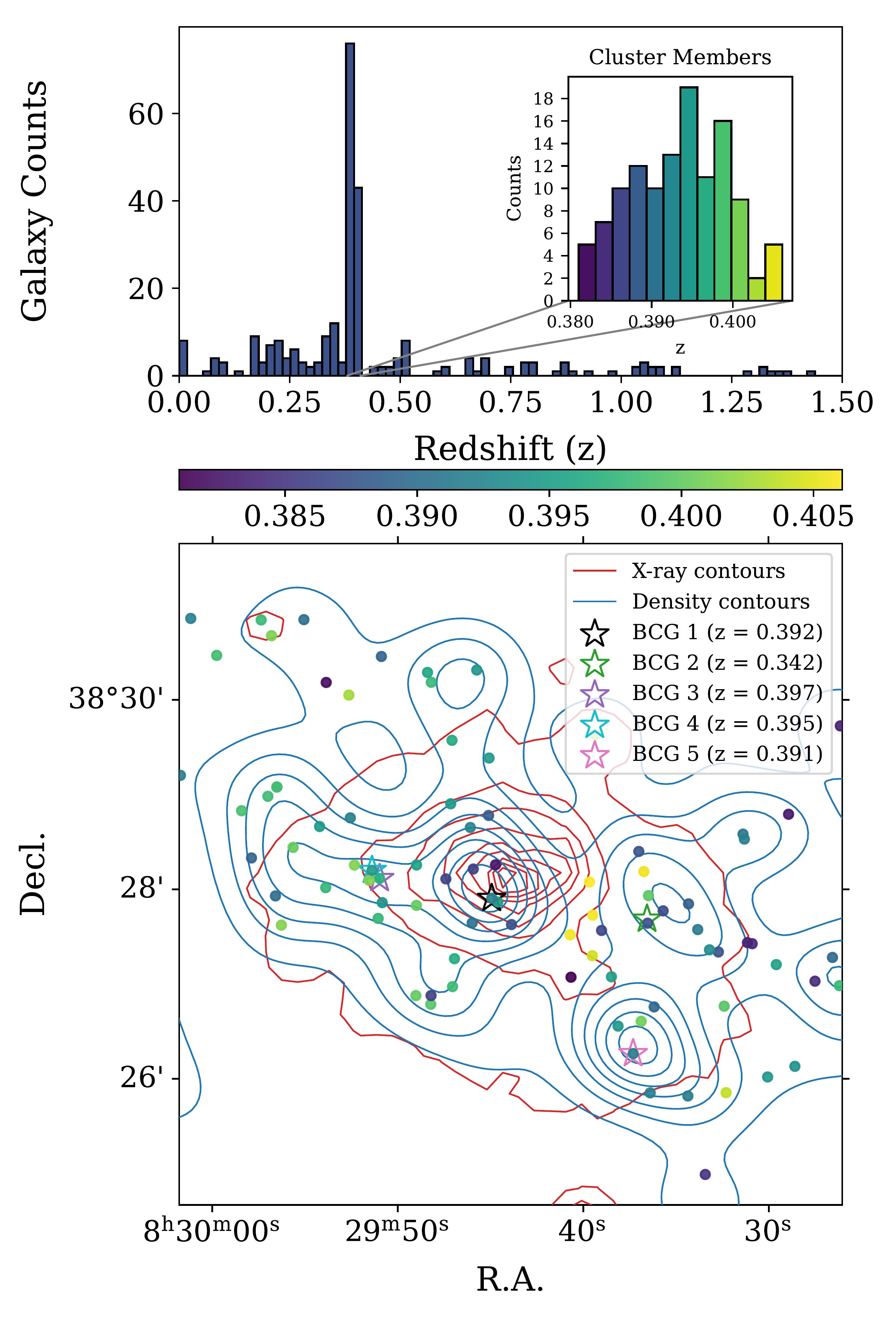}
    \caption{Redshift distribution of RMJ0829. Top panel: histogram of spectroscopic redshifts, combining archival data within 10\arcmin\ of the cluster with new observations from DEIMOS. The inset highlights galaxies in the redshift range $0.38 \leq z \leq 0.41$, which are classified as cluster members. Bottom panel: spatial distribution of those same members.}
    \label{fig:RMJ0829_redshifts}
\end{figure}}

%%%%%%%%%%%%%%%%%%%%%%%%%%%%%%%%%        RMJ 0829 Text       %%%%%%%%%%%%%%%%%%%%%%%%%%%%%%%%%
RMJ0829 is a complicated system, with redMaPPer identifying BCGs 1, 2, and 3 all having $>10\%$ probability of being the true BCG (\cref{tab:RMJ0829_BCG}). However, we confirmed BCG 2 to be at $z=0.342$ while the other BCGs are spectroscopically confirmed at $z \sim0.39$, corresponding to a difference in BCG line-of-sight velocities of ${\sim}\SI{10000}{\kms}$. We believe this to be part of a foreground system west of the target cluster (\cref{fig:RMJ0829_optical}), with our target being located at $0.38 \leq z \leq 0.41$. BCG 5 is the brightest in the cluster, with $r$-band magnitude of 18.29 as compared to 18.42 for BCG 1. 

%%%%%%%%%%%%%%%%%%%%%%%%%% RMJ0829: Figure - Subclusters %%%%%%%%%%%%%%%%%%%%%%%%%%
\pdffig{
\begin{figure}[t]
    \centering
    \includegraphics[clip, trim=0.0cm 1.0cm 0.0cm 0.0cm, width=\columnwidth]{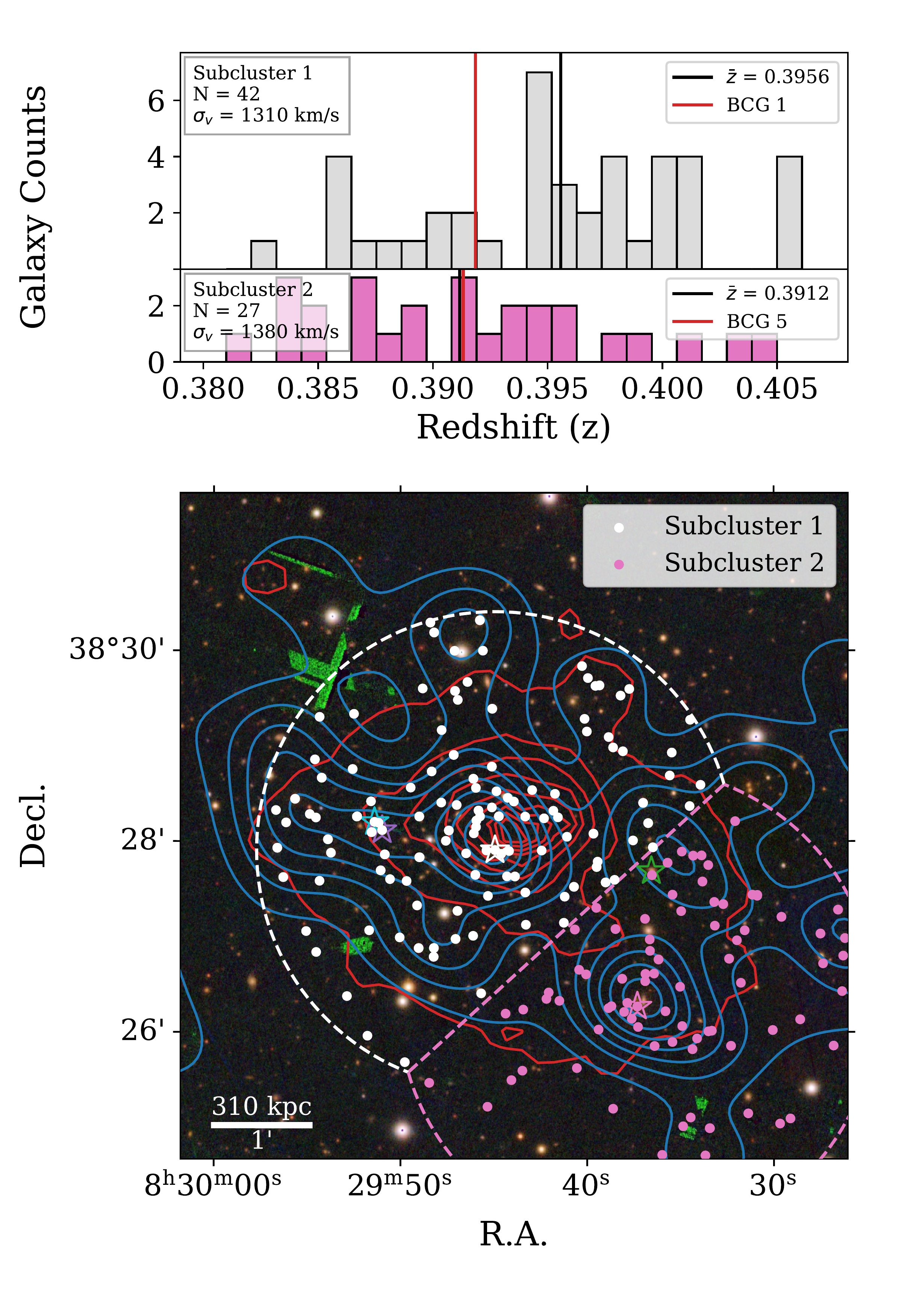}
    \caption{Subclustering analysis of RMJ0829. Bottom panel: spatial distribution of subcluster members, both spectroscopic and photometric, identified within regions defined by the bisector between BCG pairs and a radial distance of 2.5\arcmin\ from the BCG in each region. Top panel: redshift distribution of the spectroscopic members of those regions.}
    \label{fig:RMJ0829_sub_1}
\end{figure}}

The X-ray contours are elongated both along an axis between BCG 1 and BCG 3/4 and to the southwest and BCG 5, with a sharp peak $\approx\SI{100}{\kpc}$ northwest of BCG 1. This sharp peak is nearly coincident with BCG 1 but is not located between any pair of BCGs. We considered the possibility that a point source was embedded in the extended cluster emission and missed by the ESAS \texttt{cheese} routine. To rule out this possibility, we manually extracted high-energy events in a 2.5--\SI{7.5}{\keV} band using \texttt{evselect} and found no events corresponding to the bright emission. Luminosity density contours show a complex morphology, with two distinct overdensities at BCGs 1 and 5 and an extended tail towards BCGs 3 and 4. However, the extended structure to the northeast is greatly reduced in unweighted number density and is unlikely to represent any significant substructure. 

%%%%%%%%%%%%%%%%%%%%%%%%%% RMJ0829: Figure - Redshift Heatmap %%%%%%%%%%%%%%%%%%%%%%%%%%
\htmlfig{
\begin{figure}[t]
    \centering
    \includegraphics[clip, trim=0.0cm 0.75cm 0.0cm 0.0cm, width=\columnwidth]{Figures/RMJ_0829_redshifts_hist_hmap.jpg}
    \caption{Redshift distribution of RMJ0829. Top panel: histogram of spectroscopic redshifts, combining archival data within 10\arcmin\ of the cluster with new observations from DEIMOS. The inset highlights galaxies in the redshift range $0.38 \leq z \leq 0.41$, which are classified as cluster members. Bottom panel: spatial distribution of those same members.}
    \label{fig:RMJ0829_redshifts}
\end{figure}}

The redshift distribution, shown in \cref{fig:RMJ0829_redshifts}, is best-fit with a single component in the cluster region with both KS ($D=0.060$, $p=0.76$) and AD ($A^2=0.41$, $p=0.36$) tests consistent with a normal distribution. For the entire range of $z \leq 1.5$, a three-component fit is preferred with a single narrow component at the cluster redshift. Spatially, the highest redshift members of the cluster are all contained between BCG 1 and 2, indicating a possible localized line-of-sight velocity contribution. Dividing the cluster into regions defined by bisectors between BCGs 1 and 5 and a radius of 2.5\arcmin\ from each BCG (\cref{fig:RMJ0829_sub_1}) shows a slight bias towards higher redshift for Subcluster 1 and a velocity difference of \SI{950}{\kms} between subcluster means (\cref{tab:RMJ0829_subclusters}).

%%%%%%%%%%%%%%%%%%%%%%%%%% RMJ0829: Figure - Subclusters %%%%%%%%%%%%%%%%%%%%%%%%%%
\htmlfig{
\begin{figure}[t]
    \centering
    \includegraphics[clip, trim=0.0cm 1.0cm 0.0cm 0.0cm, width=\columnwidth]{Figures/RMJ_0829_subcluster_histograms_1_5.jpg}
    \caption{Subclustering analysis of RMJ0829. Bottom panel: spatial distribution of subcluster members, both spectroscopic and photometric, identified within regions defined by the bisector between BCG pairs and a radial distance of 2.5\arcmin\ from the BCG in each region. Top panel: redshift distribution of the spectroscopic members of those regions.}
    \label{fig:RMJ0829_sub_1}
\end{figure}

\begin{table}
    \centering
    \caption{RMJ0829 Subcluster Properties}
    \label{tab:RMJ0829_subclusters}
\begin{tabular*}{\columnwidth}{@{\extracolsep{\fill}}cccccc} \toprule\toprule
    Subcluster & $N$ & BCG & BCG $z$ & Mean $z$ & $\sigma_v$ [km s$^{-1}$] \\ \toprule
All & 119 & ... & ... & 0.3934 & 1330 \\
1 & 42 & 1 & 0.392 & 0.3956 & 1310  \\
2 & 27 & 5 & 0.391 & 0.3912 & 1380  \\
\bottomrule
\end{tabular*}
\end{table}}

Temperature ($\SI{6.95}{\keV}$) and X-ray luminosity ($\SI{5.50e44}{\ergs}$) are both elevated ${>}1\sigma$ above expected values (\cref{tab:XMM}). Additionally, RMJ0829 (PSZ2 G183.30+34.98) has been identified as hosting a candidate radio halo by \citet{botteon2022planck}. The halo is reported as a newly claimed detection and is accompanied by diffuse radio emission of unknown origin elongated in the north-south direction to the east of the cluster center. The combination of disturbed X-ray morphology and diffuse radio emission strongly suggest a dynamically active system. The optical and X-ray data are broadly consistent with a classification as a likely merger between subclusters associated with BCGs 1 and 5, although the location of the X-ray peak likely disfavors a simple head-on configuration. This appears to be a binary system with a non-zero impact parameter, but we are unable to definitively rule out contributions from substructure in the northeast so classify it as possibly binary. Weak-lensing constraints on the mass distribution would be valuable in clarifying the merger geometry and assessing whether the system can be modeled reliably as a two-body interaction.

%%%%%%%%%%%%%%%%%%%%%%%%%% RMJ0829: Table - Subclusters %%%%%%%%%%%%%%%%%%%%%%%%%%
\pdffig{
\begin{table}
    \centering
    \caption{RMJ0829 Subcluster Properties}
    \label{tab:RMJ0829_subclusters}
\begin{tabular*}{\columnwidth}{@{\extracolsep{\fill}}cccccc} \toprule\toprule
    Subcluster & $N$ & BCG & BCG $z$ & Mean $z$ & $\sigma_v$ [km s$^{-1}$] \\ \toprule
All & 119 & ... & ... & 0.3934 & 1330 \\
1 & 42 & 1 & 0.392 & 0.3956 & 1310  \\
2 & 27 & 5 & 0.391 & 0.3912 & 1380  \\
\bottomrule
\end{tabular*}
\end{table}}

%%%%%%%%%%%%%%%%%%%%%%%%%%%%%%%%%%%%%%%%%%%%%%%%%%%%%%%%%%%%%%%%%%%%%%%%%%%%%%%%%%%%%%%%%%%%%%
%%%%%%%%%%%%%%%%%%%%%%%%%%%%%%%%%          RMJ 0926          %%%%%%%%%%%%%%%%%%%%%%%%%%%%%%%%%
%%%%%%%%%%%%%%%%%%%%%%%%%%%%%%%%%%%%%%%%%%%%%%%%%%%%%%%%%%%%%%%%%%%%%%%%%%%%%%%%%%%%%%%%%%%%%%

\subsubsection{RM J092647.3+050004.0}\label{subsubsec:RMJ0926}

\htmlfig{
\begin{table}
    \centering
    \caption{RMJ0926 BCG Information}
    \label{tab:RMJ0926_BCG}
    \tablenotetext{a}{SDSS DR18 \citep{almeida2023eighteenth}}
    \tablenotetext{b}{DEIMOS (This work)}
    \tablenotetext{c}{DESI DR1 \citep{abdul2025data}}
    \tablenotetext{d}{DESI Legacy Survey DR10 \citep{dey2019overview}}
\begin{tabular*}{\columnwidth}{@{\extracolsep{\fill}}cccccc} \toprule\toprule
    BCG & Probability & Redshift & $r$-mag\tablenotemark{\footnotesize{d}} & RA {[}deg{]} & Dec {[}deg{]} \\ \toprule
    1  & 0.7821 & 0.462\tablenotemark{\footnotesize{a}} & 18.65 & 141.69696 & 5.00110 \\ \midrule
    2  & 0.0746 & 0.465\tablenotemark{\footnotesize{b}} & 19.14 & 141.71086 & 5.02208 \\ \midrule
    3  & 0.0511 & 0.460\tablenotemark{\footnotesize{c}} & 19.30 & 141.69759 & 5.02156 \\ \midrule
    4  & 0.0479 & ...                                   & 19.47 & 141.69760 & 5.01306 \\ \midrule
    5  & 0.0444 & 0.462\tablenotemark{\footnotesize{b}} & 18.88 & 141.69397 & 5.02389 \\ \bottomrule
\printtablenotes{6}
\end{tabular*}
  \tablenotesreset
\end{table}}

%%%%%%%%%%%%%%%%%%%%%%%%%% RMJ0926: Figure - Optical %%%%%%%%%%%%%%%%%%%%%%%%%%
\begin{figure}[t]
    \centering
    \includegraphics[clip, trim=0.0cm 1.25cm 0.0cm 0.0cm, width=\columnwidth]{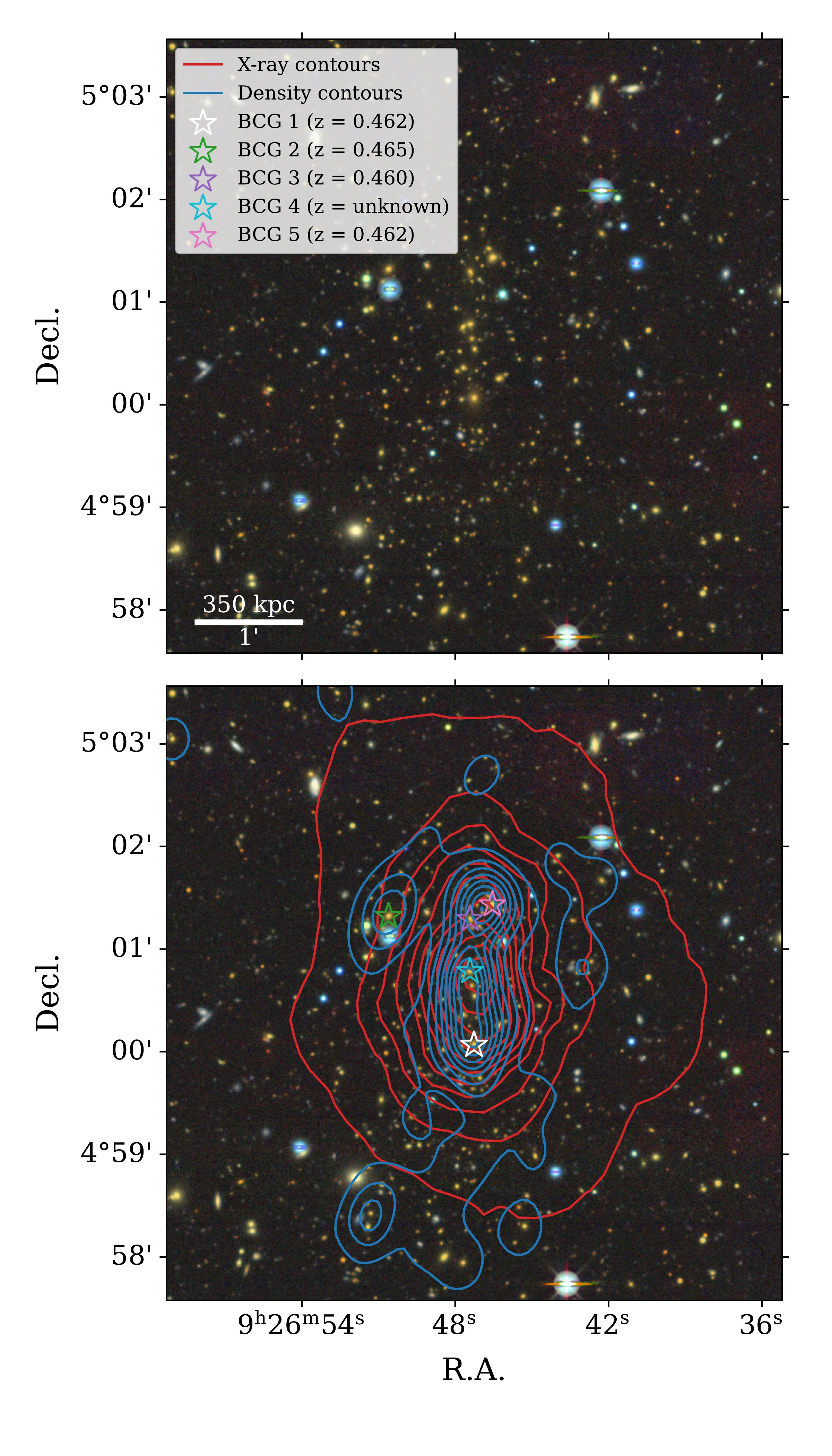}
    \caption{Optical image of RMJ0926. Top panel: 6\arcmin\ by 6\arcmin\ (2.10 by 2.10 Mpc) Legacy Survey image. Bottom panel: red sequence density contours (blue), X-ray surface brightness contours (red), and BCG candidates identified by redMaPPer.}
    \label{fig:RMJ0926_optical}
\end{figure}

%%%%%%%%%%%%%%%%%%%%%%%%%% RMJ0926: Table - BCGs %%%%%%%%%%%%%%%%%%%%%%%%%%
\pdffig{
\begin{table}
    \centering
    \caption{RMJ0926 BCG Information}
    \label{tab:RMJ0926_BCG}
    \tablenotetext{a}{SDSS DR18 \citep{almeida2023eighteenth}}
    \tablenotetext{b}{DEIMOS (This work)}
    \tablenotetext{c}{DESI DR1 \citep{abdul2025data}}
    \tablenotetext{d}{DESI Legacy Survey DR10 \citep{dey2019overview}}
\begin{tabular*}{\columnwidth}{@{\extracolsep{\fill}}cccccc} \toprule\toprule
    BCG & Probability & Redshift & $r$-mag\tablenotemark{\footnotesize{d}} & RA {[}deg{]} & Dec {[}deg{]} \\ \toprule
    1  & 0.7821 & 0.462\tablenotemark{\footnotesize{a}} & 18.65 & 141.69696 & 5.00110 \\ \midrule
    2  & 0.0746 & 0.465\tablenotemark{\footnotesize{b}} & 19.14 & 141.71086 & 5.02208 \\ \midrule
    3  & 0.0511 & 0.460\tablenotemark{\footnotesize{c}} & 19.30 & 141.69759 & 5.02156 \\ \midrule
    4  & 0.0479 & ...                                   & 19.47 & 141.69760 & 5.01306 \\ \midrule
    5  & 0.0444 & 0.462\tablenotemark{\footnotesize{b}} & 18.88 & 141.69397 & 5.02389 \\ \bottomrule
\printtablenotes{6}
\end{tabular*}
  \tablenotesreset
\end{table}}

%%%%%%%%%%%%%%%%%%%%%%%%%%%%%%%%%        RMJ 0926 Text       %%%%%%%%%%%%%%%%%%%%%%%%%%%%%%%%%
In contrast with RMJ0829, RMJ0926 appears to be a relatively simple system. The highest probability BCG in redMaPPer is at $0.7821$ for BCG 1, while all other BCGs have roughly the same probability of $\sim0.05$ (\cref{tab:RMJ0926_BCG}). We were unable to get a confirmed redshift on BCG 4, but it was selected through a fit of the red sequence, and we believe it to be a cluster member. The X-ray surface brightness (\cref{fig:RMJ0926_optical}) has a single peak at BCG 4 and is elongated along a north-south axis, coinciding with an axis that runs through all but BCG 2. The single peak located with BCG 4 may suggest a relaxed cluster at that location. However, BCG 4 is the faintest of the BCG candidates with an $r$-band magnitude of 19.47 and is unlikely to be the single BCG of a relaxed system. The galaxy luminosity density has two primary peaks, one encompassing BCGs 1 and 4 and another with BCGs 3 and 5. The contours are elongated along the same axis as the X-ray surface brightness, but also extend to the east and enclose BCG 2. The galaxy density at BCG 2 is likely underreported due to contamination from point sources, but it does not appear to be a separate subcluster.

%%%%%%%%%%%%%%%%%%%%%%%%%% RMJ0926: Figure - Redshift Heatmap %%%%%%%%%%%%%%%%%%%%%%%%%%
\begin{figure}[b]
    \centering
    \includegraphics[clip, trim=0.0cm 0.75cm 0.0cm 0.0cm, width=\columnwidth]{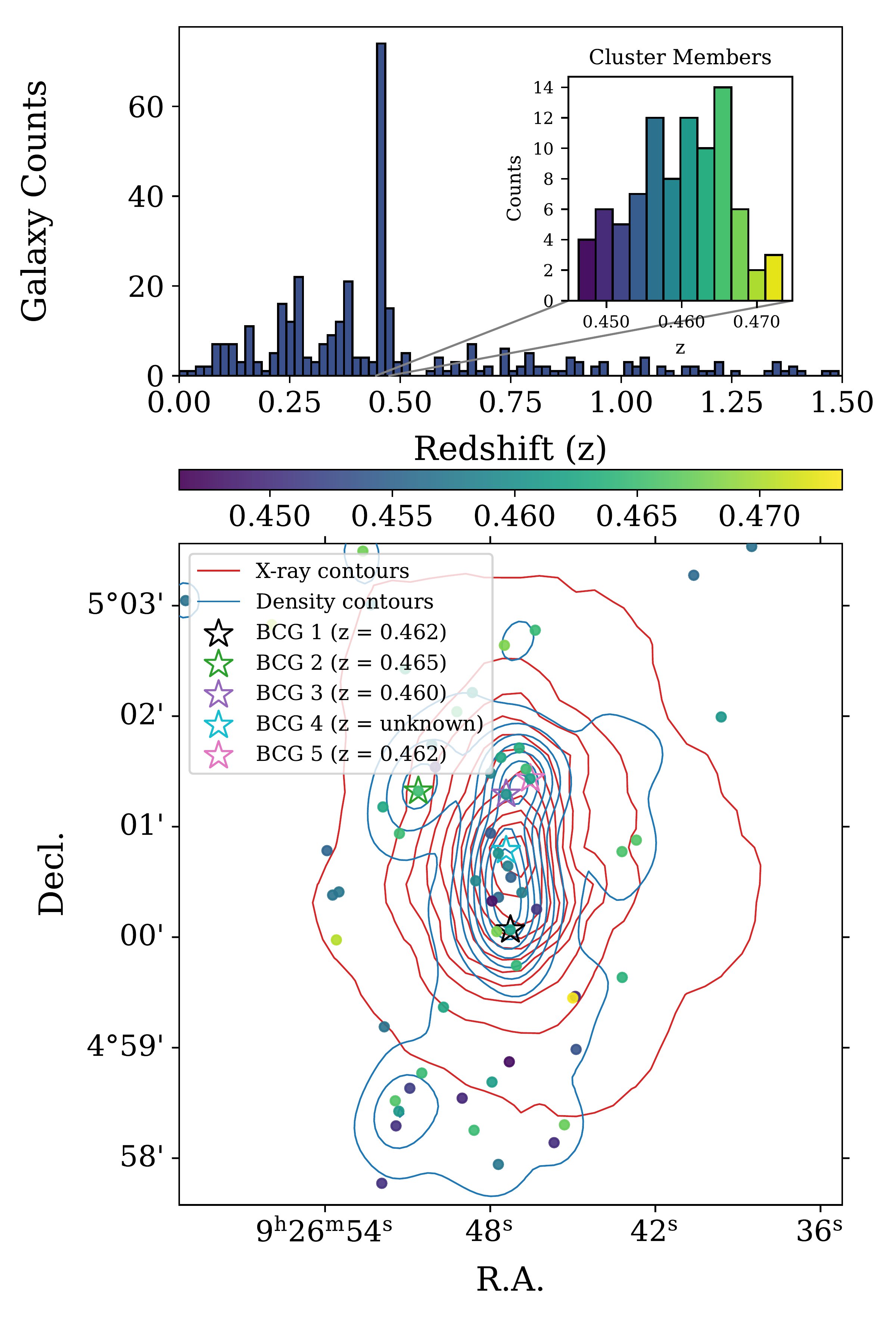}
    \caption{Redshift distribution of RMJ0926. Top panel: histogram of spectroscopic redshifts, combining archival data within 10\arcmin\ of the cluster with new observations from DEIMOS. The inset highlights galaxies in the redshift range $0.445 \leq z \leq 0.475$, which are classified as cluster members. Bottom panel: spatial distribution of those same members.}
    \label{fig:RMJ0926_redshifts}
\end{figure}

Two foreground structures were identified in a GMM fit of the redshift distribution (\cref{fig:RMJ0926_redshifts}) both with ${\sim}45$ members. The first, at $z=0.25$, is widely scattered through the eastern half of the full field of view, with only ${\sim}5$ members in the cluster region. The second, at $z=0.36$, is largely concentrated to the southeast of the cluster, but has ${\sim}10$ members overlapping along the cluster's eastern edge. That group could contribute some to the widening of the X-ray surface brightness south of BCG 2, but does not affect the general observed morphology. In the cluster region, $0.445 \leq z \leq 0.475$, a single-component GMM fit is preferred and consistent with Gaussianity ($D=0.060$, $p=0.88$; $A^2=0.33$, $p=0.51$). The spatial distribution of redshifts, shown in the bottom panel of \cref{fig:RMJ0926_redshifts}, shows some clustering of lower redshift members near BCG 1, which could indicate a local LOS velocity component. Dividing the field into northern and southern subclusters centered on BCGs 1 and 3 (\cref{fig:RMJ0926_sub_1}) shows the northern (southern) subcluster with a velocity dispersion of \SI{670}{\kms} (\SI{1410}{\kms}) (\cref{tab:RMJ0926_subclusters}) and a velocity difference of \SI{740}{\kms} between subcluster means.

\htmlfig{
\begin{figure}[b]
    \centering
    \includegraphics[clip, trim=0.0cm 1.0cm 0.0cm 0.0cm, width=\columnwidth]{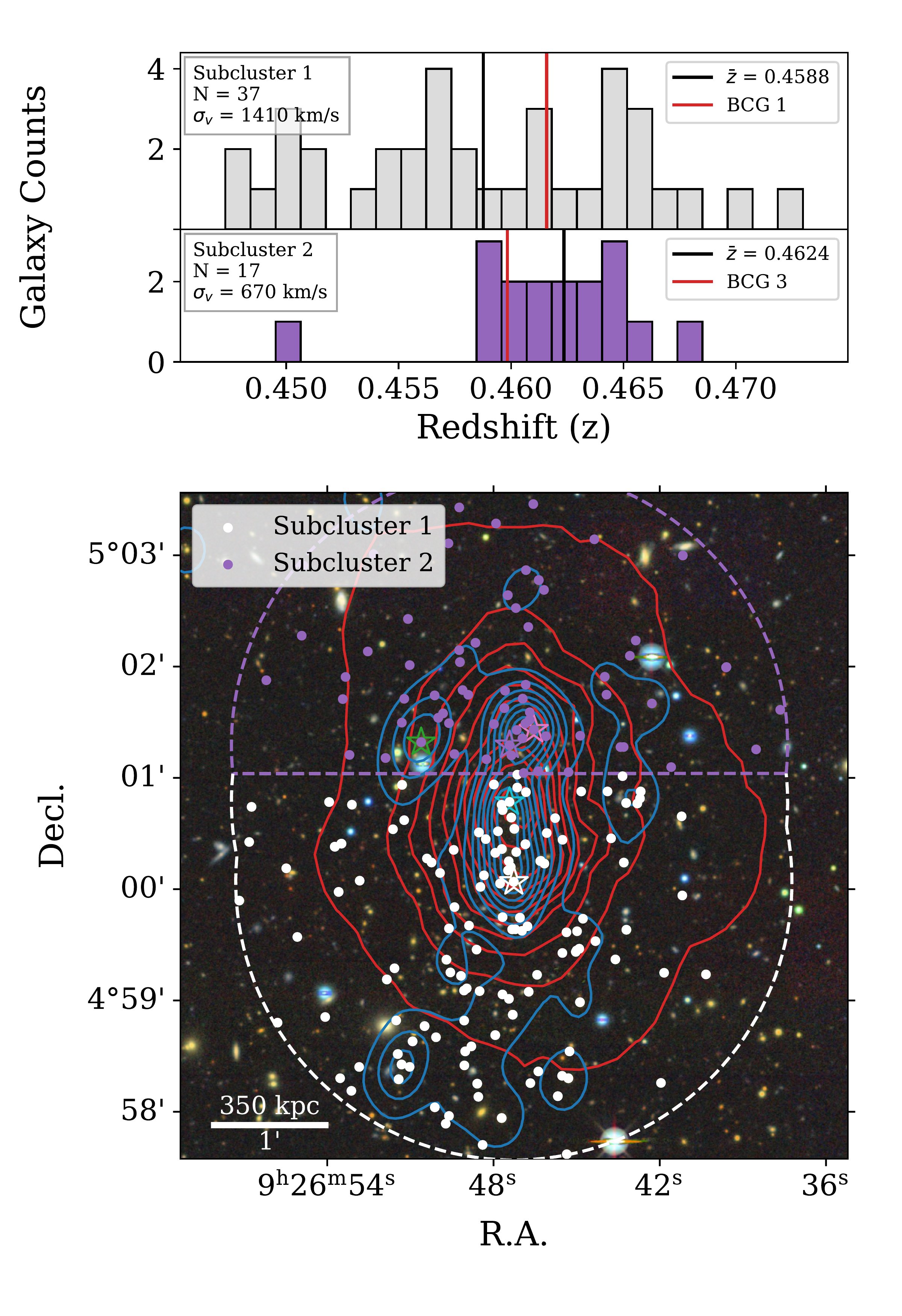}
    \caption{Subclustering analysis of RMJ0926. Bottom panel: spatial distribution of subcluster members, both spectroscopic and photometric, identified within regions defined by the bisector between BCG pairs and a radial distance of 2.5\arcmin\ from the BCG in each region. Top panel: redshift distribution of the spectroscopic members of those regions.}
    \label{fig:RMJ0926_sub_1}
\end{figure}}

%%%%%%%%%%%%%%%%%%%%%%%%%% RMJ0926: Table - Subclusters %%%%%%%%%%%%%%%%%%%%%%%%%%
\begin{table}[t]
    \centering
    \caption{RMJ0926 Subcluster Properties}
    \label{tab:RMJ0926_subclusters}
\begin{tabular*}{\columnwidth}{@{\extracolsep{\fill}}cccccc} \toprule\toprule
    Subcluster & $N$ & BCG & BCG $z$ & Mean $z$ & $\sigma_v$ [km s$^{-1}$] \\ \toprule
All & 89 & ... & ... & 0.4600 & 1358 \\
1 & 37 & 1 & 0.462 & 0.4588 & 1410  \\
2 & 17 & 3 & 0.460 & 0.4624 & \phantom{0}670  \\
\bottomrule
\end{tabular*}
\end{table}

%%%%%%%%%%%%%%%%%%%%%%%%%% RMJ0926: Figure - Subclusters %%%%%%%%%%%%%%%%%%%%%%%%%%
\pdffig{
\begin{figure}[b]
    \centering
    \includegraphics[clip, trim=0.0cm 1.0cm 0.0cm 0.0cm, width=\columnwidth]{Figures/RMJ_0926_subcluster_histograms_1_4_3.jpg}
    \caption{Subclustering analysis of RMJ0926. Bottom panel: spatial distribution of subcluster members, both spectroscopic and photometric, identified within regions defined by the bisector between BCG pairs and a radial distance of 2.5\arcmin\ from the BCG in each region. Top panel: redshift distribution of the spectroscopic members of those regions.}
    \label{fig:RMJ0926_sub_1}
\end{figure}}
 
%%%%%%%%%%%%%%%%%%%%%%%%%% RMJ1043: Figure - Optical %%%%%%%%%%%%%%%%%%%%%%%%%%
\pdffig{
\begin{figure}[t]
    \centering
    \includegraphics[clip, trim=0.0cm 1.0cm 0.0cm 0.0cm, width=\columnwidth]{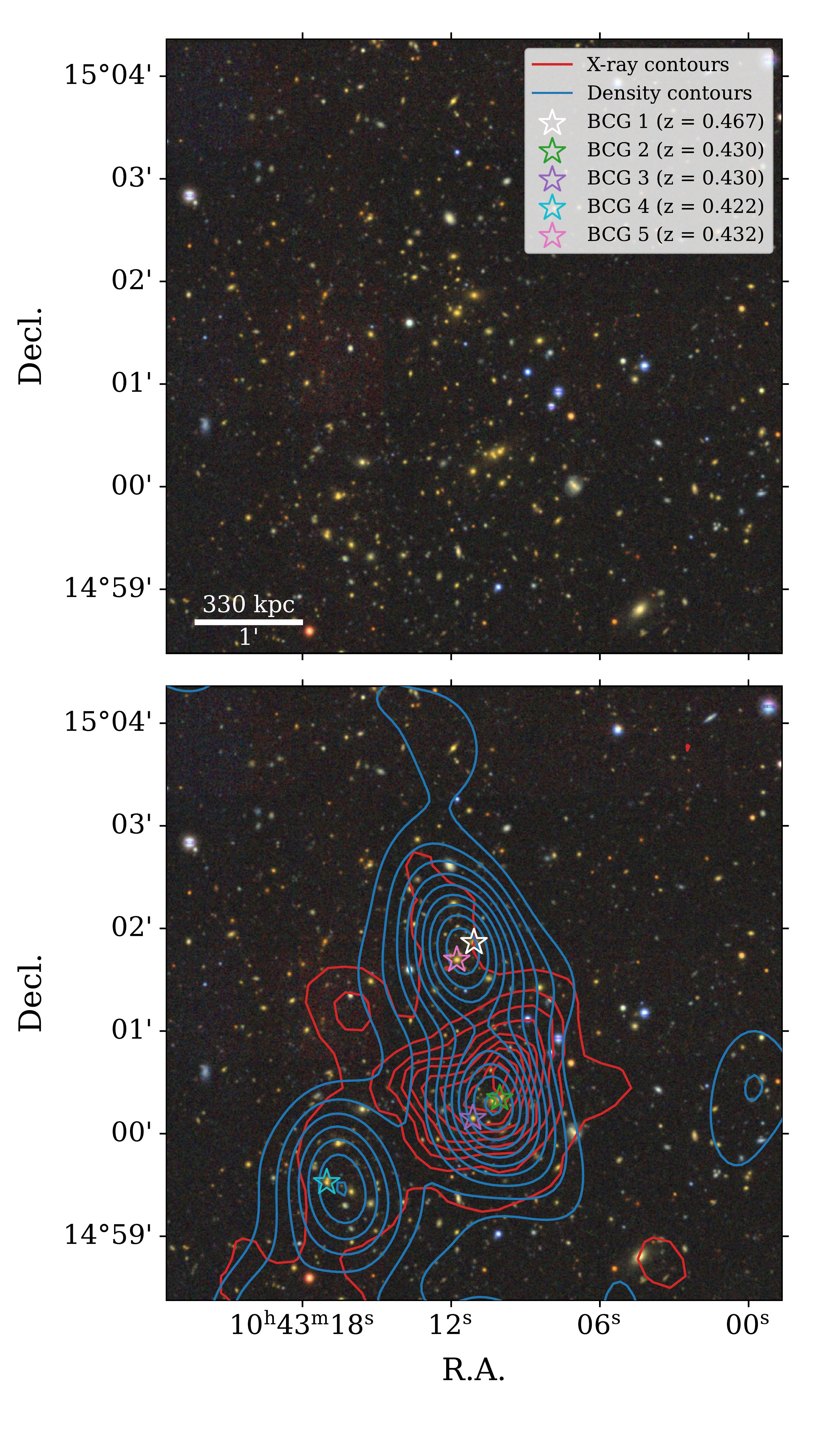}
    \caption{Optical image of RMJ1043. Top panel: 6\arcmin\ by 6\arcmin\ (1.98 by 1.98 Mpc) Legacy Survey image. Bottom panel: red sequence density contours (blue), X-ray surface brightness contours (red), and BCG candidates identified by redMaPPer.}
    \label{fig:RMJ1043_optical}
\end{figure}}

RMJ0926 has the highest temperature of any cluster in the sample at $\SI{10.55}{\keV}$, which is $2.7\sigma$ above the expected value of $\SI{5.65}{\keV}$. The X-ray luminosity is similarly elevated at $\SI{6.84e44}{\ergs}$ (\cref{tab:XMM}), compared to the expected value of $\SI{2.35e44}{\ergs}$ for a cluster of richness 140 (\cref{tab:clusters}), corresponding to a $1.8\sigma$ excess. This, along with a separation between northern and southern subclusters in both galaxy density and redshift distribution, supports a binary merger classification. Subclusters at BCG 1 and BCG 3/5 are separated by ${\approx}\SI{480}{\kpc}$, with the X-ray peak directly between the subclusters, ${\approx}\SI{240}{\kpc}$ from each. This is one of the cleanest binary mergers in the sample and a top-priority for follow-up, particularly with weak lensing. 

%%%%%%%%%%%%%%%%%%%%%%%%%%%%%%%%%%%%%%%%%%%%%%%%%%%%%%%%%%%%%%%%%%%%%%%%%%%%%%%%%%%%%%%%%%%%%%
%%%%%%%%%%%%%%%%%%%%%%%%%%%%%%%%%          RMJ 1043          %%%%%%%%%%%%%%%%%%%%%%%%%%%%%%%%%
%%%%%%%%%%%%%%%%%%%%%%%%%%%%%%%%%%%%%%%%%%%%%%%%%%%%%%%%%%%%%%%%%%%%%%%%%%%%%%%%%%%%%%%%%%%%%%
\subsubsection{RM J104311.1+150151.9}\label{subsubsec:RMJ1043}
%%%%%%%%%%%%%%%%%%%%%%%%%% RMJ1043: Table - BCGs %%%%%%%%%%%%%%%%%%%%%%%%%%
\begin{table}[b]
    \centering
    \caption{RMJ1043 BCG Information}
    \label{tab:RMJ1043_BCG}
    \tablenotetext{a}{DEIMOS (This work)}
    \tablenotetext{b}{SDSS DR18 \citep{almeida2023eighteenth}}
    \tablenotetext{c}{DESI Legacy Survey DR10 \citep{dey2019overview}}
\begin{tabular*}{\columnwidth}{@{\extracolsep{\fill}}cccccc} \toprule\toprule
    BCG & Probability & Redshift & $r$-mag\tablenotemark{\footnotesize{c}} & RA {[}deg{]} & Dec {[}deg{]} \\ \toprule
    1  & 0.3327 & 0.467\tablenotemark{\footnotesize{a}} & 19.08 & 160.79619 & 15.03108 \\ \midrule
    2  & 0.2613 & 0.430\tablenotemark{\footnotesize{b}} & 18.66 & 160.79184 & 15.00572 \\ \midrule
    3  & 0.1514 & 0.430\tablenotemark{\footnotesize{a}} & 19.68 & 160.79639 & 15.00251 \\ \midrule
    4  & 0.1506 & 0.422\tablenotemark{\footnotesize{a}} & 19.31 & 160.82095 & 14.99200 \\ \midrule
    5  & 0.1041 & 0.432\tablenotemark{\footnotesize{b}} & 19.05 & 160.79907 & 15.02822 \\ \bottomrule
\printtablenotes{6}
\end{tabular*}
  \tablenotesreset
\end{table}

%%%%%%%%%%%%%%%%%%%%%%%%%% RMJ1043: Figure - Optical %%%%%%%%%%%%%%%%%%%%%%%%%%
\htmlfig{
\begin{figure}[t]
    \centering
    \includegraphics[clip, trim=0.0cm 1.0cm 0.0cm 0.0cm, width=\columnwidth]{Figures/RMJ_1043_2panel_optical_contours_vert.jpg}
    \caption{Optical image of RMJ1043. Top panel: 6\arcmin\ by 6\arcmin\ (1.98 by 1.98 Mpc) Legacy Survey image. Bottom panel: red sequence density contours (blue), X-ray surface brightness contours (red), and BCG candidates identified by redMaPPer.}
    \label{fig:RMJ1043_optical}
\end{figure}}

%%%%%%%%%%%%%%%%%%%%%%%%%% RMJ1043: Figure - Redshift Heatmap %%%%%%%%%%%%%%%%%%%%%%%%%%
\pdffig{
\begin{figure}[t]
    \centering
    \includegraphics[clip, trim=0.0cm 0.75cm 0.0cm 0.0cm, width=\columnwidth]{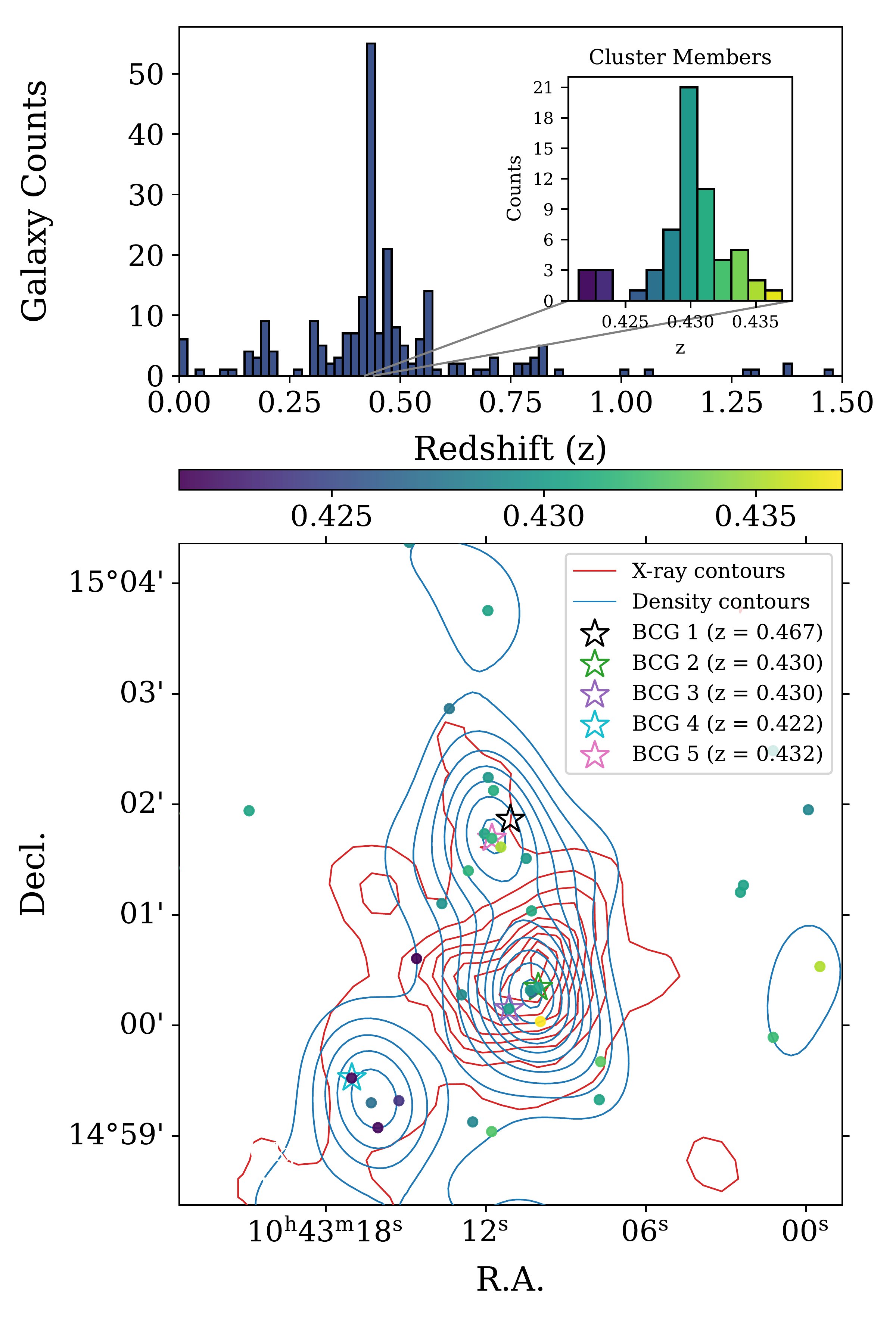}
    \caption{Redshift distribution of RMJ1043. Top panel: histogram of spectroscopic redshifts, combining archival data within 10\arcmin\ of the cluster with new observations from DEIMOS. The inset highlights galaxies in the redshift range $0.42 \leq z \leq 0.44$, which are classified as cluster members. Bottom panel: spatial distribution of those same members.}
    \label{fig:RMJ1043_redshifts}
\end{figure}

%%%%%%%%%%%%%%%%%%%%%%%%%% RMJ1043: Figure - Subclusters %%%%%%%%%%%%%%%%%%%%%%%%%%
\begin{figure}[t]
    \centering
    \includegraphics[clip, trim=0.0cm 1.0cm 0.0cm 0.0cm, width=\columnwidth]{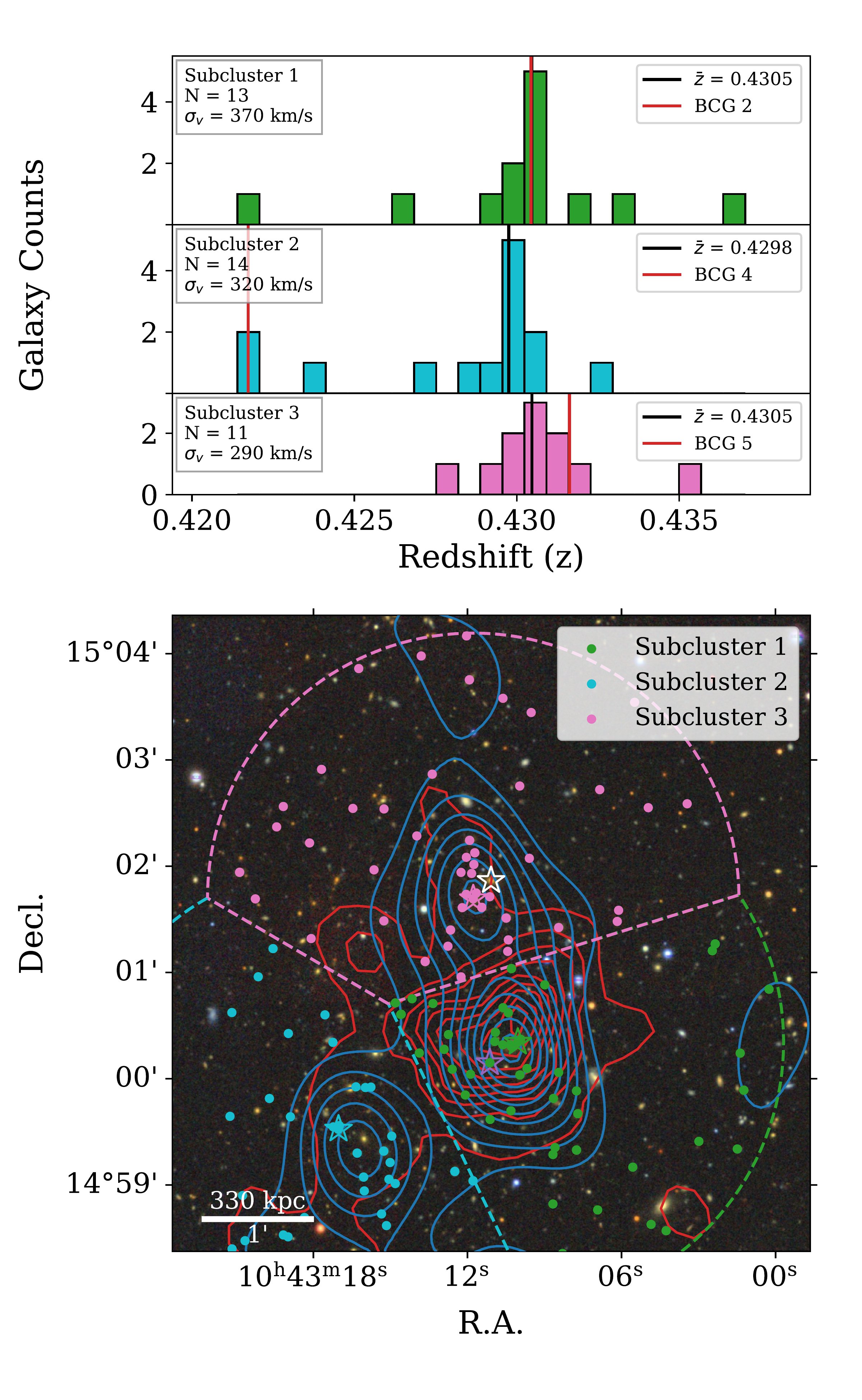}
    \caption{Subclustering analysis of RMJ1043. Bottom panel: spatial distribution of subcluster members, both spectroscopic and photometric, identified within regions defined by the bisector between BCG pairs and a radial distance of 2.5\arcmin\ from the BCG in each region. Top panel: redshift distribution of the spectroscopic members of those regions.}
    \label{fig:RMJ1043_sub_1}
\end{figure}}
%%%%%%%%%%%%%%%%%%%%%%%%%%%%%%%%%        RMJ 1043 Text       %%%%%%%%%%%%%%%%%%%%%%%%%%%%%%%%%
RMJ1043 is another complicated system, with all BCGs having a redMaPPer probability ${>}0.10$ (\cref{tab:RMJ1043_BCG}). BCG 1, with probability $0.3327$, is at a higher redshift than the rest of the BCGs at $z=0.467$ versus $z\sim0.43$, representing a velocity difference of $>\SI{7500}{\kms}$. Galaxy density contours show three distinct peaks localized on BCG 1/5, BCG  2/3, and BCG 4 (\cref{fig:RMJ1043_optical}). The X-ray surface brightness is primarily localized around BCGs 2 and 3, with a peak north of BCG 2. The bulk of the X-ray emission is also extended to the east, roughly between all three galaxy density peaks. 

In the range of $0.42 \leq z \leq 0.48$, the redshift distribution (\cref{fig:RMJ1043_redshifts}) is best-fit with a two-component GMM. The primary peak, with 56 members at $z=0.43$, encompasses BCGs 2--5 while BCG 1 is located with a secondary peak containing 27 members at $z=0.47$. Given the difference in redshift ($\sim\SI{7500}{\kms}$ between BCGs 1 and 2), we consider the cluster redshift to be $0.42 \leq z \leq 0.44$, and the cluster associated with BCG 1 to be in the background. Both luminosity-weighted and unweighted density contours of spectroscopic members show this background group has two primary concentrations, one just north of BCG 1 and the other near BCG 4. Neither of these regions overlap with the observed X-ray emission, and contributions to the cluster luminosity density are minimal. Within the cluster redshift range, the redshift distribution is consistent with a normal distribution by both KS ($D=0.21$, $p=0.85$) and AD ($A^2=0.37$, $p=0.34$) tests. A GMM fit across the entire redshift range of $z \leq 1.5$ identifies an additional background group with 20 members at $z=0.56$. This group is again divided into two primary concentrations, with one north of the cluster and one south, neither of which contaminate the observed X-rays. 

%%%%%%%%%%%%%%%%%%%%%%%%%% RMJ1043: Figure - Redshift Heatmap %%%%%%%%%%%%%%%%%%%%%%%%%%
\htmlfig{
\begin{figure}[t]
    \centering
    \includegraphics[clip, trim=0.0cm 0.75cm 0.0cm 0.0cm, width=\columnwidth]{Figures/RMJ_1043_redshifts_hist_hmap.jpg}
    \caption{Redshift distribution of RMJ1043. Top panel: histogram of spectroscopic redshifts, combining archival data within 10\arcmin\ of the cluster with new observations from DEIMOS. The inset highlights galaxies in the redshift range $0.42 \leq z \leq 0.44$, which are classified as cluster members. Bottom panel: spatial distribution of those same members.}
    \label{fig:RMJ1043_redshifts}
\end{figure}}

Dividing the cluster into three regions centered on BCGs 2, 4, and 5 (\cref{fig:RMJ1043_sub_1}) shows all three groups with small velocity dispersions (${\leq}\SI{370}{\kms}$) and ${<}\SI{150}{\kms}$ velocity difference between subcluster means (\cref{tab:RMJ1043_subclusters}), suggesting minimal interaction along the line of sight. However, the picture is complicated by a small handful of galaxies in Subcluster 2, including BCG 4, which are at a lower redshift than the primary distribution (a velocity difference of $\SI{1600}{\kms}$ between BCG 4 and the subcluster mean). We can see in the bottom panel of \cref{fig:RMJ1043_redshifts} that the lowest redshift members of the cluster are exclusively concentrated near BCG 4. If the cluster redshift range is defined $0.425 \leq z \leq 0.44$, removing the four low-redshift galaxies from the luminosity density eliminates the peak seen in the southeast. 

%%%%%%%%%%%%%%%%%%%%%%%%%% RMJ1043: Figure - Subclusters %%%%%%%%%%%%%%%%%%%%%%%%%%
\htmlfig{
\begin{figure}[t]
    \centering
    \includegraphics[clip, trim=0.0cm 1.0cm 0.0cm 0.0cm, width=\columnwidth]{Figures/RMJ_1043_subcluster_histograms_2_4_5.jpg}
    \caption{Subclustering analysis of RMJ1043. Bottom panel: spatial distribution of subcluster members, both spectroscopic and photometric, identified within regions defined by the bisector between BCG pairs and a radial distance of 2.5\arcmin\ from the BCG in each region. Top panel: redshift distribution of the spectroscopic members of those regions.}
    \label{fig:RMJ1043_sub_1}
\end{figure}

\begin{table}
    \centering
    \caption{RMJ1043 Subcluster Properties}
    \label{tab:RMJ1043_subclusters}
\begin{tabular*}{\columnwidth}{@{\extracolsep{\fill}}cccccc} \toprule\toprule
    Subcluster & $N$ & BCG & BCG $z$ & Mean $z$ & $\sigma_v$ [km s$^{-1}$] \\ \toprule
All & 61 & ... & ... & 0.4302 & 470 \\
1 & 13 & 2 & 0.430 & 0.4305 & 370  \\
2 & 14 & 4 & 0.422 & 0.4298 & 320  \\
3 & 11 & 5 & 0.432 & 0.4305 & 290  \\
\bottomrule
\end{tabular*}
\end{table}}

Given the disturbed X-ray morphology and peak offset, we classify this as a likely merger, possibly binary between Subclusters 1 and 3 with Subcluster 3 being stripped of its gas. Subcluster 2 could also be involved, and further spectroscopy would help in determining if there is a unique group at a lower redshift associated with BCG 4. Of the 12 clusters studied, RMJ1043 has the lowest richness, X-ray temperature, and X-ray luminosity at $121$ (\cref{tab:clusters}), $\SI{2.77}{\keV}$, and $\SI{0.56e44}{\ergs}$ (\cref{tab:XMM}), respectively. Both temperature and luminosity lie $2\sigma$ below expected values. Such low $T_X$ and $L_X$ disfavor the low impact parameter suggested by the gas-stripping of Subcluster 2. However, in addition to the lowest $\lambda$, $T_X$, and $L_X$ in our sample, RMJ1043 also has the lowest velocity dispersion at $\SI{470}{\kms}$ (\cref{tab:RMJ1043_subclusters}), indicating a less massive system than the richness would suggest.

RedMaPPer computes richness by first assigning galaxies a probability of being a cluster member based on a fit of the red sequence. The richness is then defined as a sum over those probabilities, weighted by radius and luminosity-dependent factors. The radius-weighting selects for members using a ``soft'' cut which penalizes galaxies outside of a richness-dependent scale radius of $\sim\SI{1}{\mpc}$. With BCG 1 having a probability of only 0.3327, again, the lowest in the sample, it is possible that the redMaPPer richness was influenced by an ambiguity in cluster location as well as interlopers near Subclusters 1 and 3. A lower richness would help to explain the discrepancy in $T_X$ and $L_X$, as well as the low velocity dispersion.

This is a complicated system that would benefit from further spectroscopy to untangle any structure along the line-of-sight. While it is likely to be a binary dissociative merger, these interlopers could prove prohibitive to building accurate mass models with weak lensing. 

%%%%%%%%%%%%%%%%%%%%%%%%%% RMJ1043: Table - Subclusters %%%%%%%%%%%%%%%%%%%%%%%%%%
\pdffig{
\begin{table}
    \centering
    \caption{RMJ1043 Subcluster Properties}
    \label{tab:RMJ1043_subclusters}
\begin{tabular*}{\columnwidth}{@{\extracolsep{\fill}}cccccc} \toprule\toprule
    Subcluster & $N$ & BCG & BCG $z$ & Mean $z$ & $\sigma_v$ [km s$^{-1}$] \\ \toprule
All & 61 & ... & ... & 0.4302 & 470 \\
1 & 13 & 2 & 0.430 & 0.4305 & 370  \\
2 & 14 & 4 & 0.422 & 0.4298 & 320  \\
3 & 11 & 5 & 0.432 & 0.4305 & 290  \\
\bottomrule
\end{tabular*}
\end{table}}
%%%%%%%%%%%%%%%%%%%%%%%%%%%%%%%%%%%%%%%%%%%%%%%%%%%%%%%%%%%%%%%%%%%%%%%%%%%%%%%%%%%%%%%%%%%%%%
%%%%%%%%%%%%%%%%%%%%%%%%%%%%%%%%%          RMJ 1219          %%%%%%%%%%%%%%%%%%%%%%%%%%%%%%%%%
%%%%%%%%%%%%%%%%%%%%%%%%%%%%%%%%%%%%%%%%%%%%%%%%%%%%%%%%%%%%%%%%%%%%%%%%%%%%%%%%%%%%%%%%%%%%%%

%%%%%%%%%%%%%%%%%%%%%%%%%% RMJ1219: Figure - Optical %%%%%%%%%%%%%%%%%%%%%%%%%%
\pdffig{
\begin{figure}[ht]
    \centering
    \includegraphics[clip, trim=0.0cm 1.25cm 0.0cm 0.0cm, width=\columnwidth]{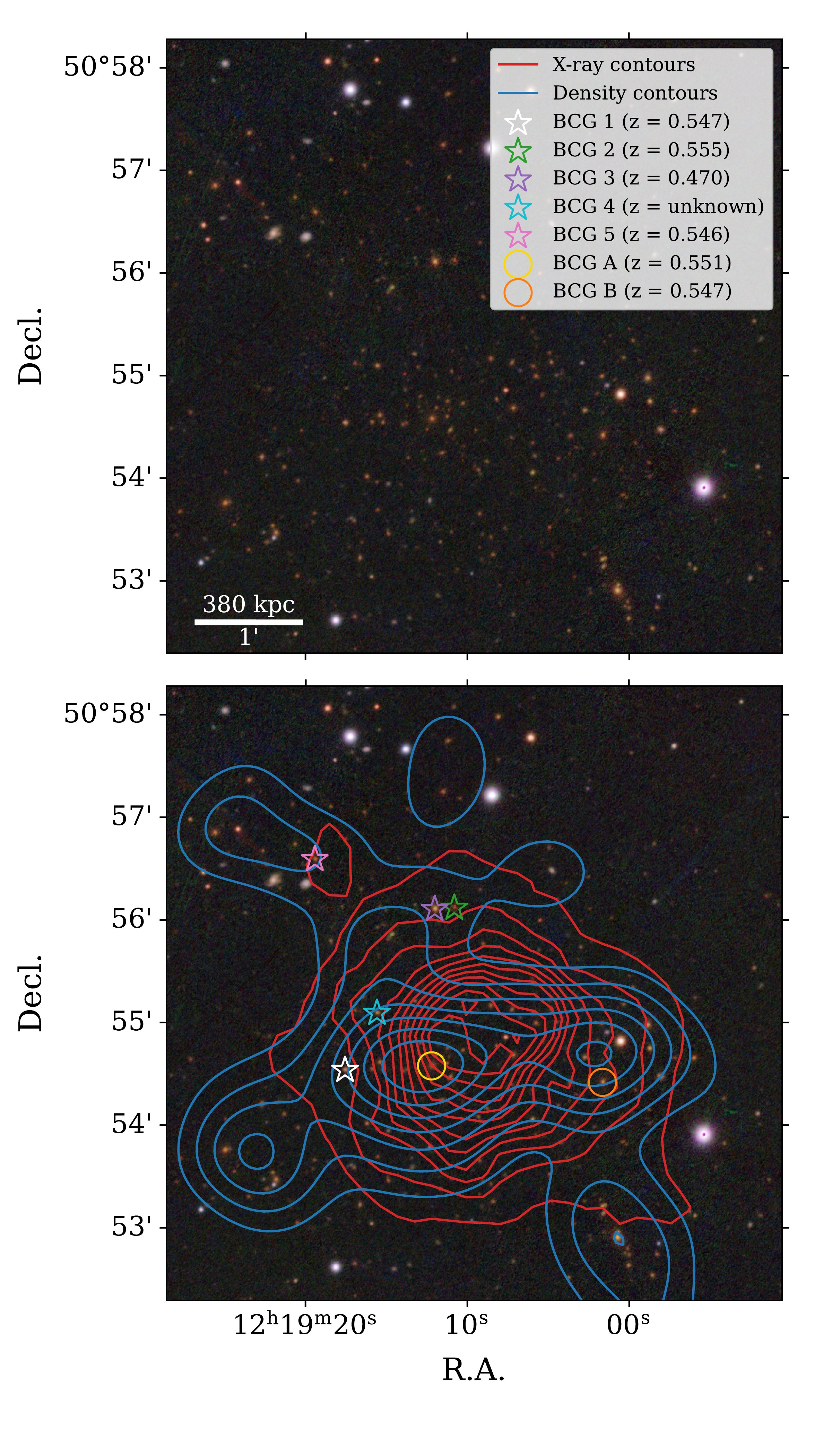}
    \caption{Optical image of RMJ1219. Top panel: 6\arcmin\ by 6\arcmin\ (2.28 by 2.28 Mpc) Pan-STARRS image. Bottom panel: red sequence density contours (blue), X-ray surface brightness contours (red), and BCG candidates identified by redMaPPer.}
    \label{fig:RMJ1219_optical}
\end{figure}}

%%%%%%%%%%%%%%%%%%%%%%%%%% RMJ1219: Table - BCGs %%%%%%%%%%%%%%%%%%%%%%%%%%
\pdffig{
\begin{table}[hb]
    \centering
    \caption{RMJ1219 BCG Information}
    \label{tab:RMJ1219_BCG}
    \tablenotetext{a}{DEIMOS (This work)}
    \tablenotetext{b}{SDSS DR18 \citep{almeida2023eighteenth}}
    \tablenotetext{c}{DESI Legacy Survey DR10 \citep{dey2019overview}}
\begin{tabular*}{\columnwidth}{@{\extracolsep{\fill}}cccccc} \toprule\toprule
    BCG & Probability & Redshift & $r$-mag\tablenotemark{\footnotesize{c}} & RA {[}deg{]} & Dec {[}deg{]} \\ \toprule
    1  & 0.5124 & 0.547\tablenotemark{\footnotesize{a}} & 20.07 & 184.82321 & 50.90911 \\ \midrule
    2  & 0.2050 & 0.555\tablenotemark{\footnotesize{b}} & 20.56 & 184.79510 & 50.93541 \\ \midrule
    3  & 0.1742 & 0.470\tablenotemark{\footnotesize{b}} & 18.80 & 184.80006 & 50.93522 \\ \midrule
    4  & 0.1032 & ...                                   & 20.37 & 184.81497 & 50.91838 \\ \midrule
    5  & 0.0052 & 0.546\tablenotemark{\footnotesize{b}} & 20.14 & 184.83102 & 50.94331 \\ \midrule
    A  & ...    & 0.551\tablenotemark{\footnotesize{a}} & 19.49 & 184.80083 & 50.90971 \\ \midrule
    B  & ...    & 0.547\tablenotemark{\footnotesize{a}} & 19.88 & 184.75687 & 50.90707 \\ \bottomrule
\printtablenotes{6}
\end{tabular*}
  \tablenotesreset
\end{table}}

%%%%%%%%%%%%%%%%%%%%%%%%%% RMJ1219: Figure - Gauss %%%%%%%%%%%%%%%%%%%%%%%%%%
\pdffig{
\begin{figure}[ht]
    \centering
    \includegraphics[clip, trim=0.0cm 0.25cm 0.0cm 0.0cm, width=\columnwidth]{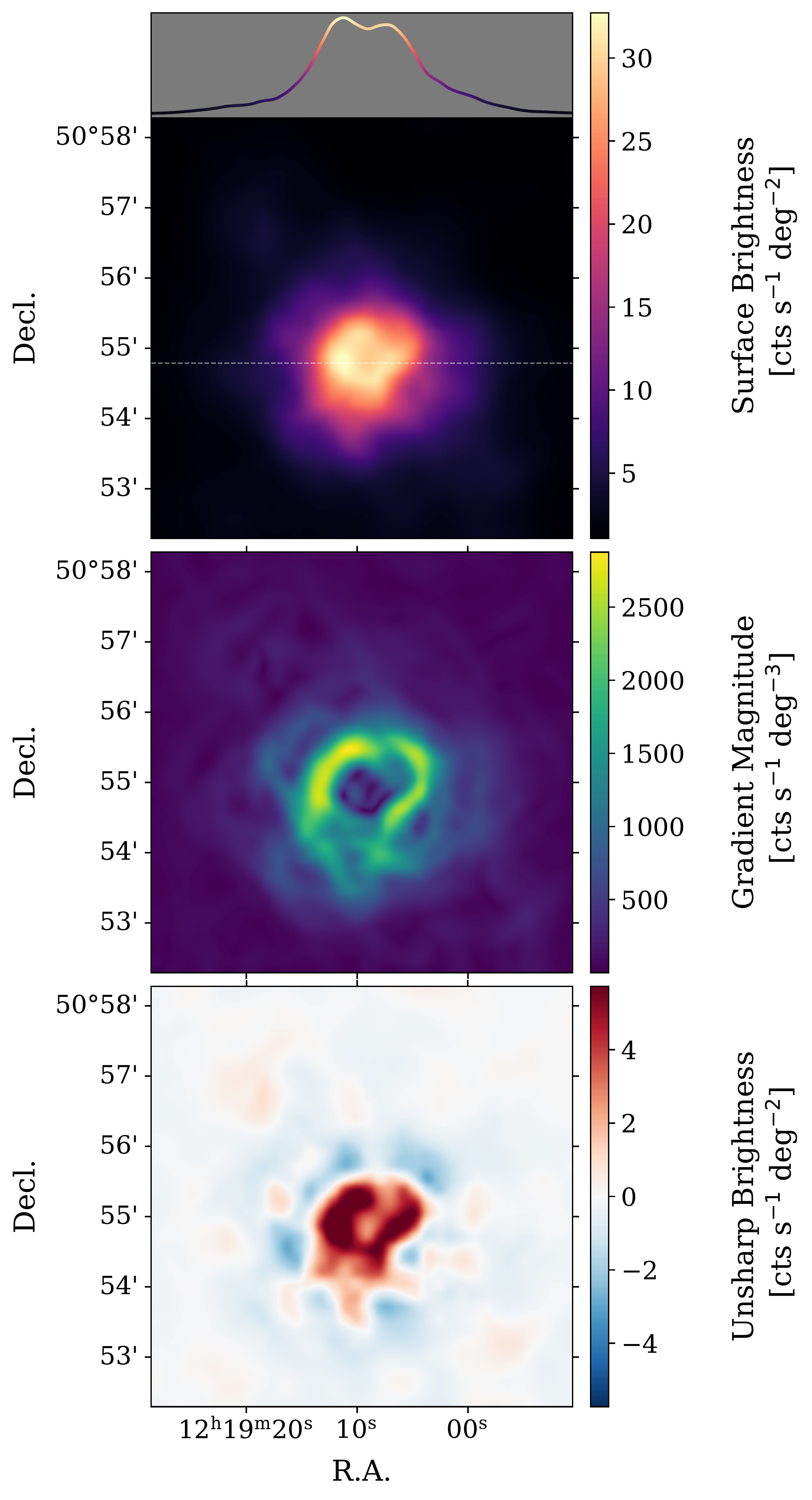}
    \caption{RMJ1219 X-ray features. Top panels: X-ray surface brightness smoothed with a 10\arcsec\ kernel and reprojected on an $800\times800$ pixel grid with the extracted 1D profile. The dashed line shows where the profile was extracted. Center panel: gradient magnitude of the brightness. Bottom panel: residual features after unsharp masking, where an image smoothed with a kernel of 50\arcsec\ is subtracted from the image smoothed at 10\arcsec.}
    \label{fig:RMJ1219_gauss}
\end{figure}}

%%%%%%%%%%%%%%%%%%%%%%%%%% RMJ1219: Figure - Redshift Heatmap %%%%%%%%%%%%%%%%%%%%%%%%%%
\pdffig{
\begin{figure}[t]
    \centering
    \includegraphics[clip, trim=0.0cm 0.75cm 0.0cm 0.0cm, width=\columnwidth]{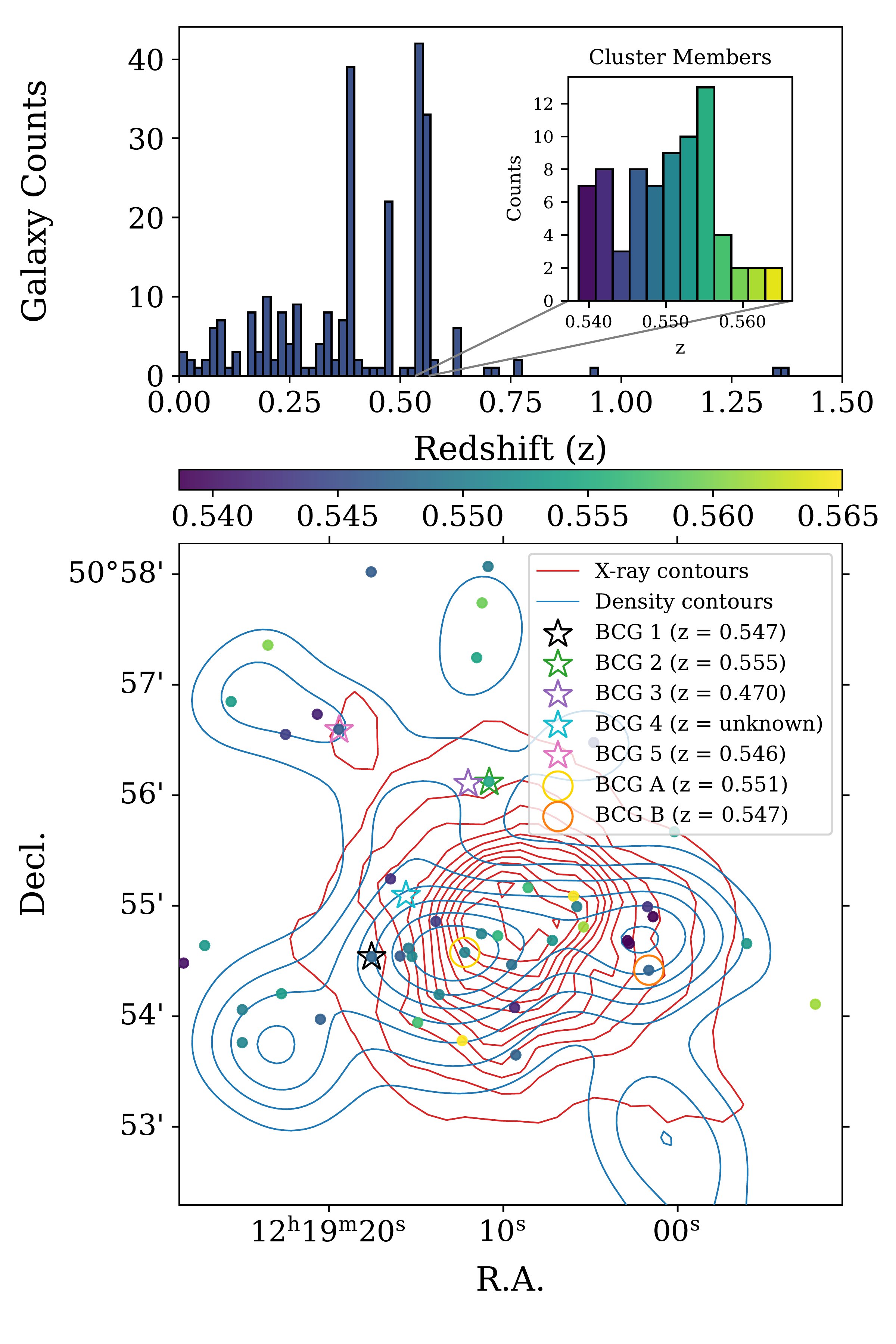}
    \caption{Redshift distribution of RMJ1219. Top panel: histogram of spectroscopic redshifts, combining archival data within 10\arcmin\ of the cluster with new observations from DEIMOS. The inset highlights galaxies in the redshift range $0.535 \leq z \leq 0.566$, which are classified as cluster members. Bottom panel: spatial distribution of those same members.}
    \label{fig:RMJ1219_redshifts}
\end{figure}}

\subsubsection{RM J121917.6+505432.8}\label{subsubsec:RMJ1219}

%%%%%%%%%%%%%%%%%%%%%%%%%% RMJ1219: Table - BCGs %%%%%%%%%%%%%%%%%%%%%%%%%%
\htmlfig{
\begin{table}[hb]
    \centering
    \caption{RMJ1219 BCG Information}
    \label{tab:RMJ1219_BCG}
    \tablenotetext{a}{DEIMOS (This work)}
    \tablenotetext{b}{SDSS DR18 \citep{almeida2023eighteenth}}
    \tablenotetext{c}{DESI Legacy Survey DR10 \citep{dey2019overview}}
\begin{tabular*}{\columnwidth}{@{\extracolsep{\fill}}cccccc} \toprule\toprule
    BCG & Probability & Redshift & $r$-mag\tablenotemark{\footnotesize{c}} & RA {[}deg{]} & Dec {[}deg{]} \\ \toprule
    1  & 0.5124 & 0.547\tablenotemark{\footnotesize{a}} & 20.07 & 184.82321 & 50.90911 \\ \midrule
    2  & 0.2050 & 0.555\tablenotemark{\footnotesize{b}} & 20.56 & 184.79510 & 50.93541 \\ \midrule
    3  & 0.1742 & 0.470\tablenotemark{\footnotesize{b}} & 18.80 & 184.80006 & 50.93522 \\ \midrule
    4  & 0.1032 & ...                                   & 20.37 & 184.81497 & 50.91838 \\ \midrule
    5  & 0.0052 & 0.546\tablenotemark{\footnotesize{b}} & 20.14 & 184.83102 & 50.94331 \\ \midrule
    A  & ...    & 0.551\tablenotemark{\footnotesize{a}} & 19.49 & 184.80083 & 50.90971 \\ \midrule
    B  & ...    & 0.547\tablenotemark{\footnotesize{a}} & 19.88 & 184.75687 & 50.90707 \\ \bottomrule
\printtablenotes{6}
\end{tabular*}
  \tablenotesreset
\end{table}}

%%%%%%%%%%%%%%%%%%%%%%%%%% RMJ1219: Figure - Optical %%%%%%%%%%%%%%%%%%%%%%%%%%
\htmlfig{
\begin{figure}[ht]
    \centering
    \includegraphics[clip, trim=0.0cm 1.25cm 0.0cm 0.0cm, width=\columnwidth]{Figures/RMJ_1219_2panel_optical_contours_vert.jpg}
    \caption{Optical image of RMJ1219. Top panel: 6\arcmin\ by 6\arcmin\ (2.28 by 2.28 Mpc) Pan-STARRS image. Bottom panel: red sequence density contours (blue), X-ray surface brightness contours (red), and BCG candidates identified by redMaPPer.}
    \label{fig:RMJ1219_optical}
\end{figure}}

%%%%%%%%%%%%%%%%%%%%%%%%%%%%%%%%%        RMJ 1219 Text       %%%%%%%%%%%%%%%%%%%%%%%%%%%%%%%%%

RMJ1219 has archival \CXO\ data (Observation Id. 21715, P.I. Kraft) and LOFAR analysis \citep{van2021lofar,botteon2022planck}, which were combined with this work as well as weak lensing from HST by \citet[][in preparation]{stancioli2025inprep} in a comprehensive analysis. All five BCGs identified by redMaPPer (four with a probability of $>0.10$) (\cref{tab:RMJ1219_BCG}) are located northeast of the cluster, making this largely a serendipitous discovery. The brightest of these, BCG 3, is at a significantly lower redshift than the cluster (0.47 versus ${\sim}0.55$ for the cluster). BCG 4 does not have a confirmed spectroscopic redshift and was not selected as cluster member in our fit of the red sequence. The true BCGs, labeled A and B, have $r$-band magnitudes of 19.49 and 19.88, respectively, and are each associated with one of the two overdensities of galaxies, shown in \cref{fig:RMJ1219_optical}, in the east and the west.

%%%%%%%%%%%%%%%%%%%%%%%%%% RMJ1219: Figure - Foregrounds %%%%%%%%%%%%%%%%%%%%%%%%%%
\pdffig{
\begin{figure}[t]
    \centering
    \includegraphics[clip, trim=0.0cm 0.25cm 0.0cm 0.0cm, width=\columnwidth]{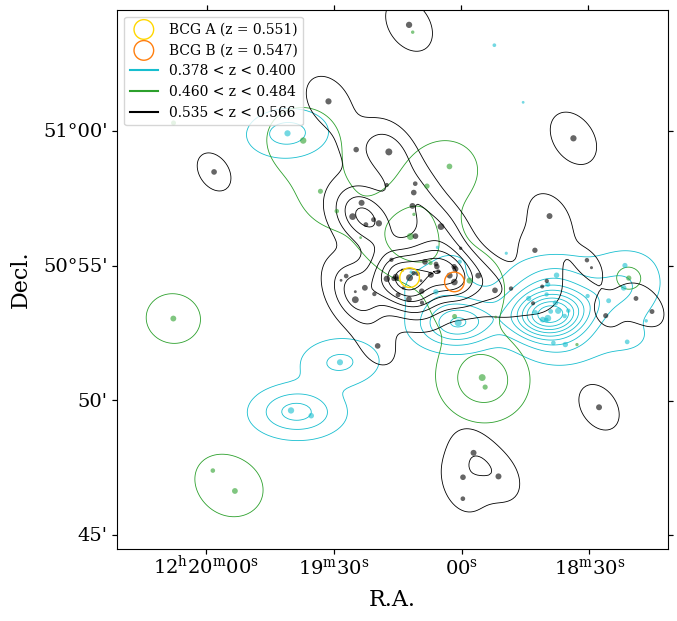}
    \caption{Distribution of foreground clusters in the full 10\arcmin\ field of RMJ1219. Contours show the luminosity-weighted density of spectroscopic members in each redshift bin. Individual galaxies are indicated by markers sized proportional to their luminosity.}
    \label{fig:RMJ1219_foregrounds}
\end{figure}

%%%%%%%%%%%%%%%%%%%%%%%%%% RMJ1219: Figure - Subclusters %%%%%%%%%%%%%%%%%%%%%%%%%%
\begin{figure}[h]
    \centering
    \includegraphics[clip, trim=0.0cm 1.0cm 0.0cm 0.0cm, width=0.99\columnwidth]{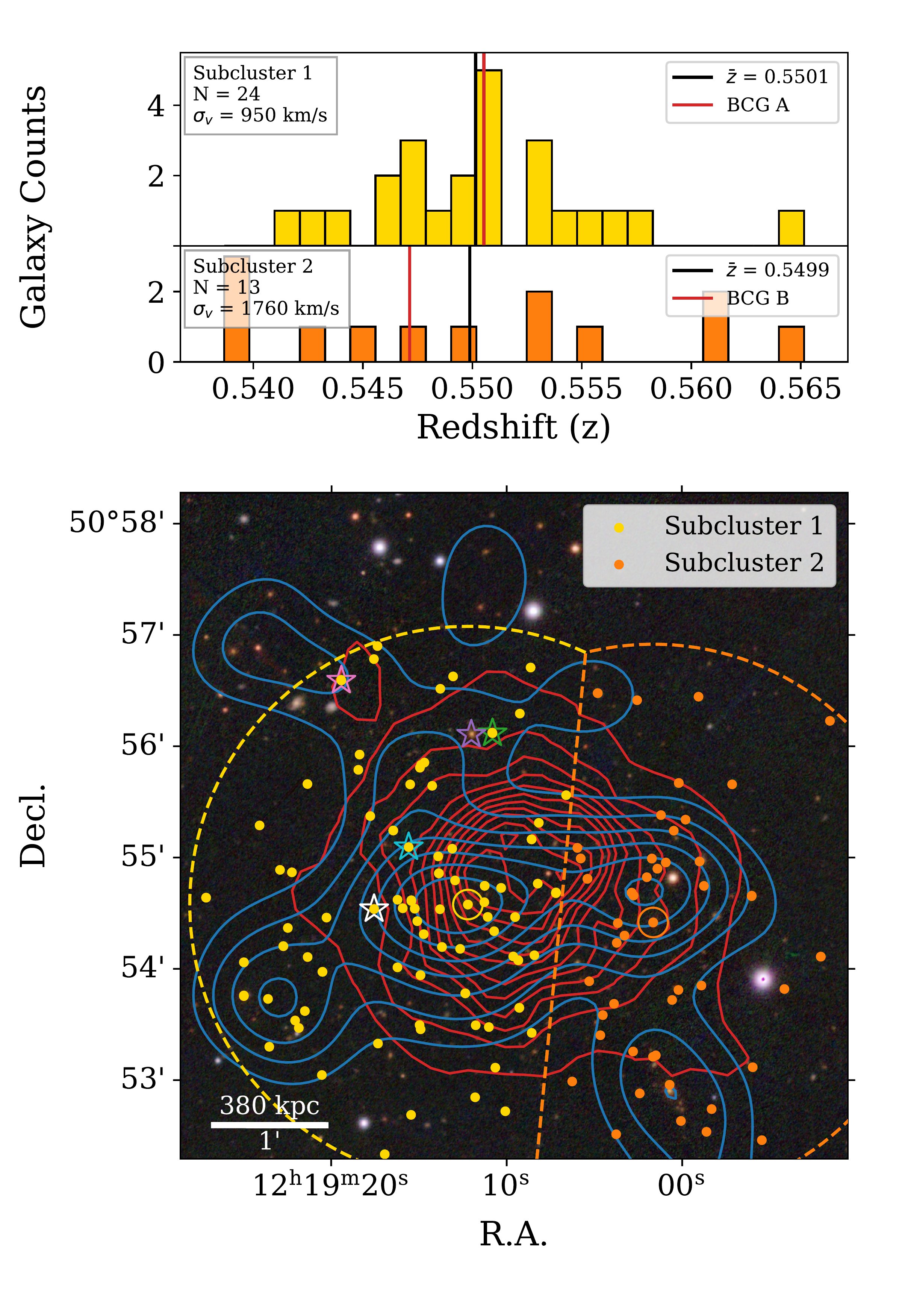}
    \caption{Subclustering analysis of RMJ1219. Bottom panel: spatial distribution of subcluster members, both spectroscopic and photometric, identified within regions defined by the bisector between BCG pairs and a radial distance of 2.5\arcmin\ from the BCG in each region. Top panel: redshift distribution of the spectroscopic members of those regions.}
    \label{fig:RMJ1219_sub_1}
\end{figure}}

The X-ray surface brightness is strongly disturbed, with a crescent-shaped central structure between the two optical subclusters. This peak is roughly bimodal, as seen in the one-dimensional profile extracted across the cluster's center (top panel of \cref{fig:RMJ1219_gauss}), where two local maxima are resolved within the broader emission envelope. The gradient map (center panel) highlights sharp brightness discontinuities along the eastern and western edges of the emission. These features are further emphasized in the unsharp-mask image (bottom panel). Both the location and morphology of the X-ray emission provide strong evidence of a dissociative merger.

%%%%%%%%%%%%%%%%%%%%%%%%%% RMJ1219: Figure - Gauss %%%%%%%%%%%%%%%%%%%%%%%%%%
\htmlfig{
\begin{figure}[ht]
    \centering
    \includegraphics[clip, trim=0.0cm 0.25cm 0.0cm 0.0cm, width=\columnwidth]{Figures/RMJ1219_gaussian_gradient.jpg}
    \caption{RMJ1219 X-ray features. Top panels: X-ray surface brightness smoothed with a 10\arcsec\ kernel and reprojected on an $800\times800$ pixel grid with the extracted 1D profile. The dashed line shows where the profile was extracted. Center panel: gradient magnitude of the brightness. Bottom panel: residual features after unsharp masking, where an image smoothed with a kernel of 50\arcsec\ is subtracted from the image smoothed at 10\arcsec.}
    \label{fig:RMJ1219_gauss}
\end{figure}}

The redshift histogram, shown in the top panel of \cref{fig:RMJ1219_redshifts}, exhibits two large foreground structures, both confirmed with a GMM fit. The spatial distribution of these structures is given in \cref{fig:RMJ1219_foregrounds}, with luminosity-weighted density contours created from spectroscopic members with \texttt{scipy.stats.gaussian\_kde} using a kernel factor of 0.2. The first, with 55 spectroscopic members at $0.378 \leq z \leq 0.40$, is concentrated southwest of the majority of the X-ray emission, 1.2 Mpc southwest of BCG B. The observed X-ray and luminosity density are not affected by this cluster, but analysis further from the cluster center could be impacted. The second, with 23 members at $0.46 \leq z \leq 0.484$, is distributed broadly across the cluster region. This cluster could produce some foreground contamination, particularly in the galaxy density where the luminosity peaks in the southwest and northeast are likely artifacts of interlopers in the fit of the CMD. However, with only 22 spectroscopic members across such a broad region, the location of the X-ray surface brightness peak is likely not materially affected.

 %%%%%%%%%%%%%%%%%%%%%%%%%% RMJ1219: Table - Subclusters %%%%%%%%%%%%%%%%%%%%%%%%%%
 \pdffig{
\begin{table}[b]
    \centering
    \caption{RMJ1219 Subcluster Properties}
    \label{tab:RMJ1219_subclusters}
\begin{tabular*}{\columnwidth}{@{\extracolsep{\fill}}cccccc} \toprule\toprule
    Subcluster & $N$ & BCG & BCG $z$ & Mean $z$ & $\sigma_v$ [km s$^{-1}$] \\ \toprule
All & 75 & ... & ... & 0.5503 & 1290 \\
1 & 24 & A & 0.551 & 0.5501 & 950 \\
2 & 13 & B & 0.547 & 0.5499 & 1760 \\
\bottomrule
\end{tabular*}
\end{table}}

%%%%%%%%%%%%%%%%%%%%%%%%%% RMJ1219: Figure - Redshift Heatmap %%%%%%%%%%%%%%%%%%%%%%%%%%
\htmlfig{
\begin{figure}[t]
    \centering
    \includegraphics[clip, trim=0.0cm 0.75cm 0.0cm 0.0cm, width=\columnwidth]{Figures/RMJ_1219_redshifts_hist_hmap.jpg}
    \caption{Redshift distribution of RMJ1219. Top panel: histogram of spectroscopic redshifts, combining archival data within 10\arcmin\ of the cluster with new observations from DEIMOS. The inset highlights galaxies in the redshift range $0.535 \leq z \leq 0.566$, which are classified as cluster members. Bottom panel: spatial distribution of those same members.}
    \label{fig:RMJ1219_redshifts}
\end{figure}

%%%%%%%%%%%%%%%%%%%%%%%%%% RMJ1219: Figure - Foregrounds %%%%%%%%%%%%%%%%%%%%%%%%%%
\begin{figure}[t]
    \centering
    \includegraphics[clip, trim=0.0cm 0.25cm 0.0cm 0.0cm, width=\columnwidth]{Figures/RMJ_1219_foreground_gals_weight.jpg}
    \caption{Distribution of foreground clusters in the full 10\arcmin\ field of RMJ1219. Contours show the luminosity-weighted density of spectroscopic members in each redshift bin. Individual galaxies are indicated by markers sized proportional to their luminosity.}
    \label{fig:RMJ1219_foregrounds}
\end{figure}}

Within the cluster redshift range, $0.535 \leq z \leq 0.566$, there appears to be an excess of galaxies at the low end of the redshift distribution. However, a one-component fit is preferred and consistent with a Gaussian distribution ($D=0.078$, $p=0.73$; $A^2=0.59$, $p=0.12$). The spatial distribution of redshifts within the cluster region (\cref{fig:RMJ1219_redshifts}) shows some grouping of low redshift members near BCG B. We divide the cluster into regions defined by the bisector between BCGs A and B and a 2.5\arcmin\ radius from each (\cref{fig:RMJ1219_sub_1}). Subcluster 1 has a velocity dispersion of $\SI{950}{\kms}$ while Subcluster 2 is more widely distributed across the full redshift range with a dispersion of $\SI{1760}{\kms}$ (falsely elevated due to the low-number statistics)(\cref{tab:RMJ1219_subclusters}). Both are at nearly the same redshift with a LOS velocity difference of only $\SI{40}{\kms}$, indicating a merger near the plane of the sky.

%%%%%%%%%%%%%%%%%%%%%%%%%% RMJ1219: Figure - Subclusters %%%%%%%%%%%%%%%%%%%%%%%%%%
\htmlfig{
\begin{figure}[h]
    \centering
    \includegraphics[clip, trim=0.0cm 1.0cm 0.0cm 0.0cm, width=0.99\columnwidth]{Figures/RMJ_1219_subcluster_histograms_6_7.jpg}
    \caption{Subclustering analysis of RMJ1219. Bottom panel: spatial distribution of subcluster members, both spectroscopic and photometric, identified within regions defined by the bisector between BCG pairs and a radial distance of 2.5\arcmin\ from the BCG in each region. Top panel: redshift distribution of the spectroscopic members of those regions.}
    \label{fig:RMJ1219_sub_1}
\end{figure}}

 %%%%%%%%%%%%%%%%%%%%%%%%%% RMJ1219: Table - Subclusters %%%%%%%%%%%%%%%%%%%%%%%%%%
\htmlfig{
\begin{table}[b]
    \centering
    \caption{RMJ1219 Subcluster Properties}
    \label{tab:RMJ1219_subclusters}
\begin{tabular*}{\columnwidth}{@{\extracolsep{\fill}}cccccc} \toprule\toprule
    Subcluster & $N$ & BCG & BCG $z$ & Mean $z$ & $\sigma_v$ [km s$^{-1}$] \\ \toprule
All & 75 & ... & ... & 0.5503 & 1290 \\
1 & 24 & A & 0.551 & 0.5501 & 950 \\
2 & 13 & B & 0.547 & 0.5499 & 1760 \\
\bottomrule
\end{tabular*}
\end{table}}

Given a richness of 165 (\cref{tab:clusters}), the X-ray temperature and luminosity of $ \SI{7.27}{\keV}$ and $\SI{5.32e44}{\ergs}$ (\cref{tab:XMM}) are both slightly elevated from expected scaling relations. Weak lensing results from \citet[][in preparation]{stancioli2025inprep} confirm a dissociative merger between east and west subclusters. Consistent with this interpretation, LOFAR observations indicate the presence of both a radio relic and radio halo \citep{van2021lofar,botteon2022planck}. Overall, this is an interesting system with clear evidence of merging activity and a great candidate for additional study.

%%%%%%%%%%%%%%%%%%%%%%%%%%%%%%%%%%%%%%%%%%%%%%%%%%%%%%%%%%%%%%%%%%%%%%%%%%%%%%%%%%%%%%%%%%%%%%
%%%%%%%%%%%%%%%%%%%%%%%%%%%%%%%%%          RMJ 1257          %%%%%%%%%%%%%%%%%%%%%%%%%%%%%%%%%
%%%%%%%%%%%%%%%%%%%%%%%%%%%%%%%%%%%%%%%%%%%%%%%%%%%%%%%%%%%%%%%%%%%%%%%%%%%%%%%%%%%%%%%%%%%%%%

\htmlfig{
\subsubsection{RM J125725.9+365429.4}\label{subsubsec:RMJ1257}}

%%%%%%%%%%%%%%%%%%%%%%%%%% RMJ1257: Table - BCGs %%%%%%%%%%%%%%%%%%%%%%%%%%
\begin{table}[hb]
    \centering
    \caption{RMJ1257 BCG Information}
    \label{tab:RMJ1257_BCG}
    \tablenotetext{a}{SDSS DR18 \citep{almeida2023eighteenth}}
    \tablenotetext{b}{DEIMOS (This work)}
    \tablenotetext{c}{DESI DR1 \citep{abdul2025data}}
    \tablenotetext{d}{DESI Legacy Survey DR10 \citep{dey2019overview}}
\begin{tabular*}{\columnwidth}{@{\extracolsep{\fill}}cccccc} \toprule\toprule
    BCG & Probability & Redshift & $r$-mag\tablenotemark{\footnotesize{d}} & RA {[}deg{]} & Dec {[}deg{]} \\ \toprule
    1  & 0.9473 & 0.518\tablenotemark{\footnotesize{a}} & 19.13 & 194.35800 & 36.90816 \\ \midrule
    2  & 0.0367 & 0.527\tablenotemark{\footnotesize{b}} & 19.81 & 194.36916 & 36.92155 \\ \midrule
    3  & 0.0129 & 0.504\tablenotemark{\footnotesize{a}} & 19.76 & 194.36880 & 36.91972 \\ \midrule
    4  & 0.0029 & 0.531\tablenotemark{\footnotesize{b}} & 19.83 & 194.33660 & 36.90151 \\ \midrule
    5  & 0.0002 & 0.522\tablenotemark{\footnotesize{c}} & 20.30 & 194.35435 & 36.90672 \\ \midrule
    A  & ... & 0.527\tablenotemark{\footnotesize{a}} & 19.71 & 194.34135 & 36.91943 \\ \bottomrule
\printtablenotes{6}
\end{tabular*}
  \tablenotesreset
\end{table}

\pdffig{
\subsubsection{RM J125725.9+365429.4}\label{subsubsec:RMJ1257}}

%%%%%%%%%%%%%%%%%%%%%%%%%% RMJ1257: Figure - Optical %%%%%%%%%%%%%%%%%%%%%%%%%%
\begin{figure}[t]
    \centering
    \includegraphics[clip, trim=0.0cm 1.25cm 0.0cm 0.0cm, width=\columnwidth]{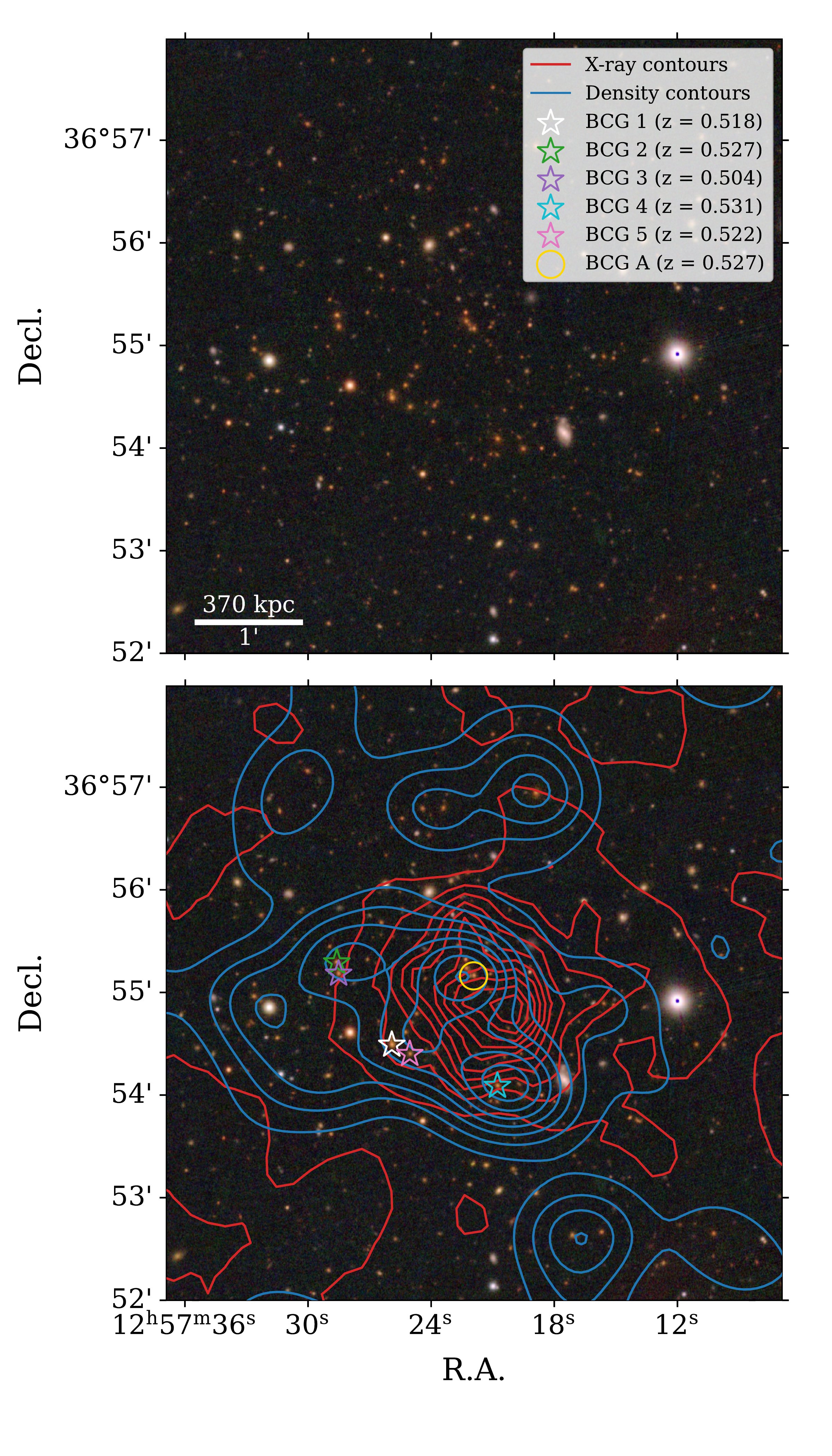}
    \caption{Optical image of RMJ1257. Top panel: 6\arcmin\ by 6\arcmin\ (2.22 by 2.22 Mpc) Pan-STARRS image. Bottom panel: red sequence density contours (blue), X-ray surface brightness contours (red), and BCG candidates identified by redMaPPer.}
    \label{fig:RMJ1257_optical}
\end{figure}

%%%%%%%%%%%%%%%%%%%%%%%%%% RMJ1257: Figure - Redshift Heatmap %%%%%%%%%%%%%%%%%%%%%%%%%%
\pdffig{
\begin{figure}[t]
    \centering
    \includegraphics[clip, trim=0.0cm 0.75cm 0.0cm 0.0cm, width=\columnwidth]{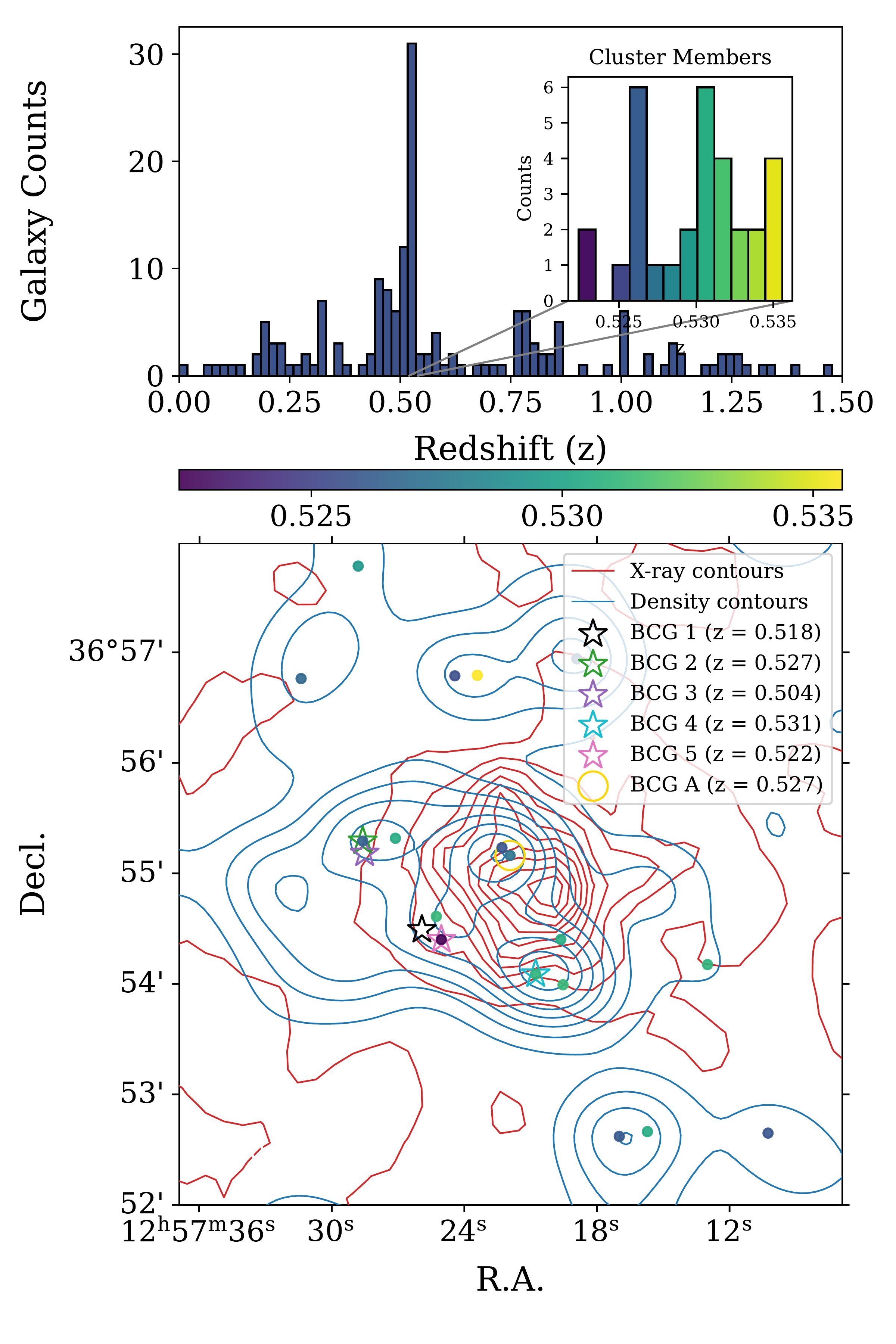}
    \caption{Redshift distribution of RMJ1257. Top panel: histogram of spectroscopic redshifts, combining archival data within 10\arcmin\ of the cluster with new observations from DEIMOS. The inset highlights galaxies in the redshift range $0.52 \leq z \leq 0.54$, which are classified as cluster members. Bottom panel: spatial distribution of those same members.}
    \label{fig:RMJ1257_redshifts}
\end{figure}

%%%%%%%%%%%%%%%%%%%%%%%%%% RMJ1257: Figure - Subclusters %%%%%%%%%%%%%%%%%%%%%%%%%%
\begin{figure}[t]
    \centering
    \includegraphics[clip, trim=0.0cm 1.0cm 0.0cm 0.0cm, width=\columnwidth]{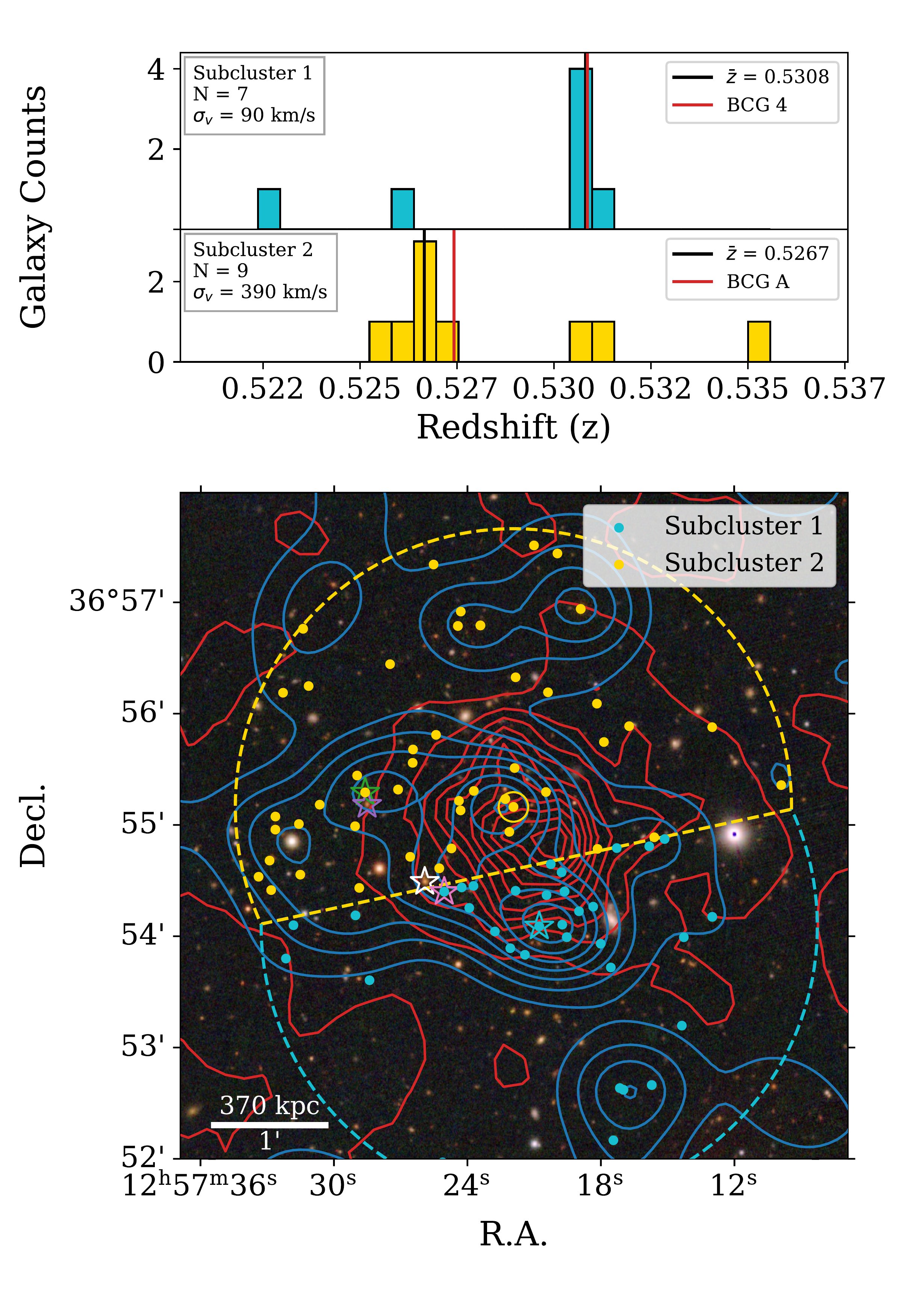}
    \caption{Subclustering analysis of RMJ1257. Bottom panel: spatial distribution of subcluster members, both spectroscopic and photometric, identified within regions defined by the bisector between BCG pairs and a radial distance of 2.5\arcmin\ from each BCG. Top panel: redshift distribution of the spectroscopic members of those regions.}
    \label{fig:RMJ1257_sub_1}
\end{figure}}

%%%%%%%%%%%%%%%%%%%%%%%%%%%%%%%%%        RMJ 1257 Text       %%%%%%%%%%%%%%%%%%%%%%%%%%%%%%%%%
RMJ1257 is another complicated cluster. The BCGs identified by redMaPPer form roughly along a line ${\sim}\SI{400}{\kpc}$ east of the peak in the X-ray emission, \cref{fig:RMJ1257_optical}. We identified an additional bright cluster member, which we label BCG A, northeast of the X-ray peak, with an r-magnitude of 19.71, making it brighter than all but BCG 1 (\cref{tab:RMJ1257_BCG}). Luminosity density contours show two primary peaks located on BCGs A and 4, with the central peak extending towards BCGs 1 and 5. There is also an additional, smaller, peak located at BCGs 2/3 (\cref{fig:RMJ1257_optical}).

The redshift distribution, \cref{fig:RMJ1257_redshifts}, defines cluster members from $0.52 \leq z \leq 0.54$. In this range, a one-component fit is preferred with a BIC of -254 (two-component BIC of -247) and consistent with Gaussianity ($D=0.12$, $p=0.72$; $A^2=0.45$, $p=0.26$). However, this excludes BCGs 1 and 3. If the range is extended down to $z=0.50$ to capture those members, a two-component GMM fit ($\text{BIC}=-289$, $\Delta\text{BIC}=18$) is preferred with 13 galaxies at $z=0.51$ and 30 at $0.53$. The lower redshift group is spread widely through the entire 10\arcmin\ field of view, but does have a handful of members concentrated near BCG 3. With a velocity difference of ${\sim}\SI{5000}{\kms}$ between BCG 3 and the primary cluster, we consider this to be a foreground group. There is an additional foreground group at $z=0.49$ with 16 members, also broadly covering the region, with some clustering near BCG 4. 

%%%%%%%%%%%%%%%%%%%%%%%%%% RMJ1257: Figure - Redshift Heatmap %%%%%%%%%%%%%%%%%%%%%%%%%%
\htmlfig{
\begin{figure}[t]
    \centering
    \includegraphics[clip, trim=0.0cm 0.75cm 0.0cm 0.0cm, width=\columnwidth]{Figures/RMJ_1257_redshifts_hist_hmap.jpg}
    \caption{Redshift distribution of RMJ1257. Top panel: histogram of spectroscopic redshifts, combining archival data within 10\arcmin\ of the cluster with new observations from DEIMOS. The inset highlights galaxies in the redshift range $0.52 \leq z \leq 0.54$, which are classified as cluster members. Bottom panel: spatial distribution of those same members.}
    \label{fig:RMJ1257_redshifts}
\end{figure}}

Given the location of the X-ray peak and primary density peaks at BCG A and BCG 4, we divide the cluster into subclusters centered on those two BCGs (\cref{fig:RMJ1257_sub_1}). There is differentiation in redshift between the two subclusters, with a velocity difference of \SI{820}{\kms} between subcluster means (\cref{tab:RMJ1257_subclusters}). However, with only 7 and 9 members in each group, we cannot draw any strong conclusions. New DEIMOS spectroscopy was limited to a single mask, adding only 39 new redshifts. RMJ1257 has the lowest number of both spectroscopic and total cluster members in the sample.

%%%%%%%%%%%%%%%%%%%%%%%%%% RMJ1257: Figure - Subclusters %%%%%%%%%%%%%%%%%%%%%%%%%%
\htmlfig{
\begin{figure}[t]
    \centering
    \includegraphics[clip, trim=0.0cm 1.0cm 0.0cm 0.0cm, width=\columnwidth]{Figures/RMJ_1257_subcluster_histograms_4_6.jpg}
    \caption{Subclustering analysis of RMJ1257. Bottom panel: spatial distribution of subcluster members, both spectroscopic and photometric, identified within regions defined by the bisector between BCG pairs and a radial distance of 2.5\arcmin\ from each BCG. Top panel: redshift distribution of the spectroscopic members of those regions.}
    \label{fig:RMJ1257_sub_1}
\end{figure}

\begin{table}
    \centering
    \caption{RMJ1257 Subcluster Properties}
    \label{tab:RMJ1257_subclusters}
\begin{tabular*}{\columnwidth}{@{\extracolsep{\fill}}cccccc} \toprule\toprule
    Subcluster & $N$ & BCG & BCG $z$ & Mean $z$ & $\sigma_v$ [km s$^{-1}$] \\ \toprule
All & 31 & ... & ... & 0.5301 & 720 \\
1 & 7 & 4 & 0.531 & 0.5308 & \phantom{0}90  \\
2 & 9 & A & 0.527 & 0.5267 & 390  \\
\bottomrule
\end{tabular*}
\end{table}}

RMJ1257 has the second lowest X-ray temperature in our sample at $\SI{4.04}{\keV}$, $1.2\sigma$ below the expected value of $\SI{5.71}{\keV}$, while its X-ray luminosity of $\SI{3.04e44}{\ergs}$ slightly exceeds the expectation of $\SI{2.44e44}{\ergs}$ (\cref{tab:XMM}). RMJ1257 had the lowest \XMM\ exposure time of any cluster in our sample with 14 ks, which was further diminished by extensive proton flaring, especially on the PN detector (only 1.08 ks of usable exposure, which we drop from our spectral fitting).

Given the displaced X-ray peak between BCGs 4 and A, we classify this as a post-collision merger, with the BCGs separated by $\SI{410}{\kpc}$. The X-ray peak is $\SI{300}{\kpc}$ from BCG 4 and \SI{225}{\kpc} from BCG A, offset from the presumed merger axis by ${\sim}\SI{150}{\kpc}$. However, the lack of data gives us low confidence in this assessment and we classify this as likely binary.  Additional spectroscopy (or weak lensing) would help to better define the subclusters involved while additional X-ray imaging is needed to compensate for the high flaring in this data. If the observed morphology is confirmed, RMJ1257 has the potential to be a promising dissociative merger.

%%%%%%%%%%%%%%%%%%%%%%%%%% RMJ1257: Table - Subclusters %%%%%%%%%%%%%%%%%%%%%%%%%%
\pdffig{
\begin{table}
    \centering
    \caption{RMJ1257 Subcluster Properties}
    \label{tab:RMJ1257_subclusters}
\begin{tabular*}{\columnwidth}{@{\extracolsep{\fill}}cccccc} \toprule\toprule
    Subcluster & $N$ & BCG & BCG $z$ & Mean $z$ & $\sigma_v$ [km s$^{-1}$] \\ \toprule
All & 31 & ... & ... & 0.5301 & 720 \\
1 & 7 & 4 & 0.531 & 0.5308 & \phantom{0}90  \\
2 & 9 & A & 0.527 & 0.5267 & 390  \\
\bottomrule
\end{tabular*}
\end{table}}

%%%%%%%%%%%%%%%%%%%%%%%%%%%%%%%%%%%%%%%%%%%%%%%%%%%%%%%%%%%%%%%%%%%%%%%%%%%%%%%%%%%%%%%%%%%%%%
%%%%%%%%%%%%%%%%%%%%%%%%%%%%%%%%%          RMJ 1327          %%%%%%%%%%%%%%%%%%%%%%%%%%%%%%%%%
%%%%%%%%%%%%%%%%%%%%%%%%%%%%%%%%%%%%%%%%%%%%%%%%%%%%%%%%%%%%%%%%%%%%%%%%%%%%%%%%%%%%%%%%%%%%%%
\subsubsection{RM J132724.2+534656.5}\label{subsubsec:RMJ1327}

%%%%%%%%%%%%%%%%%%%%%%%%%% RMJ1327: Table - BCGs %%%%%%%%%%%%%%%%%%%%%%%%%%
\htmlfig{
\begin{table}
    \centering
    \caption{RMJ1327 BCG Information}
    \label{tab:RMJ1327_BCG}
    \tablenotetext{a}{DESI DR1 \citep{abdul2025data}}
    \tablenotetext{b}{SDSS DR18 \citep{almeida2023eighteenth}}
    \tablenotetext{c}{DEIMOS (This work)}
    \tablenotetext{d}{DESI Legacy Survey DR10 \citep{dey2019overview}}
\begin{tabular*}{\columnwidth}{@{\extracolsep{\fill}}cccccc} \toprule\toprule
    BCG & Probability & Redshift & $r$-mag\tablenotemark{\footnotesize{d}} & RA {[}deg{]} & Dec {[}deg{]} \\ \toprule
    1  & 0.9436 & 0.411\tablenotemark{\footnotesize{a}} & 18.90 & 201.85063 & 53.78235 \\ \midrule
    2  & 0.0484 & 0.390\tablenotemark{\footnotesize{a}} & 19.04 & 201.73846 & 53.81966 \\ \midrule
    3  & 0.0029 & 0.393\tablenotemark{\footnotesize{a}} & 18.26 & 201.76472 & 53.83118 \\ \midrule
    4  & 0.0026 & 0.396\tablenotemark{\footnotesize{b}} & 18.82 & 201.68665 & 53.86797 \\ \midrule
    5  & 0.0025 & 0.409\tablenotemark{\footnotesize{a}} & 19.12 & 201.85095 & 53.77650 \\ \midrule
    A  & ... & 0.414\tablenotemark{\footnotesize{a}} & 18.57 & 201.87647 & 53.79540 \\ \midrule
    B  & ... & 0.394\tablenotemark{\footnotesize{c}} & 19.28 & 201.73490 & 53.86985 \\ \bottomrule
\printtablenotes{6}
\end{tabular*}
  \tablenotesreset
\end{table}}

%%%%%%%%%%%%%%%%%%%%%%%%%% RMJ1327: Figure - Optical %%%%%%%%%%%%%%%%%%%%%%%%%%
\begin{figure}[t]
    \centering
    \includegraphics[clip, trim=0.0cm 1.25cm 0.0cm 0.0cm, width=\columnwidth]{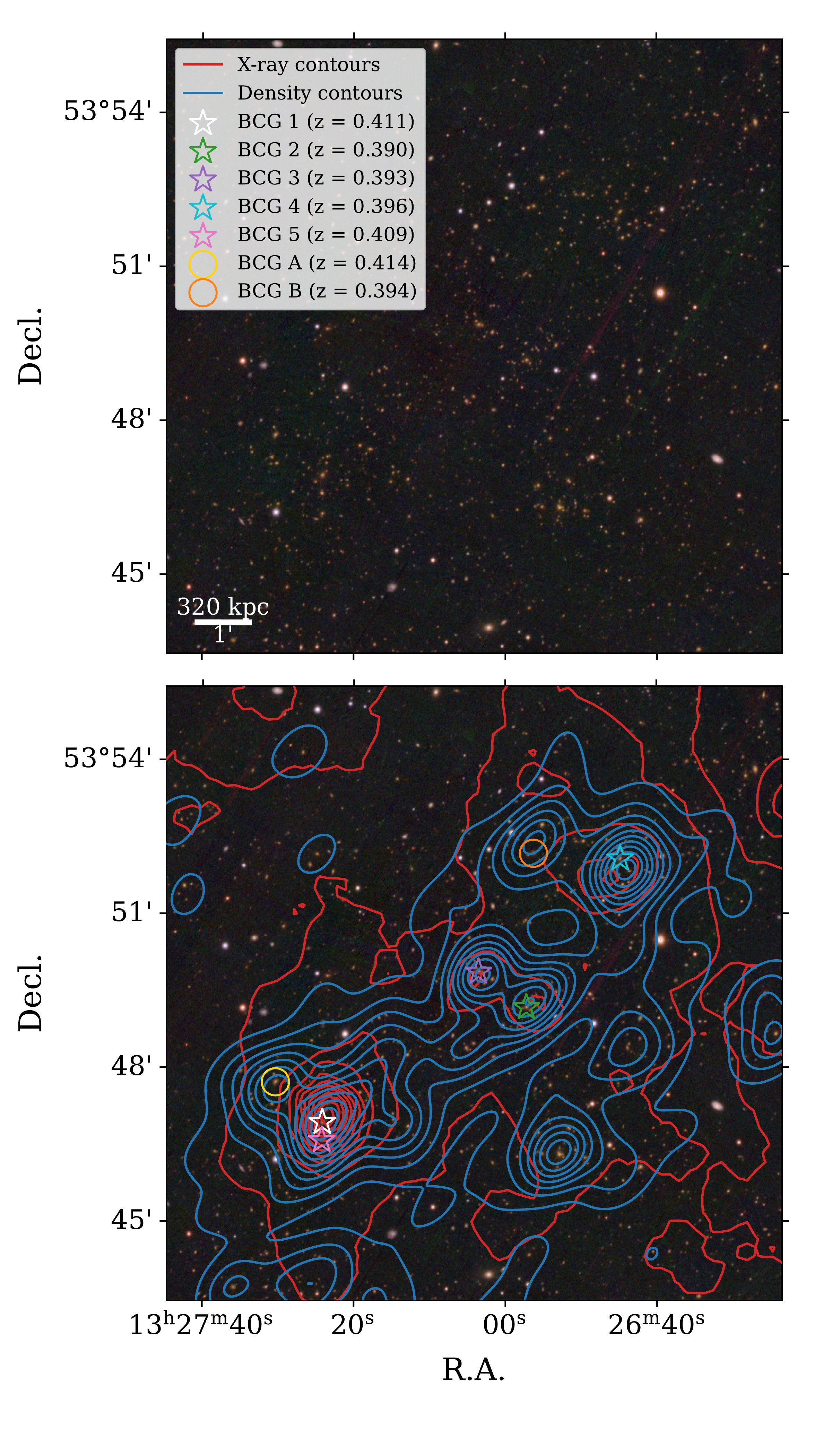}
    \caption{Optical image of RMJ1327. Top panel: 12\arcmin\ by 12\arcmin\ (3.84 by 3.84 Mpc) Pan-STARRS image. Bottom panel: red sequence density contours (blue), X-ray surface brightness contours (red), and BCG candidates identified by redMaPPer.}
    \label{fig:RMJ1327_optical}
\end{figure}

%%%%%%%%%%%%%%%%%%%%%%%%%% RMJ1327: Figure - Redshift Heatmap %%%%%%%%%%%%%%%%%%%%%%%%%%
\pdffig{
\begin{figure}[t]
    \centering
    \includegraphics[clip, trim=0.0cm 0.75cm 0.0cm 0.0cm, width=\columnwidth]{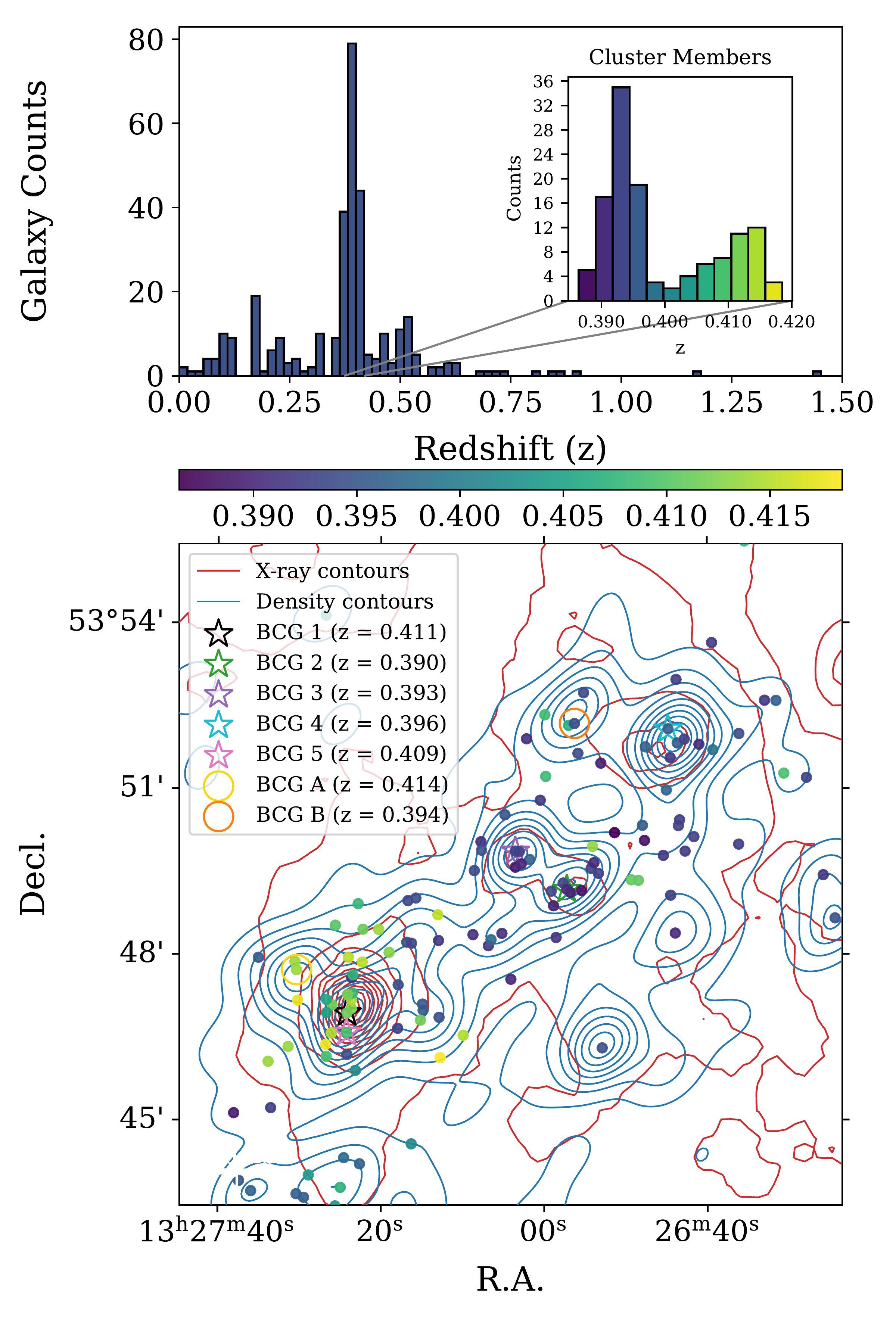}
    \caption{Redshift distribution of RMJ1327. Top panel: histogram of spectroscopic redshifts, combining archival data within 10\arcmin\ of the cluster with new observations from DEIMOS. The inset highlights galaxies in the redshift range $0.38 \leq z \leq 0.42$, which are classified as cluster members. Bottom panel: spatial distribution of those same members.}
    \label{fig:RMJ1327_redshifts}
\end{figure}

%%%%%%%%%%%%%%%%%%%%%%%%%% RMJ1327: Table - BCGs %%%%%%%%%%%%%%%%%%%%%%%%%%
\begin{table}
    \centering
    \caption{RMJ1327 BCG Information}
    \label{tab:RMJ1327_BCG}
    \tablenotetext{a}{DESI DR1 \citep{abdul2025data}}
    \tablenotetext{b}{SDSS DR18 \citep{almeida2023eighteenth}}
    \tablenotetext{c}{DEIMOS (This work)}
    \tablenotetext{d}{DESI Legacy Survey DR10 \citep{dey2019overview}}
\begin{tabular*}{\columnwidth}{@{\extracolsep{\fill}}cccccc} \toprule\toprule
    BCG & Probability & Redshift & $r$-mag\tablenotemark{\footnotesize{d}} & RA {[}deg{]} & Dec {[}deg{]} \\ \toprule
    1  & 0.9436 & 0.411\tablenotemark{\footnotesize{a}} & 18.90 & 201.85063 & 53.78235 \\ \midrule
    2  & 0.0484 & 0.390\tablenotemark{\footnotesize{a}} & 19.04 & 201.73846 & 53.81966 \\ \midrule
    3  & 0.0029 & 0.393\tablenotemark{\footnotesize{a}} & 18.26 & 201.76472 & 53.83118 \\ \midrule
    4  & 0.0026 & 0.396\tablenotemark{\footnotesize{b}} & 18.82 & 201.68665 & 53.86797 \\ \midrule
    5  & 0.0025 & 0.409\tablenotemark{\footnotesize{a}} & 19.12 & 201.85095 & 53.77650 \\ \midrule
    A  & ... & 0.414\tablenotemark{\footnotesize{a}} & 18.57 & 201.87647 & 53.79540 \\ \midrule
    B  & ... & 0.394\tablenotemark{\footnotesize{c}} & 19.28 & 201.73490 & 53.86985 \\ \bottomrule
\printtablenotes{6}
\end{tabular*}
  \tablenotesreset
\end{table}}

%%%%%%%%%%%%%%%%%%%%%%%%%%%%%%%%%        RMJ 1327 Text       %%%%%%%%%%%%%%%%%%%%%%%%%%%%%%%%%
RMJ1327 is perhaps the most interesting system in our sample, with multiple well-separated clusters, highlighting the ability to identify subclusters from redMaPPer BCG separations. RedMaPPer assigns a probability of $0.9436$ to BCG 1 (\cref{tab:RMJ1327_BCG}) making it the dominant candidate, but still within our $0.98$ threshold. We also identified two additional bright galaxies, listed as BCGs A and B, with magnitudes comparable to the BCGs identified by redMaPPer. The luminosity density and the XSB contours (\cref{fig:RMJ1327_optical}) show three primary subclusters: southeast, containing BCGs 1, 5, and A; center, containing BCGs 2 and 3; and northwest, with BCGs 4 and B. The undetected BCGs are associated with more minor structures than BCGs 1--5, justifying their exclusion from the top five candidates identified by redMaPPer. Additionally, the true brightest galaxy in the field ($r=18.09$ from DESI Legacy Survey DR10 \citep{dey2019overview} and $z=0.3945$ from DESI DR1 \citep{abdul2025data}), located ${\sim}3.5\arcmin$ south of BCG 2 and ${\sim}4\arcmin$ west of BCG 1, was not identified. However, this galaxy was separately classified as a BCG using an optical selection algorithm utilizing a fit of the red sequence on SDSS DR7 data by \citet{hao2010gmbcg}. While this galaxy has been previously identified as a BCG, it is relatively isolated and not associated with any extended X-ray emission, thus, we choose to omit it from our list of potential BCG candidates. Of note, the furthest separation between BCG candidates is ${\sim}8\arcmin$ (${\sim}2.6$ Mpc) between BCGs 1/5 and BCG 4, the largest in our sample, which could offer a way to identify similar extended systems. 

The XSB contours are dominated by the southeastern subcluster. The central and northwestern subclusters have elevated X-ray emission, but are easily overwhelmed by the contour scale of the primary subcluster. Notably, there is no enhanced X-ray emission in the regions between the three subclusters. In addition to the three largest subclusters, there are several smaller structures that could themselves be involved in independent mergers. Each subcluster contains two BCG candidates, each associated with an overdensity in galaxies. The peak in the luminosity density containing a single galaxy west of BCG 1 is the previously identified BCG mentioned above.

%%%%%%%%%%%%%%%%%%%%%%%%%% RMJ1327: Figure - Primary Subclusters %%%%%%%%%%%%%%%%%%%%%%%%%%
\pdffig{
\begin{figure}
    \centering
    \includegraphics[clip, trim=0.0cm 0.55cm 0.0cm 0.5cm, width=0.99\columnwidth]{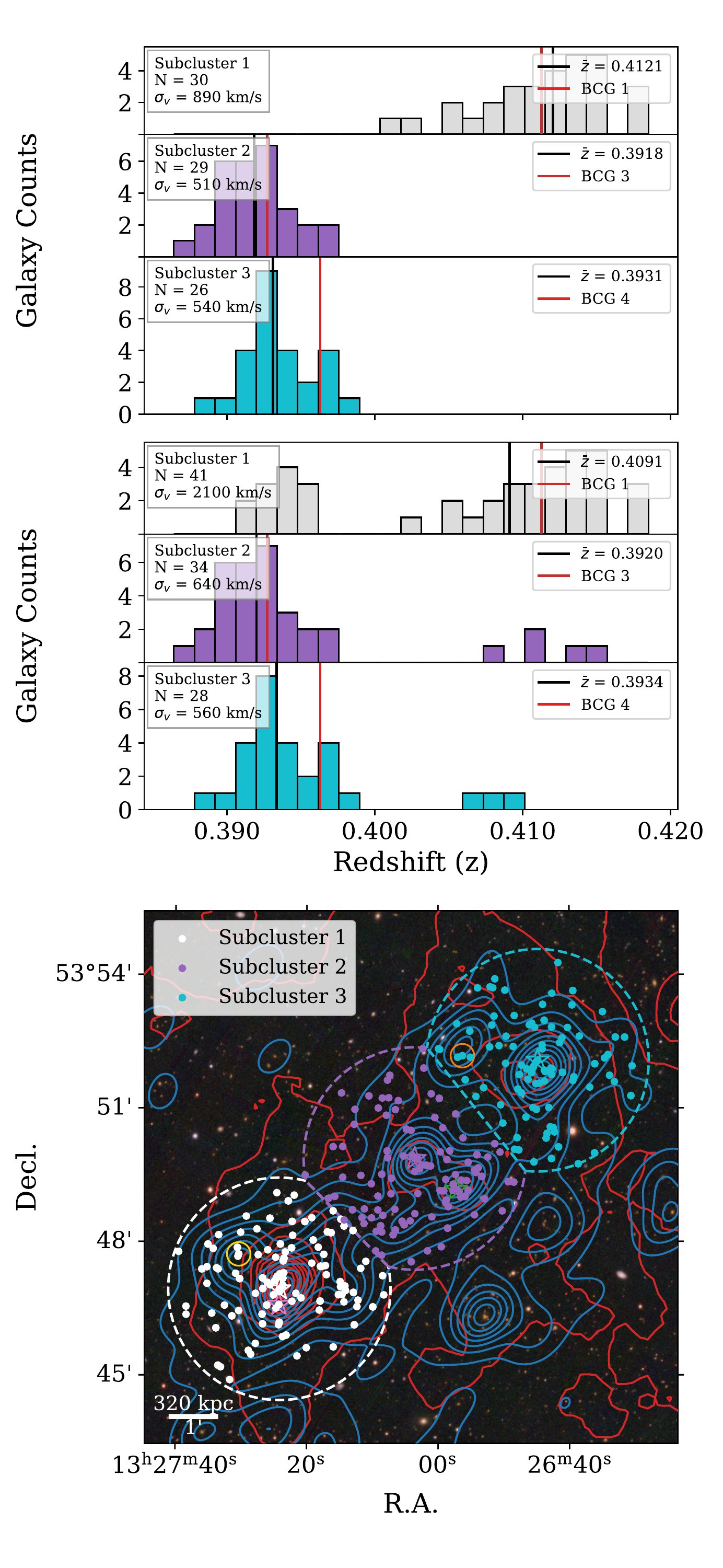}
    \caption{Subclustering analysis of the primary subclusters of RMJ1327. Bottom panel: full 12\arcmin\ field with the three primary subclusters defined by bisectors between BCGs 1, 3, and 4 and a 2.75\arcmin\ radius from each. The center set of histograms show the redshift distribution when considering the full range $0.38 \leq z \leq 0.42$, while the bottom histograms limit Subcluster 1 to the range $0.40 \leq z \leq 0.42$ and Subclusters 2 and 3 to the range $0.38 \leq z \leq 0.40$.}
    \label{fig:RMJ1327_sub_1}
\end{figure}}

\htmlfig{
\begin{figure}[t]
    \centering
    \includegraphics[clip, trim=0.0cm 0.75cm 0.0cm 0.0cm, width=\columnwidth]{Figures/RMJ_1327_redshifts_hist_hmap.jpg}
    \caption{Redshift distribution of RMJ1327. Top panel: histogram of spectroscopic redshifts, combining archival data within 10\arcmin\ of the cluster with new observations from DEIMOS. The inset highlights galaxies in the redshift range $0.38 \leq z \leq 0.42$, which are classified as cluster members. Bottom panel: spatial distribution of those same members.}
    \label{fig:RMJ1327_redshifts}
\end{figure}}

The redshift distribution in the cluster redshift range of $0.38 \leq z \leq 0.42$ is best-fit with a two-component GMM corresponding to the two prominent peaks in \cref{fig:RMJ1327_redshifts}. Both are consistent with a Gaussian distribution ($D=0.069$, $p=0.83$; $A^2=0.25$, $p=0.74$ at $\bar{z}=0.393$ and $D=0.10$, $p=0.68$; $A^2=0.38$, $p=0.40$ at $\bar{z}=0.410$). The bottom panel of \cref{fig:RMJ1327_redshifts} shows the southeastern subcluster at a distinctly higher redshift than the two other primary subclusters. To explore this further, we begin by dividing the field into regions separated by the bisectors between BCGs 1--3 and 3--4 and a $2.75\arcmin$ radius from each, shown in the bottom panel of \cref{fig:RMJ1327_sub_1}. The redshift distribution in the center panel shows the bimodality and the difference in redshifts between Subcluster 1 and Subclusters 2/3. Given such a clean separation, we elect to enforce a redshift range of $0.40 \leq z \leq 0.42$ for Subcluster 1, and $0.38 \leq z \leq 0.40$ for Subclusters 2 and 3. The resulting distributions are shown in the top panel of \cref{fig:RMJ1327_sub_1} and given in \cref{tab:RMJ1327_subclusters}. Subclusters 2 and 3 have similar distributions, with velocity dispersions of $\SI{510}{\kms}$ and $\SI{540}{\kms}$, respectively, and a LOS velocity difference of $\SI{280}{\kms}$ between the subcluster means. Subcluster 1 is more broad, with $\sigma_v=\SI{890}{\kms}$. The ultimate LOS velocity difference between Subcluster 1 and Subclusters 2/3 is then $\SI{4200}{\kms}$, too large for an active merger between Subclusters 1 and 2 at a projected separation of 1.5 Mpc.

%%%%%%%%%%%%%%%%%%%%%%%%%% RMJ1327: Figure - Primary Subclusters %%%%%%%%%%%%%%%%%%%%%%%%%%
\htmlfig{
\begin{figure}
    \centering
    \includegraphics[clip, trim=0.0cm 0.55cm 0.0cm 0.5cm, width=0.99\columnwidth]{Figures/RMJ_1327_full.jpg}
    \caption{Subclustering analysis of the primary subclusters of RMJ1327. Bottom panel: full 12\arcmin\ field with the three primary subclusters defined by bisectors between BCGs 1, 3, and 4 and a 2.75\arcmin\ radius from each. The center set of histograms show the redshift distribution when considering the full range $0.38 \leq z \leq 0.42$, while the bottom histograms limit Subcluster 1 to the range $0.40 \leq z \leq 0.42$ and Subclusters 2 and 3 to the range $0.38 \leq z \leq 0.40$.}
    \label{fig:RMJ1327_sub_1}
\end{figure}}

%%%%%%%%%%%%%%%%%%%%%%%%%% RMJ1327: Table - Subclusters %%%%%%%%%%%%%%%%%%%%%%%%%%
\begin{table}
    \centering
    \caption{RMJ1327 Subcluster Properties}
    \label{tab:RMJ1327_subclusters}
\begin{tabular*}{\columnwidth}{@{\extracolsep{\fill}}cccccc} \toprule\toprule
    Subcluster & $N$ & BCG & BCG $z$ & Mean $z$ & $\sigma_v$ [km s$^{-1}$] \\ \toprule
All & 124 & ... & ... & 0.3961 & 2140 \\
1 & 30 & 1 & 0.411 & 0.4121 & 890  \\
2 & 29 & 3 & 0.393 & 0.3918 & 510  \\
3 & 26 & 4 & 0.396 & 0.3931 & 540  \\
\bottomrule
\end{tabular*}
\end{table}

%%%%%%%%%%%%%%%%%%%%%%%%%% RMJ1327: Figure - Subclusters %%%%%%%%%%%%%%%%%%%%%%%%%%
\begin{figure*}[t]
    \centering
    \includegraphics[clip, trim=0.0cm 0.5cm 0.25cm 0.5cm, width=\textwidth]{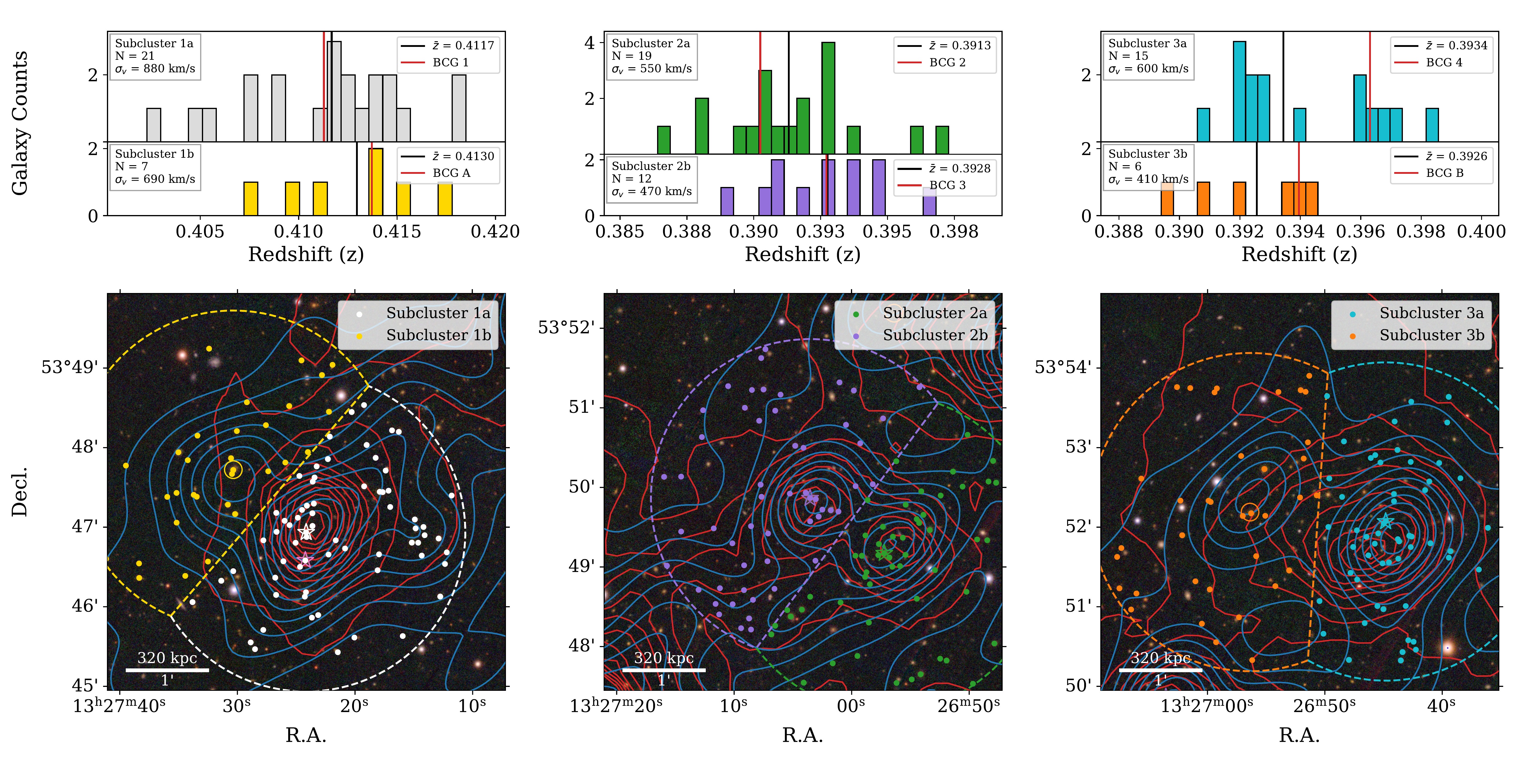}
    \caption{Subclustering analysis of the individual subclusters of RMJ1327. Each panel shows a 5\arcmin\ field about one of the primary subclusters shown in \cref{fig:RMJ1327_sub_1}, arranged from southeast to northwest, left to right. Regions are defined by the bisectors between relevant BCGs and a 1.75\arcmin\ radius from each BCG.}
    \label{fig:RMJ1327_sub_2}
\end{figure*}

Next, we take a closer look at each subcluster individually. We divide each subcluster into two regions defined by a bisector between BCGs and a radius of 1.75\arcmin\ from each (\cref{fig:RMJ1327_sub_2}). 
\begin{itemize}
    \item Subcluster 1, shown in the left panel, has two peaks in the luminosity density, but X-ray emission is almost entirely centered on BCG 1. This could indicate gas-stripping from a recent merger, but there appears to be minimal disruption to the X-ray morphology. More likely, this is a minor merger or possibly even a relaxed cluster. The luminosity density in the northeast is dominated by BCG A and unweighted galaxy density contours are centered almost exclusively on BCG 1. This is also supported by X-ray temperatures and luminosities, which were extracted from a region centered on BCG 1, both being ${\sim}0.5\sigma$ lower than predicted at $\SI{4.31}{\keV}$ and $\SI{1.21e44}{\ergs}$ (\cref{tab:XMM}). There was a fair amount of proton flaring, but all detectors had at least 10 ks of exposure.
    \item Subcluster 2, shown in the center panel, features two distinct luminosity density and XSB peaks, with both groups having similar redshifts. The X-ray morphology is somewhat disturbed, with a bridge in the XSB between Subclusters 2a and 2b and each X-ray peak offset counterclockwise from their BCG, possibly indicating rotation. Of the three subclusters, this is the best candidate for a binary merger, possibly at a large impact parameter.
    \item Subcluster 3, shown in the far right panel, again has two luminosity density peaks, centered on BCGs 4 and B. The XSB is largely centered on BCG 4, but is extended towards BCG B with the peak offset from BCG 4 by ${\sim}\SI{100}{\kpc}$, possibly indicating a merger. However, the luminosity density peak near BCG B is largely dominated by the BCG and largely vanishes when considering unweighted galaxy density, calling into question the existence of a unique group.
\end{itemize}
While it does not appear that any of the three primary subclusters are actively merging with each other, each of the three shows some evidence of being involved in binary mergers themselves. While not a clean binary system, RMJ1327 appears to be an excellent example of a dynamic system along a filament, warranting further observations. We classify RMJ1327 as a possible merger and not binary.

%%%%%%%%%%%%%%%%%%%%%%%%%%%%%%%%%%%%%%%%%%%%%%%%%%%%%%%%%%%%%%%%%%%%%%%%%%%%%%%%%%%%%%%%%%%%%%
%%%%%%%%%%%%%%%%%%%%%%%%%%%%%%%%%          RMJ 1635          %%%%%%%%%%%%%%%%%%%%%%%%%%%%%%%%%
%%%%%%%%%%%%%%%%%%%%%%%%%%%%%%%%%%%%%%%%%%%%%%%%%%%%%%%%%%%%%%%%%%%%%%%%%%%%%%%%%%%%%%%%%%%%%%
\subsubsection{RM J163509.2+152951.5}\label{subsubsec:RMJ1635}
%%%%%%%%%%%%%%%%%%%%%%%%%% RMJ1635: Table - BCGs %%%%%%%%%%%%%%%%%%%%%%%%%%

\begin{table}[b]
    \centering
    \caption{RMJ1635 BCG Information}
    \label{tab:RMJ1635_BCG}
    \tablenotetext{a}{SDSS DR18 \citep{almeida2023eighteenth}}
    \tablenotetext{b}{DEIMOS (This work)}
    \tablenotetext{c}{DESI Legacy Survey DR10 \citep{dey2019overview}}
\begin{tabular*}{\columnwidth}{@{\extracolsep{\fill}}cccccc} \toprule\toprule
    BCG & Probability & Redshift & $r$-mag\tablenotemark{\footnotesize{c}} & RA {[}deg{]} & Dec {[}deg{]} \\ \toprule
    1  & 0.5402 & 0.474\tablenotemark{\footnotesize{a}} & 19.43 & 248.78848 & 15.49764 \\ \midrule
    2  & 0.3727 & 0.472\tablenotemark{\footnotesize{a}} & 19.04 & 248.76398 & 15.48361 \\ \midrule
    3  & 0.0762 & 0.481\tablenotemark{\footnotesize{b}} & 18.99 & 248.77975 & 15.48494 \\ \midrule
    4  & 0.0069 & 0.478\tablenotemark{\footnotesize{b}} & 19.85 & 248.79898 & 15.49102 \\ \midrule
    5  & 0.0040 & 0.478\tablenotemark{\footnotesize{a}} & 19.81 & 248.77858 & 15.48468 \\ \bottomrule
\printtablenotes{6}
\end{tabular*}
  \tablenotesreset
\end{table}

%%%%%%%%%%%%%%%%%%%%%%%%%% RMJ1635: Figure - Optical %%%%%%%%%%%%%%%%%%%%%%%%%%
\begin{figure}[t]
    \centering
    \includegraphics[clip, trim=0.0cm 1.0cm 0.0cm 0.0cm, width=\columnwidth]{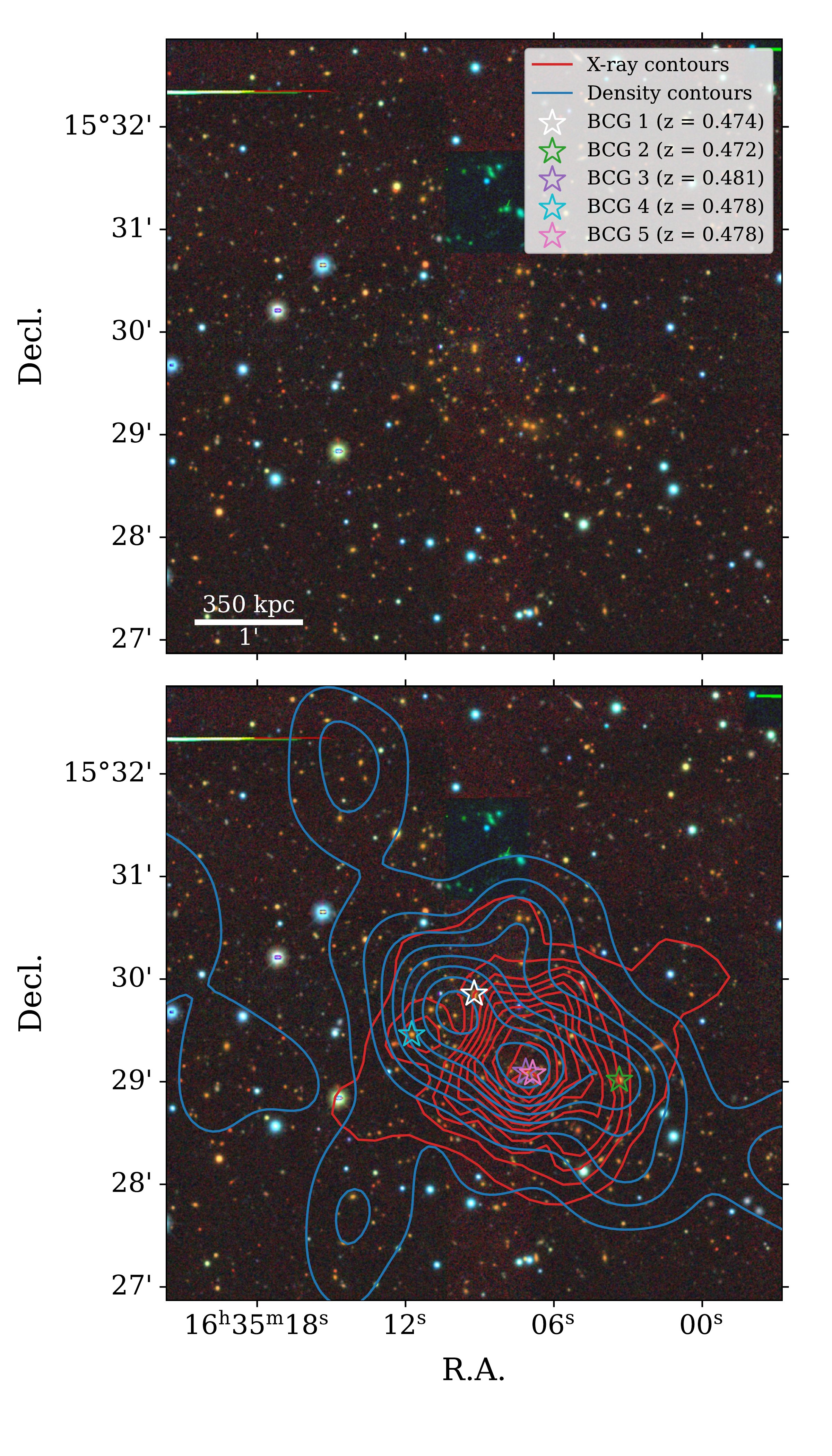}
    \caption{Optical image of RMJ1635. Top panel: 6\arcmin\ by 6\arcmin\ (2.10 by 2.10 Mpc) Legacy Survey image. Bottom panel: red sequence density contours (blue), X-ray surface brightness contours (red), and BCG candidates identified by redMaPPer.}
    \label{fig:RMJ1635_optical}
\end{figure}

\begin{figure}[t]
    \centering
    \includegraphics[clip, trim=0.0cm 1.0cm 0.0cm 0.0cm, width=\columnwidth]{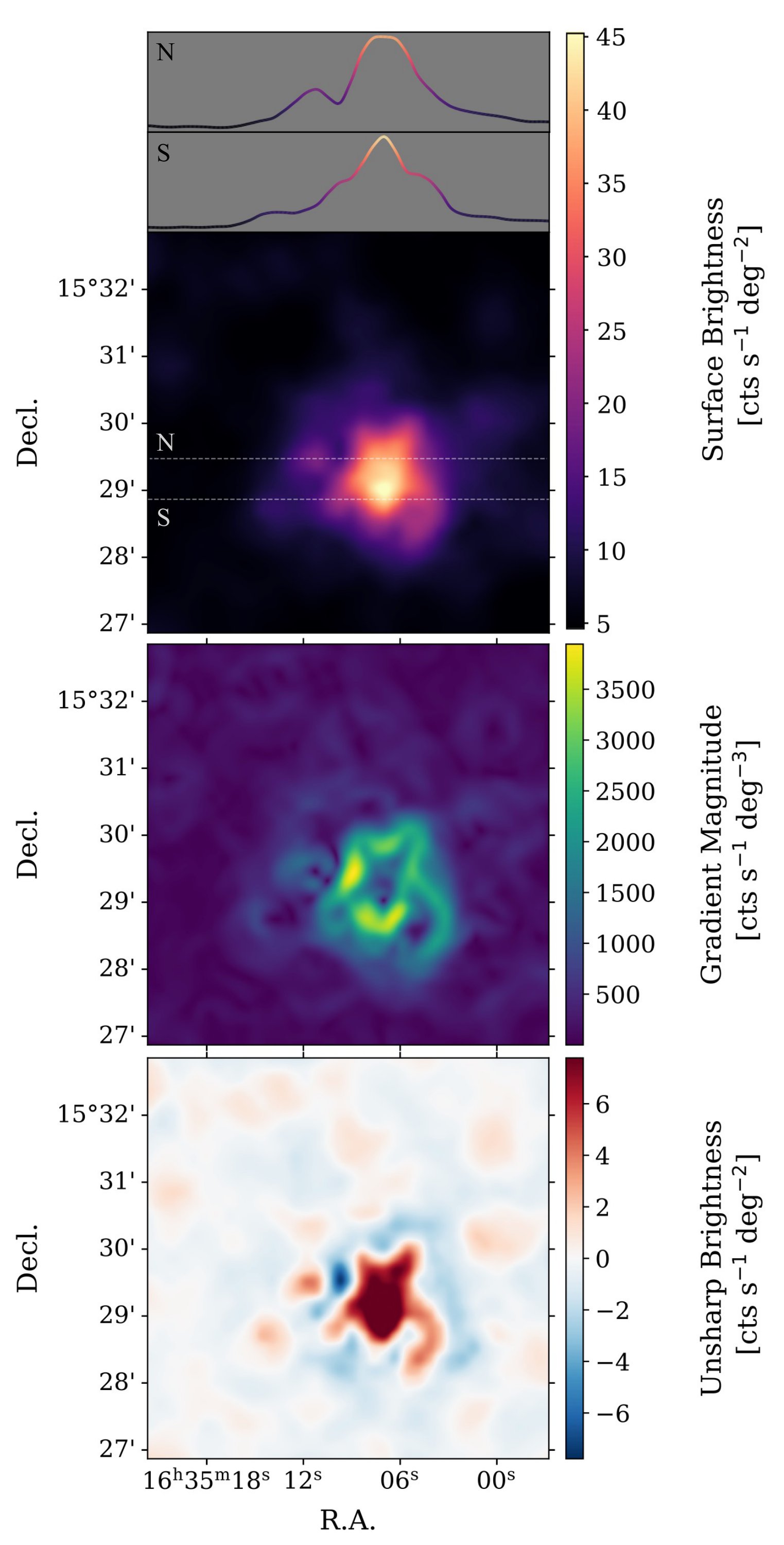}
    \caption{RMJ1635 X-ray features. The top panel shows the X-ray surface brightness smoothed with a 10\arcsec\ kernel and reprojected on an $800\times800$ pixel grid with the extracted one-dimensional profiles. The dashed lines show where the profiles were extracted across the southern and northeastern edges. The center panel displays the gradient magnitude of the brightness. The bottom panel shows residual features after unsharp masking, where an image smoothed with a kernel of 50\arcsec\ is subtracted from the image smoothed at 10\arcsec.}
    \label{fig:RMJ1635_gauss}
\end{figure}

%%%%%%%%%%%%%%%%%%%%%%%%%%%%%%%%%        RMJ 1635 Text       %%%%%%%%%%%%%%%%%%%%%%%%%%%%%%%%%
The top two most likely BCG candidates for RMJ1635 each have relatively large probabilities ($0.5402$ and $0.3727$), with the third-ranked BCG at $0.0762$ (\cref{tab:RMJ1635_BCG}). The luminosity density contours show two distinct peaks, one centered on BCG 3/5 and the other located between BCGs 1 and 4 (\cref{fig:RMJ1635_optical}). The X-ray surface brightness exhibits a disturbed morphology, with a primary peak coincident with BCG 3/5, and a secondary peak at BCG 4. The XSB is asymmetric, with pronounced edges south of BCG 3/5 as well as south of BCG 1, indicative of dynamic activity. These features are clearly visible in the smoothed XSB map (top panel of \cref{fig:RMJ1635_gauss}) and are further emphasized in the gradient magnitude (center) and unsharp-masked (bottom) images. 

A one-dimensional surface brightness profile extracted across the southern edge of the cluster (marked `S' in the top panel of \cref{fig:RMJ1635_gauss}) shows a clear, abrupt change in surface brightness on both the eastern and western sides of the feature. From the gradient magnitude and unsharp-mask, it is evident that the northeastern feature (marked `N') has an even more severe gradient than the southern edge. Extracting a one-dimensional profile across that feature shows the rapid change in XSB and also resolves the secondary XSB peak near BCG 4. 

%%%%%%%%%%%%%%%%%%%%%%%%%% RMJ1635: Figure - Gauss %%%%%%%%%%%%%%%%%%%%%%%%%%

%%%%%%%%%%%%%%%%%%%%%%%%%% RMJ1635: Figure - Redshift Heatmap %%%%%%%%%%%%%%%%%%%%%%%%%%
\begin{figure}[ht]
    \centering
    \includegraphics[clip, trim=0.0cm 0.75cm 0.0cm 0.25cm, width=\columnwidth]{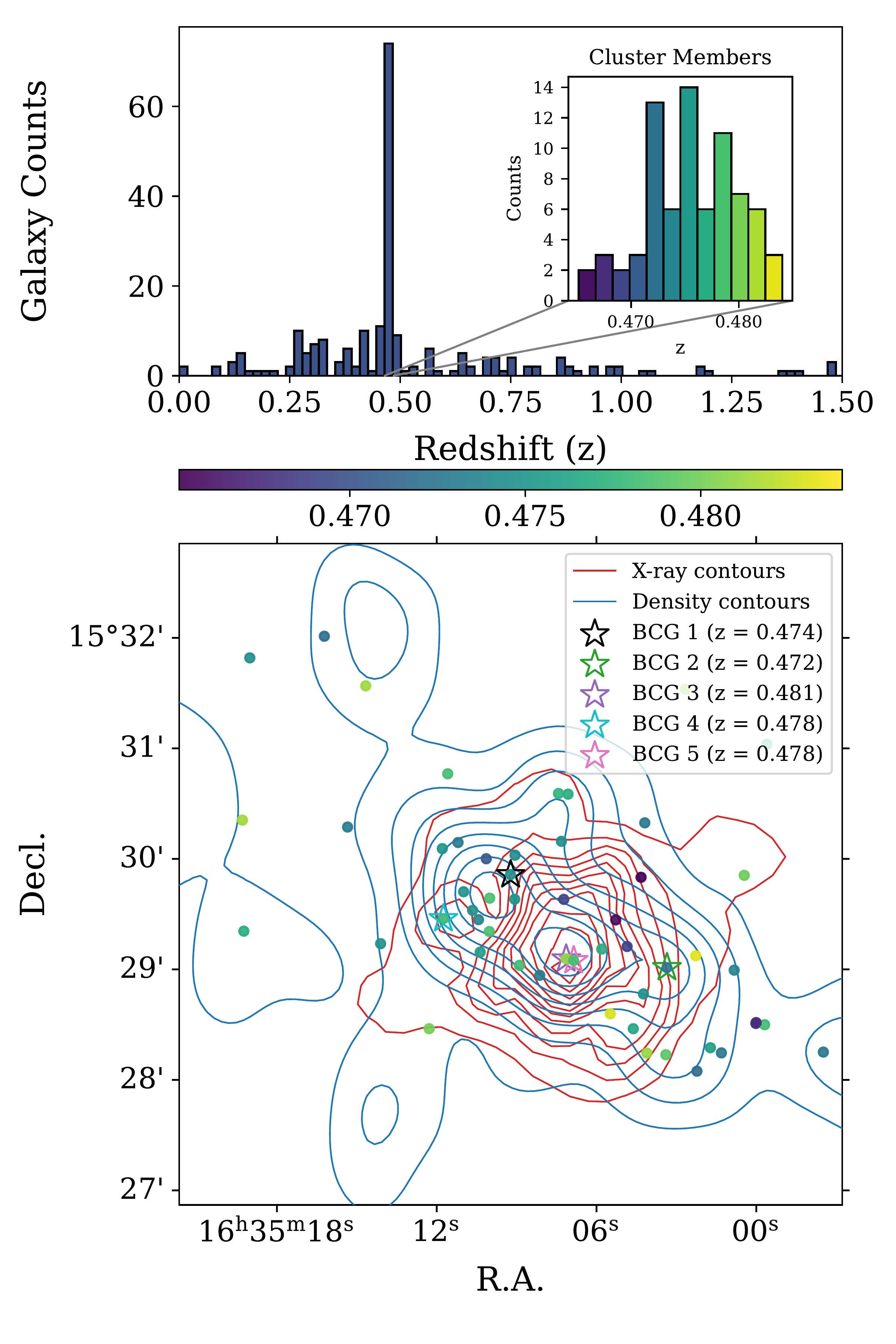}
    \caption{Redshift distribution of RMJ1635. Top panel: histogram of spectroscopic redshifts, combining archival data within 10\arcmin\ of the cluster with new observations from DEIMOS. The inset highlights cluster members in the redshift range $0.465 \leq z \leq 0.485$.}
    \label{fig:RMJ1635_redshifts}
\end{figure}

In the cluster redshift range of $0.465 \leq z \leq 0.485$, the redshift distribution (\cref{fig:RMJ1635_redshifts}) is well fit by a single Gaussian ($D=0.067$, $p=0.87$; $A^2=0.31$, $p=0.57$), and no foreground or background structures contaminate the field. Spatially, the highest redshift galaxies in the cluster are preferentially located in the south near BCG 2 (bottom panel of \cref{fig:RMJ1635_redshifts}). Dividing the cluster into subclusters centered on BCGs 1 and 3 (\cref{fig:RMJ1635_sub_1}) shows a relatively tight distribution of redshifts around BCG 1, with a velocity dispersion of \SI{640}{\kms}. Subcluster 2 is more broadly distributed, with $\sigma_v=\SI{1110}{\kms}$. Both subclusters have approximately the same mean redshift, with a velocity difference of ${<}\SI{50}{\kms}$ (\cref{tab:RMJ1635_subclusters}).

\htmlfig{
\begin{figure}[t]
    \centering
    \includegraphics[clip, trim=0.0cm 1.0cm 0.0cm 0.0cm, width=\columnwidth]{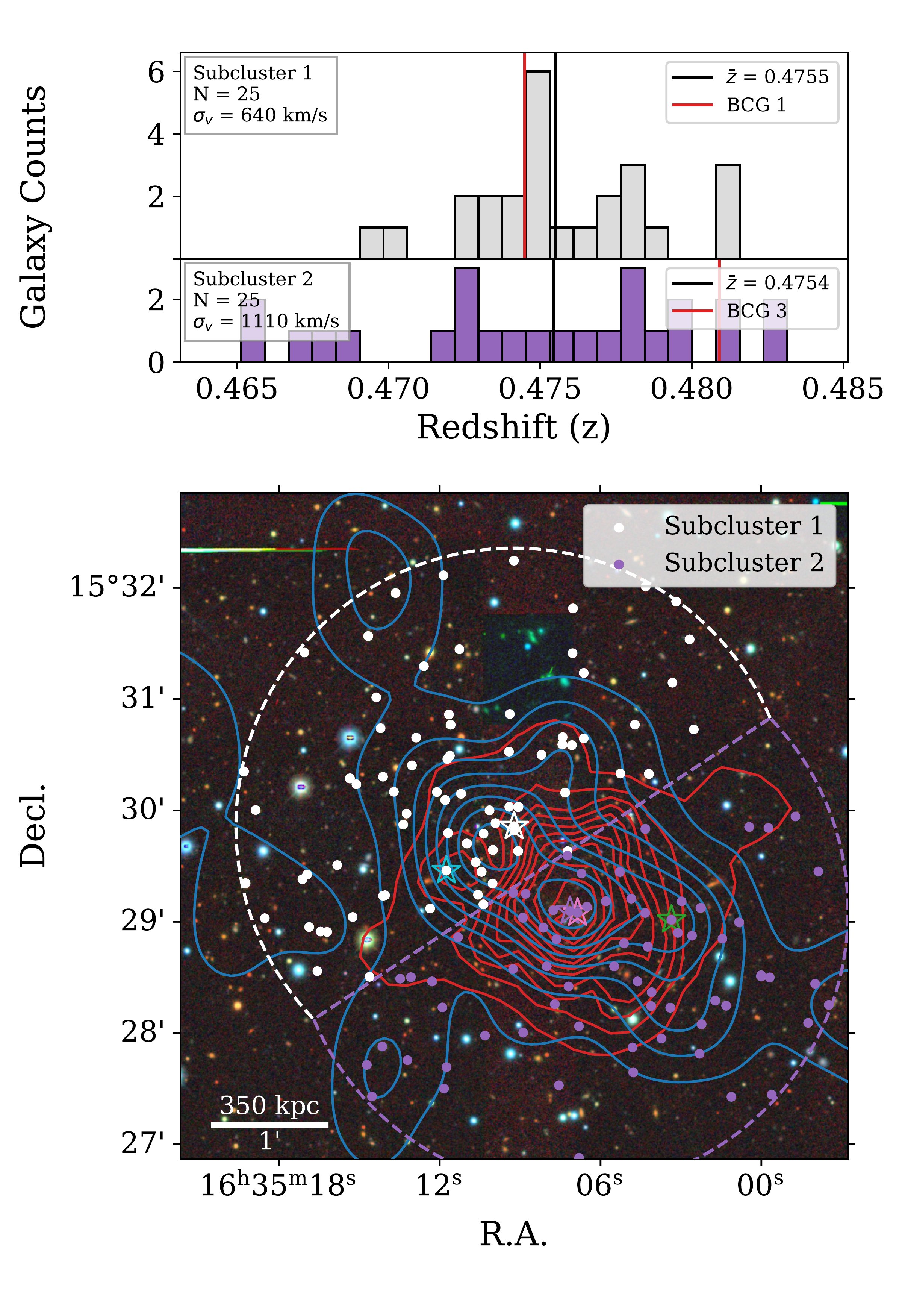}
    \caption{Subclustering analysis of RMJ1635. Bottom panel: spatial distribution of subcluster members, both spectroscopic and photometric, identified within regions defined by the bisector between BCG pairs and a radial distance of 2.5\arcmin\ from the BCG in each region. Top panel: redshift distribution of the spectroscopic members of those regions.}
    \label{fig:RMJ1635_sub_1}
\end{figure}}

%%%%%%%%%%%%%%%%%%%%%%%%%% RMJ1635: Table - Subclusters %%%%%%%%%%%%%%%%%%%%%%%%%%
\begin{table}
    \centering
    \caption{RMJ1635 Subcluster Properties}
    \label{tab:RMJ1635_subclusters}
\begin{tabular*}{\columnwidth}{@{\extracolsep{\fill}}cccccc} \toprule\toprule
    Subcluster & $N$ & BCG & BCG $z$ & Mean $z$ & $\sigma_v$ [km s$^{-1}$] \\ \toprule
All & 76 & ... & ... & 0.4756 & 870 \\
1 & 25 & 1 & 0.474 & 0.4755 & 640  \\
2 & 25 & 3 & 0.481 & 0.4754 & 1110  \\
\bottomrule
\end{tabular*}
\end{table}

The X-ray temperature of $\SI{5.74}{\keV}$ is consistent with the predicted value, while the X-ray luminosity is elevated by ${\sim}1\sigma$ at $\SI{4.09e44}{\ergs}$ (\cref{tab:XMM}). The bimodal galaxy distribution, disturbed X-ray morphology, and presence of sharp surface brightness variations collectively indicate a dynamically active system. The small LOS velocity difference between the two galaxy concentrations indicates this velocity axis is near the plane of the sky and we classify it as a merger, and likely a binary configuration. While the XSB peak is not displaced from the primary cluster, this is a relatively clean system and an excellent candidate for follow-up with weak lensing studies.

%%%%%%%%%%%%%%%%%%%%%%%%%% RMJ1635: Figure - Subclusters %%%%%%%%%%%%%%%%%%%%%%%%%%
\pdffig{
\begin{figure}[t]
    \centering
    \includegraphics[clip, trim=0.0cm 1.0cm 0.0cm 0.0cm, width=\columnwidth]{Figures/RMJ_1635_subcluster_histograms_1_3.jpg}
    \caption{Subclustering analysis of RMJ1635. Bottom panel: spatial distribution of subcluster members, both spectroscopic and photometric, identified within regions defined by the bisector between BCG pairs and a radial distance of 2.5\arcmin\ from the BCG in each region. Top panel: redshift distribution of the spectroscopic members of those regions.}
    \label{fig:RMJ1635_sub_1}
\end{figure}}

%%%%%%%%%%%%%%%%%%%%%%%%%%%%%%%%%%%%%%%%%%%%%%%%%%%%%%%%%%%%%%%%%%%%%%%%%%%%%%%%%%%%%%%%%%%%%%
%%%%%%%%%%%%%%%%%%%%%%%%%%%%%%%%%          RMJ 2321          %%%%%%%%%%%%%%%%%%%%%%%%%%%%%%%%%
%%%%%%%%%%%%%%%%%%%%%%%%%%%%%%%%%%%%%%%%%%%%%%%%%%%%%%%%%%%%%%%%%%%%%%%%%%%%%%%%%%%%%%%%%%%%%%
\subsubsection{RM J232104.1+291134.5}\label{subsubsec:RMJ2321}

%%%%%%%%%%%%%%%%%%%%%%%%%% RMJ2321: Table - BCGs %%%%%%%%%%%%%%%%%%%%%%%%%%

\begin{table}[b]
    \centering
    \caption{RMJ2321 BCG Information}
    \label{tab:RMJ2321_BCG}
    \tablenotetext{a}{SDSS DR18 \citep{almeida2023eighteenth}}
    \tablenotetext{b}{DEIMOS (This work)}
    \tablenotetext{c}{DESI Legacy Survey DR10 \citep{dey2019overview}}
\begin{tabular*}{\columnwidth}{@{\extracolsep{\fill}}cccccc} \toprule\toprule
    BCG & Probability & Redshift & $r$-mag\tablenotemark{\footnotesize{c}} & RA {[}deg{]} & Dec {[}deg{]} \\ \toprule
    1  & 0.7827 & 0.485\tablenotemark{\footnotesize{a}} & 18.64 & 350.26727 & 29.19291 \\ \midrule
    2  & 0.1679 & 0.501\tablenotemark{\footnotesize{a}} & 19.13 & 350.24640 & 29.21685 \\ \midrule
    3  & 0.0460 & 0.494\tablenotemark{\footnotesize{a}} & 19.27 & 350.24790 & 29.20312 \\ \midrule
    4  & 0.0020 & 0.489\tablenotemark{\footnotesize{a}} & 20.72 & 350.26820 & 29.19231 \\ \midrule
    5  & 0.0015 & 0.497\tablenotemark{\footnotesize{b}} & 19.70 & 350.24000 & 29.19139 \\ \bottomrule
\printtablenotes{6}
\end{tabular*}
  \tablenotesreset
\end{table}

%%%%%%%%%%%%%%%%%%%%%%%%%% RMJ2321: Figure - Optical %%%%%%%%%%%%%%%%%%%%%%%%%%
\htmlfig{
\begin{figure}[t]
    \centering
    \includegraphics[clip, trim=0.0cm 1.0cm 0.0cm 0.0cm, width=\columnwidth]{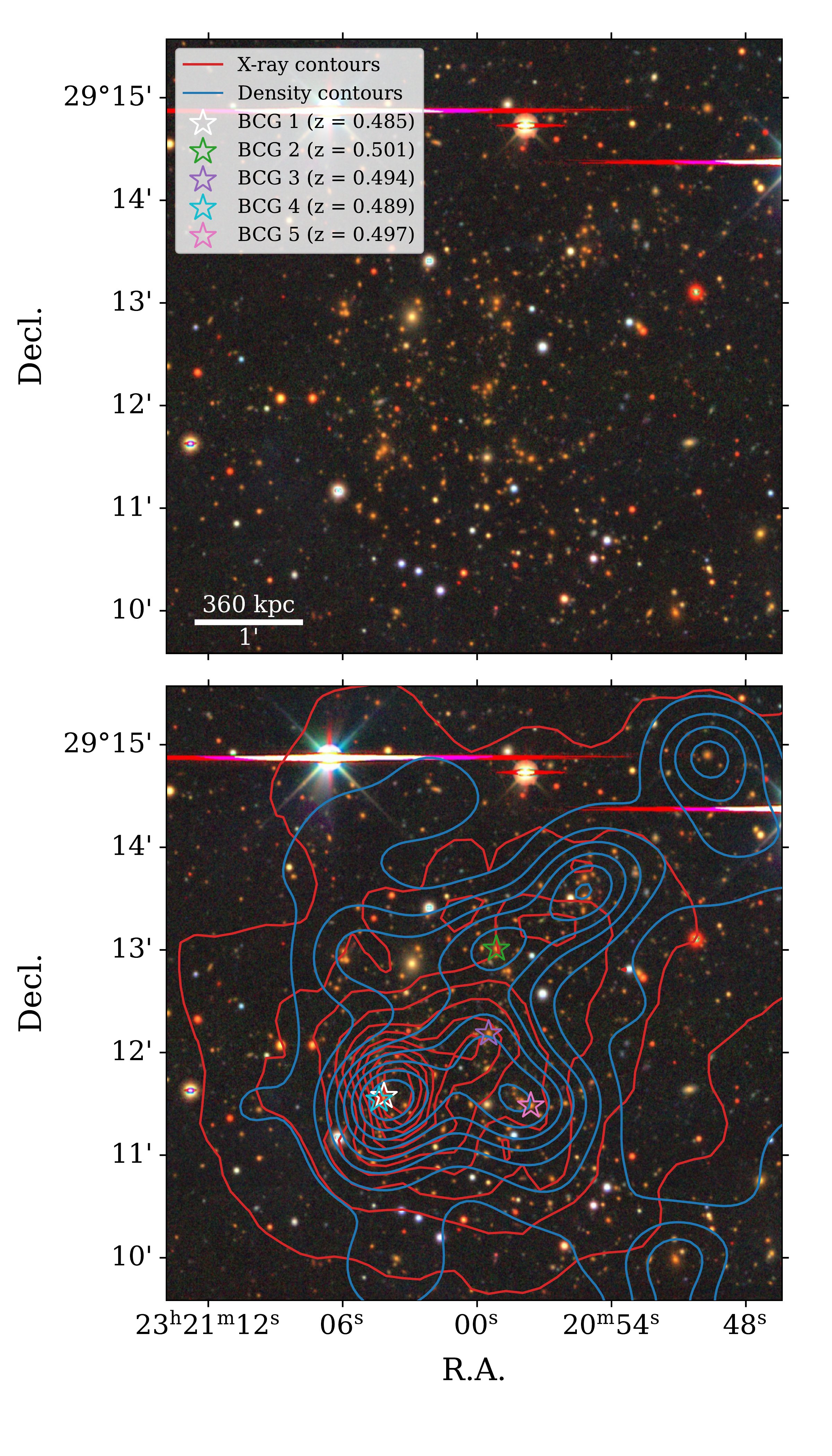}
    \caption{Optical image of RMJ2321. Top panel: 6\arcmin\ by 6\arcmin\ (2.16 by 2.16 Mpc) Legacy Survey image. Bottom panel: red sequence density contours (blue), X-ray surface brightness contours (red), and BCG candidates identified by redMaPPer.}
    \label{fig:RMJ2321_optical}
\end{figure}}

%%%%%%%%%%%%%%%%%%%%%%%%%%%%%%%%%        RMJ 2321 Text       %%%%%%%%%%%%%%%%%%%%%%%%%%%%%%%%%

Our final cluster, RMJ2321, is another complicated system. The top three BCGs are listed with probabilities of $0.7827$, $0.1679$, and $0.0460$ (\cref{tab:RMJ2321_BCG}). However, BCGs 1 and 2 have a large LOS velocity difference of ${\sim}\SI{3200}{\kms}$, indicating either a significant LOS component to a potential merger or foreground/background group. There are multiple peaks in the luminosity density contours: one located near BCG 1/4, another at BCG 5, and two in the north near BCG 2 (\cref{fig:RMJ2321_optical}). The X-ray surface brightness is peaked just northwest of BCG 1/4, with a secondary peak near BCG 3. The X-ray contours also extend out towards BCG 2.

%%%%%%%%%%%%%%%%%%%%%%%%%% RMJ2321: Figure - Optical %%%%%%%%%%%%%%%%%%%%%%%%%%
\pdffig{
\begin{figure}[t]
    \centering
    \includegraphics[clip, trim=0.0cm 1.0cm 0.0cm 0.0cm, width=\columnwidth]{Figures/RMJ_2321_2panel_optical_contours_vert.jpg}
    \caption{Optical image of RMJ2321. Top panel: 6\arcmin\ by 6\arcmin\ (2.16 by 2.16 Mpc) Legacy Survey image. Bottom panel: red sequence density contours (blue), X-ray surface brightness contours (red), and BCG candidates identified by redMaPPer.}
    \label{fig:RMJ2321_optical}
\end{figure}}

%%%%%%%%%%%%%%%%%%%%%%%%%% RMJ2321: Figure - Redshift Heatmap %%%%%%%%%%%%%%%%%%%%%%%%%%
\begin{figure}[t]
    \centering
    \includegraphics[clip, trim=0.0cm 0.75cm 0.0cm 0.0cm, width=\columnwidth]{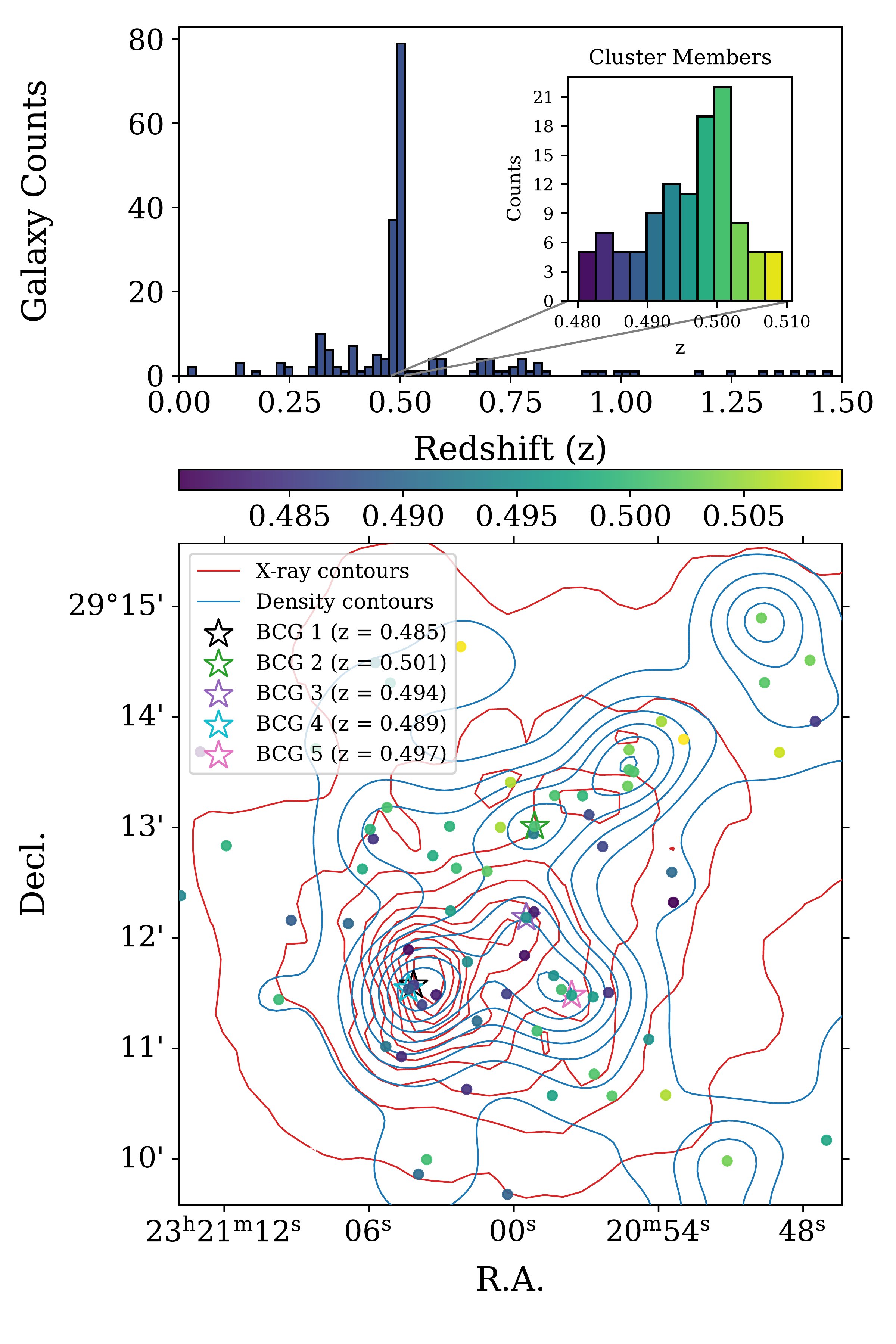}
    \caption{Redshift distribution of RMJ2321. Top panel: histogram of spectroscopic redshifts, combining archival data within 10\arcmin\ of the cluster with new observations from DEIMOS. The inset highlights galaxies in the redshift range $0.48 \leq z \leq 0.51$, which are classified as cluster members. Bottom panel: spatial distribution of those same members.}
    \label{fig:RMJ2321_redshifts}
\end{figure}

%%%%%%%%%%%%%%%%%%%%%%%%%% RMJ2321: Figure - Subclusters %%%%%%%%%%%%%%%%%%%%%%%%%%
\pdffig{
\begin{figure*}[t]
    \centering
    \includegraphics[clip, trim=0.0cm 0.75cm 0.5cm 0.5cm, width=\textwidth]{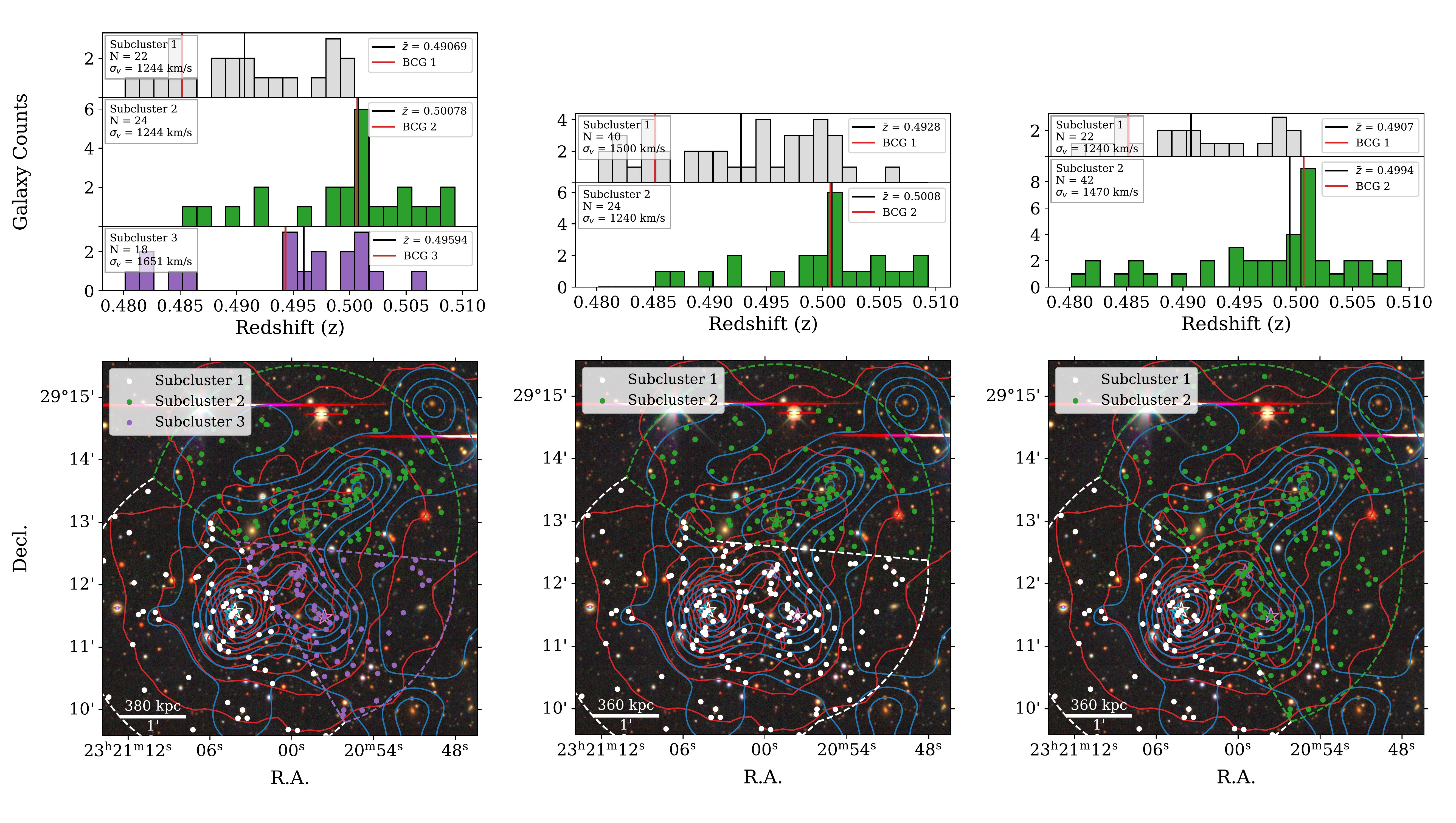}
    \caption{Subclustering analysis of RMJ2321. Bottom panel: spatial distribution of subcluster members, both spectroscopic and photometric, identified within regions defined by the bisector between BCG pairs and a radial distance of 2.5' from the BCG in each region. Top panel: redshift distribution of the spectroscopic members of those regions.}
    \label{fig:RMJ2321_sub_1}
\end{figure*}}

A GMM fit of the full redshift distribution is best-fit with three total components but contains only a single narrow Gaussian and shows no foreground or background structures (\cref{fig:RMJ2321_redshifts}). In the cluster redshift range of $0.48 \leq z \leq 0.51$, the distribution is consistent with Gaussianity by the KS test ($D=0.104$, $p=0.15$), but not the AD test ($A^2=1.60$, $p=0.0004$). There is a gradient in the spatial distribution of redshifts, following the trend from BCGs 1 and 2, with the southeast subcluster at a lower redshift. To investigate further, we divided the cluster into regions centered on BCGs 1, 2, and 3 (left panel of \cref{fig:RMJ2321_sub_1}). If there is a distinct subcluster associated with BCG 3, it also likely includes BCG 5. However, it is unclear which would be the primary BCG of the subcluster. BCG 3 is brighter ($r=19.27$ versus $19.70$) and associated with a peak in the XSB. However, BCG 5 is associated with a peak in the luminosity density. A merger axis between BCGs 1 and 3 places both XSB peaks between the BCGs. Subclusters 1 and 2 have noticeable differences in redshift, with a velocity difference of \SI{2000}{\kms} between subcluster means (\cref{tab:RMJ2321_subclusters}). Subcluster 3 has members across both redshift ranges.

%%%%%%%%%%%%%%%%%%%%%%%%%% RMJ2321: Figure - Subclusters %%%%%%%%%%%%%%%%%%%%%%%%%%
\htmlfig{
\begin{figure*}[t]
    \centering
    \includegraphics[clip, trim=0.0cm 0.75cm 0.5cm 0.5cm, width=\textwidth]{Figures/RMJ_2321_all.jpg}
    \caption{Subclustering analysis of RMJ2321. Bottom panel: spatial distribution of subcluster members, both spectroscopic and photometric, identified within regions defined by the bisector between BCG pairs and a radial distance of 2.5' from the BCG in each region. Top panel: redshift distribution of the spectroscopic members of those regions.}
    \label{fig:RMJ2321_sub_1}
\end{figure*}

\begin{table}
    \centering
    \caption{RMJ2321 Subcluster Properties}
    \label{tab:RMJ2321_subclusters}
\begin{tabular*}{\columnwidth}{@{\extracolsep{\fill}}cccccc} \toprule\toprule
    Subcluster & $N$ & BCG & BCG $z$ & Mean $z$ & $\sigma_v$ [km s$^{-1}$] \\ \toprule
All & 113 & ... & ... & 0.4970 & 1440 \\
1 & 22 & 1 & 0.485 & 0.4907 & 1240  \\
2 & 24 & 2 & 0.501 & 0.5008 & 1240  \\
3 & 18 & 3 & 0.494 & 0.4959 & 1650  \\
\bottomrule
\end{tabular*}
\end{table}}

We then combine Subclusters 1 and 3 (center panel of \cref{fig:RMJ2321_sub_1}) which increases the dispersion on Subcluster 1 to \SI{1500}{\kms}, but decreases the LOS velocity difference between means to \SI{1600}{\kms}. We also consider a scenario combining Subclusters 2 and 3. This increases the dispersion on Subcluster 2 to \SI{1500}{\kms} with a velocity difference between means of \SI{1740}{\kms}.

While the cluster redshift range is best fit with a single component Gaussian ($\mathrm{BIC}=-794$), a two-component GMM yields a nearly identical BIC of $-793$, indicating no strong statistical preference between the models. To explore these two distributions further, we divide the cluster into two redshift bins, $0.480 \leq z \leq 0.495$ and $0.495 \leq z \leq 0.510$, and examine their spatial distributions in \cref{fig:RMJ2321_foregrounds}. From this, Subclusters 1 and 2 appear distinct, whereas the region of Subcluster 3 remains ambiguous. It contains approximately equal numbers of galaxies from each redshift bin, with BCGs 3 and 5 divided between the bins. Adjusting the redshift boundary slightly, from $0.495$ to $0.494$ or $0.497$, to capture both BCGs in the same bin biases the observed density contours somewhat, but the general morphology of an extension in the low redshift contours towards BCG 3 and a concentration in high redshift members near BCG 5 remains. 

%%%%%%%%%%%%%%%%%%%%%%%%%% RMJ2321: Table - Subclusters %%%%%%%%%%%%%%%%%%%%%%%%%%
\pdffig{
\begin{table}
    \centering
    \caption{RMJ2321 Subcluster Properties}
    \label{tab:RMJ2321_subclusters}
\begin{tabular*}{\columnwidth}{@{\extracolsep{\fill}}cccccc} \toprule\toprule
    Subcluster & $N$ & BCG & BCG $z$ & Mean $z$ & $\sigma_v$ [km s$^{-1}$] \\ \toprule
All & 113 & ... & ... & 0.4970 & 1440 \\
1 & 22 & 1 & 0.485 & 0.4907 & 1240  \\
2 & 24 & 2 & 0.501 & 0.5008 & 1240  \\
3 & 18 & 3 & 0.494 & 0.4959 & 1650  \\
\bottomrule
\end{tabular*}
\end{table}}

\htmlfig{
\begin{figure}[t]
    \centering
    \includegraphics[clip, trim=0.0cm 0.2cm 0.0cm 0.2cm, width=0.9\columnwidth]{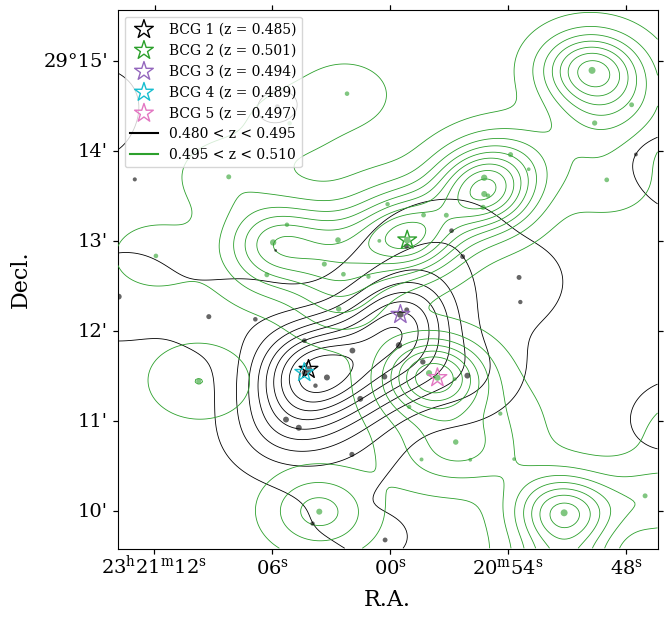}
    \caption{Distribution of galaxies in RMJ2321 separated by redshift. Contours show the luminosity-weighted density of spectroscopic members in each redshift bin. Individual galaxies are indicated by markers sized proportional to their luminosity.}
    \label{fig:RMJ2321_foregrounds}
\end{figure}}

RMJ2321 has the highest X-ray luminosity in the sample at $\SI{7.20e44}{\ergs}$, and second-highest X-ray temperature, $\SI{8.62}{\keV}$ (\cref{tab:XMM}). However, it also has the highest richness in our sample, at $\lambda = 239$ (\cref{tab:clusters}). This places both the X-ray temperature and luminosity only ${\lesssim}0.5\sigma$ above the expected values. RMJ2321 (PSZ2 G100.22-29.64) was also studied as one of the LOFAR--Planck SZ clusters, where \citet{botteon2022planck} report no diffuse emission.

The disturbed X-ray and luminosity density morphologies are strongly suggestive of merging activity, but the underlying substructure remains uncertain. One possible configuration is a merger between Subclusters 1 and 3 at the lower redshift, with Subcluster 2 in the background; however, the extent to which Subcluster 3 represents a distinct system, rather than a mixture of interlopers from Subclusters 1 and 2, is unclear. Any active merger between Subclusters 1 and 2 would require a large LOS velocity component, complicating modeling. Additional spectroscopy focused on Subclusters 2 and 3 would be particularly useful in determining the substructure of the cluster. We classify this system as a likely merger, and possibly a binary system.

%%%%%%%%%%%%%%%%%%%%%%%%%% RMJ2321: Figure - Foregrounds %%%%%%%%%%%%%%%%%%%%%%%%%%
\pdffig{
\begin{figure}[t]
    \centering
    \includegraphics[clip, trim=0.0cm 0.2cm 0.0cm 0.2cm, width=0.9\columnwidth]{Figures/RMJ_2321_foregrounds.png}
    \caption{Distribution of galaxies in RMJ2321 separated by redshift. Contours show the luminosity-weighted density of spectroscopic members in each redshift bin. Individual galaxies are indicated by markers sized proportional to their luminosity.}
    \label{fig:RMJ2321_foregrounds}
\end{figure}}

\subsection{Archival Results}\label{subsec:archival results}

While the bulk of this work focuses on the 12 X-SORTER targets, the selection method also yields 27 candidates with existing archival X-ray data, providing additional context for evaluating how the optical selection criteria perform. We classify each of these archival systems using the framework of \S\ref{subsec:Classification}, drawing on published multiwavelength results, with the resulting classifications recorded in \cref{tab:CXO_and_XMM,tab:CXO,tab:XMM_archive}. Of the 27 archival systems, 14 are Mergers, seven are Likely or Possible Mergers, two are Indeterminate, and four are Relaxed or Likely Relaxed. Roughly 78\% of the archival sample shows evidence of merger activity at some level. 

We note that the archival sample is biased toward bright, easily detected X-ray clusters. Because mergers can transiently boost X-ray luminosity \citep{ricker2001off,poole2007impact}, X-ray bright samples may inflate the apparent merger fraction relative to a general cluster population. However, the cut on RASS excess serves a similar function for the 12 clusters studied here, making the two samples roughly comparable. The archival results, combined with the 12 new X-SORTER targets, none of which are definitively relaxed, demonstrate that the selection method consistently produces a high fraction of dynamically active clusters. Both populations are dominated by mergers and merger candidates, supporting the conclusion that ambiguity in the redMaPPer BCG identification, paired with a non-trivial BCG separation, is an effective tracer of dynamical activity. Several of the archival mergers exhibit features that illustrate the diversity of merger phases and configurations recovered by the X-SORTER selection:
\begin{itemize}
    \item \textbf{Abell 98}: Early-stage merger involving multiple subclusters along a common axis; deep Chandra observations have revealed both a merger shock and cold-front features in the ICM \citep{Paterno-Mahler2014, sarkar2022shock, Sarkar2023}.
    \item \textbf{Abell 781}: Cluster-chain complex composed of multiple aligned components; combined LOFAR, GMRT, and \XMM\ analysis identifies a merger shock and cold fronts within the main cluster \citep{Sehgal2008Abell781, WittmanA781, BotteonA781}.
    \item \textbf{Abell 1319}: Pre-merger configuration with three converging subclusters; the lack of detected radio halo or relic emission is consistent with a pre-pericenter state \citep{Wilber2019A}.
    \item \textbf{Abell 1430}: Off-axis binary merger; the more massive component hosts a low-power radio halo, while the secondary is associated with a diffuse radio source \citep{Hoeft2021}.
    \item \textbf{MACS J0032.1+1808}: Massive bimodal merger characterized by a particularly large effective Einstein radius of $\theta_E \approx 40\arcsec$ \citep{acebron2020relics, gargantua2026inprep}.
    \item \textbf{MACS J1149.6+2223}: Hubble Frontier Fields cluster identified as a complex multi-component merger; deep Chandra observations have revealed one of the most distant cold fronts known to date \citep{ogrean2016frontier,golovichDynamical2016,Finner2025}.
    \item \textbf{MACS J2243.3$-$0935}: Massive merging cluster with $M_{200} = (1.54 \pm 0.29) \times 10^{15}~M_\odot$, hosting a radio halo and a candidate radio relic \citep{schirmer2011macsj2243, Cantwell2016}.
\end{itemize}

The X-SORTER approach has also produced new confirmed dissociative merger discoveries. Abell 56 \citep{wittman2023new, albuquerque2024unravelling} and RMJ1508 \citep{stancioli2024new} were identified through this selection method and characterized prior to the present \XMM\ survey.

The archival mergers above span a range of merger configurations, from pre-pericenter (Abell 98, Abell 1319) to post-pericenter dissociative (Abell 56, RMJ1508), illustrating the diversity of dynamical states recovered by the X-SORTER selection. \cref{fig:results} summarizes the classification outcomes across the full program.

\section{Summary and Conclusion}\label{sec:Conclusion}

%%%%%%%%%%%%%%%%%%%%%%%%%% Summary Figure %%%%%%%%%%%%%%%%%%%%%%%%%%
\begin{figure*}
    \centering
    \includegraphics[width=\textwidth]{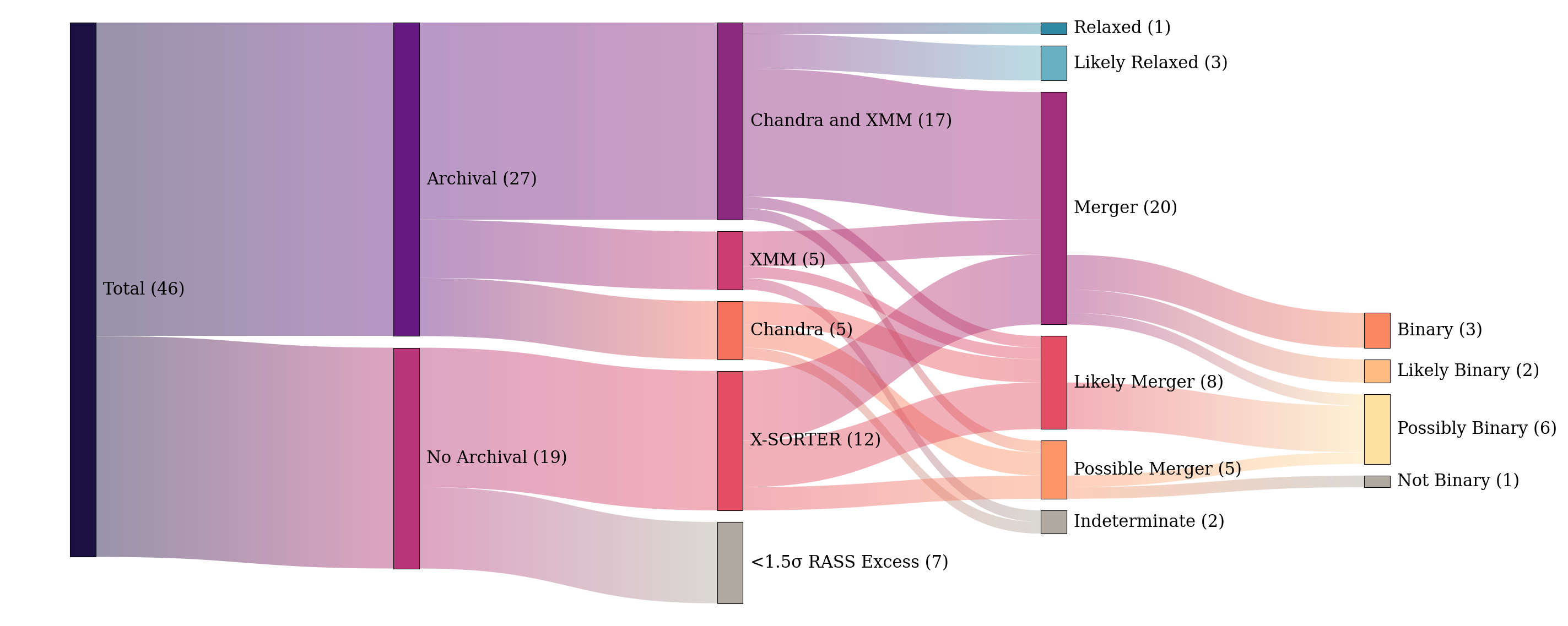}
    \caption{Flow diagram summarizing the classification of cluster candidates across the full program. Clusters with archival X-ray data are categorized according to classifications given in \cref{tab:CXO_and_XMM,tab:CXO,tab:XMM_archive} while new designations are assigned to the X-SORTER sample. Binarity classifications are limited to the X-SORTER systems presented in this work.}
    \label{fig:results}
\end{figure*}

We have presented X-ray and spectroscopic follow-up of 12 galaxy clusters selected from the redMaPPer catalog using optical indicators of binarity, forming the core sample of the X-SORTER program. Our selection aimed to identify massive, dissociative mergers by requiring low BCG probability ($ < 0.98$), projected BCG separations exceeding $0.95\arcmin$, and richness $\lambda \geq 120$. These criteria were designed to preferentially select post-pericenter, binary mergers observable near the plane of the sky---ideal conditions for constraining dark matter self-interactions through the spatial offsets between galaxies, gas, and dark matter.

While only a handful of the 12 observed clusters in this sample can be modeled as clean binary mergers, the success rate in identifying disturbed and dynamically active systems remains high. The majority of clusters show clear evidence of merging activity, even if the underlying substructure is somewhat ambiguous. No clusters from the sample are definitively relaxed systems, while multiple appear to be pre-merger or multi-body systems, illustrating the diversity of dynamical states that can arise from a selection based on optical BCG properties.

To provide a consistent framework for comparison, we assigned qualitative classifications based on the combined X-ray morphology, optical luminosity density, and redshift distribution as laid out in \S\ref{subsec:Classification}.  These classifications, along with classifications from the full program, including archival results, are summarized in \cref{fig:results}.

Below, we summarize the results for the 12 X-SORTER targets presented in this work, grouped into qualitative tiers reflecting our overall confidence that each system meets the goals of the X-SORTER program, namely, isolating clean binary, post-pericenter mergers near the plane of the sky; candidates are listed by ascending RA in each category:
\paragraph{\textbf{Top Binary Merger Candidates}}
\begin{itemize}
    \item \textbf{RMJ0219}: Binary merger, likely post-pericenter and near the plane of the sky; X-ray contours show dual peaks with a bullet-like feature.
    \item \textbf{RMJ0926}: Binary merger. Highest $T_X$ in the sample and $L_X$ is ${>}1.5\sigma$ greater than expected; XSB is extended along the presumed merger axis with a single peak between merging groups.
    \item \textbf{RMJ1219}: Binary dissociative merger confirmed with multi-wavelength follow-up and weak lensing by \citet{stancioli2025inprep}.
    \item \textbf{RMJ1635}: Merger, likely binary; two clear peaks in luminosity density and sharp features in the X-ray emission.
\end{itemize}
\paragraph{\textbf{Intermediate Candidates}}
\begin{itemize}
    \item \textbf{RMJ0003}: Likely merger, possibly binary. X-ray emission is primarily located with the top BCG candidate and is elongated along a presumed merger axis, but also exhibits an offset X-ray peak.
    \item \textbf{RMJ0109}: Merger, possibly binary. X-ray peak is offset between two galaxy overdensities. Role of BCG A, which was not identified by redMaPPer, is unclear.
    \item \textbf{RMJ0829}: Likely merger, possibly binary. Disturbed X-ray morphology and two dominant luminosity density peaks, but possible additional substructure; high $L_X$ and candidate radio halo by \citet{botteon2022planck}.
    \item \textbf{RMJ1257}: Merger, likely binary. XSB peak offset between two peaks in the luminosity density; low $L_X$, $T_X$, X-ray exposure time, and spectroscopic members.
\end{itemize}
\paragraph{\textbf{Marginal Candidates}}
\begin{itemize}
    \item \textbf{RMJ0801}: Possible merger, possibly binary; Elongated gas and galaxy contours with three coaxial galaxy density peaks; X-ray peak along the presumed merger axis and slightly offset towards the southern luminosity density peak. $T_X$ exceeds $1.6\sigma$ above the value predicted from richness scaling relation. 
    \item \textbf{RMJ1043}: Likely merger, possibly binary. Moderately disturbed system with three luminosity density peaks; possible dissociative merger between two of the three subclusters.
\end{itemize}
\paragraph{\textbf{Complex Systems}}
\begin{itemize}
    \item \textbf{RMJ1327}: Possible merger, not binary. Low temperature system with three well-separated subclusters along a filamentary structure. Possible binary merger within central subcluster.
    \item \textbf{RMJ2321}: Likely merger, possibly binary. Complicated, disturbed system, with ambiguous substructure. Possibly binary with background contamination.
\end{itemize}

Taken together, these results show that redMaPPer-based selection is effective at identifying disturbed and dynamically complex clusters. None of the 12 systems is definitively relaxed, and several are strong candidates for binary, post-pericenter mergers conducive to simple modeling, including RMJ1219 which has already been confirmed as a binary dissociative merger. This is even more apparent when considering the full program of 46 targets meeting our selection criteria, which includes 27 clusters with archival X-ray data (22 \CXO, 5 \XMM). Among those with archival data, 12 are previously known mergers, two have been confirmed as newly identified dissociative mergers (Abell 56 and RMJ1508) and seven are disturbed systems.

However, our X-SORTER sample also demonstrates that BCG-based selection alone does not uniquely isolate simple binary-merger geometries; while many of the systems are dynamically active, few are clean two-body mergers. In addition, several systems (RMJ0109, RMJ0801, RMJ1219, RMJ1257, and RMJ1327) have merger scenarios involving BCGs that were not among redMaPPer's top candidates, making these partly serendipitous discoveries. These results motivate combining BCG probability and separation with additional optical information. Incorporating BCG properties into broader optical selection frameworks, such as that of \citet{wen2024catalogue}, may improve the yield of clean binary systems.

Even with these limitations, BCG probabilities and separations provide a useful and inexpensive first step toward constructing the larger, controlled samples needed for statistical studies of cluster mergers. The sample presented here provides a foundation for targeted multi-wavelength follow-up. Weak lensing will be essential for constraining subcluster masses and confirming merger geometries, while radio observations can identify relics and halos diagnostic of the merger stage. Together with the archival systems, this program contributes to a growing ensemble of merging clusters, a resource valuable not only for constraining dark matter self-interactions, but also for studies of ICM dynamics, structure formation, and galaxy evolution in disturbed environments.

\section*{Acknowledgments}

Support for this work was provided by the National Science Foundation under grant NSF 2308383.

This work is based in part on observations obtained with \XMM, an ESA science mission with instruments and contributions directly funded by ESA Member States and NASA. We thank the \XMM\ operations team for their support of this program. The X-ray analysis was supported by NASA grants 80NSSC23K0496, 80NSSC24K0332, and 80NSSC23K1462.

Spectroscopic data were obtained at the W. M. Keck Observatory using the DEIMOS instrument, which is operated as a scientific partnership among the California Institute of Technology, the University of California and the National Aeronautics and Space Administration. The Observatory was made possible by the generous financial support of the W.M. Keck Foundation. We are grateful to the Keck Observatory staff and observing assistants for their expert support. The authors wish to recognize and acknowledge the very significant cultural role and reverence that the summit of Mauna Kea has always had within the indigenous Hawaiian community. We are most fortunate to have the opportunity to conduct observations from this mountain. 

This research has made use of data obtained from the Chandra Data Archive and the \textit{ROSAT} All-Sky Survey in the target selection process. 

This research has made use of the NASA/IPAC Extragalactic Database (NED), which is funded by the National Aeronautics and Space Administration and operated by the California Institute of Technology.

This work has made use of data from the Dark Energy Spectroscopic Instrument (DESI) Data Release 1 and from the Sloan Digital Sky Survey (SDSS) Data Release 18. Funding for the DESI Project has been provided by the U.S. Department of Energy, the National Science Foundation, and participating institutions. Funding for the SDSS has been provided by the Alfred P. Sloan Foundation, the U.S. Department of Energy Office of Science, and participating institutions.

This research has made use of the Pan-STARRS1 Surveys (PS1) and imaging data from the DESI Legacy Imaging Surveys (DECaLS, BASS, and MzLS).

\subsection*{Facilities}{XMM (EPIC), Keck:II (DEIMOS)}
\subsection*{Software}
SAS \citep{gabriel2004astronomical},
ESAS \citep{snowden2014cookbook},
XSPEC \citep{arnaud1996xspec},
PypeIt \citep{prochaska2020pypeit},
Astropy \citep{robitaille2013astropy,astropy2018,collaboration2022astropy},
Astroquery \citep{ginsburg2019astroquery},
SciPy \citep{virtanen2020scipy},
scikit-image \citep{van2014scikit},
scikit-learn \citep{pedregosa2011scikit},
reproject \citep{robitaille2020reproject}

\begin{appendix}
\section{Redshift Catalog}\label{sec:Catalog}

\Cref{tab:redshift_table} lists newly obtained spectroscopic redshifts for galaxies in the 12 X-SORTER cluster fields observed with Keck/DEIMOS. The observations, mask design, and reduction procedures are described in \S\ref{sec:deimos}. Redshifts were measured via cross-correlation with galaxy templates following the method of \citet{wittman2023new}, with typical statistical uncertainties of $\lesssim10^{-4}$.

\begin{table}[h]
    \centering
    \caption{New Spectroscopic Redshifts from Keck/DEIMOS}
    \label{tab:redshift_table}
\begin{tabular*}{0.7\textwidth}{@{\extracolsep{\fill}}ccccc} \toprule\toprule
Cluster & \hspace{1em}RA [deg]\hspace{1em} & \hspace{1em}Dec [deg]\hspace{1em} & \hspace{1em}Redshift\hspace{1em} & Error \\ \toprule
RMJ000343.8+100123.8 &  0.854225 & 9.982806 & 0.376286 & 7.8e-05 \\
RMJ000343.8+100123.8 &  0.874092 & 9.985403 & 0.376233 & 8.1e-05 \\
RMJ000343.8+100123.8 &  0.874900 & 9.982392 & 0.375552 & 7.3e-05 \\
RMJ000343.8+100123.8 &  0.884100 & 9.992076 & 0.266945 & 1.6e-05 \\
RMJ000343.8+100123.8 &  0.886304 & 9.989572 & 0.375704 & 3.0e-05 \\ \bottomrule
\tablenoterow{5}{This table is published in its entirety in machine-readable format at \citet{hopp2026redshifts}.}
\tablenotelastrow{5}{A portion is shown here for guidance regarding its form and content.}
\end{tabular*}
  \tablenotesreset
\end{table}

\end{appendix}

\bibliography{bib}{}

\begin{thebibliography}{}
\expandafter\ifx\csname natexlab\endcsname\relax\def\natexlab#1{#1}\fi
\providecommand{\url}[1]{\href{#1}{#1}}
\providecommand{\dodoi}[1]{doi:~\href{http://doi.org/#1}{\nolinkurl{#1}}}
\providecommand{\doeprint}[1]{\href{http://ascl.net/#1}{\nolinkurl{http://ascl.net/#1}}}
\providecommand{\doarXiv}[1]{\href{https://arxiv.org/abs/#1}{\nolinkurl{https://arxiv.org/abs/#1}}}

\bibitem[{{Acebron} {et~al.}(2020){Acebron}, {Zitrin}, {Coe}, {Mahler}, {Sharon}, {Oguri}, {Brada{\v{c}}}, {Bradley}, {Frye}, {Forman}, {Strait}, {Su}, {Umetsu}, {Andrade-Santos}, {Avila}, {Carrasco}, {Cerny}, {Czakon}, {Dawson}, {Fox}, {Hoag}, {Huang}, {Johnson}, {Kikuchihara}, {Lam}, {Lovisari}, {Mainali}, {Nonino}, {Oesch}, {Ogaz}, {Ouchi}, {Past}, {Paterno-Mahler}, {Peterson}, {Ryan}, {Salmon}, {Stark}, {Toft}, {Trenti}, {Vulcani}, \& {Welch}}]{acebron2020relics}
{Acebron}, A., {Zitrin}, A., {Coe}, D., {et~al.} 2020, \apj, 898, 6, \dodoi{10.3847/1538-4357/ab929d}

\bibitem[{{Albuquerque} {et~al.}(2024){Albuquerque}, {Machado}, \& {Monteiro-Oliveira}}]{albuquerque2024unravelling}
{Albuquerque}, R.~P., {Machado}, R. E.~G., \& {Monteiro-Oliveira}, R. 2024, \mnras, 530, 2146, \dodoi{10.1093/mnras/stae1004}

\bibitem[{{Almeida} {et~al.}(2023){Almeida}, {Anderson}, {Argudo-Fern{\'a}ndez}, {Badenes}, {Barger}, {Barrera-Ballesteros}, {Bender}, {Benitez}, {Besser}, {Bird}, {Bizyaev}, {Blanton}, {Bochanski}, {Bovy}, {Brandt}, {Brownstein}, {Buchner}, {Bulbul}, {Burchett}, {Cano D{\'\i}az}, {Carlberg}, {Casey}, {Chandra}, {Cherinka}, {Chiappini}, {Coker}, {Comparat}, {Conroy}, {Contardo}, {Cortes}, {Covey}, {Crane}, {Cunha}, {Dabbieri}, {Davidson}, {Davis}, {de Andrade Queiroz}, {De Lee}, {M{\'e}ndez Delgado}, {Demasi}, {Di Mille}, {Donor}, {Dow}, {Dwelly}, {Eracleous}, {Eriksen}, {Fan}, {Farr}, {Frederick}, {Fries}, {Frinchaboy}, {G{\"a}nsicke}, {Ge}, {Gonz{\'a}lez {\'A}vila}, {Grabowski}, {Grier}, {Guiglion}, {Gupta}, {Hall}, {Hawkins}, {Hayes}, {Hermes}, {Hern{\'a}ndez-Garc{\'\i}a}, {Hogg}, {Holtzman}, {Ibarra-Medel}, {Ji}, {Jofre}, {Johnson}, {Jones}, {Kinemuchi}, {Kluge}, {Koekemoer}, {Kollmeier}, {Kounkel}, {Krishnarao}, {Krumpe}, {Lacerna}, {Lago}, {Laporte}, {Liu}, {Liu}, {Liu}, {Lopes}, {Macktoobian},
  {Majewski}, {Malanushenko}, {Maoz}, {Masseron}, {Masters}, {Matijevic}, {McBride}, {Medan}, {Merloni}, {Morrison}, {Myers}, {M{\'e}sz{\'a}ros}, {Negrete}, {Nidever}, {Nitschelm}, {Oravetz}, {Oravetz}, {Pan}, {Peng}, {Pinsonneault}, {Pogge}, {Qiu}, {Ramirez}, {Rix}, {Fern{\'a}ndez Rosso}, {Runnoe}, {Salvato}, {Sanchez}, {Santana}, {Saydjari}, {Sayres}, {Schlaufman}, {Schneider}, {Schwope}, {Serna}, {Shen}, {Sobeck}, {Song}, {Souto}, {Spoo}, {Stassun}, {Steinmetz}, {Straumit}, {Stringfellow}, {S{\'a}nchez-Gallego}, {Taghizadeh-Popp}, {Tayar}, {Thakar}, {Tissera}, {Tkachenko}, {Hernandez Toledo}, {Trakhtenbrot}, {Fern{\'a}ndez-Trincado}, {Troup}, {Trump}, {Tuttle}, {Ulloa}, {Vazquez-Mata}, {Vera Alfaro}, {Villanova}, {Wachter}, {Weijmans}, {Wheeler}, {Wilson}, {Wojno}, {Wolf}, {Xue}, {Ybarra}, {Zari}, \& {Zasowski}}]{almeida2023eighteenth}
{Almeida}, A., {Anderson}, S.~F., {Argudo-Fern{\'a}ndez}, M., {et~al.} 2023, \apjs, 267, 44, \dodoi{10.3847/1538-4365/acda98}

\bibitem[{{Arendt} {et~al.}(2024){Arendt}, {Perrott}, {Contreras-Santos}, {de Andres}, {Cui}, \& {Rennehan}}]{arendt2024identifying}
{Arendt}, A.~R., {Perrott}, Y.~C., {Contreras-Santos}, A., {et~al.} 2024, \mnras, 530, 20, \dodoi{10.1093/mnras/stae568}

\bibitem[{{Arnaud}(1996)}]{arnaud1996xspec}
{Arnaud}, K.~A. 1996, in Astronomical Society of the Pacific Conference Series, Vol. 101, Astronomical Data Analysis Software and Systems V, ed. G.~H. {Jacoby} \& J.~{Barnes}, 17

\bibitem[{{Arnaud} {et~al.}(2005){Arnaud}, {Pointecouteau}, \& {Pratt}}]{arnaud2005structural}
{Arnaud}, M., {Pointecouteau}, E., \& {Pratt}, G.~W. 2005, \aap, 441, 893, \dodoi{10.1051/0004-6361:20052856}

\bibitem[{{Arnaud} {et~al.}(2002){Arnaud}, {Majerowicz}, {Lumb}, {Neumann}, {Aghanim}, {Blanchard}, {Boer}, {Burke}, {Collins}, {Giard}, {Nevalainen}, {Nichol}, {Romer}, \& {Sadat}}]{arnaud2002xmm}
{Arnaud}, M., {Majerowicz}, S., {Lumb}, D., {et~al.} 2002, \aap, 390, 27, \dodoi{10.1051/0004-6361:20020669}

\bibitem[{{Astropy Collaboration} {et~al.}(2013){Astropy Collaboration}, {Robitaille}, {Tollerud}, {Greenfield}, {Droettboom}, {Bray}, {Aldcroft}, {Davis}, {Ginsburg}, {Price-Whelan}, {Kerzendorf}, {Conley}, {Crighton}, {Barbary}, {Muna}, {Ferguson}, {Grollier}, {Parikh}, {Nair}, {Unther}, {Deil}, {Woillez}, {Conseil}, {Kramer}, {Turner}, {Singer}, {Fox}, {Weaver}, {Zabalza}, {Edwards}, {Azalee Bostroem}, {Burke}, {Casey}, {Crawford}, {Dencheva}, {Ely}, {Jenness}, {Labrie}, {Lim}, {Pierfederici}, {Pontzen}, {Ptak}, {Refsdal}, {Servillat}, \& {Streicher}}]{robitaille2013astropy}
{Astropy Collaboration}, {Robitaille}, T.~P., {Tollerud}, E.~J., {et~al.} 2013, \aap, 558, A33, \dodoi{10.1051/0004-6361/201322068}

\bibitem[{{Astropy Collaboration} {et~al.}(2018){Astropy Collaboration}, {Price-Whelan}, {Sip{\H{o}}cz}, {G{\"u}nther}, {Lim}, {Crawford}, {Conseil}, {Shupe}, {Craig}, {Dencheva}, {Ginsburg}, {VanderPlas}, {Bradley}, {P{\'e}rez-Su{\'a}rez}, {de Val-Borro}, {Aldcroft}, {Cruz}, {Robitaille}, {Tollerud}, {Ardelean}, {Babej}, {Bach}, {Bachetti}, {Bakanov}, {Bamford}, {Barentsen}, {Barmby}, {Baumbach}, {Berry}, {Biscani}, {Boquien}, {Bostroem}, {Bouma}, {Brammer}, {Bray}, {Breytenbach}, {Buddelmeijer}, {Burke}, {Calderone}, {Cano Rodr{\'\i}guez}, {Cara}, {Cardoso}, {Cheedella}, {Copin}, {Corrales}, {Crichton}, {D'Avella}, {Deil}, {Depagne}, {Dietrich}, {Donath}, {Droettboom}, {Earl}, {Erben}, {Fabbro}, {Ferreira}, {Finethy}, {Fox}, {Garrison}, {Gibbons}, {Goldstein}, {Gommers}, {Greco}, {Greenfield}, {Groener}, {Grollier}, {Hagen}, {Hirst}, {Homeier}, {Horton}, {Hosseinzadeh}, {Hu}, {Hunkeler}, {Ivezi{\'c}}, {Jain}, {Jenness}, {Kanarek}, {Kendrew}, {Kern}, {Kerzendorf}, {Khvalko}, {King}, {Kirkby}, {Kulkarni},
  {Kumar}, {Lee}, {Lenz}, {Littlefair}, {Ma}, {Macleod}, {Mastropietro}, {McCully}, {Montagnac}, {Morris}, {Mueller}, {Mumford}, {Muna}, {Murphy}, {Nelson}, {Nguyen}, {Ninan}, {N{\"o}the}, {Ogaz}, {Oh}, {Parejko}, {Parley}, {Pascual}, {Patil}, {Patil}, {Plunkett}, {Prochaska}, {Rastogi}, {Reddy Janga}, {Sabater}, {Sakurikar}, {Seifert}, {Sherbert}, {Sherwood-Taylor}, {Shih}, {Sick}, {Silbiger}, {Singanamalla}, {Singer}, {Sladen}, {Sooley}, {Sornarajah}, {Streicher}, {Teuben}, {Thomas}, {Tremblay}, {Turner}, {Terr{\'o}n}, {van Kerkwijk}, {de la Vega}, {Watkins}, {Weaver}, {Whitmore}, {Woillez}, {Zabalza}, \& {Astropy Contributors}}]{astropy2018}
{Astropy Collaboration}, {Price-Whelan}, A.~M., {Sip{\H{o}}cz}, B.~M., {et~al.} 2018, \aj, 156, 123, \dodoi{10.3847/1538-3881/aabc4f}

\bibitem[{{Astropy Collaboration} {et~al.}(2022){Astropy Collaboration}, {Price-Whelan}, {Lim}, {Earl}, {Starkman}, {Bradley}, {Shupe}, {Patil}, {Corrales}, {Brasseur}, {N{\"o}the}, {Donath}, {Tollerud}, {Morris}, {Ginsburg}, {Vaher}, {Weaver}, {Tocknell}, {Jamieson}, {van Kerkwijk}, {Robitaille}, {Merry}, {Bachetti}, {G{\"u}nther}, {Aldcroft}, {Alvarado-Montes}, {Archibald}, {B{\'o}di}, {Bapat}, {Barentsen}, {Baz{\'a}n}, {Biswas}, {Boquien}, {Burke}, {Cara}, {Cara}, {Conroy}, {Conseil}, {Craig}, {Cross}, {Cruz}, {D'Eugenio}, {Dencheva}, {Devillepoix}, {Dietrich}, {Eigenbrot}, {Erben}, {Ferreira}, {Foreman-Mackey}, {Fox}, {Freij}, {Garg}, {Geda}, {Glattly}, {Gondhalekar}, {Gordon}, {Grant}, {Greenfield}, {Groener}, {Guest}, {Gurovich}, {Handberg}, {Hart}, {Hatfield-Dodds}, {Homeier}, {Hosseinzadeh}, {Jenness}, {Jones}, {Joseph}, {Kalmbach}, {Karamehmetoglu}, {Ka{\l}uszy{\'n}ski}, {Kelley}, {Kern}, {Kerzendorf}, {Koch}, {Kulumani}, {Lee}, {Ly}, {Ma}, {MacBride}, {Maljaars}, {Muna}, {Murphy}, {Norman},
  {O'Steen}, {Oman}, {Pacifici}, {Pascual}, {Pascual-Granado}, {Patil}, {Perren}, {Pickering}, {Rastogi}, {Roulston}, {Ryan}, {Rykoff}, {Sabater}, {Sakurikar}, {Salgado}, {Sanghi}, {Saunders}, {Savchenko}, {Schwardt}, {Seifert-Eckert}, {Shih}, {Jain}, {Shukla}, {Sick}, {Simpson}, {Singanamalla}, {Singer}, {Singhal}, {Sinha}, {Sip{\H{o}}cz}, {Spitler}, {Stansby}, {Streicher}, {{\v{S}}umak}, {Swinbank}, {Taranu}, {Tewary}, {Tremblay}, {de Val-Borro}, {Van Kooten}, {Vasovi{\'c}}, {Verma}, {de Miranda Cardoso}, {Williams}, {Wilson}, {Winkel}, {Wood-Vasey}, {Xue}, {Yoachim}, {Zhang}, {Zonca}, \& {Astropy Project Contributors}}]{collaboration2022astropy}
{Astropy Collaboration}, {Price-Whelan}, A.~M., {Lim}, P.~L., {et~al.} 2022, \apj, 935, 167, \dodoi{10.3847/1538-4357/ac7c74}

\bibitem[{{B{\^{\i}}rzan} {et~al.}(2019){B{\^{\i}}rzan}, {Rafferty}, {Cassano}, {Brunetti}, {van Weeren}, {Br{\"u}ggen}, {Intema}, {de Gasperin}, {Andrade-Santos}, {Botteon}, {R{\"o}ttgering}, \& {Shimwell}}]{Birzan2019Abell959}
{B{\^{\i}}rzan}, L., {Rafferty}, D.~A., {Cassano}, R., {et~al.} 2019, \mnras, 487, 4775, \dodoi{10.1093/mnras/stz1456}

\bibitem[{{Boschin} {et~al.}(2009){Boschin}, {Barrena}, \& {Girardi}}]{Boschin2009Abell959}
{Boschin}, W., {Barrena}, R., \& {Girardi}, M. 2009, \aap, 495, 15, \dodoi{10.1051/0004-6361:200811043}

\bibitem[{{Botteon} {et~al.}(2019){Botteon}, {Shimwell}, {Bonafede}, {Dallacasa}, {Gastaldello}, {Eckert}, {Brunetti}, {Venturi}, {van Weeren}, {Mandal}, {Br{\"u}ggen}, {Cassano}, {de Gasperin}, {Drabent}, {Dumba}, {Intema}, {Hoang}, {Rafferty}, {R{\"o}ttgering}, {Savini}, {Shulevski}, {Stroe}, \& {Wilber}}]{BotteonA781}
{Botteon}, A., {Shimwell}, T.~W., {Bonafede}, A., {et~al.} 2019, \aap, 622, A19, \dodoi{10.1051/0004-6361/201833861}

\bibitem[{{Botteon} {et~al.}(2022){Botteon}, {Shimwell}, {Cassano}, {Cuciti}, {Zhang}, {Bruno}, {Camillini}, {Natale}, {Jones}, {Gastaldello}, {Simionescu}, {Rossetti}, {Akamatsu}, {van Weeren}, {Brunetti}, {Br{\"u}ggen}, {Groeneveld}, {Hoang}, {Hardcastle}, {Ignesti}, {Di Gennaro}, {Bonafede}, {Drabent}, {R{\"o}ttgering}, {Hoeft}, \& {de Gasperin}}]{botteon2022planck}
{Botteon}, A., {Shimwell}, T.~W., {Cassano}, R., {et~al.} 2022, \aap, 660, A78, \dodoi{10.1051/0004-6361/202143020}

\bibitem[{{Bouhrik} {et~al.}(2025){Bouhrik}, {Stancioli}, \& {Wittman}}]{Bouhrik2025Champagne}
{Bouhrik}, F., {Stancioli}, R., \& {Wittman}, D. 2025, \apj, 988, 166, \dodoi{10.3847/1538-4357/ade67c}

\bibitem[{{Brada{\v{c}}} {et~al.}(2008){Brada{\v{c}}}, {Allen}, {Treu}, {Ebeling}, {Massey}, {Morris}, {von der Linden}, \& {Applegate}}]{bradavc2008revealing}
{Brada{\v{c}}}, M., {Allen}, S.~W., {Treu}, T., {et~al.} 2008, \apj, 687, 959, \dodoi{10.1086/591246}

\bibitem[{{Burchett} {et~al.}(2018){Burchett}, {Tripp}, {Wang}, {Willmer}, {Bowen}, \& {Jenkins}}]{Burchett2018Abell1095}
{Burchett}, J.~N., {Tripp}, T.~M., {Wang}, Q.~D., {et~al.} 2018, \mnras, 475, 2067, \dodoi{10.1093/mnras/stx3170}

\bibitem[{{Cantwell} {et~al.}(2016){Cantwell}, {Scaife}, {Oozeer}, {Wen}, \& {Han}}]{Cantwell2016}
{Cantwell}, T.~M., {Scaife}, A.~M.~M., {Oozeer}, N., {Wen}, Z.~L., \& {Han}, J.~L. 2016, \mnras, 458, 1803, \dodoi{10.1093/mnras/stw419}

\bibitem[{{Carter} \& {Read}(2007)}]{carter2007xmm}
{Carter}, J.~A., \& {Read}, A.~M. 2007, \aap, 464, 1155, \dodoi{10.1051/0004-6361:20065882}

\bibitem[{{Cerny} {et~al.}(2018){Cerny}, {Sharon}, {Andrade-Santos}, {Avila}, {Brada{\v{c}}}, {Bradley}, {Carrasco}, {Coe}, {Czakon}, {Dawson}, {Frye}, {Hoag}, {Huang}, {Johnson}, {Jones}, {Lam}, {Lovisari}, {Mainali}, {Oesch}, {Ogaz}, {Past}, {Paterno-Mahler}, {Peterson}, {Riess}, {Rodney}, {Ryan}, {Salmon}, {Sendra-Server}, {Stark}, {Strolger}, {Trenti}, {Umetsu}, {Vulcani}, \& {Zitrin}}]{cerny2018relics}
{Cerny}, C., {Sharon}, K., {Andrade-Santos}, F., {et~al.} 2018, \apj, 859, 159, \dodoi{10.3847/1538-4357/aabe7b}

\bibitem[{{Cha} {et~al.}(2025){Cha}, {Cho}, {Joo}, {Lee}, {HyeongHan}, {Scofield}, {Finner}, \& {Jee}}]{cha2025high}
{Cha}, S., {Cho}, B.~Y., {Joo}, H., {et~al.} 2025, \apjl, 987, L15, \dodoi{10.3847/2041-8213/add2f0}

\bibitem[{{Chatterjee} {et~al.}(2025){Chatterjee}, {Pillay}, {Datta}, {Raja}, {Knowles}, {Rahaman}, \& {Sikhosana}}]{chatterjee2025exploring}
{Chatterjee}, S., {Pillay}, D., {Datta}, A., {et~al.} 2025, \mnras, 539, 981, \dodoi{10.1093/mnras/staf480}

\bibitem[{{Cibirka} {et~al.}(2018){Cibirka}, {Acebron}, {Zitrin}, {Coe}, {Agulli}, {Andrade-Santos}, {Brada{\v{c}}}, {Frye}, {Livermore}, {Mahler}, {Salmon}, {Sharon}, {Trenti}, {Umetsu}, {Avila}, {Bradley}, {Carrasco}, {Cerny}, {Czakon}, {Dawson}, {Hoag}, {Huang}, {Johnson}, {Jones}, {Kikuchihara}, {Lam}, {Lovisari}, {Mainali}, {Oesch}, {Ogaz}, {Ouchi}, {Past}, {Paterno-Mahler}, {Peterson}, {Ryan}, {Sendra-Server}, {Stark}, {Strait}, {Toft}, \& {Vulcani}}]{cibirka2018relics}
{Cibirka}, N., {Acebron}, A., {Zitrin}, A., {et~al.} 2018, \apj, 863, 145, \dodoi{10.3847/1538-4357/aad2d3}

\bibitem[{{Clowe} {et~al.}(2006){Clowe}, {Brada{\v{c}}}, {Gonzalez}, {Markevitch}, {Randall}, {Jones}, \& {Zaritsky}}]{clowe2006direct}
{Clowe}, D., {Brada{\v{c}}}, M., {Gonzalez}, A.~H., {et~al.} 2006, \apjl, 648, L109, \dodoi{10.1086/508162}

\bibitem[{{Dark Energy Survey Collaboration} {et~al.}(2016){Dark Energy Survey Collaboration}, {Abbott}, {Abdalla}, {Aleksi{\'c}}, {Allam}, {Amara}, {Bacon}, {Balbinot}, {Banerji}, {Bechtol}, {Benoit-L{\'e}vy}, {Bernstein}, {Bertin}, {Blazek}, {Bonnett}, {Bridle}, {Brooks}, {Brunner}, {Buckley-Geer}, {Burke}, {Caminha}, {Capozzi}, {Carlsen}, {Carnero-Rosell}, {Carollo}, {Carrasco-Kind}, {Carretero}, {Castander}, {Clerkin}, {Collett}, {Conselice}, {Crocce}, {Cunha}, {D'Andrea}, {da Costa}, {Davis}, {Desai}, {Diehl}, {Dietrich}, {Dodelson}, {Doel}, {Drlica-Wagner}, {Estrada}, {Etherington}, {Evrard}, {Fabbri}, {Finley}, {Flaugher}, {Foley}, {Fosalba}, {Frieman}, {Garc{\'\i}a-Bellido}, {Gaztanaga}, {Gerdes}, {Giannantonio}, {Goldstein}, {Gruen}, {Gruendl}, {Guarnieri}, {Gutierrez}, {Hartley}, {Honscheid}, {Jain}, {James}, {Jeltema}, {Jouvel}, {Kessler}, {King}, {Kirk}, {Kron}, {Kuehn}, {Kuropatkin}, {Lahav}, {Li}, {Lima}, {Lin}, {Maia}, {Makler}, {Manera}, {Maraston}, {Marshall}, {Martini}, {McMahon},
  {Melchior}, {Merson}, {Miller}, {Miquel}, {Mohr}, {Morice-Atkinson}, {Naidoo}, {Neilsen}, {Nichol}, {Nord}, {Ogando}, {Ostrovski}, {Palmese}, {Papadopoulos}, {Peiris}, {Peoples}, {Percival}, {Plazas}, {Reed}, {Refregier}, {Romer}, {Roodman}, {Ross}, {Rozo}, {Rykoff}, {Sadeh}, {Sako}, {S{\'a}nchez}, {Sanchez}, {Santiago}, {Scarpine}, {Schubnell}, {Sevilla-Noarbe}, {Sheldon}, {Smith}, {Smith}, {Soares-Santos}, {Sobreira}, {Soumagnac}, {Suchyta}, {Sullivan}, {Swanson}, {Tarle}, {Thaler}, {Thomas}, {Thomas}, {Tucker}, {Vieira}, {Vikram}, {Walker}, {Wechsler}, {Weller}, {Wester}, {Whiteway}, {Wilcox}, {Yanny}, {Zhang}, \& {Zuntz}}]{dark2016dark}
{Dark Energy Survey Collaboration}, {Abbott}, T., {Abdalla}, F.~B., {et~al.} 2016, \mnras, 460, 1270, \dodoi{10.1093/mnras/stw641}

\bibitem[{{Dawson} {et~al.}(2013){Dawson}, {Schlegel}, {Ahn}, {Anderson}, {Aubourg}, {Bailey}, {Barkhouser}, {Bautista}, {Beifiori}, {Berlind}, {Bhardwaj}, {Bizyaev}, {Blake}, {Blanton}, {Blomqvist}, {Bolton}, {Borde}, {Bovy}, {Brandt}, {Brewington}, {Brinkmann}, {Brown}, {Brownstein}, {Bundy}, {Busca}, {Carithers}, {Carnero}, {Carr}, {Chen}, {Comparat}, {Connolly}, {Cope}, {Croft}, {Cuesta}, {da Costa}, {Davenport}, {Delubac}, {de Putter}, {Dhital}, {Ealet}, {Ebelke}, {Eisenstein}, {Escoffier}, {Fan}, {Filiz Ak}, {Finley}, {Font-Ribera}, {G{\'e}nova-Santos}, {Gunn}, {Guo}, {Haggard}, {Hall}, {Hamilton}, {Harris}, {Harris}, {Ho}, {Hogg}, {Holder}, {Honscheid}, {Huehnerhoff}, {Jordan}, {Jordan}, {Kauffmann}, {Kazin}, {Kirkby}, {Klaene}, {Kneib}, {Le Goff}, {Lee}, {Long}, {Loomis}, {Lundgren}, {Lupton}, {Maia}, {Makler}, {Malanushenko}, {Malanushenko}, {Mandelbaum}, {Manera}, {Maraston}, {Margala}, {Masters}, {McBride}, {McDonald}, {McGreer}, {McMahon}, {Mena}, {Miralda-Escud{\'e}}, {Montero-Dorta},
  {Montesano}, {Muna}, {Myers}, {Naugle}, {Nichol}, {Noterdaeme}, {Nuza}, {Olmstead}, {Oravetz}, {Oravetz}, {Owen}, {Padmanabhan}, {Palanque-Delabrouille}, {Pan}, {Parejko}, {P{\^a}ris}, {Percival}, {P{\'e}rez-Fournon}, {P{\'e}rez-R{\`a}fols}, {Petitjean}, {Pfaffenberger}, {Pforr}, {Pieri}, {Prada}, {Price-Whelan}, {Raddick}, {Rebolo}, {Rich}, {Richards}, {Rockosi}, {Roe}, {Ross}, {Ross}, {Rossi}, {Rubi{\~n}o-Martin}, {Samushia}, {S{\'a}nchez}, {Sayres}, {Schmidt}, {Schneider}, {Sc{\'o}ccola}, {Seo}, {Shelden}, {Sheldon}, {Shen}, {Shu}, {Slosar}, {Smee}, {Snedden}, {Stauffer}, {Steele}, {Strauss}, {Streblyanska}, {Suzuki}, {Swanson}, {Tal}, {Tanaka}, {Thomas}, {Tinker}, {Tojeiro}, {Tremonti}, {Vargas Maga{\~n}a}, {Verde}, {Viel}, {Wake}, {Watson}, {Weaver}, {Weinberg}, {Weiner}, {West}, {White}, {Wood-Vasey}, {Yeche}, {Zehavi}, {Zhao}, \& {Zheng}}]{dawson2012baryon}
{Dawson}, K.~S., {Schlegel}, D.~J., {Ahn}, C.~P., {et~al.} 2013, \aj, 145, 10, \dodoi{10.1088/0004-6256/145/1/10}

\bibitem[{{Dawson} {et~al.}(2012){Dawson}, {Wittman}, {Jee}, {Gee}, {Hughes}, {Tyson}, {Schmidt}, {Thorman}, {Brada{\v{c}}}, {Miyazaki}, {Lemaux}, {Utsumi}, \& {Margoniner}}]{dawson2012discovery}
{Dawson}, W.~A., {Wittman}, D., {Jee}, M.~J., {et~al.} 2012, \apjl, 747, L42, \dodoi{10.1088/2041-8205/747/2/L42}

\bibitem[{{DESI Collaboration} {et~al.}(2025){DESI Collaboration}, {Abdul-Karim}, {Adame}, {Aguado}, {Aguilar}, {Ahlen}, {Alam}, {Aldering}, {Alexander}, {Alfarsy}, {Allen}, {Allende Prieto}, {Alves}, {Anand}, {Andrade}, {Armengaud}, {Avila}, {Aviles}, {Awan}, {Bailey}, {Baleato Lizancos}, {Ballester}, {Bault}, {Bautista}, {BenZvi}, {Beraldo e Silva}, {Bermejo-Climent}, {Beutler}, {Bianchi}, {Blake}, {Blum}, {Bolton}, {Bonici}, {Brieden}, {Brodzeller}, {Brooks}, {Buckley-Geer}, {Burtin}, {Canning}, {Carnero Rosell}, {Carr}, {Carrilho}, {Casas}, {Castander}, {Cereskaite}, {Cervantes-Cota}, {Chaussidon}, {Chaves-Montero}, {Chen}, {Chen}, {Claybaugh}, {Cole}, {Cooper}, {Cousinou}, {Cuceu}, {Davis}, {Dawson}, {de Belsunce}, {de la Cruz}, {de la Macorra}, {de Mattia}, {Deiosso}, {Della Costa}, {Demina}, {Demirbozan}, {DeRose}, {Dey}, {Dey}, {Ding}, {Ding}, {Doel}, {Douglass}, {Dowicz}, {Ebina}, {Edelstein}, {Eisenstein}, {Elbers}, {Emas}, {Escoffier}, {Fagrelius}, {Fan}, {Fanning}, {Fawcett},
  {Fern\'andez-Garc\'ia}, {Ferraro}, {Findlay}, {Font-Ribera}, {Forero-Romero}, {Forero-S\'anchez}, {Frenk}, {G\''ansicke}, {Galbany}, {Garc\'ia-Bellido}, {Garcia-Quintero}, {Garrison}, {Gaztatop\~naga}, {Gil-Martop'in}, {Gnedin}, {Gontcho}, {Gonzalez-Morales}, {Gonzalez-Perez}, {Gordon}, {Graur}, {Green}, {Gruen}, {Gsponer}, {Guandalin}, {Gutierrez}, {Guy}, {Hahn}, {Han}, {Han}, {He}, {Herrera-Alcantar}, {Honscheid}, {Hou}, {Howlett}, {Huterer}, {Irtopv\{s\}itopv\{c\}}, {Ishak}, {Jacques}, {Jimenez}, {Jing}, {Joachimi}, {Joudaki}, {Joyce}, {Jullo}, {Juneau}, {Karatopc\{c\}ayl\{topi\}}, {Karim}, {Kehoe}, {Kent}, {Khederlarian}, {Kirkby}, {Kisner}, {Kitaura}, {Kizhuprakkat}, {Kong}, {Koposov}, {Kremin}, {Krolewski}, {Lahav}, {Lai}, {Lamman}, {Lan}, {Landriau}, {Lang}, {Lange}, {Lasker}, {Le Goff}, {Le Guillou}, {Leauthaud}, {Levi}, {Li}, {Li}, {Lodha}, {Lokken}, {Luo}, {Magneville}, {Manera}, {Manser}, {Margala}, {Martini}, {Maus}, {McCullough}, {McDonald}, {Medina}, {Medina-Varela}, {Meisner},
  {Mena-Ferntop'andez}, {Menegas}, {Mezcua}, {Miquel}, {Montero-Camacho}, {Moon}, {Moustakas}, {Mutop\~noz-Gutitop'errez}, {Mutop\~noz-Santos}, {Myers}, {Myles}, {Nadathur}, {Najita}, {Napolitano}, {Newman}, {Nikakhtar}, {Nikutta}, {Niz}, {Noriega}, {Padmanabhan}, {Paillas}, {Palanque-Delabrouille}, {Palmese}, {Pan}, {Pan}, {Parkinson}, {Peacock}, {Percival}, {Ptop'erez-Ferntop'andez}, {Ptop'erez-Rtop`afols}, \& {Peterson}}]{abdul2025data}
{DESI Collaboration}, {Abdul-Karim}, M., {Adame}, A.~G., {et~al.} 2025, arXiv e-prints, arXiv:2503.14745, \dodoi{10.48550/arXiv.2503.14745}

\bibitem[{{Dey} {et~al.}(2019){Dey}, {Schlegel}, {Lang}, {Blum}, {Burleigh}, {Fan}, {Findlay}, {Finkbeiner}, {Herrera}, {Juneau}, {Landriau}, {Levi}, {McGreer}, {Meisner}, {Myers}, {Moustakas}, {Nugent}, {Patej}, {Schlafly}, {Walker}, {Valdes}, {Weaver}, {Y{\`e}che}, {Zou}, {Zhou}, {Abareshi}, {Abbott}, {Abolfathi}, {Aguilera}, {Alam}, {Allen}, {Alvarez}, {Annis}, {Ansarinejad}, {Aubert}, {Beechert}, {Bell}, {BenZvi}, {Beutler}, {Bielby}, {Bolton}, {Brice{\~n}o}, {Buckley-Geer}, {Butler}, {Calamida}, {Carlberg}, {Carter}, {Casas}, {Castander}, {Choi}, {Comparat}, {Cukanovaite}, {Delubac}, {DeVries}, {Dey}, {Dhungana}, {Dickinson}, {Ding}, {Donaldson}, {Duan}, {Duckworth}, {Eftekharzadeh}, {Eisenstein}, {Etourneau}, {Fagrelius}, {Farihi}, {Fitzpatrick}, {Font-Ribera}, {Fulmer}, {G{\"a}nsicke}, {Gaztanaga}, {George}, {Gerdes}, {Gontcho}, {Gorgoni}, {Green}, {Guy}, {Harmer}, {Hernandez}, {Honscheid}, {Huang}, {James}, {Jannuzi}, {Jiang}, {Joyce}, {Karcher}, {Karkar}, {Kehoe}, {Kneib}, {Kueter-Young}, {Lan},
  {Lauer}, {Le Guillou}, {Le Van Suu}, {Lee}, {Lesser}, {Perreault Levasseur}, {Li}, {Mann}, {Marshall}, {Mart{\'\i}nez-V{\'a}zquez}, {Martini}, {du Mas des Bourboux}, {McManus}, {Meier}, {M{\'e}nard}, {Metcalfe}, {Mu{\~n}oz-Guti{\'e}rrez}, {Najita}, {Napier}, {Narayan}, {Newman}, {Nie}, {Nord}, {Norman}, {Olsen}, {Paat}, {Palanque-Delabrouille}, {Peng}, {Poppett}, {Poremba}, {Prakash}, {Rabinowitz}, {Raichoor}, {Rezaie}, {Robertson}, {Roe}, {Ross}, {Ross}, {Rudnick}, {Safonova}, {Saha}, {S{\'a}nchez}, {Savary}, {Schweiker}, {Scott}, {Seo}, {Shan}, {Silva}, {Slepian}, {Soto}, {Sprayberry}, {Staten}, {Stillman}, {Stupak}, {Summers}, {Sien Tie}, {Tirado}, {Vargas-Maga{\~n}a}, {Vivas}, {Wechsler}, {Williams}, {Yang}, {Yang}, {Yapici}, {Zaritsky}, {Zenteno}, {Zhang}, {Zhang}, {Zhou}, \& {Zhou}}]{dey2019overview}
{Dey}, A., {Schlegel}, D.~J., {Lang}, D., {et~al.} 2019, \aj, 157, 168, \dodoi{10.3847/1538-3881/ab089d}

\bibitem[{{Duchesne} {et~al.}(2020){Duchesne}, {Johnston-Hollitt}, {Zhu}, {Wayth}, \& {Line}}]{duchesne2020murchison}
{Duchesne}, S.~W., {Johnston-Hollitt}, M., {Zhu}, Z., {Wayth}, R.~B., \& {Line}, J.~L.~B. 2020, \pasa, 37, e037, \dodoi{10.1017/pasa.2020.29}

\bibitem[{{Dutta} {et~al.}(2025){Dutta}, {Peterson}, {Cianfaglione}, \& {Sembroski}}]{dutta2025weak}
{Dutta}, A., {Peterson}, J.~R., {Cianfaglione}, M., \& {Sembroski}, G. 2025, \pasp, 137, 064102, \dodoi{10.1088/1538-3873/adcbd5}

\bibitem[{{Faber} {et~al.}(2003){Faber}, {Phillips}, {Kibrick}, {Alcott}, {Allen}, {Burrous}, {Cantrall}, {Clarke}, {Coil}, {Cowley}, {Davis}, {Deich}, {Dietsch}, {Gilmore}, {Harper}, {Hilyard}, {Lewis}, {McVeigh}, {Newman}, {Osborne}, {Schiavon}, {Stover}, {Tucker}, {Wallace}, {Wei}, {Wirth}, \& {Wright}}]{faber2003deimos}
{Faber}, S.~M., {Phillips}, A.~C., {Kibrick}, R.~I., {et~al.} 2003, in Society of Photo-Optical Instrumentation Engineers (SPIE) Conference Series, Vol. 4841, Instrument Design and Performance for Optical/Infrared Ground-based Telescopes, ed. M.~{Iye} \& A.~F.~M. {Moorwood}, 1657--1669, \dodoi{10.1117/12.460346}

\bibitem[{{Finner} {et~al.}(2025){Finner}, {Jee}, {Cho}, {HyeongHan}, {Lee}, {van Weeren}, {Wittman}, \& {Yoon}}]{Finner2025}
{Finner}, K., {Jee}, M.~J., {Cho}, H., {et~al.} 2025, \apjs, 277, 28, \dodoi{10.3847/1538-4365/adb0b6}

\bibitem[{{Ge} {et~al.}(2019){Ge}, {Sun}, {Rozo}, {Sehgal}, {Vikhlinin}, {Forman}, {Jones}, \& {Nagai}}]{geXray2019}
{Ge}, C., {Sun}, M., {Rozo}, E., {et~al.} 2019, \mnras, 484, 1946, \dodoi{10.1093/mnras/stz088}

\bibitem[{{Ge} {et~al.}(2016){Ge}, {Wang}, {Tripp}, {Li}, {Gu}, \& {Ji}}]{ge2016baryon}
{Ge}, C., {Wang}, Q.~D., {Tripp}, T.~M., {et~al.} 2016, \mnras, 459, 366, \dodoi{10.1093/mnras/stw599}

\bibitem[{{George} {et~al.}(2021){George}, {Kale}, \& {Wadadekar}}]{george2021imaging}
{George}, L.~T., {Kale}, R., \& {Wadadekar}, Y. 2021, \mnras, 507, 4487, \dodoi{10.1093/mnras/stab2309}

\bibitem[{{Ginsburg} {et~al.}(2019){Ginsburg}, {Sip{\H{o}}cz}, {Brasseur}, {Cowperthwaite}, {Craig}, {Deil}, {Guillochon}, {Guzman}, {Liedtke}, {Lian Lim}, {Lockhart}, {Mommert}, {Morris}, {Norman}, {Parikh}, {Persson}, {Robitaille}, {Segovia}, {Singer}, {Tollerud}, {de Val-Borro}, {Valtchanov}, {Woillez}, {Astroquery Collaboration}, \& {a subset of astropy Collaboration}}]{ginsburg2019astroquery}
{Ginsburg}, A., {Sip{\H{o}}cz}, B.~M., {Brasseur}, C.~E., {et~al.} 2019, \aj, 157, 98, \dodoi{10.3847/1538-3881/aafc33}

\bibitem[{{Golovich} {et~al.}(2016){Golovich}, {Dawson}, {Wittman}, {Ogrean}, {van Weeren}, \& {Bonafede}}]{golovichDynamical2016}
{Golovich}, N., {Dawson}, W.~A., {Wittman}, D., {et~al.} 2016, \apj, 831, 110, \dodoi{10.3847/0004-637X/831/1/110}

\bibitem[{{Golovich} {et~al.}(2019{\natexlab{a}}){Golovich}, {Dawson}, {Wittman}, {van Weeren}, {Andrade-Santos}, {Jee}, {Benson}, {de Gasperin}, {Venturi}, {Bonafede}, {Sobral}, {Ogrean}, {Lemaux}, {Brada{\v{c}}}, {Br{\"u}ggen}, \& {Peter}}]{golovich2019merging1}
{Golovich}, N., {Dawson}, W.~A., {Wittman}, D.~M., {et~al.} 2019{\natexlab{a}}, \apj, 882, 69, \dodoi{10.3847/1538-4357/ab2f90}

\bibitem[{{Golovich} {et~al.}(2019{\natexlab{b}}){Golovich}, {Dawson}, {Wittman}, {Jee}, {Benson}, {Lemaux}, {van Weeren}, {Andrade-Santos}, {Sobral}, {de Gasperin}, {Br{\"u}ggen}, {Brada{\v{c}}}, {Finner}, \& {Peter}}]{golovich2019merging2}
---. 2019{\natexlab{b}}, \apjs, 240, 39, \dodoi{10.3847/1538-4365/aaf88b}

\bibitem[{{Groeneveld} {et~al.}(2025){Groeneveld}, {van Weeren}, {Botteon}, {Cassano}, {de Gasperin}, {Osinga}, {Brunetti}, \& {R{\"o}ttgering}}]{groeneveld2025serendipitous}
{Groeneveld}, C., {van Weeren}, R.~J., {Botteon}, A., {et~al.} 2025, \aap, 693, A99, \dodoi{10.1051/0004-6361/202452687}

\bibitem[{{Hao} {et~al.}(2010){Hao}, {McKay}, {Koester}, {Rykoff}, {Rozo}, {Annis}, {Wechsler}, {Evrard}, {Siegel}, {Becker}, {Busha}, {Gerdes}, {Johnston}, \& {Sheldon}}]{hao2010gmbcg}
{Hao}, J., {McKay}, T.~A., {Koester}, B.~P., {et~al.} 2010, \apjs, 191, 254, \dodoi{10.1088/0067-0049/191/2/254}

\bibitem[{{Harvey} {et~al.}(2015){Harvey}, {Massey}, {Kitching}, {Taylor}, \& {Tittley}}]{harvey2015nongravitational}
{Harvey}, D., {Massey}, R., {Kitching}, T., {Taylor}, A., \& {Tittley}, E. 2015, Science, 347, 1462, \dodoi{10.1126/science.1261381}

\bibitem[{{Hoeft} {et~al.}(2021){Hoeft}, {Dumba}, {Drabent}, {Rajpurohit}, {Rossetti}, {Nuza}, {van Weeren}, {Meusinger}, {Botteon}, {Brunetti}, {Shimwell}, {Cassano}, {Br{\"u}ggen}, {R{\"o}ttgering}, {Gastaldello}, {Lovisari}, {Yepes}, {Andrade-Santos}, \& {Eckert}}]{Hoeft2021}
{Hoeft}, M., {Dumba}, C., {Drabent}, A., {et~al.} 2021, \aap, 654, A68, \dodoi{10.1051/0004-6361/202039725}

\bibitem[{{Hopp} {et~al.}(2026){Hopp}, {Wittman}, {Stancioli}, {Gao}, {Bouhrik}, \& {Adler}}]{hopp2026redshifts}
{Hopp}, C., {Wittman}, D., {Stancioli}, R., {et~al.} 2026, {Spectroscopic Redshifts of 1,561 Galaxies in 12 Merging Galaxy Clusters from the X-SORTER Survey},  Zenodo, \dodoi{10.5281/zenodo.18879235}

\bibitem[{{Ivezi{\'c}} {et~al.}(2019){Ivezi{\'c}}, {Kahn}, {Tyson}, {Abel}, {Acosta}, {Allsman}, {Alonso}, {AlSayyad}, {Anderson}, {Andrew}, {Angel}, {Angeli}, {Ansari}, {Antilogus}, {Araujo}, {Armstrong}, {Arndt}, {Astier}, {Aubourg}, {Auza}, {Axelrod}, {Bard}, {Barr}, {Barrau}, {Bartlett}, {Bauer}, {Bauman}, {Baumont}, {Bechtol}, {Bechtol}, {Becker}, {Becla}, {Beldica}, {Bellavia}, {Bianco}, {Biswas}, {Blanc}, {Blazek}, {Blandford}, {Bloom}, {Bogart}, {Bond}, {Booth}, {Borgland}, {Borne}, {Bosch}, {Boutigny}, {Brackett}, {Bradshaw}, {Brandt}, {Brown}, {Bullock}, {Burchat}, {Burke}, {Cagnoli}, {Calabrese}, {Callahan}, {Callen}, {Carlin}, {Carlson}, {Chandrasekharan}, {Charles-Emerson}, {Chesley}, {Cheu}, {Chiang}, {Chiang}, {Chirino}, {Chow}, {Ciardi}, {Claver}, {Cohen-Tanugi}, {Cockrum}, {Coles}, {Connolly}, {Cook}, {Cooray}, {Covey}, {Cribbs}, {Cui}, {Cutri}, {Daly}, {Daniel}, {Daruich}, {Daubard}, {Daues}, {Dawson}, {Delgado}, {Dellapenna}, {de Peyster}, {de Val-Borro}, {Digel}, {Doherty}, {Dubois},
  {Dubois-Felsmann}, {Durech}, {Economou}, {Eifler}, {Eracleous}, {Emmons}, {Fausti Neto}, {Ferguson}, {Figueroa}, {Fisher-Levine}, {Focke}, {Foss}, {Frank}, {Freemon}, {Gangler}, {Gawiser}, {Geary}, {Gee}, {Geha}, {Gessner}, {Gibson}, {Gilmore}, {Glanzman}, {Glick}, {Goldina}, {Goldstein}, {Goodenow}, {Graham}, {Gressler}, {Gris}, {Guy}, {Guyonnet}, {Haller}, {Harris}, {Hascall}, {Haupt}, {Hernandez}, {Herrmann}, {Hileman}, {Hoblitt}, {Hodgson}, {Hogan}, {Howard}, {Huang}, {Huffer}, {Ingraham}, {Innes}, {Jacoby}, {Jain}, {Jammes}, {Jee}, {Jenness}, {Jernigan}, {Jevremovi{\'c}}, {Johns}, {Johnson}, {Johnson}, {Jones}, {Juramy-Gilles}, {Juri{\'c}}, {Kalirai}, {Kallivayalil}, {Kalmbach}, {Kantor}, {Karst}, {Kasliwal}, {Kelly}, {Kessler}, {Kinnison}, {Kirkby}, {Knox}, {Kotov}, {Krabbendam}, {Krughoff}, {Kub{\'a}nek}, {Kuczewski}, {Kulkarni}, {Ku}, {Kurita}, {Lage}, {Lambert}, {Lange}, {Langton}, {Le Guillou}, {Levine}, {Liang}, {Lim}, {Lintott}, {Long}, {Lopez}, {Lotz}, {Lupton}, {Lust}, {MacArthur}, {Mahabal},
  {Mandelbaum}, {Markiewicz}, {Marsh}, {Marshall}, {Marshall}, {May}, {McKercher}, {McQueen}, {Meyers}, {Migliore}, {Miller}, \& {Mills}}]{ivezic2019lsst}
{Ivezi{\'c}}, {\v{Z}}., {Kahn}, S.~M., {Tyson}, J.~A., {et~al.} 2019, \apj, 873, 111, \dodoi{10.3847/1538-4357/ab042c}

\bibitem[{{Jee} {et~al.}(2014){Jee}, {Hughes}, {Menanteau}, {Sif{\'o}n}, {Mandelbaum}, {Barrientos}, {Infante}, \& {Ng}}]{jee2014weighing}
{Jee}, M.~J., {Hughes}, J.~P., {Menanteau}, F., {et~al.} 2014, \apj, 785, 20, \dodoi{10.1088/0004-637X/785/1/20}

\bibitem[{{Kaiser} {et~al.}(2002){Kaiser}, {Aussel}, {Burke}, {Boesgaard}, {Chambers}, {Chun}, {Heasley}, {Hodapp}, {Hunt}, {Jedicke}, {Jewitt}, {Kudritzki}, {Luppino}, {Maberry}, {Magnier}, {Monet}, {Onaka}, {Pickles}, {Rhoads}, {Simon}, {Szalay}, {Szapudi}, {Tholen}, {Tonry}, {Waterson}, \& {Wick}}]{kaiser2002pan}
{Kaiser}, N., {Aussel}, H., {Burke}, B.~E., {et~al.} 2002, in Society of Photo-Optical Instrumentation Engineers (SPIE) Conference Series, Vol. 4836, Survey and Other Telescope Technologies and Discoveries, ed. J.~A. {Tyson} \& S.~{Wolff}, 154--164, \dodoi{10.1117/12.457365}

\bibitem[{{Kale} {et~al.}(2015){Kale}, {Venturi}, {Giacintucci}, {Dallacasa}, {Cassano}, {Brunetti}, {Cuciti}, {Macario}, \& {Athreya}}]{kale2015extended}
{Kale}, R., {Venturi}, T., {Giacintucci}, S., {et~al.} 2015, \aap, 579, A92, \dodoi{10.1051/0004-6361/201525695}

\bibitem[{{Kim} {et~al.}(2022){Kim}, {Ko}, {Smith}, {Kim}, {Hwang}, {Song}, {Shin}, \& {Yoo}}]{kim2022a2261}
{Kim}, H., {Ko}, J., {Smith}, R., {et~al.} 2022, \apj, 928, 170, \dodoi{10.3847/1538-4357/ac510e}

\bibitem[{{Kim} {et~al.}(2017){Kim}, {Peter}, \& {Wittman}}]{kim2017wake}
{Kim}, S.~Y., {Peter}, A. H.~G., \& {Wittman}, D. 2017, \mnras, 469, 1414, \dodoi{10.1093/mnras/stx896}

\bibitem[{{Kollmeier} {et~al.}(2019){Kollmeier}, {Anderson}, {Blanc}, {Blanton}, {Covey}, {Crane}, {Drory}, {Frinchaboy}, {Froning}, {Johnson}, {Kneib}, {Kreckel}, {Merloni}, {Pellegrini}, {Pogge}, {Ramirez}, {Rix}, {Sayres}, {S{\'a}nchez-Gallego}, {Shen}, {Tkachenko}, {Trump}, {Tuttle}, {Weijmans}, {Zasowski}, {Barbuy}, {Beaton}, {Bergemann}, {Bochanski}, {Brandt}, {Casey}, {Cherinka}, {Eracleous}, {Fan}, {Garc{\'\i}a}, {Green}, {Hekker}, {Lane}, {Longa-Pe{\~n}a}, {Mathur}, {Meza}, {Minchev}, {Myers}, {Nidever}, {Nitschelm}, {O'Connell}, {Price-Whelan}, {Raddick}, {Rossi}, {Sankrit}, {Simon}, {Stutz}, {Ting}, {Trakhtenbrot}, {Weaver}, {Willmer}, \& {Weinberg}}]{kollmeier2019sdss}
{Kollmeier}, J., {Anderson}, S.~F., {Blanc}, G.~A., {et~al.} 2019, in Bulletin of the American Astronomical Society, Vol.~51, 274

\bibitem[{{Liu} {et~al.}(2018){Liu}, {Yu}, {Diaferio}, {Tozzi}, {Hwang}, {Umetsu}, {Okabe}, \& {Yang}}]{Liu2018Abell2142}
{Liu}, A., {Yu}, H., {Diaferio}, A., {et~al.} 2018, \apj, 863, 102, \dodoi{10.3847/1538-4357/aad090}

\bibitem[{{Mann} \& {Ebeling}(2012)}]{mann2012x}
{Mann}, A.~W., \& {Ebeling}, H. 2012, \mnras, 420, 2120, \dodoi{10.1111/j.1365-2966.2011.20170.x}

\bibitem[{{Markevitch} {et~al.}(2004){Markevitch}, {Gonzalez}, {Clowe}, {Vikhlinin}, {Forman}, {Jones}, {Murray}, \& {Tucker}}]{markevitch2004direct}
{Markevitch}, M., {Gonzalez}, A.~H., {Clowe}, D., {et~al.} 2004, \apj, 606, 819, \dodoi{10.1086/383178}

\bibitem[{{McDonald} {et~al.}(2022){McDonald}, {Obreschkow}, \& {Garratt-Smithson}}]{McDonald2022}
{McDonald}, W., {Obreschkow}, D., \& {Garratt-Smithson}, L. 2022, \mnras, 516, 5289, \dodoi{10.1093/mnras/stac2276}

\bibitem[{{Ochsenbein} {et~al.}(2004){Ochsenbein}, {Allen}, \& {Egret}}]{gabriel2004astronomical}
{Ochsenbein}, F., {Allen}, M.~G., \& {Egret}, D., eds. 2004, Astronomical Society of the Pacific Conference Series, Vol. 314, {Astronomical Data Analysis Software and Systems (ADASS) XIII}

\bibitem[{Ogrean {et~al.}(2016)Ogrean, Weeren, Jones, Forman, Dawson, Golovich, Andrade-Santos, Murray, Nulsen, Roediger, Zitrin, Bulbul, Kraft, Goulding, Umetsu, Mroczkowski, Bonafede, Randall, Sayers, Churazov, David, Merten, Donahue, Mason, Rosati, Vikhlinin, \& Ebeling}]{ogrean2016frontier}
Ogrean, G.~A., Weeren, R. J.~v., Jones, C., {et~al.} 2016, The Astrophysical Journal, 819, 113, \dodoi{10.3847/0004-637X/819/2/113}

\bibitem[{{Paterno-Mahler} {et~al.}(2014){Paterno-Mahler}, {Randall}, {Bulbul}, {Andrade-Santos}, {Blanton}, {Jones}, {Murray}, \& {Johnson}}]{Paterno-Mahler2014}
{Paterno-Mahler}, R., {Randall}, S.~W., {Bulbul}, E., {et~al.} 2014, \apj, 791, 104, \dodoi{10.1088/0004-637X/791/2/104}

\bibitem[{{Paul} {et~al.}(2021){Paul}, {Gupta}, {Salunkhe}, {Bhagat}, {Sonkamble}, {Hiray}, {Dabhade}, \& {Raychaudhury}}]{paul2021ugmrt}
{Paul}, S., {Gupta}, P., {Salunkhe}, S., {et~al.} 2021, \mnras, 506, 5389, \dodoi{10.1093/mnras/stab1965}

\bibitem[{{Pedregosa} {et~al.}(2011){Pedregosa}, {Varoquaux}, {Gramfort}, {Michel}, {Thirion}, {Grisel}, {Blondel}, {M{\"u}ller}, {Nothman}, {Louppe}, {Prettenhofer}, {Weiss}, {Dubourg}, {Vanderplas}, {Passos}, {Cournapeau}, {Brucher}, {Perrot}, \& {Duchesnay}}]{pedregosa2011scikit}
{Pedregosa}, F., {Varoquaux}, G., {Gramfort}, A., {et~al.} 2011, Journal of Machine Learning Research, 12, 2825, \dodoi{10.48550/arXiv.1201.0490}

\bibitem[{{Pillay} {et~al.}(2021){Pillay}, {Turner}, {Hilton}, {Knowles}, {Kesebonye}, {Moodley}, {Mroczkowski}, {Oozeer}, {Pfrommer}, {Sikhosana}, \& {Wollack}}]{pillay2021multiwavelength}
{Pillay}, D.~S., {Turner}, D.~J., {Hilton}, M., {et~al.} 2021, Galaxies, 9, 97, \dodoi{10.3390/galaxies9040097}

\bibitem[{{Poole} {et~al.}(2007){Poole}, {Babul}, {McCarthy}, {Fardal}, {Bildfell}, {Quinn}, \& {Mahdavi}}]{poole2007impact}
{Poole}, G.~B., {Babul}, A., {McCarthy}, I.~G., {et~al.} 2007, \mnras, 380, 437, \dodoi{10.1111/j.1365-2966.2007.12107.x}

\bibitem[{{Prochaska} {et~al.}(2020){Prochaska}, {Hennawi}, {Westfall}, {Cooke}, {Wang}, {Hsyu}, {Davies}, {Farina}, \& {Pelliccia}}]{prochaska2020pypeit}
{Prochaska}, J., {Hennawi}, J., {Westfall}, K., {et~al.} 2020, The Journal of Open Source Software, 5, 2308, \dodoi{10.21105/joss.02308}

\bibitem[{{Randall} {et~al.}(2008){Randall}, {Markevitch}, {Clowe}, {Gonzalez}, \& {Brada{\v{c}}}}]{randall2008constraints}
{Randall}, S.~W., {Markevitch}, M., {Clowe}, D., {Gonzalez}, A.~H., \& {Brada{\v{c}}}, M. 2008, \apj, 679, 1173, \dodoi{10.1086/587859}

\bibitem[{{Ricker} \& {Sarazin}(2001)}]{ricker2001off}
{Ricker}, P.~M., \& {Sarazin}, C.~L. 2001, \apj, 561, 621, \dodoi{10.1086/323365}

\bibitem[{{Rines} {et~al.}(2022){Rines}, {Sohn}, {Geller}, \& {Diaferio}}]{rines2022spectroscopic}
{Rines}, K.~J., {Sohn}, J., {Geller}, M.~J., \& {Diaferio}, A. 2022, \apj, 930, 156, \dodoi{10.3847/1538-4357/ac67a8}

\bibitem[{{Robertson} {et~al.}(2019){Robertson}, {Harvey}, {Massey}, {Eke}, {McCarthy}, {Jauzac}, {Li}, \& {Schaye}}]{robertson2019observable}
{Robertson}, A., {Harvey}, D., {Massey}, R., {et~al.} 2019, \mnras, 488, 3646, \dodoi{10.1093/mnras/stz1815}

\bibitem[{{Robertson} {et~al.}(2017){Robertson}, {Massey}, \& {Eke}}]{robertson2016does}
{Robertson}, A., {Massey}, R., \& {Eke}, V. 2017, \mnras, 465, 569, \dodoi{10.1093/mnras/stw2670}

\bibitem[{{Robitaille} {et~al.}(2020){Robitaille}, {Deil}, \& {Ginsburg}}]{robitaille2020reproject}
{Robitaille}, T., {Deil}, C., \& {Ginsburg}, A. 2020, {reproject: Python-based astronomical image reprojection}, Astrophysics Source Code Library, record ascl:2011.023.
\newblock \doeprint{2011.023}

\bibitem[{{Rykoff} {et~al.}(2014){Rykoff}, {Rozo}, {Busha}, {Cunha}, {Finoguenov}, {Evrard}, {Hao}, {Koester}, {Leauthaud}, {Nord}, {Pierre}, {Reddick}, {Sadibekova}, {Sheldon}, \& {Wechsler}}]{rykoff2014redmapper}
{Rykoff}, E.~S., {Rozo}, E., {Busha}, M.~T., {et~al.} 2014, \apj, 785, 104, \dodoi{10.1088/0004-637X/785/2/104}

\bibitem[{{Rykoff} {et~al.}(2016){Rykoff}, {Rozo}, {Hollowood}, {Bermeo-Hernandez}, {Jeltema}, {Mayers}, {Romer}, {Rooney}, {Saro}, {Vergara Cervantes}, {Wechsler}, {Wilcox}, {Abbott}, {Abdalla}, {Allam}, {Annis}, {Benoit-L{\'e}vy}, {Bernstein}, {Bertin}, {Brooks}, {Burke}, {Capozzi}, {Carnero Rosell}, {Carrasco Kind}, {Castander}, {Childress}, {Collins}, {Cunha}, {D'Andrea}, {da Costa}, {Davis}, {Desai}, {Diehl}, {Dietrich}, {Doel}, {Evrard}, {Finley}, {Flaugher}, {Fosalba}, {Frieman}, {Glazebrook}, {Goldstein}, {Gruen}, {Gruendl}, {Gutierrez}, {Hilton}, {Honscheid}, {Hoyle}, {James}, {Kay}, {Kuehn}, {Kuropatkin}, {Lahav}, {Lewis}, {Lidman}, {Lima}, {Maia}, {Mann}, {Marshall}, {Martini}, {Melchior}, {Miller}, {Miquel}, {Mohr}, {Nichol}, {Nord}, {Ogando}, {Plazas}, {Reil}, {Sahl{\'e}n}, {Sanchez}, {Santiago}, {Scarpine}, {Schubnell}, {Sevilla-Noarbe}, {Smith}, {Soares-Santos}, {Sobreira}, {Stott}, {Suchyta}, {Swanson}, {Tarle}, {Thomas}, {Tucker}, {Uddin}, {Viana}, {Vikram}, {Walker}, {Zhang}, \& {DES
  Collaboration}}]{rykoff2016redmapper}
{Rykoff}, E.~S., {Rozo}, E., {Hollowood}, D., {et~al.} 2016, \apjs, 224, 1, \dodoi{10.3847/0067-0049/224/1/1}

\bibitem[{{Sarkar} {et~al.}(2022){Sarkar}, {Randall}, {Su}, {Alvarez}, {Sarazin}, {Nulsen}, {Blanton}, {Forman}, {Jones}, {Bulbul}, {Zuhone}, {Andrade-Santos}, {Johnson}, \& {Chakraborty}}]{sarkar2022shock}
{Sarkar}, A., {Randall}, S., {Su}, Y., {et~al.} 2022, \apjl, 935, L23, \dodoi{10.3847/2041-8213/ac86d4}

\bibitem[{{Sarkar} {et~al.}(2023){Sarkar}, {Randall}, {Su}, {Alvarez}, {Sarazin}, {Jones}, {Blanton}, {Nulsen}, {Chakraborty}, {Bulbul}, {Zuhone}, {Andrade-Santos}, \& {Johnson}}]{Sarkar2023}
---. 2023, \apj, 944, 132, \dodoi{10.3847/1538-4357/acae9f}

\bibitem[{{Schirmer} {et~al.}(2011){Schirmer}, {Hildebrandt}, {Kuijken}, \& {Erben}}]{schirmer2011macsj2243}
{Schirmer}, M., {Hildebrandt}, H., {Kuijken}, K., \& {Erben}, T. 2011, \aap, 532, A57, \dodoi{10.1051/0004-6361/201016348}

\bibitem[{{Sehgal} {et~al.}(2008){Sehgal}, {Hughes}, {Wittman}, {Margoniner}, {Tyson}, {Gee}, \& {dell'Antonio}}]{Sehgal2008Abell781}
{Sehgal}, N., {Hughes}, J.~P., {Wittman}, D., {et~al.} 2008, \apj, 673, 163, \dodoi{10.1086/523840}

\bibitem[{Snowden \& Kuntz(2024)}]{snowden2014cookbook}
Snowden, S., \& Kuntz, K. 2024, Cookbook for analysis procedures for XMM-Newton EPIC observations of extended objects and the diffuse background,  Tech. report

\bibitem[{{Soucail} {et~al.}(2015){Soucail}, {Fo{\"e}x}, {Pointecouteau}, {Arnaud}, \& {Limousin}}]{Soucail2015}
{Soucail}, G., {Fo{\"e}x}, G., {Pointecouteau}, E., {Arnaud}, M., \& {Limousin}, M. 2015, \aap, 581, A31, \dodoi{10.1051/0004-6361/201322689}

\bibitem[{{Stancioli} {et~al.}(2024){Stancioli}, {Wittman}, {Finner}, \& {Bouhrik}}]{stancioli2024new}
{Stancioli}, R., {Wittman}, D., {Finner}, K., \& {Bouhrik}, F. 2024, \apj, 966, 49, \dodoi{10.3847/1538-4357/ad3249}

\bibitem[{Stancioli {et~al.}(2026)Stancioli, Wittman, Finner, Bouhrik, \& Gao}]{stancioli2025inprep}
Stancioli, R., Wittman, D., Finner, K., Bouhrik, F., \& Gao, Z. 2026

\bibitem[{Stephens(1974)}]{stephens1974edf}
Stephens, M.~A. 1974, Journal of the American statistical Association, 69, 730

\bibitem[{{Stuardi} {et~al.}(2025){Stuardi}, {Botteon}, {Sereno}, {Umetsu}, {Gavazzi}, {Bonafede}, \& {Gheller}}]{stuardi2025radio}
{Stuardi}, C., {Botteon}, A., {Sereno}, M., {et~al.} 2025, \aap, 695, L16, \dodoi{10.1051/0004-6361/202452581}

\bibitem[{{Szabo} {et~al.}(2011){Szabo}, {Pierpaoli}, {Dong}, {Pipino}, \& {Gunn}}]{szabo2011optical}
{Szabo}, T., {Pierpaoli}, E., {Dong}, F., {Pipino}, A., \& {Gunn}, J. 2011, \apj, 736, 21, \dodoi{10.1088/0004-637X/736/1/21}

\bibitem[{{Tempel} {et~al.}(2017){Tempel}, {Tuvikene}, {Kipper}, \& {Libeskind}}]{tempel2017merging}
{Tempel}, E., {Tuvikene}, T., {Kipper}, R., \& {Libeskind}, N.~I. 2017, \aap, 602, A100, \dodoi{10.1051/0004-6361/201730499}

\bibitem[{{Tulin} \& {Yu}(2018)}]{tulin2018dark}
{Tulin}, S., \& {Yu}, H.-B. 2018, \physrep, 730, 1, \dodoi{10.1016/j.physrep.2017.11.004}

\bibitem[{{Upsdell} {et~al.}(2023){Upsdell}, {Giles}, {Romer}, {Wilkinson}, {Turner}, {Hilton}, {Rykoff}, {Farahi}, {Bhargava}, {Jeltema}, {Klein}, {Bermeo}, {Collins}, {Ebrahimpour}, {Hollowood}, {Mann}, {Manolopoulou}, {Miller}, {Rooney}, {Sahl{\'e}n}, {Stott}, {Viana}, {Allam}, {Alves}, {Bacon}, {Bertin}, {Bocquet}, {Brooks}, {Burke}, {Carrasco Kind}, {Carretero}, {Costanzi}, {da Costa}, {Pereira}, {De Vicente}, {Desai}, {Diehl}, {Dietrich}, {Everett}, {Ferrero}, {Frieman}, {Garc{\'\i}a-Bellido}, {Gerdes}, {Gutierrez}, {Hinton}, {Honscheid}, {James}, {Kuehn}, {Kuropatkin}, {Lima}, {Marshall}, {Mena-Fern{\'a}ndez}, {Menanteau}, {Miquel}, {Mohr}, {Ogando}, {Pieres}, {Raveri}, {Rodriguez-Monroy}, {Sanchez}, {Scarpine}, {Sevilla-Noarbe}, {Smith}, {Suchyta}, {Swanson}, {Tarle}, {To}, {Weaverdyck}, {Weller}, \& {Wiseman}}]{upsdell2023xmm}
{Upsdell}, E.~W., {Giles}, P.~A., {Romer}, A.~K., {et~al.} 2023, \mnras, 522, 5267, \dodoi{10.1093/mnras/stad1220}

\bibitem[{{van der Walt} {et~al.}(2014){van der Walt}, {Sch{\"o}nberger}, {Nunez-Iglesias}, {Boulogne}, {Warner}, {Yager}, {Gouillart}, {Yu}, \& {scikit-image Contributors}}]{van2014scikit}
{van der Walt}, S., {Sch{\"o}nberger}, J.~L., {Nunez-Iglesias}, J., {et~al.} 2014, PeerJ, 2, e453, \dodoi{10.7717/peerj.453}

\bibitem[{{van Weeren} {et~al.}(2021){van Weeren}, {Shimwell}, {Botteon}, {Brunetti}, {Br{\"u}ggen}, {Boxelaar}, {Cassano}, {Di Gennaro}, {Andrade-Santos}, {Bonnassieux}, {Bonafede}, {Cuciti}, {Dallacasa}, {de Gasperin}, {Gastaldello}, {Hardcastle}, {Hoeft}, {Kraft}, {Mandal}, {Rossetti}, {R{\"o}ttgering}, {Tasse}, \& {Wilber}}]{van2021lofar}
{van Weeren}, R.~J., {Shimwell}, T.~W., {Botteon}, A., {et~al.} 2021, \aap, 651, A115, \dodoi{10.1051/0004-6361/202039826}

\bibitem[{{Virtanen} {et~al.}(2020){Virtanen}, {Gommers}, {Oliphant}, {Haberland}, {Reddy}, {Cournapeau}, {Burovski}, {Peterson}, {Weckesser}, {Bright}, {van der Walt}, {Brett}, {Wilson}, {Millman}, {Mayorov}, {Nelson}, {Jones}, {Kern}, {Larson}, {Carey}, {Polat}, {Feng}, {Moore}, {VanderPlas}, {Laxalde}, {Perktold}, {Cimrman}, {Henriksen}, {Quintero}, {Harris}, {Archibald}, {Ribeiro}, {Pedregosa}, {van Mulbregt}, \& {SciPy 1. 0 Contributors}}]{virtanen2020scipy}
{Virtanen}, P., {Gommers}, R., {Oliphant}, T.~E., {et~al.} 2020, Nature Medicine, 17, 261, \dodoi{10.1038/s41592-019-0686-2}

\bibitem[{{Voges} {et~al.}(1999){Voges}, {Aschenbach}, {Boller}, {Br{\"a}uninger}, {Briel}, {Burkert}, {Dennerl}, {Englhauser}, {Gruber}, {Haberl}, {Hartner}, {Hasinger}, {K{\"u}rster}, {Pfeffermann}, {Pietsch}, {Predehl}, {Rosso}, {Schmitt}, {Tr{\"u}mper}, \& {Zimmermann}}]{voges1999rosat}
{Voges}, W., {Aschenbach}, B., {Boller}, T., {et~al.} 1999, \aap, 349, 389, \dodoi{10.48550/arXiv.astro-ph/9909315}

\bibitem[{{Wen} {et~al.}(2024){Wen}, {Han}, \& {Yuan}}]{wen2024catalogue}
{Wen}, Z.~L., {Han}, J.~L., \& {Yuan}, Z.~S. 2024, \mnras, 532, 1849, \dodoi{10.1093/mnras/stae1614}

\bibitem[{{Wilber} {et~al.}(2019){Wilber}, {Br{\"u}ggen}, {Bonafede}, {Rafferty}, {Shimwell}, {van Weeren}, {Akamatsu}, {Botteon}, {Savini}, {Intema}, {Heino}, {Cuciti}, {Cassano}, {Brunetti}, {R{\"o}ttgering}, \& {de Gasperin}}]{Wilber2019A}
{Wilber}, A., {Br{\"u}ggen}, M., {Bonafede}, A., {et~al.} 2019, \aap, 622, A25, \dodoi{10.1051/0004-6361/201833884}

\bibitem[{{Wittman} {et~al.}(2014){Wittman}, {Dawson}, \& {Benson}}]{WittmanA781}
{Wittman}, D., {Dawson}, W., \& {Benson}, B. 2014, \mnras, 437, 3578, \dodoi{10.1093/mnras/stt2151}

\bibitem[{{Wittman} {et~al.}(2018){Wittman}, {Golovich}, \& {Dawson}}]{wittman2018mismeasure}
{Wittman}, D., {Golovich}, N., \& {Dawson}, W.~A. 2018, \apj, 869, 104, \dodoi{10.3847/1538-4357/aaee77}

\bibitem[{{Wittman} {et~al.}(2025){Wittman}, {Stancioli}, {Bouhrik}, {van Weeren}, \& {Botteon}}]{gargantua2026inprep}
{Wittman}, D., {Stancioli}, R., {Bouhrik}, F., {van Weeren}, R., \& {Botteon}, A. 2025, arXiv e-prints, arXiv:2512.14945, \dodoi{10.48550/arXiv.2512.14945}

\bibitem[{{Wittman} {et~al.}(2023){Wittman}, {Stancioli}, {Finner}, {Bouhrik}, {van Weeren}, \& {Botteon}}]{wittman2023new}
{Wittman}, D., {Stancioli}, R., {Finner}, K., {et~al.} 2023, \apj, 954, 36, \dodoi{10.3847/1538-4357/acdb73}

\bibitem[{{York} {et~al.}(2000){York}, {Adelman}, {Anderson}, {Anderson}, {Annis}, {Bahcall}, {Bakken}, {Barkhouser}, {Bastian}, {Berman}, {Boroski}, {Bracker}, {Briegel}, {Briggs}, {Brinkmann}, {Brunner}, {Burles}, {Carey}, {Carr}, {Castander}, {Chen}, {Colestock}, {Connolly}, {Crocker}, {Csabai}, {Czarapata}, {Davis}, {Doi}, {Dombeck}, {Eisenstein}, {Ellman}, {Elms}, {Evans}, {Fan}, {Federwitz}, {Fiscelli}, {Friedman}, {Frieman}, {Fukugita}, {Gillespie}, {Gunn}, {Gurbani}, {de Haas}, {Haldeman}, {Harris}, {Hayes}, {Heckman}, {Hennessy}, {Hindsley}, {Holm}, {Holmgren}, {Huang}, {Hull}, {Husby}, {Ichikawa}, {Ichikawa}, {Ivezi{\'c}}, {Kent}, {Kim}, {Kinney}, {Klaene}, {Kleinman}, {Kleinman}, {Knapp}, {Korienek}, {Kron}, {Kunszt}, {Lamb}, {Lee}, {Leger}, {Limmongkol}, {Lindenmeyer}, {Long}, {Loomis}, {Loveday}, {Lucinio}, {Lupton}, {MacKinnon}, {Mannery}, {Mantsch}, {Margon}, {McGehee}, {McKay}, {Meiksin}, {Merelli}, {Monet}, {Munn}, {Narayanan}, {Nash}, {Neilsen}, {Neswold}, {Newberg}, {Nichol}, {Nicinski},
  {Nonino}, {Okada}, {Okamura}, {Ostriker}, {Owen}, {Pauls}, {Peoples}, {Peterson}, {Petravick}, {Pier}, {Pope}, {Pordes}, {Prosapio}, {Rechenmacher}, {Quinn}, {Richards}, {Richmond}, {Rivetta}, {Rockosi}, {Ruthmansdorfer}, {Sandford}, {Schlegel}, {Schneider}, {Sekiguchi}, {Sergey}, {Shimasaku}, {Siegmund}, {Smee}, {Smith}, {Snedden}, {Stone}, {Stoughton}, {Strauss}, {Stubbs}, {SubbaRao}, {Szalay}, {Szapudi}, {Szokoly}, {Thakar}, {Tremonti}, {Tucker}, {Uomoto}, {Vanden Berk}, {Vogeley}, {Waddell}, {Wang}, {Watanabe}, {Weinberg}, {Yanny}, {Yasuda}, \& {SDSS Collaboration}}]{york2000sloan}
{York}, D.~G., {Adelman}, J., {Anderson}, Jr., J.~E., {et~al.} 2000, \aj, 120, 1579, \dodoi{10.1086/301513}

\bibitem[{{Yuan} \& {Han}(2020)}]{yuan2020dynamical}
{Yuan}, Z.~S., \& {Han}, J.~L. 2020, \mnras, 497, 5485, \dodoi{10.1093/mnras/staa2363}

\bibitem[{{Yuan} {et~al.}(2022){Yuan}, {Han}, \& {Wen}}]{yuan2022dynamical}
{Yuan}, Z.~S., {Han}, J.~L., \& {Wen}, Z.~L. 2022, \mnras, 513, 3013, \dodoi{10.1093/mnras/stac1037}

\bibitem[{{Zhang} {et~al.}(2023){Zhang}, {Simionescu}, {Gastaldello}, {Eckert}, {Camillini}, {Natale}, {Rossetti}, {Brunetti}, {Akamatsu}, {Botteon}, {Cassano}, {Cuciti}, {Bruno}, {Shimwell}, {Jones}, {Kaastra}, {Ettori}, {Br{\"u}ggen}, {de Gasperin}, {Drabent}, {van Weeren}, \& {R{\"o}ttgering}}]{zhang2023planck}
{Zhang}, X., {Simionescu}, A., {Gastaldello}, F., {et~al.} 2023, \aap, 672, A42, \dodoi{10.1051/0004-6361/202244761}

\bibitem[{{Zitrin} {et~al.}(2020){Zitrin}, {Acebron}, {Coe}, {Kelly}, {Koekemoer}, {Nonino}, {Windhorst}, {Frye}, {Pascale}, {Broadhurst}, {Cohen}, {Diego}, {Finkelstein}, {Jansen}, {Larson}, {Yan}, {Alpaslan}, {Bhatawdekar}, {Conselice}, {Griffiths}, {Strolger}, \& {Wyithe}}]{zitrin2020strong}
{Zitrin}, A., {Acebron}, A., {Coe}, D., {et~al.} 2020, \apj, 903, 137, \dodoi{10.3847/1538-4357/abb8dd}

\end{thebibliography}
\bibliographystyle{aasjournal}

\end{document}